\documentclass[11pt,letterpaper,twoside,leqno]{article}
\usepackage[margin=1.05in]{geometry}
\usepackage[T1]{fontenc}
\usepackage{lmodern,microtype}
\usepackage{etoolbox}
\usepackage{amsmath,amssymb,amsthm,mathtools}
\usepackage{longtable,booktabs,tabularx}
\usepackage{xcolor}
\usepackage{tikz}
\usetikzlibrary{arrows.meta}
\usepackage[hidelinks,hypertexnames=false,linktoc=all]{hyperref}

\usepackage{fancyhdr}
\hypersetup{pdftitle={Sharp High-Dimensional Gaussian Approximations over Rectangles without Covariance Restrictions},pdfsubject={Independent sums with bounded, sub-exponential, or polynomial-moment envelopes},pdfauthor={Victor Chernozhukov, Denis Chetverikov, Kengo Kato, Yuta Koike}}
\allowdisplaybreaks[2]
\newtheoremstyle{bookplain}{6pt plus 2pt minus 1pt}{6pt plus 2pt minus 1pt}
  {\itshape}{}{\bfseries}{.}{.5em}
  {\thmname{#1}\thmnumber{ #2}\thmnote{ (#3)}}
\newtheoremstyle{bookremark}{6pt plus 2pt minus 1pt}{6pt plus 2pt minus 1pt}
  {\normalfont}{}{\bfseries}{.}{.5em}
  {\thmname{#1}\thmnumber{ #2}\thmnote{ (#3)}}
\theoremstyle{bookplain}
\newtheorem{theorem}{Theorem}
\newtheorem{proposition}{Proposition}
\newtheorem{lemma}{Lemma}
\newtheorem{corollary}{Corollary}
\newtheorem{conjecture}{Conjecture}
\theoremstyle{bookremark}
\newtheorem{remark}{Remark}
\makeatletter
\renewenvironment{proof}[1][\proofname]{\par
  \pushQED{\qed}\normalfont\topsep6pt plus6pt\relax
  \trivlist\item[\hskip\labelsep\bfseries #1\@addpunct{.}]\ignorespaces
}{\popQED\endtrivlist\@endpefalse}
\makeatother
\newcommand{\E}{\mathbb E}
\newcommand{\Pp}{\mathbb P}
\newcommand{\R}{\mathbb R}

\newcommand{\Cov}{\operatorname{Cov}}
\newcommand{\Var}{\operatorname{Var}}
\newcommand{\tr}{\operatorname{tr}}
\newcommand{\op}{\mathrm{op}}
\newcommand{\norm}[1]{\left\lVert#1\right\rVert}
\newcommand{\maxnorm}[1]{\lVert#1\rVert_{\max}}
\newcommand{\rh}{\rho_{\mathcal R}}
\newcommand{\ind}{\mathbf 1}
\newcommand{\dd}{\,d}

\title{\Large Sharp High-Dimensional Gaussian Approximations\\[0.15em]
over Rectangles without Covariance Restrictions}
\author{\normalsize Victor Chernozhukov, Denis Chetverikov, Kengo Kato, Yuta
Koike\thanks{This paper
was developed in genuine collaboration with the AI system GPT-Astra.
Lean formalization and verification of the archived version were performed
with GPT-6 Sol
(see Section~\ref{sec:lean-introduction}). The human authors take full
responsibility for the content of this research.}}
\date{\normalsize 16 November 2026}

\makeatletter
\renewcommand{\@maketitle}{%
  \begin{center}
    {\Large\@title\par}\vskip.85em
    {\normalsize\lineskip.5em\begin{tabular}[t]{c}\@author\end{tabular}\par}
    \vskip.65em{\normalsize\@date\par}
  \end{center}}
\makeatother

\begin{document}
\maketitle

\begin{abstract}

We study how well a sum of \(n\) independent random vectors with \(p\)
coordinates can be approximated by a Gaussian vector with the same
mean and covariance. We measure the error over rectangles, so the bounds apply
to maxima and simultaneous confidence regions. The coordinates may
be arbitrarily dependent, even exactly linearly dependent, and the
bounds do not deteriorate as the covariance matrix becomes singular.
After standardizing the coordinate variances, bounded or
sub-exponential coordinates give the \(n^{-1/2}\) sample-size
rate up to logarithmic factors, with the size or tail bounds held
fixed. {The main upper bounds retain the Gaussian expected maximum,
improving the rates when the coordinates are strongly dependent.
Fourth-moment refinements give further improvements when large
observations contribute little to individual coordinate variances.} If we instead impose a fixed moment bound of order
\(q>4\) on the largest absolute coordinate, together with a coordinate
fourth-moment bound, the rate becomes \(n^{-1/2+1/q}\) up to
logarithmic factors. At \(q=4\), the slower \(n^{-1/4}\) rate is
unavoidable, confirming our earlier conjecture. We also allow the
bounds on observation size to grow with \(n\). Lower-bound examples
show that the dependence on {sample size, observation size, and
Gaussian width} is sharp
up to logarithmic factors in the regimes studied, and in several
regimes the upper and lower bounds match including the logarithmic
terms.
\end{abstract}

\begingroup
\setcounter{tocdepth}{1}
\small
\makeatletter
\patchcmd{\l@section}{1.0em \@plus\p@}{0.5em \@plus\p@}{}{}
\makeatother
\tableofcontents
\endgroup
\clearpage

\section{\texorpdfstring{Introduction and main results}{Introduction and main results}}\label{sec:results}
Gaussian approximation over rectangles provides a common foundation for
approximating maxima and constructing simultaneous confidence regions, with
applications to a wide variety of statistical inference problems reviewed in~\cite{CCKK2023}.
In many of these applications the {covariance} matrix may be nearly or
exactly singular. {In this paper we obtain} approximation bounds that remain uniform
as its eigenvalues approach zero. {These bounds are sharp up to
logarithmic factors, bringing our research program on rectangular
high-dimensional central limit theorems close to completion.}

Let \(W\) be a sum of independent centered bounded random vectors in
\(\R^p\). Suppose that each coordinate of the sum has unit variance,
and let \(Z\) be a centered Gaussian vector with the same covariance.
For the bounded-observation results, our assumptions and notation are
\begin{equation}
 \begin{gathered}
 W\coloneqq (W_j)_{j=1}^p=\sum_{i=1}^n\xi_i,\qquad
 \E\xi_i=0,\qquad \norm{\xi_i}_\infty\le\delta\quad\text{a.s.},\\
 \Sigma\coloneqq \Cov(W),\qquad\Sigma_{jj}=1\ (j\le p),
 \qquad Z\sim N(0,\Sigma),
 \end{gathered}
 \label{eq:setup}
\end{equation}
where $\|\cdot\|_\infty$ denotes the maximum norm of a vector. 
The summands need not have the same distribution, and coordinates within
a summand may be dependent. The unit marginal variances fix the scale
of each coordinate, and they can be obtained by coordinatewise rescaling
whenever all marginal variances are positive. There is no
restriction on the rank of \(\Sigma\).

The parameter \(\delta\) bounds the contribution of any one summand
to any coordinate of \(W\), and such a uniform bound is called an
\emph{envelope}. To state the
bound compactly, {define}
\[
 L\coloneqq \log(ep),\qquad
 D_L\coloneqq 1+\log_+\frac1{\delta\sqrt L},\qquad
 \log_+x\coloneqq \max\{0,\log x\}.
\]
Here \(L\) encodes the dependence on dimension, while \(D_L\)
contributes an additional logarithmic factor. Constants denoted by
\(C\) are absolute unless a subscript records their dependence.

Let \(\mathcal R\) denote the class of axis-aligned rectangles in
\(\R^p\). We measure the approximation error by the \emph{Kolmogorov distance over rectangles}
\[
 \rh(W,Z)\coloneqq \sup_{A\in\mathcal R}
       |\Pp(W\in A)-\Pp(Z\in A)|.
\]
We use the same notation for any two random vectors in \(\R^p\). For
example, taking \(A\coloneqq (-\infty,s]^p\) shows that this distance controls the
error in approximating the distribution of \(\max_{j\le p}W_j\), the maximum
coordinate of \(W\).

{
In our proof, the envelope \(\delta\) sets a lower bound on the width
used to smooth rectangle indicators. To sharpen the comparison of the
smoothed expectations, we also retain information about the summands'
third and fourth moments. We collect this information in matrices that
will be compared with the covariance matrix \(\Sigma\). For
\(1\le j\le p\), define
\[
 T_j:=\sum_i\E[\xi_{ij}\,\xi_i\xi_i^\top],\qquad
 Q_j:=\sum_i\E[\xi_{ij}^2\,\xi_i\xi_i^\top],\qquad
 \widetilde Q_j:=\sum_i\E[\xi_{ij}^2]\,\E[\xi_i\xi_i^\top].
\]
We measure these moments relative to the variance in each direction.
Specifically, we call \(s_3,s_4\ge0\) \emph{relative moment bounds} if,
for every \(1\le j\le p\) and \(u\in\R^p\),
\begin{equation}\label{eq:relative-moments}
 |u^\top T_j u|\le s_3\,u^\top\Sigma u,\qquad
 u^\top Q_j u\le s_4\,u^\top\Sigma u,\qquad
 u^\top\widetilde Q_j u\le s_4\,u^\top\Sigma u.
\end{equation}
Here \(u^\top\Sigma u\) is the variance of the linear contrast
\(u^\top W\). The first bound controls the aggregate third moment
\(\sum_i\E[\xi_{ij}(u^\top\xi_i)^2]\), allowing cancellation across
summands. The remaining bounds control mixed fourth moments: the
coordinate and the contrast come from the same summand in \(Q_j\),
and from independent copies in \(\widetilde Q_j\). These comparisons
remain meaningful when \(\Sigma\) is singular. The envelope bound
\(|\xi_{ij}|\le\delta\) always permits \(s_3=\delta\) and
\(s_4=\delta^2\).
}

The number of coordinates can overstate the size of their maximum
when they are strongly dependent. To retain this information, define
the \emph{Gaussian width} of the coordinates by
\[
 g_\Sigma:=\E\|Z\|_\infty.
\]
The unit marginal variances imply
\(\sqrt{2/\pi}\le g_\Sigma\le C\sqrt L\). For example, if all
coordinates agree up to sign, then \(g_\Sigma=\sqrt{2/\pi}\),
regardless of \(p\). Our first result retains this width in the
anti-concentration estimates used to compare rectangle probabilities.

\begin{theorem}[Unrestricted covariance]\label{thm:unrestricted}
{Under~\eqref{eq:setup}, for every admissible $s_4$ in \eqref{eq:relative-moments},}
\begin{equation}\label{eq:intrinsic-unrestricted}
{\rh(W,Z)\le C\min\{1,\,g_\Sigma L(\delta+\sqrt{s_4D_L})\}.}
\end{equation}
{In particular, the envelope alone gives}
\begin{equation}\label{eq:intrinsic-envelope}
{\rh(W,Z)\le C\min\{1,\,\delta g_\Sigma L\sqrt{D_L}\}.}
\end{equation}
\end{theorem}
{
The factor \(g_\Sigma\) measures boundary concentration. Using
\(g_\Sigma\le C\sqrt L\) gives the worst-case dimension dependence,
whereas a smaller width improves the bound for strongly dependent
coordinates. The remaining factor \(L\) cannot be removed in general,
even when \(g_\Sigma\) is bounded. Theorem~\ref{opt:symmetric-lower}
constructs centrally symmetric observations with error of order at
least \(\delta g_\Sigma L\), allowing the width to range from a
constant to order \(\sqrt L\). These examples do not establish
the necessity of the additional factor \(\sqrt{D_L}\).
Section~\ref{sec:proof-theorem-one} proves the theorem.
}

{
The rate in \eqref{eq:intrinsic-unrestricted} is \(O(\delta g_\Sigma L)\)
whenever \({s_4D_L\le C\delta^2}\). In the derivation of
\eqref{eq:intrinsic-unrestricted}, aggregate third moments contribute a
term of order \(s_3g_\Sigma L\), which is absorbed into
\(C\delta g_\Sigma L\) because we may always take \(s_3\le\delta\).
Even under exact cancellation, \(s_3=0\), smoothing still costs order
\(\delta g_\Sigma L\) in our argument. {The lower bounds show
that this order is necessary in general, including when the width
stays bounded;} see
Theorem~\ref{opt:symmetric-lower} in Section~\ref{opt:section}.
}

{To illustrate the gain from the fourth-moment
refinement, let \(R_1,\ldots,R_n\) be independent copies of a centered
random vector \(R\in\{-1,1\}^p\) with covariance \(\Sigma\). The
coordinates of \(R\) may be dependent. For deterministic weights
\(a_1,\ldots,a_n\) satisfying \(\sum_i a_i^2=1\), set
\(\xi_i:=a_iR_i\).} Since $R_{ij}^2=1$, both fourth-moment matrices equal $(\sum_i a_i^4)\Sigma$. {Give one summand a weight proportional to $\sqrt{\log n}$ and each of the other $n-1$ summands a weight proportional to $1$.} After normalization, $\delta\asymp\sqrt{(\log n)/n}$ and $s_4\asymp1/n$. For $p\asymp n^a$ with fixed $a>0$, \eqref{eq:intrinsic-unrestricted} gains a factor of order $\sqrt{\log n}$ over \eqref{eq:intrinsic-envelope}, for every covariance of the sign vector.

Proposition~G.1 of the Supplemental Material~\cite{Supplement2026}
replaces the factor \(\sqrt{D_L}\) in the envelope bound
\eqref{eq:intrinsic-envelope} by \(\sqrt{\min\{D_L,L\}}\),
while retaining the Gaussian-width factor \(g_\Sigma\).
This refinement is useful when dimension grows slowly relative to the
inverse summand bound. It uses the same interpolation and an additional
integrated remainder estimate.

For normalized observations, Theorem~\ref{thm:unrestricted} gives
the following dependence on the sample size.

\begin{corollary}[Normalized observations]\label{cor:observations}
{
Let \(X_1,\ldots,X_n\in\R^p\) be independent and centered. Suppose
\(\norm{X_i}_\infty\le B\) almost surely and
\(n^{-1}\sum_i\E X_{ij}^2=1\) for every \(j\). Define
\(W_n:=n^{-1/2}\sum_iX_i\), let \(Z_n\sim N(0,\Cov(W_n))\),
and define \(g_n:=\E\|Z_n\|_\infty\). Then
\begin{equation}\label{eq:observations-intrinsic}
 \rh(W_n,Z_n)\le C\min\!\left\{1,
 \frac{B g_n L}{\sqrt n}
 \sqrt{1+\log_+\frac{\sqrt n}{B\sqrt L}}\right\}.
\end{equation}
}
\end{corollary}

{
Corollary~\ref{cor:observations} follows from Theorem~\ref{thm:unrestricted}
by setting \(\xi_i:=X_i/\sqrt n\), which has envelope
\(\delta=B/\sqrt n\). For fixed \(B\) and \(p\asymp n^a\), with
fixed \(a>0\), the bound \(g_n\le C\sqrt L\) gives the rate
\(n^{-1/2}(\log n)^2\). If \(g_n\) remains bounded, the rate improves
to \(n^{-1/2}(\log n)^{3/2}\).

Theorem~\ref{opt:symmetric-lower} gives centrally symmetric examples
whose error is at least of order \(B g_nL/\sqrt n\), whenever
this rate tends to zero, with widths ranging from a constant to
order \(\sqrt L\). Thus the corollary is sharp in these regimes
up to its additional square-root logarithm. Theorem~\ref{opt:skew-lower}
attains the same rate with independent skewed coordinates, for which
\(g_n\asymp\sqrt L\). This example extends that of
Chernozhukov, Chetverikov, and Koike~\cite[Proposition~2.1]{CCK}.
Conjecture~\ref{conj:sharp-unrestricted} asks whether the additional
factor can be removed while retaining the Gaussian width.

The sub-exponential and polynomial-moment upper bounds below also
retain Gaussian width. Section~\ref{opt:section} states the lower
bounds in the same notation and specifies the dimension and width
sequences for which the constructions apply.
}

{
Corollary~\ref{cor:observations} is stated for a normalized sum \(W_n\)
whose coordinates have unit variance. For unequal positive variances, let
\[
 \sigma_j^2\coloneqq n^{-1}\sum_i\E X_{ij}^2,
 \qquad
 \underline\sigma\coloneqq \min_j\sigma_j>0.
\]
We then apply Theorem~\ref{thm:unrestricted} to the summands
\(\xi_{ij}\coloneqq X_{ij}/(\sigma_j\sqrt n)\). If
\(\max_i\norm{X_i}_\infty\le K_n\), these summands have
envelope \(\delta=K_n/(\underline\sigma\sqrt n)\).
Positive diagonal scaling preserves the class of rectangles and hence
the distance \(\rh\), so the resulting bound also holds in the
original coordinates.
}

\subsection{Sub-exponential coordinates}\label{sec:subexponential}
{
We next extend the bounded-observation result to observations whose
coordinates have sub-exponential tails. We measure tail size by the
\emph{Orlicz norm}
\[
 \|V\|_{\psi_1}:=\inf\{s>0:\E\exp(|V|/s)\le2\}.
\]
A uniform bound \(\|X_{ij}\|_{\psi_1}\le K\) gives
\(\Pp(|X_{ij}|>t)\le2e^{-t/K}\) for \(t\ge0\).
We cap each coordinate's magnitude at a deterministic threshold \(T\),
then recenter and rescale the coordinates so that their normalized
sum has unit marginal variances. We can then apply
Theorem~\ref{thm:unrestricted} to this bounded sum. To transfer the
resulting bound to the original sum, we must also control the residual
introduced by capping. The exponential tail bound allows us to do so
with a threshold of order \(K\log(en)\). This choice yields an
effective envelope of order \(K\log(en)/\sqrt n\), introducing an
additional logarithmic factor in the bound.
}

\begin{corollary}[Sub-exponential transfer]\label{cor:subexponential}
Let \(X_1,\ldots,X_n\in\R^p\) be independent and centered, with
\(n^{-1}\sum_i\E X_{ij}^2=1\) and
\(\max_{i,j}\|X_{ij}\|_{\psi_1}\le K\), where \(K\ge1\).
For \(W_n\coloneqq n^{-1/2}\sum_iX_i\) and a centered Gaussian \(Z_n\)
with the same covariance, {define \(g_n:=\E\|Z_n\|_\infty\) and}
\[
 \ell_n\coloneqq \log(en),\qquad b_n\coloneqq \frac{K\ell_n}{\sqrt n},\qquad
 D_n\coloneqq 1+\log_+\frac1{b_n\sqrt L}.
\]
Then
\begin{equation}
 {\rh(W_n,Z_n)\le C\min\{1,b_ng_nL\sqrt{D_n}\}.}
 \label{eq:subexponential}
\end{equation}

\end{corollary}

{
{
Capping with a common threshold does not increase the Gaussian
width before marginal standardization. The standardization changes
it by at most an absolute factor, so the transfer preserves the
width improvement in Theorem~\ref{thm:unrestricted}.
For fixed \(K\) and \(p\asymp n^a\), \(a>0\), the bound is
\(O(n^{-1/2}(\log n)^3)\) in the worst case and improves to
\(O(n^{-1/2}(\log n)^{5/2})\) when \(g_n\) remains bounded.
Appendix~\ref{app:subexponential} gives the proof.
}
}

\subsection{Polynomial moment envelopes}\label{sec:polynomial-main}

{
We next consider observations whose largest absolute coordinate has a
finite moment of order \(q>2\), allowing heavier tails than the
sub-exponential condition. We combine two Gaussian approximation
bounds. The first follows from truncation and
Theorem~\ref{thm:unrestricted} under this moment condition alone. The
second also uses a separate bound on the coordinate fourth moments.
This additional condition can improve the treatment of large
observations because an observation with a large maximum need not
contribute much to any given coordinate.
Theorem~\ref{thm:polynomial} states the two estimates separately,
so that we can use whichever applies and gives the smaller bound.
}

Let \(X_1,\ldots,X_n\in\R^p\) be independent and centered, with
\(n^{-1}\sum_i\E X_{ij}^2=1\) for each \(j\). For this setting, write
\begin{equation}
 W:=\frac1{\sqrt n}\sum_{i=1}^nX_i,\qquad
 \Sigma:=\Cov(W),\qquad Z\sim N(0,\Sigma),\qquad L:=\log(ep).
 \label{poly:setup}
\end{equation}
{As above, \(g_\Sigma:=\E\|Z\|_\infty\) measures the Gaussian
width of the coordinates.}
{
For fixed \(q>2\) and \(B\ge1\), we consider the envelope and
coordinate moment conditions
\begin{align}
 \max_{i\le n}\E\norm{X_i}_\infty^q&\le B^q,
 \label{poly:envelope}\\
 \max_{j\le p}\frac1n\sum_{i=1}^n\E X_{ij}^4&\le B^2.
 \label{poly:fourth}
\end{align}
The first condition controls the \emph{random envelope}
\(\norm{X_i}_\infty\), including when the largest coordinate occurs
at a different location in each observation. The second condition is
an additional assumption on the averaged coordinate fourth moments,
including when \(2<q<4\). At \(q=4\), the moment bounds for
the envelope and for each coordinate are \(B^4\) and \(B^2\),
respectively. We write \(\mathcal C_q(B,p)\) for the class of arrays
satisfying both conditions, suppressing its dependence on \(n\).
}

{
\begin{theorem}[Polynomial moment envelopes]\label{thm:polynomial}
{
Let \(X_1,\ldots,X_n\in\R^p\) be independent and centered, with
\(n^{-1}\sum_i\E X_{ij}^2=1\) for every \(j\), and let \(W,Z,L\)
be as in~\eqref{poly:setup}, with \(g_\Sigma:=\E\|Z\|_\infty\).
For every fixed \(q>2\), there is a
constant \(C_q>0\), depending only on \(q\), such that the following
statements hold for all \(n,p\ge1\) and \(B\ge1\).
\begin{subequations}\label{poly:combined}
\par\noindent\textup{(i)} If the envelope condition~\eqref{poly:envelope}
holds, then
\begin{equation}
 \rh(W,Z)\le C_q\min\left\{1,
 \frac{B g_\Sigma L^{1-1/q}\{\log(en)\}^{1/2-1/q}}
 {n^{1/2-1/q}}\right\}.
 \label{poly:main-envelope}
\end{equation}
\noindent\textup{(ii)} If the coordinate fourth-moment
condition~\eqref{poly:fourth} also holds, then
\begin{equation}
 \rh(W,Z)\le C_q\min\left\{1,
 g_\Sigma\left[
 \frac{B^{1/2}L^{3/4}}{n^{1/4}}
 +\frac{B L^{1-1/q}}
 {n^{1/2-1/q}\{\log(e+B(n/L)^{2/q-1/2})\}^{1-1/q}}\right]\right\}.
 \label{poly:main-fourth}
\end{equation}
\end{subequations}
When both moment conditions hold, we may take the smaller of
\eqref{poly:main-envelope} and~\eqref{poly:main-fourth}.
}
\end{theorem}
}

{
The bounds allow \(B\) to grow with \(n\).
Bound~\eqref{poly:main-envelope} requires only~\eqref{poly:envelope},
while~\eqref{poly:main-fourth} uses~\eqref{poly:fourth} to exploit
smaller coordinate fourth moments. For example, at \(q=4\),
\eqref{poly:envelope} alone bounds their averages by \(B^4\),
whereas~\eqref{poly:fourth} reduces this bound to \(B^2\).
{The lower bounds below retain the same Gaussian-width factor.}
Theorem~\ref{lower:dense}\textup{(i)} matches
\eqref{poly:main-envelope} from below up to the logarithmic factor
\(\{\log(en)\}^{1/2-1/q}\) under the envelope condition alone,
for every fixed \(q>2\) and fixed or growing \(B\ge2\).

When both moment conditions hold and \(B\ge2\) is fixed, the
comparison has three regimes.
When \(2<q<4\) and \(n/L\to\infty\),
\eqref{poly:main-fourth} improves the logarithmic factors in
\eqref{poly:main-envelope}. Theorem~\ref{lower:sparse} gives a
matching lower bound when \(\log n=o(L)\) and the rate tends to zero.
At \(q=4\), \eqref{poly:main-fourth} gives the rate
{\(g_\Sigma n^{-1/4}L^{3/4}\), which
Theorem~\ref{opt:q4-lower} matches from below for widths ranging
from a constant to order \(\sqrt L\).}
This confirms our conjecture that the quartic rate in
\cite[Theorem~2.5]{CCKK2022} is optimal under fourth-moment envelopes.
When \(q>4\), \eqref{poly:main-envelope} improves the sample-size
power to \(n^{-1/2+1/q}\). Theorem~\ref{lower:dense}\textup{(ii)} matches the
rate in~\eqref{poly:main-envelope} from below up to the logarithmic factor
\(\{\log(en)\}^{1/2-1/q}\).

For growing \(B\), \eqref{poly:main-fourth} can also improve on
\eqref{poly:main-envelope} when \(q>4\). More generally, for every
fixed \(q>2\), Theorem~\ref{lower:sparse} matches the rate in
\eqref{poly:main-fourth} from below when \(B(n/L)^{2/q-1/2}\to\infty\),
provided that the logarithm of this quantity is \(o(L)\) and the
rate tends to zero. At \(q=4\), Theorem~\ref{opt:q4-growing}
matches the dependence \(B/\{\log(e+B)\}^{3/4}\) on the growing
envelope from below. These lower-bound comparisons use the dimension sequences
specified in the cited theorems. See Section~\ref{sec:poly} for
the proof of Theorem~\ref{thm:polynomial} and
Section~\ref{opt:section} for the lower bounds and their summary
in Table~\ref{opt:tab:rates}.
}

\subsection{Transfer to polyhedra}\label{sec:polyhedral-transfer}

{
A polyhedron is specified by inequalities
\(\nu_\alpha^\top x\le a_\alpha\), for \(\alpha=1,\ldots,m\),
where the \(\nu_\alpha\) are its normal vectors. Membership of
\(W\) is therefore a lower-orthant event for the vector
\((\nu_1^\top W,\ldots,\nu_m^\top W)\), as
in~\cite[Section~3.1]{CCK2017}.
The following corollary extends our Gaussian approximation bounds to
polyhedra. Its condition on the normal vectors limits the amplification
caused by standardizing these linear forms and permits singular
covariance matrices.
}

{
\begin{corollary}[Polyhedral transfer]\label{cor:polyhedral-transfer}
Let \(W\in\R^p\) be centered with covariance \(\Sigma\), and let
\(Z\sim N(0,\Sigma)\). For normal vectors \(\nu_1,\ldots,\nu_m\in\R^p\)
with \(\sigma_\alpha^2:=\nu_\alpha^\top\Sigma\nu_\alpha>0\), define
\[
 P_a:=\bigcap_{\alpha=1}^m\{x:\nu_\alpha^\top x\le a_\alpha\},
 \qquad
 \Delta_m(W,Z):=\sup_{a\in\R^m}
 \bigl|\Pp(W\in P_a)-\Pp(Z\in P_a)\bigr|.
\]
Suppose that, for some \(J\ge1\),
\begin{equation}
 \|\nu_\alpha\|_1\le J\sigma_\alpha\quad(\alpha\le m).
 \label{eq:polyhedral-normal-condition}
\end{equation}
{Define the Gaussian width of the standardized projections by}
\[
 {g_P:=\E\max_{\alpha\le m}
                  \frac{|\nu_\alpha^\top Z|}{\sigma_\alpha}.}
\]
Under the bounded-summand assumptions of Theorem~\ref{thm:unrestricted},
{bounds~\eqref{eq:intrinsic-unrestricted} and~\eqref{eq:intrinsic-envelope}
hold for \(\Delta_m(W,Z)\), with \(g_\Sigma\) replaced by \(g_P\)
and the substitutions}
\[
 p\mapsto m,\qquad \delta\mapsto J\delta,\qquad s_4\mapsto J^2s_4.
\]
Under the setup~\eqref{poly:setup} and the envelope
condition~\eqref{poly:envelope}, the bound~\eqref{poly:main-envelope}
holds for \(\Delta_m(W,Z)\) with
\[
 p\mapsto m,\qquad B\mapsto JB.
\]
If the coordinate fourth-moment condition~\eqref{poly:fourth} also holds,
then~\eqref{poly:main-fourth} holds for \(\Delta_m(W,Z)\) with
\[
 p\mapsto m,\qquad B\mapsto J^2B.
\]
Under the assumptions of Corollary~\ref{cor:subexponential},
the bound \eqref{eq:subexponential} holds for \(\Delta_m(W_n,Z_n)\) with
\[
 p\mapsto m,\qquad K\mapsto JK.
\]
{In each case, \(L\) becomes \(\log(em)\), the Gaussian width
becomes \(g_P\le C\sqrt{\log(em)}\), and the auxiliary quantities
in the cited bound are recomputed using the substituted parameters.}
The bounds are uniform over normal vectors satisfying
\eqref{eq:polyhedral-normal-condition} for the given \(J\).
Zero-variance inequalities may be deleted unless they exclude both
vectors almost surely, in which case the probability difference is zero.
\end{corollary}
}

{
The proof is given in Appendix~\ref{app:polyhedral-transfer}.
}

\subsection{Relationship to the Literature}\label{sec:literature}

Gaussian approximation for high-dimensional maxima and rectangles was
developed in~\cite{CCK2013,CCK2017}. Under coordinatewise sub-exponential
tails and an averaged fourth-moment bound, Chernozhukov, Chetverikov,
Kato, and Koike~\cite[Theorem~2.1]{CCKK2022} obtained a rate of
\(n^{-1/4}\) up to logarithmic factors, without a covariance restriction.
For fixed tail scale and polynomially growing dimension, the bound
in~\cite[Theorem~1]{GaoZhang} improves this to
\(O(n^{-1/3}(\log n)^{7/3})\). In the same regime, our
Corollaries~\ref{cor:observations} and~\ref{cor:subexponential}
achieve the optimal rate of \(n^{-1/2}\) up to logarithmic factors
for bounded and sub-exponential coordinates, respectively.
The covariance may be singular, and the constants do not depend on
its smallest positive eigenvalue.

Stronger covariance assumptions permit sharper logarithmic factors.
When the smallest eigenvalue is bounded away from zero,
Chernozhukov, Chetverikov, and Koike~\cite{CCK} obtain the
bounded-observation rate \(Bn^{-1/2}\{\log(ep)\}^{3/2}\log n\)
by combining exchangeable pairs with Slepian--Stein interpolation
and mixed smoothing. The bound in~\cite{AL} removes the final
\(\log n\) under the same type of nondegeneracy condition.
Corollary~\ref{cor:positive-correlation-gap}, proved in
the Supplemental Material~\cite[Appendix~H]{Supplement2026}, attains the same rate under a
pairwise correlation gap bounded away from zero, a condition that allows
singular covariance matrices. Its bound also retains $g_\Sigma$, giving
an improvement when the Gaussian width is smaller than its worst-case
order. The companion paper~\cite{CorrelationGap} develops broader
moment-sensitive bounds under this pairwise condition.
Theorem~\ref{thm:unrestricted} imposes no such condition and identifies
the relative fourth moments that govern its remaining logarithmic factor.

Our main approximation bounds also retain the Gaussian width
\(g_\Sigma=\E\|Z\|_\infty\). This yields a further improvement
when dependence among the coordinates makes their width smaller
than its worst-case order \(\sqrt{\log(ep)}\). For bounded,
sub-exponential, and polynomial-moment envelopes, the gain over
the corresponding worst-case bounds is of order
\(\sqrt{\log(ep)}/g_\Sigma\), reaching order
\(\sqrt{\log(ep)}\) when the width remains bounded.
The lower-bound constructions in Section~\ref{opt:section}
retain the same width factor in the regimes studied.

To obtain the improved power of \(n\), we exploit the fact that a
smooth rectangle indicator varies appreciably only near the boundary.
We use a logistic approximation and estimate its derivatives in
expectation, retaining the small probability of lying near that
boundary. As in~\cite{GaoZhang}, Gaussian matrix estimates control
the derivative directions without introducing a factor depending on
rank. We derive these estimates directly from the logistic weights.
The key step bounds the required derivatives of Gaussian rectangle
probabilities with respect to their thresholds by probabilities of
thin layers between a rectangle and a slightly enlarged rectangle.
This construction retains the geometry underlying Nazarov's
inequality~\cite{Nazarov2003,CCK17N}, while avoiding surface integrals
over faces and their intersections. Appendices~\ref{appx-an:sec:matrix}
and~\ref{appx-an:sec:threshold} give these estimates.
The positive-gap refinement proved in the Supplemental
Material~\cite[Appendix~H]{Supplement2026}
additionally uses angle-weighted surface estimates on facets and ridges.

Boundary concentration improves the averaged derivative estimates
by one power of the smoothing width. The main difficulty is
preserving this gain as we replace the original summands by Gaussian
noise. Two Gaussian integrations by parts replace two derivatives
of the logistic approximation by a centered quadratic Gaussian
expression, a second Hermite polynomial. This expression and the
remaining logistic weights depend on the same Gaussian vector, so
we must estimate their weighted expectation jointly. We also sum
the moment matrices before taking operator norms, allowing the
relative moment assumptions to control their aggregate contribution
without introducing a factor depending on the covariance rank.
Together, the boundary-concentration estimate and these weighted
Gaussian and aggregate moment estimates preserve the gain of one
power of the smoothing width, at the cost of an additional factor
\(L\).

To obtain the logarithmic refinement, we use the calibrated dropout
path defined in~\cite[Section~4]{AL}, preceded by a short Slepian
interpolation that introduces a small Gaussian component. This
\emph{Gaussian seed} permits two Gaussian integrations by parts
throughout the main comparison, including at its starting point,
which in our view simplifies the proof. We control the short initial
Slepian step directly, before applying Gaussian integration by parts
along the dropout path.

The dropout calibration controls the cubic contribution near the
initial endpoint. The retention probabilities and amplitudes are
chosen so that the summands' contribution to covariance decreases
linearly while the time derivative of their third moments vanishes
at that endpoint. The calibration makes the cubic coefficient vanish linearly
at the start of the comparison. This offsets the inverse
Gaussian-variance factor produced by two integrations by parts,
making the cubic contribution integrable uniformly as the Gaussian
seed tends to zero. With the same derivative estimate, ordinary
Slepian interpolation alone would yield a logarithmic time integral
because the third-moment derivative does not vanish at the initial
endpoint. The calibrated path removes this logarithm from the cubic
contribution.

The fourth-order remainder still contributes a logarithmic time
integral. Balancing this contribution against the smoothing error
produces the factor \(\sqrt{D_L}\) in the envelope
bound~\eqref{eq:intrinsic-envelope}. In the refined
bound~\eqref{eq:intrinsic-unrestricted}, the size of this contribution
is governed by the relative fourth-moment bound \(s_4\).
Section~\ref{sec:smoothing} explains these estimates and their balance.

For polynomial moment envelopes, Theorem~\ref{thm:polynomial}
separates two bounds according to their assumptions.
Part~\textup{(i)} requires only the envelope
condition~\eqref{poly:envelope}. Conditional truncation transfers
the bounded-observation result to the rate \(n^{-1/2+1/q}\) up to
logarithmic factors. For fixed \(B\) and \(q>4\), this improves
the power of \(n\) over the \(n^{-1/4}\) term
in~\cite[Theorem~2.5]{CCKK2022}. Part~\textup{(ii)} also assumes
the coordinate fourth-moment condition~\eqref{poly:fourth} and
sharpens the treatment of large observations in the bound
in~\cite{CCKK2022}. We give a self-contained logistic--Slepian
proof of a version of the matching-covariance comparison
estimate~\cite[Lemma~A.1]{CCKK2022}. Interchanging each summand
with an independent copy and applying a trapezoidal Taylor identity
gives a fourth-order remainder. A Bennett inequality using both
moment conditions controls the expected maximum of the large
increments and reduces the envelope contribution by the logarithmic
factor in~\eqref{poly:main-fourth}. When both conditions hold,
we may take the smaller of the two bounds.

The lower-bound constructions retain Gaussian width in these
polynomial-moment comparisons as well. At \(q=4\), the
sample-size dependence \(n^{-1/4}\) remains and is sharp.
For fixed \(B\ge2\), the centrally symmetric examples in
Theorem~\ref{opt:q4-lower} match the full rate
\(g_\Sigma n^{-1/4}L^{3/4}\) in~\eqref{poly:main-fourth}
from below, confirming our conjecture that the quartic rate
in~\cite[Theorem~2.5]{CCKK2022} is optimal under fourth-moment
envelopes. More generally, Theorem~\ref{lower:dense}\textup{(i)}
matches~\eqref{poly:main-envelope} from below up to logarithmic
factors under the envelope condition alone, while
Theorems~\ref{opt:q4-growing} and~\ref{lower:sparse}
match~\eqref{poly:main-fourth} from below in their regimes with
growing envelopes or other moment orders. These comparisons use
the dimension and vanishing-rate conditions of the cited theorems.

Our main upper bounds thus provide polynomial or logarithmic
improvements over previous bounds in the regimes described above.
Most of the lower bounds are also new, although their constructions
build in part on earlier examples in~\cite{CCK,FKLZ}. Together,
these results establish sharpness, up to logarithmic factors,
across a broad variety of moment regimes.

Table~\ref{tab:proof-comparison} summarizes the principal proof
methods, separating bounded and sub-exponential coordinates from
polynomial moment envelopes.

\begin{table}[!htb]

\centering
\small
\setlength{\tabcolsep}{4pt}
\renewcommand{\arraystretch}{1.1}
\caption{Principal proof methods for Gaussian approximation over rectangles.}
\label{tab:proof-comparison}
\smallskip
\textit{(a) Bounded or sub-exponential coordinates}\par\smallskip
{\setlength{\tabcolsep}{3pt}
\begin{tabularx}{\textwidth}{@{}>{\raggedright\arraybackslash}p{.12\textwidth}
 >{\raggedright\arraybackslash}p{.15\textwidth}
 >{\raggedright\arraybackslash}p{.15\textwidth}
 >{\raggedright\arraybackslash}X
 >{\raggedright\arraybackslash}p{.15\textwidth}
 >{\raggedright\arraybackslash}p{.16\textwidth}@{}}
\toprule
Work & Covariance assumption & Gaussian seeding & Subsequent comparison & Non-Gaussian boundary estimates & Key derivative bounds \\
\midrule
This paper & Unrestricted & Slepian regularization & Calibrated dropout & Direct & Lemma~\ref{appx-an:lem:restoration} \\[2pt]
This paper, Cor.~\ref{cor:positive-correlation-gap} & Pairwise correlation gap & Slepian regularization & Calibrated dropout & Direct & Lemma~H.3 in~\cite{Supplement2026} \\[2pt]
LM~\cite{LopesMasini2026} & Unrestricted & Gaussian smoothing & Slepian interpolation & Recursive & Prop.~3.4; Lemma~3.6 \\[2pt]
AL~\cite{AL} & Positive definite & Gaussian smoothing & Calibrated dropout & Direct & Corollary~3.2 \\[2pt]
CCK~\cite{CCK} & Positive definite & Gaussian smoothing & Slepian--Stein & Recursive & Lemmas~6.1--6.3 \\[2pt]
FKLZ~\cite{FKLZ} & Three-coordinate nondegeneracy & Gaussian smoothing & Slepian--Stein & Recursive & Lemmas~2.2--2.3 \\[2pt]
GZ~\cite{GaoZhang} & Unrestricted & Bernoulli interpolation & Slepian interpolation & Not used & Lemmas~16, 32 \\[2pt]
CCKK~\cite{CCKK2022}, Thm.~2.1 & Unrestricted & No separate stage & Iterated randomized Lindeberg; Stein kernels & Recursive & Eqs.~(C.4)--(C.8) \\
\bottomrule
\end{tabularx}}
\smallskip
{\footnotesize\raggedright
LM denotes Lopes and Masini, whose result assumes i.i.d. observations.
Boundary estimates bound the probability that a non-Gaussian sum lies in a
thin layer near a rectangle's boundary and are used to control
derivatives in expectation. A direct bound does not use the Gaussian
approximation error being proved. A recursive bound compares the sum
with a Gaussian vector, introducing an approximation error that must
itself be controlled.\par}

\medskip
\textit{(b) Polynomial moment envelopes}\par\smallskip
\begin{tabularx}{\textwidth}{@{}>{\raggedright\arraybackslash}p{.25\textwidth}
 >{\raggedright\arraybackslash}p{.30\textwidth}
 >{\raggedright\arraybackslash}X@{}}
\toprule
Result & Moment assumptions & Proof route \\
\midrule
CCKK~\cite{CCKK2022}, Thm.~2.5 & Envelope and coordinate fourth moments & Slepian--Stein and truncation \\[2pt]
This paper, Thm.~\ref{thm:polynomial}(i) & Envelope alone & Conditional truncation and the bounded-summand theorem \\[2pt]
This paper, Thm.~\ref{thm:polynomial}(ii) & Envelope and coordinate fourth moments & Logistic--Slepian comparison and Bennett's inequality \\
\bottomrule
\end{tabularx}
\smallskip
{\footnotesize\raggedright
The envelope and coordinate fourth-moment conditions are
\eqref{poly:envelope} and~\eqref{poly:fourth}, respectively.\par}
\end{table}

On 16 November 2026, while finalizing this manuscript, we became aware
of the independent and contemporaneous work of Lopes and
Masini~\cite{LopesMasini2026}. Their manuscript was submitted to arXiv
on 13 November 2026 and appeared in the new-paper listing dated
16 November 2026. They obtain Gaussian approximation bounds for i.i.d.
random vectors with sub-Gaussian coordinates and unrestricted covariance,
as well as Gaussian comparison and bootstrap results. Their approximation
bounds have an \(n^{-1/2}\) factor but a suboptimal dependence on
\(L=\log(ep)\). This distinction is consequential when \(\log p\)
grows polynomially in \(n\), because the additional powers of \(L\)
then change the polynomial rate in \(n\).

To make the comparison precise under common assumptions, first consider
centered i.i.d. observations with unit coordinate variances and a fixed
bound on their coordinate sub-Gaussian norms. These assumptions also
give a fixed bound on their sub-exponential norms, so
Corollary~\ref{cor:subexponential} applies. When
\(\log p\asymp n^a\) for fixed \(0<a<1\), using
\(g_n\le C\sqrt L\), our bound and theirs in
\cite[Theorem~2.1]{LopesMasini2026} have respective orders
\[
 n^{-1/2}L^{3/2}(\log n)^{3/2}
 \qquad\text{and}\qquad
 n^{-1/2}L^3\log n,
\]
before truncation at \(1\). Thus their bound contains an extra
factor \(L^{3/2}\) in its dimension dependence, while our stated
sub-exponential bound has an additional factor \(\sqrt{\log n}\).
The ratio of the displayed bounds is therefore of order
\(L^{3/2}/\sqrt{\log n}\asymp n^{3a/2}/\sqrt{\log n}\).

For bounded observations, a sharper comparison is available.
If \(\|X_i\|_\infty\le B\) for fixed \(B\),
Corollary~\ref{cor:observations} and their
Theorem~3.1 give respective orders
\[
 n^{-1/2}L^{3/2}\sqrt{\log n}
 \qquad\text{and}\qquad
 n^{-1/2}L^{5/2}\log n.
\]
Their orthant bound applies to rectangles by replacing \(X_i\) with
\((X_i,-X_i)\in\R^{2p}\). In this bounded setting, the ratio is of
order \(L\sqrt{\log n}\asymp n^a\sqrt{\log n}\).
Table~\ref{tab:dimension-growth} records both comparisons.
For example, when \(a=1/4\), both of our bounds tend to zero,
whereas neither of their displayed bounds does. Their dependence
on \(\log(ep)\) is therefore suboptimal even within the bounded
i.i.d. class. The additional square-root logarithm in our
bounded-observation bound is not known to be necessary.

\begin{table}[!htb]
\centering\small
\renewcommand{\arraystretch}{1.15}
\caption{Dimension growth under common i.i.d. assumptions with unit
coordinate variances, fixed envelope or tail scale, and unrestricted
covariance, when \(\log p\asymp n^a\), \(0<a<1\).}
\label{tab:dimension-growth}
\begin{tabularx}{\textwidth}{@{}>{\raggedright\arraybackslash}Xcc@{}}
\toprule
Result & Bound before truncation at \(1\) & Bound tends to zero \\
\midrule
\multicolumn{3}{@{}l}{\textit{Sub-Gaussian coordinates}} \\[2pt]
This paper, Corollary~\ref{cor:subexponential}
  & \(n^{-1/2+3a/2}(\log n)^{3/2}\) & \(0<a<1/3\) \\[3pt]
Lopes--Masini~\cite[Theorem~2.1]{LopesMasini2026}
  & \(n^{-1/2+3a}\log n\) & \(0<a<1/6\) \\[5pt]
\multicolumn{3}{@{}l}{\textit{Bounded coordinates}} \\[2pt]
This paper, Corollary~\ref{cor:observations}
  & \(n^{-1/2+3a/2}\sqrt{\log n}\) & \(0<a<1/3\) \\[3pt]
Lopes--Masini~\cite[Theorem~3.1]{LopesMasini2026}
  & \(n^{-1/2+5a/2}\log n\) & \(0<a<1/5\) \\
\bottomrule
\end{tabularx}
\smallskip
{\footnotesize\raggedright
Multiplicative constants are suppressed. The first row uses our
sub-exponential result, which applies under the stated sub-Gaussian
assumption. The last column describes when each displayed bound
guarantees convergence; it is not a necessary condition for Gaussian
approximation.\par}
\end{table}

For the main Gaussian approximation bound, both proofs use interpolation,
but their argument combines Gaussian smoothing with recursive
anti-concentration, whereas ours uses logistic smoothing and calibrated
dropout. The present paper also treats independent summands that need
not be identically distributed and develops Gaussian-width refinements,
polynomial-moment bounds, lower bounds, and refinements under pairwise
correlation-gap conditions.

\subsection{Lean verification}\label{sec:lean-introduction}
Lean 4 verification covers the 31 proved statements of the archived
7 November 2026 version and formal counterparts of its substantive
proof arguments, with the derivative-envelope estimates restricted to
integer orders \(q\ge1\). The
\href{https://github.com/VC2015/rectangle-clt-singular/blob/main/LEAN_VERIFICATION.md}{verification record}
identifies the exact manuscript and proof versions, the correspondence
between the printed and formal arguments, the treatment of external
inputs, and the successful independent checks. This certification
applies to that archived version. Subsequent revisions, including
the Gaussian-width refinements and the additional results in the
Supplemental Material~\cite{Supplement2026},
are not yet covered by this certification.
Conjecture~\ref{conj:sharp-unrestricted} is excluded.

\section*{Plan of the Paper}
{Section~\ref{sec:smoothing} proves Theorem~\ref{thm:unrestricted}
by smoothing and interpolation, retaining the dependence on Gaussian width.}
{Section~\ref{sec:poly} outlines the proofs of the two
parts of Theorem~\ref{thm:polynomial} and compares their rates.}
Section~\ref{opt:section} compares the upper bounds with examples that
give lower bounds. Section~\ref{sec:discussion} asks whether the remaining
logarithmic factor for bounded observations can be removed. The appendices
give the proofs of the auxiliary estimates and the lower bounds.
{Appendix~\ref{app:polynomial-proof} proves
Theorem~\ref{thm:polynomial}, followed by the lower-bound proofs in
Appendix~\ref{app:sharpness-proofs}.}
The Supplemental Material~\cite{Supplement2026} contains three further
appendices. Appendix~G caps the additional logarithm in the
bounded-envelope bound by \(L=\log(ep)\), retaining Gaussian width.
Appendix~H proves the positive-correlation-gap corollary, retaining
Gaussian width through its surface and matrix estimates. Appendix~I
extends that bound to an additive budget for exceptional correlation gaps.

\section{{Proof of Theorem~\ref*{thm:unrestricted}}}\label{sec:smoothing}
Throughout this section we use the bounded-summand setup~\eqref{eq:setup},
with \(W=\sum_i\xi_i\) and \(Z\sim N(0,\Sigma)\).
{We retain the expected maxima in the anti-concentration estimates
throughout the comparison, and express the resulting bound in terms
of \(g_\Sigma\).}
We replace each rectangle indicator by a smooth approximation and compare
its expectations along a path from the sum to its Gaussian counterpart.
The smoothing parameter balances the error in approximating the indicator
against the derivative bounds needed for this comparison.
We first make a change of coordinates that represents both random vectors
in \(\R^r\), where \(r:=\operatorname{rank}(\Sigma)\), and gives each
identity covariance. This allows us to carry out the comparison even
when \(\Sigma\) is singular.

{
Subsection~\ref{sec:outline-smoothing} expresses rectangles as linear
inequalities in these coordinates and bounds the error from smoothing
their indicators. Subsection~\ref{sec:outline-path} constructs the
comparison in two stages, first adding a small Gaussian component and
then gradually replacing the summands by Gaussian noise while preserving
covariance. Subsection~\ref{sec:outline-seeded} controls the change in
the smoothed expectations using aggregate covariance and Gaussian matrix
and threshold estimates. Finally, Subsection~\ref{sec:proof-theorem-one}
integrates these bounds and chooses the smoothing parameter.
Appendices~\ref{appx-sec:smoothing}--\ref{appx-app:concentration}
give the auxiliary estimates and their proofs.
}

\subsection{Support coordinates and smoothing}\label{sec:outline-smoothing}
It suffices to consider bounded closed rectangles, since the result for
general rectangles follows by monotone approximation. Choose a \(p\times r\)
matrix \(\Gamma\) of full column rank such that \(\Sigma=\Gamma\Gamma^\top\).
The two matrices have the same column space:
\[
 \operatorname{range}(\Sigma)=\operatorname{range}(\Gamma)\subseteq\R^p.
\]
Indeed, the first space is contained in the second, and both have dimension
\(r\). Each summand lies in this common subspace almost surely. To see this,
take any vector \(z\) in a basis of the null space of \(\Sigma\). Then
\[
 \sum_i\E(z^\top\xi_i)^2=z^\top\Sigma z=0,
\]
so every summand is orthogonal to that null space. The columns of \(\Gamma\)
form a basis of the common subspace, and each summand therefore has a unique
vector of coefficients in this basis. Define these \emph{support coordinates}
and the Gaussian comparison vector by
\[
 \zeta_i^0:=(\Gamma^\top\Gamma)^{-1}\Gamma^\top\xi_i\in\R^r,\qquad
 S:=\sum_i\zeta_i^0,\qquad G\sim N(0,I_r),
\]
where \(G\) is independent of the summands. Thus \(\xi_i=\Gamma\zeta_i^0\),
\(W=\Gamma S\), and we may take \(Z=\Gamma G\).
The transformed summands are independent and centered, and their
covariances sum to \(I_r\). The superscript \(0\) distinguishes them
from the independent copies used in the proofs.

A point \(\Gamma x\) belongs to the rectangle
\(B=\prod_{j=1}^p[\ell_j,b_j]\) exactly when
\((\Gamma^\top e_j)\cdot x\le b_j\) and
\(-(\Gamma^\top e_j)\cdot x\le-\ell_j\) for every \(j\), where \(e_j\)
is the \(j\)th coordinate vector and the dot denotes the usual inner product.
We collect these \(m:=2p\)
inequalities by setting
\[
 u_j:=\Gamma^\top e_j,\qquad u_{p+j}:=-u_j,\qquad
 a_j:=b_j,\qquad a_{p+j}:=-\ell_j\quad (j=1,\ldots,p).
\]
In support coordinates the rectangle is therefore the set
\[
 A_a:=\{x:u_\alpha\cdot x\le a_\alpha\ \text{for all }\alpha\le m\}.
\]
For each defining inequality \(u_\alpha\cdot x\le a_\alpha\), the vector
\(u_\alpha\) is perpendicular to the boundary hyperplane
\(u_\alpha\cdot x=a_\alpha\) and points outward from the allowed halfspace.
We call these vectors \emph{face normals}, including those associated
with redundant inequalities. They are unit vectors, since
\[
 \|u_j\|^2=e_j^\top\Gamma\Gamma^\top e_j=\Sigma_{jj}=1,
 \qquad u_{p+j}=-u_j.
\]
Some face normals may repeat, but this creates no difficulty.

The original coordinate bound also controls the components of each
transformed summand along these normals. Indeed, since
\(\Gamma\zeta_i^0=\xi_i\),
\[
 u_j\cdot\zeta_i^0=e_j^\top\Gamma\zeta_i^0=\xi_{ij},
 \qquad u_{p+j}\cdot\zeta_i^0=-\xi_{ij}.
\]
Writing \(U:=(u_\alpha)_{\alpha=1}^m\), we define the \emph{face norm} by
\[
 \|x\|_U:=\max_{\alpha\le m}|u_\alpha\cdot x|.
\]
{
For every \(x\in\R^r\), this equals \(\|\Gamma x\|_\infty\).
Since \(\Gamma\) has full column rank, \(\Gamma x=0\) implies \(x=0\),
so the face norm is indeed a norm on \(\R^r\).
}
The preceding identities give
\[
 \|\zeta_i^0\|_U=\max_{j\le p}|\xi_{ij}|
 =\|\xi_i\|_\infty\le\delta\qquad\text{almost surely},
\]
where the last inequality is the bounded-summand assumption~\eqref{eq:setup}.
Thus the transformation preserves the envelope \(\delta\), regardless
of how small the positive eigenvalues of \(\Sigma\) are. This support reduction also appears in~\cite[Section~2.3]{GaoZhang}.

{
The equivalence between \(\Gamma x\in B\) and \(x\in A_a\) also
transfers the probability comparison to \(\R^r\). Since
\(W=\Gamma S\) and \(Z=\Gamma G\),
\[
 \Pp(W\in B)=\Pp(S\in A_a),\qquad
 \Pp(Z\in B)=\Pp(G\in A_a).
\]
Every bounded closed rectangle is represented by such a threshold
vector. We allow arbitrary \(a\in\R^m\), interpreting inconsistent
inequalities as the empty set. Including these empty sets does not
change the supremum, and hence
\[
 \rho:=\sup_{a\in\R^m}
 |\Pp(S\in A_a)-\Pp(G\in A_a)|=\rh(W,Z).
\]
}

To compare these probabilities by differentiation, we introduce the
\emph{logistic approximation} to the indicator of \(A_a\). For a smoothing
parameter \(h>0\), define
\begin{equation}
 F_{a,h}(x)\coloneqq
 \left(1+\sum_{\alpha=1}^m
 e^{(u_\alpha\cdot x-a_\alpha)/h}\right)^{-1},
 \qquad x\in\R^r.
 \label{eq:selector}
\end{equation}
{
We index the approximation by the threshold vector \(a\), keeping the
full list of inequalities. A redundant inequality can change the
smoothing formula even when it does not change the set \(A_a\).
}

{
To interpret~\eqref{eq:selector}, assign each face \(\alpha\) the
\emph{violation score} \(u_\alpha\cdot x-a_\alpha\), and introduce a
\emph{baseline option} with score \(0\). The baseline score is at least
as large as every face score exactly when \(x\in A_a\). Logistic
normalization gives the baseline probability \(F_{a,h}(x)\) and the
\emph{face probabilities}
\[
 \pi_\alpha(x):=
 \frac{e^{(u_\alpha\cdot x-a_\alpha)/h}}
 {1+\sum_{\beta=1}^m e^{(u_\beta\cdot x-a_\beta)/h}},
 \qquad \sum_{\alpha=1}^m\pi_\alpha(x)=1-F_{a,h}(x).
\]
We use these probabilities to express and bound the derivatives of
the logistic approximation.
}

{
The smoother \(F_{a,h}\) in~\eqref{eq:selector} combines a smooth
maximum of the violation scores with a logistic cutoff, using the
same width \(h\) for both. By comparison,
the construction in~\cite{CCK2013} allows the cutoff and its scale to
be chosen separately from the smooth maximum. The multinomial
logistic form was used by McFadden~\cite[p.~1001]{McFadden1989} to
smooth indicators of simultaneous inequalities and also appears
in~\cite[eq.~(5)]{GaoZhang}. As \(h\) decreases to zero,
\(F_{a,h}(x)\) approaches \(\ind_{A_a}(x)\) away from the boundary
of \(A_a\), but its derivative bounds increase.
}

{The \emph{random-threshold representation} of the multinomial logit
probabilities makes the underlying random shocks explicit, with their
scale controlled by the smoothing width \(h\).} Let \(\omega_0,\ldots,\omega_m\) be independent
exponential random variables with mean \(1\), independent of \(S\) and
\(G\). Then
\begin{equation}
 \begin{split}
 F_{a,h}(x)&=\Pp_\omega(x\in A_{a,\omega}),\\
 A_{a,\omega}&\coloneqq\{x:u_\alpha\cdot x\le
 a_\alpha+h\log(\omega_\alpha/\omega_0)
                   \text{ for all }\alpha\le m\},
 \end{split}
 \label{eq:mixture}
\end{equation}
where $\Pp_\omega$ denotes probability over $\omega$ alone. {Thus smoothing replaces each threshold \(a_\alpha\) by a random threshold whose displacement is proportional to \(h\).} 

Conditioning on \(\omega_0\) gives this formula directly. This is also
the classical \emph{random-utility representation} underlying
multinomial logit probabilities, as described by
Yellott~\cite[Section~2.6]{Yellott1977}. Indeed, the variables
\(-\log\omega_\beta\), \(0\le\beta\le m\), are independent
\emph{standard Gumbel variables}, and \(F_{a,h}(x)\) is the probability
that the baseline utility \(-h\log\omega_0\) exceeds all face utilities
\(u_\alpha\cdot x-a_\alpha-h\log\omega_\alpha\).
Rearranging these utility comparisons gives the inequalities defining
\(A_{a,\omega}\). {An equivalent Gumbel-max
representation is used in~\cite[Lemma~14]{GaoZhang}.}
{
Thus the baseline records that no randomly shifted constraint is
violated, while \(\pi_\alpha(x)\) is the probability that face
\(\alpha\) has the largest positive shifted violation.
}

For the smoothing bound, we use~\eqref{eq:mixture} as a representation
by random threshold shifts with the face normals fixed.
The largest absolute threshold shift has expectation at most \(ChL\).
{
To convert this width into an error in probability, for each real \(s\) define
\[
 A_a^s:=\{x\in\R^r:u_\alpha\cdot x\le a_\alpha+s,\ \alpha\le m\}.
\]
Thus every face threshold is increased by the same amount \(s\).
Since the face normals are unit vectors, this moves each defining
hyperplane outward by distance \(s\) when \(s>0\), and inward by
distance \(|s|\) when \(s<0\). The notation denotes this common
threshold shift, not a dilation of the set \(A_a\).} The following inequality applies to the original sum and
also after independent noise has been added along the comparison path.
The lemma is of independent interest. Both parts retain the expected
largest absolute projection, which can be more informative than the
number of faces.

\begin{lemma}[Rectangle anti-concentration]\label{lem:concentration}
Let \(X_1,\ldots,X_n\in\R^r\) be independent centered random
vectors, and {define} \(V\coloneqq\sum_iX_i\). Suppose that
\[
 \norm{X_i}_U\le\eta\quad\text{almost surely for every }i,
 \qquad \Var(u_\alpha\cdot V)=v>0
                   \quad\text{for every }\alpha\le m.
\]
Let \(R\in\R^r\) be a random vector independent of the summands.
Then, for every \(a\in\R^m\) and \(s>0\),
\begin{equation}
 \Pp(V+R\in A_a^s\setminus A_a^{-s})
 \le C\min\left\{1,\frac{s+\eta}{v}\E\norm{V}_U,
                         (s+\eta)\sqrt{L/v}\right\}.
 \label{eq:bounded-intrinsic-concentration}
\end{equation}
If instead \(V\) is a centered Gaussian vector with
\(\Var(u_\alpha\cdot V)=v>0\) for every \(\alpha\), then, for
every \(R\) independent of \(V\),
\begin{equation}
 \Pp(V+R\in A_a^s\setminus A_a^{-s})
 \le\min\left\{1,\frac{2s}{v}\E\norm{V}_U\right\}
 \le C\min\{1,s\sqrt{L/v}\}.
 \label{eq:gaussian-intrinsic-concentration}
\end{equation}
The constant \(C\) is absolute.
\end{lemma}
The bounds involving \(\E\norm{V}_U\) are \emph{intrinsic}:
their dependence on the face directions enters through this expected
maximum, with no explicit dependence on the number of faces.
{The intrinsic bounds can be substantially smaller when the
coordinates are strongly dependent. For example, if all face normals
coincide up to sign}, then
\(\E\norm{V}_U\le\sqrt v\), regardless of \(p\), and in the
Gaussian case it equals \(\sqrt{2v/\pi}\).
The first inequality in~\eqref{eq:gaussian-intrinsic-concentration}
is a \emph{dimension-free alternative to Nazarov bounds} {developed in~\cite{ChernozhukovAC2026}.}

{
Appendix~\ref{appx-app:concentration} gives a self-contained proof.
The logarithmic Gaussian bound is the rectangle form of Nazarov's
inequality~\cite[Lemma~A.1]{CCK2017}. A direct integration-by-parts
argument proves the stronger geometric
bound~\eqref{eq:gaussian-intrinsic-concentration}, analogous to the
bound for centered cubes in~\cite[Theorem~3 and Comment~5]{CCK2015AC}.
}

{
The boundary estimates along the comparison involve both the Gaussian
and non-Gaussian expected maxima. To keep track of them, define
\begin{equation}\label{eq:intrinsic-boundary-coefficient}
 \kappa:=\min\left\{\sqrt L,\,
             1+\max\{\E\|W\|_\infty,g_\Sigma\}\right\}.
\end{equation}
Both alternatives come from Lemma~\ref{lem:concentration}.
For \(S\) and \(G\), the expected face norms are
\(\E\|W\|_\infty\) and \(g_\Sigma\), respectively, so their
boundary layers have probabilities at most \(C\kappa(s+\delta)\).
Lemma~\ref{intrinsic:lem:maximum} in
Appendix~\ref{appx-app:concentration} proves
\(\E\|W\|_\infty\le C(g_\Sigma+\delta L)\) without invoking
the Gaussian approximation being proved. In the nontrivial regime
\(\delta L\le g_\Sigma\), it follows that \(\kappa\le Cg_\Sigma\).
We use \(\kappa\) in the intermediate estimates and make this last
substitution when completing the proof.
}

{Let \(M_h\) denote the largest absolute difference between the
smoothed rectangle probabilities of \(S\) and \(G\), over all threshold vectors:}
\[
 M_h:=\sup_a|\E F_{a,h}(S)-\E F_{a,h}(G)|.
\]
{The expected size of the random threshold shifts and
Lemma~\ref{lem:concentration} give the following reduction.}
\begin{lemma}[Smoothing error]\label{lem:endpoint}
{Under~\eqref{eq:setup}, with \(\kappa\) defined in~\eqref{eq:intrinsic-boundary-coefficient}, for \(h\ge\delta\),} the {Kolmogorov distance over rectangles satisfies}
\begin{equation}
 \rho\le M_h+Ch{\kappa L}.
 \label{eq:endpoint-min}
\end{equation}
\end{lemma}
Appendix~\ref{appx-sec:smoothing} proves this bound and records the
derivative formulas used below. {To see why the smoothing error has
order \(h{\kappa L}\), recall that the representation~\eqref{eq:mixture}
shifts each face threshold \(a_\alpha\) by
\(h\log(\omega_\alpha/\omega_0)\). {Define}
\[
 R_h:=\max_{\alpha\le m}
 \bigl|h\log(\omega_\alpha/\omega_0)\bigr|.
\]
For each realization of the shifts,
\[
 A_a^{-R_h}\subseteq A_{a,\omega}\subseteq A_a^{R_h},
\]
so membership can change only within the shell
\(A_a^{R_h}\setminus A_a^{-R_h}\). Conditional on the shifts,
Lemma~\ref{lem:concentration} bounds the probability of this shell
by \(C(R_h+\delta){\kappa}\) {for both \(S\) and \(G\).}

The expected largest absolute shift satisfies \(\E_\omega R_h\le ChL\),
where the logarithmic factor \(L\) arises from taking the maximum over the
\(m=2p\) faces. Averaging the conditional probability bound over the shifts
therefore gives {a smoothing error for each of \(S\) and \(G\)} of at most
\[
 C(\E_\omega R_h+\delta){\kappa}
 \le C(hL+\delta){\kappa}
 \le Ch{\kappa L},
\]
since \(h\ge\delta\) and \(L\ge1\).}

\subsection{Calibrated dropout with Slepian regularization}
\label{sec:outline-path}

\begin{figure}[!htbp]
\centering
\resizebox{\textwidth}{!}{%
\begin{tikzpicture}[>=Stealth,font=\fontsize{9}{11}\selectfont]
\definecolor{seed}{RGB}{32,91,135}
\definecolor{drop}{RGB}{136,73,32}
\node[font=\Large] (S) at (0,0) {$S$};
\node[font=\Large] (U) at (6.2,0) {$U_0$};
\node[font=\Large] (G) at (12.4,0) {$\textcolor{blue}{U_1=G}$};
\draw[->,thick,seed] (S) -- node[above=9pt,align=center] {Slepian regularization} node[below=8pt] {$0\le v\le v_\varepsilon$} (U);
\draw[->,thick,drop] (U) -- node[above=9pt,align=center] {Calibrated dropout with a Gaussian seed} node[below=8pt] {$0\le t\le1$} (G);
\node[anchor=north,align=center] at (6.2,-0.85) {$\displaystyle U_0=\frac{S+\sqrt\varepsilon\,G}{\sqrt{1+\varepsilon}}$
\\[5pt] $\displaystyle v_\varepsilon=\frac{\varepsilon}{1+\varepsilon}$};
\node[anchor=north,align=center,text width=6.1cm] at (2.8,-2.9) {Rescale all summands\\and add Gaussian noise.\\[9pt]
 $\displaystyle T_v=\sqrt{1-v}\,S+\sqrt v\,G$};
\node[anchor=north,align=center,text width=7.0cm] at (10,-2.9) {Randomly delete summands, amplify survivors,\\and increase the Gaussian variance.\\[9pt]
 $\displaystyle U_t=\frac{(1+t)\sum_i\mathbb B_{i,t}\zeta_i^0+\sqrt{t+\varepsilon}\,G}{\sqrt{1+\varepsilon}}$\\[9pt]
 $\displaystyle \mathbb P(\mathbb B_{i,t}=1)=\frac{1-t}{(1+t)^2}$};
\end{tikzpicture}
}
\caption{The comparison from \(S\) to \(G\) passes through the regularized
sum \(U_0\). The first stage follows \(T_v\) up to \(v_\varepsilon\),
and the second follows \(U_t\) from \(t=0\) to \(t=1\),
\textcolor{blue}{ending at \(U_1=G\)}.}
\label{fig:two-stage-path}
\end{figure}
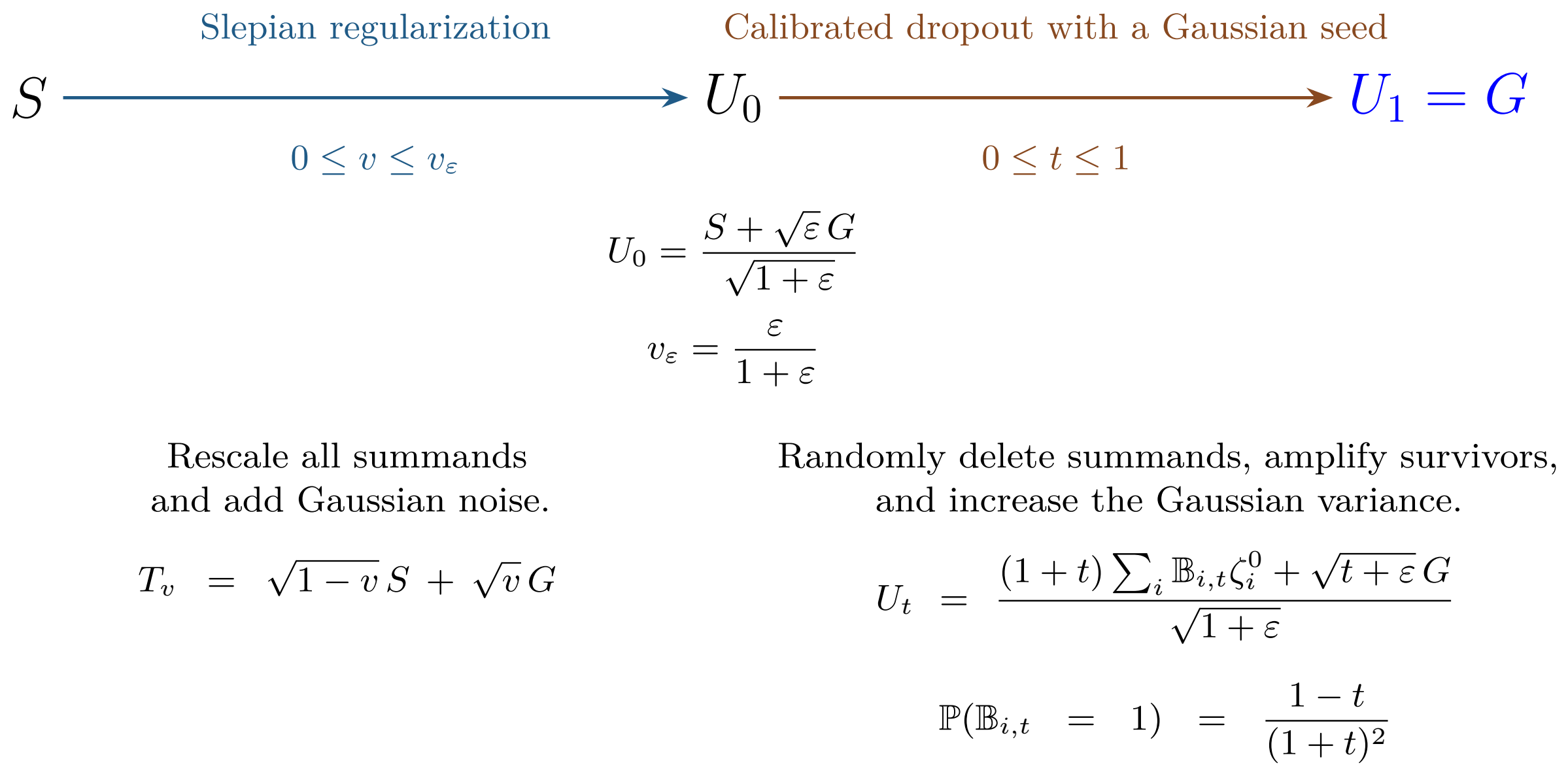

{Our goal is to bound \(M_h\), the {smoothed Kolmogorov discrepancy over rectangles} between
\(S=\sum_i\zeta_i^0\) and the independent Gaussian vector \(G\sim N(0,I_r)\).
Figure~\ref{fig:two-stage-path} shows the two stages of the comparison.
We first move from \(S\) to a regularized sum \(U_0\) by Slepian interpolation,
and then from \(U_0\) to \(U_1=G\) by calibrated dropout.
Both stages preserve the common covariance \(I_r\).}

{Fix \(\varepsilon>0\). In the first stage, we rescale \(S\)
and add a small independent Gaussian component along the
\emph{Slepian interpolation}}
\[
 T_v\coloneqq\sqrt{1-v}\,S+\sqrt v\,G,
 \qquad 0\le v\le1.
\]
{We stop at time \(v_\varepsilon\coloneqq\varepsilon/(1+\varepsilon)\),
so the first stage starts at \(T_0=S\) and ends at}
\[
 U_0\coloneqq T_{v_\varepsilon}
 =\frac{S+\sqrt\varepsilon\,G}{\sqrt{1+\varepsilon}}.
\]
{Since \(S\) and \(G\) are independent and both have covariance
\(I_r\), we have \(\Cov(T_v)=(1-v)I_r+vI_r=I_r\) throughout this stage.}

\begin{samepage}
{The initial independent Gaussian component, which we call the
\emph{Gaussian seed}, contributes \(\varepsilon I_r/(1+\varepsilon)\)
to the covariance of \(U_0\). This seed gives the second stage a positive
Gaussian variance from its starting point, permitting two Gaussian
integrations by parts throughout that stage, including at \(t=0\).
We call the initial operation \emph{Slepian regularization}.}
After a change of time it is also
\emph{Ornstein--Uhlenbeck regularization}~\cite[Theorem~4.12]{Nourdin2012}.
\par
\end{samepage}

{Starting from \(U_0\), we now delete summands at random,
rescale those retained, and increase the Gaussian variance.
For each time \(t\in[0,1]\), let \(\mathbb B_{i,t}\) be independent Bernoulli
\emph{masks}, independent of the summands and \(G\), and define}
\begin{equation}
 U_t\coloneqq
 \frac{(1+t)\sum_{i=1}^n\mathbb B_{i,t}\zeta_i^0
             +\sqrt{t+\varepsilon}\,G}{\sqrt{1+\varepsilon}},
 \qquad 0\le t\le1.
 \label{eq:path}
\end{equation}
A mask equal to \(1\) retains its summand, and a mask equal to zero deletes it.
Before the common normalization, each retained summand is multiplied by its
\emph{amplitude} \(\gamma(t)\coloneqq1+t\).

The \emph{retention probability} is
\[
 \varpi(t)\coloneqq\Pp(\mathbb B_{i,t}=1)
             =\frac{1-t}{(1+t)^2}.
\]
These choices give \(\varpi(t)\gamma(t)^2=1-t\). In the normalized
path~\eqref{eq:path}, the two independent components therefore contribute
{covariances adding up to \(I_r\):}
\[
 \underbrace{\frac{1-t}{1+\varepsilon}I_r}_{\text{retained summands}}
 +\underbrace{\frac{t+\varepsilon}{1+\varepsilon}I_r}_{\text{Gaussian component}}
 =I_r.
\]
As \(t\) increases, covariance is transferred from the retained summands
to the Gaussian component, while \(\Cov(U_t)=I_r\) throughout.
{At \(t=0\), every summand is retained, giving the regularized
sum \(U_0\) constructed in the first stage. At \(t=1\), all summands have
been deleted and \(U_1=G\).}

{The joint choice of amplitude and retention probability also
controls the third-moment contribution. Before normalization, each summand's
third-moment tensor is multiplied by}
\[
 {\varpi(t)\gamma(t)^3=1-t^2.}
\]
{The derivative of this multiplier vanishes at \(t=0\),
which helps control the third moments when the Gaussian variance is smallest.}
This random deletion is known as \emph{Bernoulli thinning}
\cite{Renyi1956,SteutelHarn1979,LastPenrose2017}. We call the
chosen amplitude and retention probability \emph{calibrated dropout},
in reference to the independent masks used in dropout regularization~\cite{Dropout2014}.
With \(\varepsilon=0\), this coincides with the calibrated path
defined in~\cite[Section~4]{AL}.
The paths describe distributions at each time, and
we do not need to couple the Bernoulli masks across different times.

{
To measure changes in the smoothed probabilities along the first stage, {define}
\[
 P_a^0(v)\coloneqq\E F_{a,h}(T_v),\qquad 0\le v\le1.
\]
The superscript \(0\) labels this initial stage. Since
\(T_0=S\) and \(T_{v_\varepsilon}=U_0\), the cost of introducing the seed
can be bounded by integrating the rate of change:
\[
 I_h\coloneqq\sup_a|P_a^0(v_\varepsilon)-P_a^0(0)|
 =\sup_a\left|\int_0^{v_\varepsilon}(P_a^0)'(v)\,dv\right|
 \le\sup_a\int_0^{v_\varepsilon}|(P_a^0)'(v)|\,dv.
\]
}

{To measure the change in the smoothed probabilities along the
second stage, {define}}
\[
 P_a(t)\coloneqq\E F_{a,h}(U_t).
\]
{
Its derivative \(P_a'(t)\) is the rate of change at time \(t\).
The two stages meet at \(P_a^0(v_\varepsilon)=P_a(0)\).
The triangle inequality and integration along the paths therefore give
\begin{equation}
\begin{aligned}
 M_h&\le I_h+\sup_a\int_0^1|P_a'(t)|\,dt\\
 &\le
 \underbrace{\sup_a\int_0^{v_\varepsilon}|(P_a^0)'(v)|\,dv}
             _{\text{Slepian regularization}}
 +\underbrace{\sup_a\int_0^1|P_a'(t)|\,dt}
             _{\text{calibrated dropout}}.
\end{aligned}
\label{eq:roadmap}
\end{equation}
It remains to bound the two derivatives over their respective time intervals.
}

\subsection{Controlling the comparison error}
\label{sec:outline-seeded}

{
We bound the two integrals in~\eqref{eq:roadmap} in the order of the
comparison. {The initial Slepian stage is controlled by direct averaged derivative estimates.} Its purpose is to introduce
the Gaussian seed needed for the main comparison. {Using this seed, we will obtain derivative bounds in the second stage.}

Let
\[
 K_3\coloneqq\sum_i\E(\zeta_i^0)^{\otimes3}
\]
denote the aggregate third-moment tensor.
{
Here \(z^{\otimes3}\) has entries \(z_jz_kz_\ell\).
}
Along the initial Slepian path,
Lemma~\ref{appx-an:lem:identity} separates its contribution from the
fourth-order remainder:
\[
 (P_a^0)'(v)
 =-\frac{\sqrt{1-v}}4\E D^3F_{a,h}(T_v):K_3+R^0_{a,h}(v),
 \qquad 0<v<1.
\]
{
The contraction \(D^3F(x):K_3\) is the sum of the third partial
derivatives of \(F\) at \(x\), each multiplied by the corresponding
entry of \(K_3\). We similarly write
\(D^qF(x)[z_1,\ldots,z_q]\) for the \(q\)th derivative evaluated in
the listed directions. The term \(R^0_{a,h}\) is the integral
remainder supplied by the lemma.
}

{
The derivative bounds improve after averaging because the logistic
 derivatives are concentrated near the smoothed boundary. Each spatial
 derivative contributes a factor \(h^{-1}\), giving the pointwise factors
 \(h^{-3}\) and \(h^{-4}\) for the third and fourth derivatives.
 However, their expansions also contain positive logistic weight products
 whose sums are bounded by a constant times
\[
 F_{a,h}(x)\bigl(1-F_{a,h}(x)\bigr).
\]
This weight is appreciable only where the smoother transitions between
\(0\) and \(1\). We will show that its expectation along the Slepian
path is at most \(Ch{\kappa}\). Multiplying by this bound improves the
 derivative factors from \(h^{-3}\) and \(h^{-4}\) to
\(h^{-2}{\kappa}\) and \(h^{-3}{\kappa}\), respectively.

{
To bound this expected weight, fix \(x\) and increase all thresholds
by the same amount \(s\). Define
\(F_s(x):=F_{a+s\mathbf1,h}(x)\), where \(\mathbf1\in\R^m\)
has all entries equal to \(1\). Direct differentiation gives
\[
 \frac{\partial}{\partial s}F_s(x)
 =h^{-1}F_s(x)\{1-F_s(x)\}.
\]
For \(|s|\le h\), the product \(F_s(x)\{1-F_s(x)\}\) is at least
\(e^{-1}F_0(x)\{1-F_0(x)\}\). Integrating over \([-h,h]\) yields
\[
 F_0(x)\{1-F_0(x)\}
 \le \frac e2\{F_h(x)-F_{-h}(x)\}.
\]
By~\eqref{eq:mixture}, \(F_h(x)-F_{-h}(x)\) is the probability,
over the random shifts \(\omega\), that \(x\) belongs to the shell
\(A_{a,\omega}^{h}\setminus A_{a,\omega}^{-h}\).

We apply this inequality at \(T_v=\sqrt{1-v}\,S+\sqrt v\,G\).
The two independent components have variances \(1-v\) and \(v\)
along every face normal. Condition on the component with the smaller
variance and on the random thresholds. The remaining component has
variance at least \(1/2\) along every face normal. {Its expected face norm is at most
\(\E\|W\|_\infty\) or \(g_\Sigma\), respectively, so the intrinsic
and dimension-based parts of Lemma~\ref{lem:concentration}} bound the conditional shell probability
by \(C(h+\delta){\kappa}\). Hence, uniformly in \(v\),
\[
 \E[F_{a,h}(T_v)\{1-F_{a,h}(T_v)\}]
 \le C(h+\delta){\kappa}\le Ch{\kappa},
 \qquad {h\ge\delta}.
\]
The smoothing error compares \(A_{a,\omega}\) with \(A_a\), and
therefore depends on the largest absolute random threshold shift.
Here we compare \(A_{a,\omega}^{h}\) with \(A_{a,\omega}^{-h}\):
each corresponding pair of thresholds differs by exactly \(2h\),
regardless of the random shifts.
}

We now combine this averaging bound with the relative moment bounds
in~\eqref{eq:relative-moments}. The cubic contribution is controlled by
\(s_3\), which measures the aggregate third moments and retains
cancellation across summands. The fourth-order remainder is controlled
by \(s_4\). To apply the same averaging bound to this remainder, we must
also account for the shifts in its derivative evaluation points.
These shifts change each scalar product \(u_\alpha\cdot x\) by at most
\(C\delta\). {Since \(h\ge\delta\)}, the shifted logistic weights
can therefore be compared with those at \(T_v\), losing only an
absolute constant. Appendix~\ref{appx-an:sec:initialization} supplies the details.
}

Combining the derivative factors with the averaged boundary weight
gives
\[
\begin{aligned}
 \left|\E D^3F_{a,h}(T_v):K_3\right|
 &\le Cs_3h^{-3}\,h{\kappa}
 =\frac{Cs_3{\kappa}}{h^2},\\
 |R^0_{a,h}(v)|
 &\le Cs_4h^{-4}\,h{\kappa}
 =\frac{Cs_4{\kappa}}{h^3}.
\end{aligned}
\]
Thus averaging supplies one factor of \(h\) before we integrate over
time. Substituting these estimates into the Slepian identity yields
\[
 |(P_a^0)'(v)|
 \le C{\kappa}\left(\frac{s_3}{h^2}+\frac{s_4}{h^3}\right),
 \qquad 0<v<1.
\]

{
The first stage lasts only until \(v_\varepsilon\le\varepsilon\),
so integration gives
\[
 I_h\le C\varepsilon{\kappa}
       \left(\frac{s_3}{h^2}+\frac{s_4}{h^3}\right).
\]
For fixed \(h\), increasing \(\varepsilon\) increases the stopping
time \(v_\varepsilon\) and the displayed bound on \(I_h\). We choose
\(\varepsilon\) after examining how the factor \(t+\varepsilon\)
enters the derivative bounds obtained by Gaussian integration by parts.
}

For the calibrated dropout path, the same lemma gives}

\begin{equation}
 P_a'(t)=
 -\frac{t}{3(1+\varepsilon)^{3/2}}\,
       \E D^3F_{a,h}(U_t):K_3+R_{4,a,h}(t).
 \label{eq:path-summary}
\end{equation}
{
{
{
The remainder \(R_{4,a,h}(t)\) consists of fourth derivatives evaluated
at shifted points \(U_t+\Delta\). Here \(\Delta\) denotes the random
displacement obtained by deleting one scaled summand and inserting a
scaled independent copy along the segment in Taylor's formula. It
depends on the summand index and the position along that segment, and
may depend on the same summands as \(U_t\) and the derivative
directions. Formula~\eqref{appx-an:translates} gives these displacements
before the normalization by \(\sqrt{1+\varepsilon}\). Every such shift
changes each face projection by at most \(C\delta\):
\[
 \max_{\alpha\le m}|u_\alpha\cdot\Delta|\le C\delta.
\]
}
} To see how the Gaussian component helps, write \(F=F_{a,h}\),
let \(g\sim N(0,I_r)\), and fix a base point \(x\in\R^r\), directions
\(z,w_1,w_2\in\R^r\), and a scale \(\sigma>0\). Two Gaussian integrations
by parts give
\[
\begin{aligned}
 &\E_g D^4F(x+\sigma g)[z,z,w_1,w_2]\\
 &\qquad=\frac1{\sigma^2}\,
 \E_g\!\left[
 \bigl\{(g\cdot z)^2-\norm z^2\bigr\}
 D^2F(x+\sigma g)[w_1,w_2]
 \right].
\end{aligned}
\]
We apply this identity after conditioning on the base and derivative
directions. It replaces two derivatives of the logistic approximation
by the \emph{Hermite term} \((g\cdot z)^2-\norm z^2\).
In our application, \(\sigma^2\) is comparable to \(t+\varepsilon\),
so these integrations contribute \((t+\varepsilon)^{-1}\), while the
two remaining derivatives contribute \(h^{-2}\).

{
To connect this identity to the fourth-moment assumption, we expand the
 two remaining derivatives. Their logistic weights are multiplied by
 face projections, such as \(u_j\cdot w_1\) and \(u_k\cdot w_2\).
When all four directions equal the same summand \(\zeta_i^0\), a term
involving \((u_j\cdot\zeta_i^0)^2\) combines with the Hermite term as
\[
\begin{aligned}
 &(u_j\cdot\zeta_i^0)^2
 \bigl\{(g\cdot\zeta_i^0)^2-\norm{\zeta_i^0}^2\bigr\}\\
 &\qquad=\tr\!\left[
 (gg^\top-I_r)(u_j\cdot\zeta_i^0)^2
 \zeta_i^0(\zeta_i^0)^\top
 \right].
\end{aligned}
\]
The sum of the expected coefficient matrices is
\[
 A_j:=\sum_i\E\!\left[
 (u_j\cdot\zeta_i^0)^2\,
 \zeta_i^0(\zeta_i^0)^\top
 \right]\preceq s_4I_r,
\]
by the fourth-moment bound in~\eqref{eq:relative-moments}, expressed in
support coordinates. Terms involving different face normals or
independent copies satisfy analogous operator-norm bounds.
{
A related use of aggregate matrix bounds after two Gaussian
integrations by parts appears in~\cite[Lemma~16]{GaoZhang}.
Here we also need to retain the boundary-concentration gain,
which requires the weighted Gaussian estimates developed in
Appendix~\ref{appx-sec:support}.
}
}

{
The inequality \(A_j\preceq s_4I_r\) controls the moment
coefficients. The remaining task is to bound the Gaussian
expectation of the Hermite term multiplied by the logistic weights.
Both factors depend on the same Gaussian vector, so we cannot
multiply their separate expectation bounds.
Estimating their absolute sizes separately would also lose the
cancellations needed to avoid a factor depending on rank.
The matrix and threshold estimates in
Appendix~\ref{appx-sec:support} control the Gaussian matrix
\(gg^\top-I_r\) averaged against the products of logistic
probabilities arising from the derivatives. Under the scale
conditions of Proposition~\ref{an:prop:derivative} {below}, these estimates
supply a factor \(Ch{\kappa L}\) in the remainder bound, with no factor
depending on \(r\). The corresponding factor without the Hermite
term was \(Ch{\kappa}\).
Consequently,
\[
 |R_{4,a,h}(t)|
 \le Cs_4\,\frac1{t+\varepsilon}\,\frac1{h^2}\,h{\kappa L}
 =\frac{Cs_4{\kappa L}}{h(t+\varepsilon)}.
\]
Thus the boundary gain of one power of \(h\) survives the two
integrations by parts, at the cost of an additional factor \(L\)
in this weighted estimate.

For the cubic contribution, two Gaussian integrations by parts
leave one derivative of \(F_{a,h}\), which contributes \(h^{-1}\).
The signed aggregate third-moment matrices supply the factor
\(s_3\). Multiplying by the weighted boundary factor \(h{\kappa L}\)
removes this remaining power of \(h^{-1}\).
Including the coefficient in~\eqref{eq:path-summary} gives a bound
\(Cs_3{\kappa L}t/(t+\varepsilon)\).
The derivative of the calibrated third-moment multiplier
\(1-t^2\) supplies the factor \(t\). Since \(t/(t+\varepsilon)\le1\),
integration over \(0\le t\le1\) contributes at most \(Cs_3{\kappa L}\),
uniformly in the seed \(\varepsilon\). The fourth-order bound
has no compensating factor \(t\), and its integral contains
\(\log(1+1/\varepsilon)\).

We choose the seed scale by comparing the errors accumulated over
both stages. The direct derivative bounds are useful during the short
initialization stage: although its fourth-order bound contains
\(h^{-3}\), it is integrated over an interval of length at most
\(\varepsilon\). Its fourth-order contribution is therefore bounded by
\(C\varepsilon s_4{\kappa}/h^3\). Choosing \(\varepsilon=Lh^2\)
makes this contribution at most \(Cs_4{\kappa L}/h\), and the full
first-stage bound becomes
\[
 I_h\le C{\kappa L}\left(s_3+\frac{s_4}{h}\right).
\]
Thus the short duration compensates for the unfavorable powers of
\(h\) in the direct derivative bounds.

With this seed choice, the fourth-order bound in the second stage
can be written as
\[
 |R_{4,a,h}(t)|
 \le \frac{Cs_4{\kappa}}{h^3}\,
       \frac{\varepsilon}{t+\varepsilon}.
\]
It has the same order as the direct derivative bound at \(t=0\),
but decreases as the Gaussian variance grows. Consequently,
\[
 \sup_a\int_0^1|R_{4,a,h}(t)|\,dt
 \le \frac{Cs_4{\kappa L}}h
       \log\!\left(1+\frac1\varepsilon\right).
\]
A larger seed would increase the initialization bound linearly while
reducing the second-stage logarithm. The choice \(\varepsilon=Lh^2\)
keeps the fourth-order initialization cost at the scale
\(s_4{\kappa L}/h\), and the second stage costs this scale times a
logarithm. The gain comes from accumulating the direct bound only
over a short interval and using the decreasing Gaussian bound for
the rest of the comparison.
The next proposition states these bounds under their precise
scale conditions.
}

}

\begin{proposition}[Initialization and seeded derivative estimates]
\label{an:prop:derivative}
{Under~\eqref{eq:setup} and \eqref{eq:relative-moments}, with
\(\kappa\) defined in~\eqref{eq:intrinsic-boundary-coefficient}, there are absolute constants} \(\mathsf A,C\)
such that, for \(h\ge\mathsf A\delta\), uniformly over
\(a\in\R^m\) and \(0<v<1\),
\begin{equation}
 |\E D^3F_{a,h}(T_v):K_3|\le\frac{Cs_3{\kappa}}{h^2},
 \qquad |R^0_{a,h}(v)|\le\frac{Cs_4{\kappa}}{h^3}.
 \label{eq:direct-initialization}
\end{equation}
If \(Lh^2<1/4\), choose \(\varepsilon=Lh^2\) in~\eqref{eq:path}.
Then
\begin{equation}
 I_h\le Cs_3 {\kappa L}+\frac{Cs_4{\kappa L}}h,
 \label{eq:initialization-bound}
\end{equation}
and, uniformly over \(a\in\R^m\) and \(0<t<1\),
\begin{equation}
 \begin{aligned}
 \left|\frac{t}{3(1+\varepsilon)^{3/2}}
          \E D^3F_{a,h}(U_t):K_3\right|
 &\le Cs_3 {\kappa L}\frac{t}{t+\varepsilon},\\
 |R_{4,a,h}(t)|&\le\frac{Cs_4{\kappa L}}{h(t+\varepsilon)}.
 \end{aligned}
 \label{eq:seeded-derivatives}
\end{equation}

\end{proposition}
\subsection{{Completion of the proof}}\label{sec:proof-theorem-one}

{
\begin{proof}[Proof of Theorem~\ref{thm:unrestricted}]
We combine the errors from smoothing and the two comparison stages,
then choose the smoothing width to balance their contributions.
The envelope condition allows us to take \(s_3\le\delta\).

First fix \(h\ge\mathsf A\delta\), where \(\mathsf A\ge1\) is the
constant in Proposition~\ref{an:prop:derivative}, and suppose that
\(Lh^2<1/4\). Set \(\varepsilon\coloneqq Lh^2\). By the smoothing
bound~\eqref{eq:endpoint-min} and the decomposition~\eqref{eq:roadmap},
\[
 \rh(W,Z)
 \le Ch\kappa L+I_h+\sup_a\int_0^1|P_a'(t)|\,dt.
\]
The first term is the smoothing error. For the initial Slepian stage,
\eqref{eq:initialization-bound} gives
\[
 I_h\le C\kappa L\left(s_3+\frac{s_4}{h}\right).
\]
For the calibrated dropout stage, we integrate the cubic and
fourth-order bounds in~\eqref{eq:seeded-derivatives}. Their time
integrals behave differently:
\[
 \int_0^1\frac{t}{t+\varepsilon}\,dt\le1,
 \qquad
 \int_0^1\frac{dt}{t+\varepsilon}
 =\log\!\left(1+\frac1\varepsilon\right).
\]
Thus the cubic contribution has a bounded time integral, whereas the
fourth-order remainder contributes a logarithm. Combining the two
stages with the smoothing error yields
\[
 \rh(W,Z)\le C\kappa L\left[
 h+s_3+\frac{s_4}{h}
 \left\{1+\log\!\left(1+\frac1{Lh^2}\right)\right\}
 \right].
\]

We now choose \(h\). Since \(h\ge\delta\),
\[
 1+\log\!\left(1+\frac1{Lh^2}\right)
 \le1+\log\!\left(1+\frac1{L\delta^2}\right)
 \le CD_L.
\]
Using also \(s_3\le\delta\), we obtain
\[
 \rh(W,Z)\le C\kappa L
 \left(h+\delta+\frac{s_4D_L}{h}\right).
\]
Increasing \(h\) enlarges the smoothing error but reduces the
fourth-order contribution. Balancing these terms, subject to
\(h\ge\mathsf A\delta\), leads to
\[
 h\coloneqq\mathsf A\max\{\delta,\sqrt{s_4D_L}\}.
\]
For this choice,
\[
 h+\delta+\frac{s_4D_L}{h}
 \le C\left(\delta+\sqrt{s_4D_L}\right).
\]
If \(Lh^2<1/4\), the preceding estimates, together with
\(\rh(W,Z)\le1\), give \(\rh(W,Z)\le C\kappa L(\delta+\sqrt{s_4D_L})\).

It remains to cover the case in which the chosen width satisfies
\(Lh^2\ge1/4\). Then
\[
 \kappa L\left(\delta+\sqrt{s_4D_L}\right)
 \ge\frac{\kappa Lh}{\mathsf A}
 \ge\frac{\kappa\sqrt L}{2\mathsf A}
 \ge\frac1{2\mathsf A}.
\]
Since \(\kappa\ge1\) and \(L\ge1\), the same bound follows
from \(\rh(W,Z)\le1\) in this case.

We now express \(\kappa\) through the Gaussian width. If
\(\delta L\le g_\Sigma\), Lemma~\ref{intrinsic:lem:maximum} and
\(g_\Sigma\ge\sqrt{2/\pi}\) give \(\kappa\le Cg_\Sigma\).
Otherwise, \(\delta g_\Sigma L>g_\Sigma^2\ge2/\pi\), so
\eqref{eq:intrinsic-unrestricted} follows directly from
\(\rh(W,Z)\le1\). This proves~\eqref{eq:intrinsic-unrestricted}
in both cases. Finally, the envelope condition permits
\(s_4=\delta^2\). Substitution and \(D_L\ge1\) give
\eqref{eq:intrinsic-envelope}.
\end{proof}
}

{
The factor \(L\) outside the final square root comes from the
random threshold shifts and the Gaussian weighted derivative estimates.
Theorem~\ref{opt:symmetric-lower} shows that a factor of this order
is necessary even when Gaussian width is bounded. The additional
logarithm inside \(D_L\) comes from integrating the fourth-order
remainder near the beginning of the second stage. Its necessity
remains open.
}

\section{Outline of the Proof of Theorem~\ref*{thm:polynomial}}\label{sec:poly}

The two parts of Theorem~\ref{thm:polynomial} use different information
about large observations. Under the envelope condition alone, we
condition on the large observations and apply the bounded-summand
theorem to the remaining sum. With the additional coordinate
fourth-moment condition, we compare smooth expectations directly and
use Bennett's inequality to control the large increments more sharply.
The second argument refines the polynomial-moment estimate
in~\cite{CCKK2022}. We explain the two mechanisms and their parameter
choices below; Appendix~\ref{app:polynomial-proof} gives the complete
proofs in Subsections~\ref{app:conditional} and~\ref{app:bennett}.

Throughout this section, $W,Z,\Sigma$, and $L$ have the meanings
in~\eqref{poly:setup}, and $g_\Sigma=\E\|Z\|_\infty$.
Let $\mathcal E_q(B,p)$ be the class of independent centered arrays
with unit average marginal variances satisfying the envelope
condition~\eqref{poly:envelope}, without requiring~\eqref{poly:fourth}.
Thus $\mathcal C_q(B,p)\subseteq\mathcal E_q(B,p)$.
To compare the rates, define
\begin{equation}
 \alpha_q:=\frac12-\frac1q,\qquad
 V_q:=\frac{B L^{3/2-1/q}}{n^{1/2-1/q}}
      =Bn^{-\alpha_q}L^{1+\alpha_q}.
 \label{poly:rates}
\end{equation}
The quantity $V_q$ records the dimension-based scale. The bounds
below retain the common factor $g_\Sigma/\sqrt L$, which improves
that scale when the Gaussian coordinates have a smaller expected
maximum. The moment order $q>2$ is fixed. Constants indexed by $q$
may depend on it, and the bounds are not asserted uniformly as $q$
approaches $2$ or infinity.

\subsection{\texorpdfstring{{Part (i): transfer from bounded observations}}{Part (i): transfer from bounded observations}}\label{poly:transfer-sec}

We separate large observations without discarding every sample in
which one occurs. For a threshold $T\ge2B$, define
\[
 I_i:=\ind\{\|X_i\|_\infty>T\},\qquad
 \mu_i:=\E[X_i\mid I_i=0].
\]
Condition on both the indicators $I_i$ and the values $I_iX_i$ of
the large observations. Given this information, the vectors
$(1-I_i)(X_i-\mu_i)$ are independent, centered, and bounded in
maximum norm by $2T$. We can therefore write
\[
 W=\frac1{\sqrt n}\sum_i(1-I_i)(X_i-\mu_i)+R,
 \qquad
 R:=\frac1{\sqrt n}\sum_i\{I_iX_i+(1-I_i)\mu_i\},
\]
where $R$ is fixed under the conditioning. This is the form needed
to apply the bounded-summand theorem to a translated rectangle.

Truncation also changes the covariance, so we must control the
conditional Gaussian target as well as the shift $R$.
Let $Q$ be the conditional covariance of the bounded sum and define
$v_T:=B^qT^{2-q}$, which bounds the second moment discarded from
each observation. Appendix~\ref{app:conditional} proves
$Q\preceq\tfrac43\Sigma$ and shows that every marginal variance
of $Q$ is at least $1/2$ outside an event of probability at most
$Cv_T$. The covariance order bounds the conditional Gaussian width
by $Cg_\Sigma$. On the event of adequate marginal variances,
Gaussian anti-concentration then controls the shift $R$.
Comparisons formed by adding independent Gaussian vectors correct
the covariance with the same width factor.

These steps yield the threshold estimate~\eqref{poly:threshold}.
Writing $\ell:=\log(en)$, its relevant terms are
\[
 \rh(W,Z)\le C_q\min\left\{1,
 g_\Sigma\left[
 \frac{TL\sqrt\ell}{\sqrt n}
 +\sqrt{Lv_T}+\frac{V_q}{\sqrt L}\right]\right\}.
\]
The first term is the approximation error for the bounded sum.
The other two control the residual shift and the Gaussian additions.
Increasing $T$ enlarges the first term and decreases
$\sqrt{Lv_T}=B^{q/2}T^{1-q/2}\sqrt L$.
When $n\ge L\ell$, the choice
\[
 T:=2B\left(\frac n{L\ell}\right)^{1/q}
\]
balances these two terms at order
$(V_q/\sqrt L)\ell^{\alpha_q}$. The remaining term is no larger.
If $n<L\ell$, the resulting rate is bounded below by a positive
absolute constant, so the bound by $1$ suffices. We obtain
\begin{equation}
 \rh(W,Z)\le C_q\min\{1,(g_\Sigma/\sqrt L)
               V_q\{\log(en)\}^{\alpha_q}\},
 \qquad (X_i)_{i=1}^n\in\mathcal E_q(B,p),
 \label{poly:transfer-bound}
\end{equation}
which is~\eqref{poly:main-envelope} in the present notation.
The extra logarithm comes from the bounded-observation theorem.
If Conjecture~\ref{conj:sharp-unrestricted} holds, the same transfer
gives $C_q\min\{1,(g_\Sigma/\sqrt L)V_q\}$ under the envelope
condition alone. Using $g_\Sigma\le C\sqrt L$ recovers
$C_q\min\{1,V_q\}$ in the worst case over covariance matrices; see
Remark~\ref{rmk:E1}.

\subsection{\texorpdfstring{{Part (ii): refinement under coordinate fourth moments}}{Part (ii): refinement under coordinate fourth moments}}\label{poly:bennett-sec}

The coordinate fourth moments can be much smaller than the fourth
moment of the random envelope. A large observation may have its
largest coordinate in a different location from other observations,
so its contribution to any fixed coordinate can remain small.
The additional condition~\eqref{poly:fourth} records this distinction
and restricts the class from $\mathcal E_q(B,p)$ to
$\mathcal C_q(B,p)$. To compare the two contributions to the error,
define
\begin{equation}
 U:=\frac{B^{1/2}L^{5/4}}{n^{1/4}},\qquad
 \mathcal K_q:=\left(\frac{V_q}{U}\right)^2
     =B\left(\frac nL\right)^{2/q-1/2},\qquad
 \mathcal L_q:=\log(e+\mathcal K_q).
 \label{poly:bennett-parameters}
\end{equation}
The scale $U$ comes from the coordinate fourth moments, while $V_q$
comes from the envelope moment. Under both assumptions, the bound of
Chernozhukov, Chetverikov, Kato, and
Koike~\cite[Theorem~2.5]{CCKK2022} is
$\rh(W,Z)\le C_q\min\{1,U+V_q\}$.
Its rectangle form follows by applying their lower-orthant bound to
$(X_i,-X_i)$, which preserves the moment bounds and replaces the
dimension logarithm by a comparable quantity.
The ratio $\mathcal K_q$ indicates whether the envelope contribution
is larger than the fourth-moment contribution. Our refinement reduces
that contribution by a power of $\mathcal L_q$ and retains Gaussian
width:
\begin{equation}
 \rh(W,Z)\le C_q\min\left\{1,
 \frac{g_\Sigma}{\sqrt L}
 \left(U+\frac{V_q}{\mathcal L_q^{1-1/q}}\right)\right\},
 \qquad (X_i)_{i=1}^n\in\mathcal C_q(B,p).
 \label{poly:bennett-bound}
\end{equation}
This is the conclusion~\eqref{poly:main-fourth} of
Theorem~\ref{thm:polynomial}\textup{(ii)}.

The proof compares logistic approximations along a Slepian path.
Take mutually independent copies $\widetilde X_i$, independent of
the original array, and write $Y_i:=(\widetilde X_i-X_i)/\sqrt n$
for the paired increments in this argument. Interchanging $X_i$ and
$\widetilde X_i$ reverses $Y_i$ and exchanges the endpoints of the
corresponding segment. Averaging the Hessian at these endpoints
turns the Taylor error into a fourth-order remainder, without
requiring the original observations to be symmetric or to have zero
third moments. Lemma~\ref{ben:paired-interpolation} gives the identity.

We use a common scale $t$ for the logistic width and the cutoff
between small and large increments. For the latter, the comparison
in Lemma~\ref{ben:comparison} keeps two quantities separate:
\[
 v_t:=\max_j\sum_i\E[Y_{ij}^2\ind\{\|Y_i\|_\infty>t\}],
 \qquad
 M_t:=\E\left\|\sum_iY_i\ind\{\|Y_i\|_\infty>t\}\right\|_\infty.
\]
The first is the largest coordinate variance sum; the second is the
expected maximum norm of the large-increment sum. Their distinction
is essential for the logarithmic improvement.
Write $A:=Bn^{-\alpha_q}$. The envelope condition gives
$\sum_i\E\|Y_i\|_\infty^q\le C_qA^q$, while the coordinate
condition gives $\max_j\sum_i\E Y_{ij}^4\le16B^2/n$.
Cauchy--Schwarz and Markov's inequality combine these bounds into
\[
 v_t\le C_q\frac B{\sqrt n}\left(\frac At\right)^{q/2}.
\]
Thus both moment assumptions enter the same tail-variance estimate.

To control $M_t$, truncate the large increments again at a level
$u>0$. These increments are independent and centrally symmetric,
so the truncated part remains centered. Bennett's inequality
controls that part, while the envelope moment controls the omitted
part. Lemma~\ref{ben:maximum} yields
\[
 M_t\le C_q\left\{
       \frac{uL}{\log(1+u^2L/v_t)}+\frac{A^q}{u^{q-1}}\right\}
 \quad\text{if }v_t>0;
 \qquad M_t=0\quad\text{if }v_t=0.
\]
Keeping the denominator in this Bennett bound is what produces the
gain over the envelope scale $V_q$.
When $\mathcal K_q\le1$, choosing $t:=U/L^{3/2}$ controls the
full comparison by $C_q(g_\Sigma/\sqrt L)U$.
When $\mathcal K_q>1$, choose
\[
 t:=A L^{-1/q}\mathcal L_q^{-(1-1/q)},\qquad
 u:=\mathcal L_q t=A(\mathcal L_q/L)^{1/q}.
\]
The mixed moment estimate above makes the Bennett denominator at
least $c_q\mathcal L_q$ when $v_t>0$. The two terms in its bound
then have the same order, giving
\[
 \sqrt L\,M_t\le C_q\frac{V_q}{\mathcal L_q^{1-1/q}}.
\]
Appendix~\ref{app:bennett} checks that this choice also controls
the remaining smoothing and fourth-order terms, proving
\eqref{poly:bennett-bound}. Gaussian anti-concentration contributes
the common factor $g_\Sigma/\sqrt L$ to the comparison in its
dimension-based normalization, so it does not change these parameter
choices or $\mathcal K_q$.
Basu and Kuchibhotla use a related truncation argument for maxima
of independent sums~\cite[Proposition~4.2]{BK2025}.

Both arguments apply on $\mathcal C_q(B,p)$. Taking the larger of
their constants as $C_q$ makes the two bounds simultaneous, so we may
use their minimum. Table~\ref{poly:summary} summarizes the assumptions,
proof routes, and lower-bound comparisons.

\begin{table}[htbp]
\centering

\small
\renewcommand{\arraystretch}{1.15}
\begin{tabularx}{\textwidth}{@{}>{\raggedright\arraybackslash}p{0.23\textwidth}
 >{\raggedright\arraybackslash}p{0.27\textwidth}
 >{\raggedright\arraybackslash}X@{}}
\toprule
Result, assumptions, and upper rate & Proof route & Lower bounds and remaining gap\\
\midrule
Theorem~\ref{thm:polynomial}\textup{(i)}\newline
Condition~\eqref{poly:envelope} alone\newline
\((g_\Sigma/\sqrt L)V_q\{\log(en)\}^{\alpha_q}\)
& Conditional truncation, followed by the bounded-summand theorem and
Gaussian anti-concentration.\newline Appendix~\ref{app:conditional}.
& Theorem~\ref{lower:dense}\textup{(i)} gives a lower bound of order
\((g_\Sigma/\sqrt L)V_q\) for every fixed \(q>2\),
with fixed or growing \(B\ge2\).
The remaining factor is \(\{\log(en)\}^{\alpha_q}\), whose
necessity is unresolved.\\[0.6em]
Theorem~\ref{thm:polynomial}\textup{(ii)}\newline
Conditions~\eqref{poly:envelope} and~\eqref{poly:fourth}\newline
\((g_\Sigma/\sqrt L)(U+V_q/\mathcal L_q^{1-1/q})\)
& Logistic smoothing, Slepian interpolation with paired increments, and
a Bennett inequality using both moment conditions.\newline
Appendix~\ref{app:bennett}.
& Matching lower bounds for fixed \(B\ge2\) at \(q=4\)
(Theorem~\ref{opt:q4-lower}) and when \(\mathcal K_q\to\infty\), \(\log\mathcal K_q=o(L)\)
(Theorem~\ref{lower:sparse}). At \(q=4\),
Theorem~\ref{opt:q4-growing} also matches the dependence on growing \(B\).\\
\bottomrule
\end{tabularx}
\caption{The two parts of Theorem~\ref{thm:polynomial}, their proofs,
and their sharpness. Constants are suppressed and both upper bounds
are capped at 1. The lower-bound examples have widths
ranging from a constant to order \(\sqrt L\), subject to the
dimension and vanishing-rate conditions in Section~\ref{opt:section}.}
\label{poly:summary}
\end{table}

\subsection{\texorpdfstring{{Comparison of the bounds under the two moment assumptions}}{Comparison of the bounds under the two moment assumptions}}\label{poly:transition}

Both upper bounds contain the factor $g_\Sigma/\sqrt L$, so their
relative sizes are determined by $U,V_q,\mathcal K_q$, and
$\mathcal L_q$. The lower-bound examples retain the same width
factor. All sharpness comparisons below use the dimension and width
sequences specified in Section~\ref{opt:section}, with the relevant
rate tending to zero.

Under the envelope condition alone,
Theorem~\ref{lower:dense}\textup{(i)} gives a lower bound of order
$(g_\Sigma/\sqrt L)V_q$ for every fixed $q>2$ and fixed or growing
$B\ge2$. Thus the remaining gap in~\eqref{poly:transfer-bound} is
the factor $\{\log(en)\}^{\alpha_q}$.

When the coordinate fourth-moment condition also holds, we may use
the smaller upper bound. First fix $B\ge2$ and suppose $n/L\to\infty$.
For $2<q<4$, the ratio $\mathcal K_q$ tends to infinity, and
\[
 \frac{V_q/\mathcal L_q^{1-1/q}}U
 =\frac{\sqrt{\mathcal K_q}}{\mathcal L_q^{1-1/q}}\to\infty.
\]
The envelope term therefore determines the order of
\eqref{poly:bennett-bound}, giving
$(g_\Sigma/\sqrt L)V_q/\mathcal L_q^{1-1/q}$ and improving the
logarithmic factors in~\eqref{poly:transfer-bound}.
Theorem~\ref{lower:sparse} matches this rate when also
$\log\mathcal K_q=o(L)$.

At $q=4$, we have $\mathcal K_4=B$. The second term in
\eqref{poly:bennett-bound} absorbs the first uniformly in $B\ge1$,
because $\{\log(e+B)\}^{3/4}/\sqrt B$ is bounded. Consequently,
\begin{equation}
 \rh(W,Z)\le C\min\left\{1,
 \frac{g_\Sigma Bn^{-1/4}L^{3/4}}
 {\{\log(e+B)\}^{3/4}}\right\}.
 \label{poly:q4}
\end{equation}
For fixed $B\ge2$, Theorem~\ref{opt:q4-lower} matches
$g_\Sigma n^{-1/4}L^{3/4}$ from below, including when the width
stays bounded. For growing $B$, Theorem~\ref{opt:q4-growing} also
matches the factor $B/\{\log(e+B)\}^{3/4}$ under its condition
$\log B=o(L)$.

For $q>4$ and fixed $B$, the ratio $\mathcal K_q$ tends to zero.
The bound~\eqref{poly:bennett-bound} retains
$(g_\Sigma/\sqrt L)U$, whereas~\eqref{poly:transfer-bound} gives
$(g_\Sigma/\sqrt L)V_q\{\log(en)\}^{\alpha_q}$.
For example, if $p\asymp n^a$ with fixed $a>0$, the latter rate is
\[
 O_{q,B,a}\!\left(g_\Sigma n^{-\alpha_q}
                    (\log n)^{1/2+2\alpha_q}\right).
\]
Since $\alpha_q>1/4$, it improves the power of $n$.
Theorem~\ref{lower:dense}\textup{(ii)} gives a lower bound of order
$(g_\Sigma/\sqrt L)V_q$ within $\mathcal C_q(B,p)$, leaving the
factor $\{\log(en)\}^{\alpha_q}$. At $q=6$, for instance, the
lower rate is $Bg_\Sigma n^{-1/3}L^{5/6}$.

Growing envelopes can change this comparison: $\mathcal K_q$ may
diverge even when $q>4$. For every fixed $q>2$, the same ratio
calculation shows that~\eqref{poly:bennett-bound} has order
$(g_\Sigma/\sqrt L)V_q/\mathcal L_q^{1-1/q}$ whenever
$\mathcal K_q\to\infty$.
Theorem~\ref{lower:sparse} matches this rate if also
$\log\mathcal K_q=o(L)$ and the intrinsic rate tends to zero.

\section{Sharpness of the bounds}\label{opt:section}

We now ask which factors in the upper bounds are necessary. We
construct observations attaining the lower rates in
Table~\ref{opt:tab:rates}, including their dependence on Gaussian
width. Some examples match the corresponding upper bounds, while
others leave the logarithmic multipliers displayed in the table.
The distribution and dimension may depend on the sample size.
For independent centered observations
$X_{n,1},\ldots,X_{n,n}\in\R^{p_n}$, define
\begin{equation}
\begin{gathered}
 W_n:=\frac1{\sqrt n}\sum_{i=1}^nX_{n,i},\qquad
 \Sigma_n:=\frac1n\sum_{i=1}^n\E X_{n,i}X_{n,i}^{\top},\\
 Z_n\sim N(0,\Sigma_n),\qquad
 L_n:=\log(ep_n),\qquad g_{\Sigma_n}:=\E\|Z_n\|_\infty.
\end{gathered}
\label{lower:eq:setup}
\end{equation}
Each example below has identically distributed observations with unit
marginal variances.

There are two mechanisms. With independent skewed coordinates,
small scalar tail errors accumulate in the distribution of the
maximum. This example has $\Sigma_n=I_{p_n}$ and hence
$g_{\Sigma_n}\asymp\sqrt{L_n}$.
The symmetric examples use a different construction: each maximum
is a leading component plus an independent random shift. A difference
between the mean shifts for the sum and its Gaussian counterpart
creates a probability discrepancy near a suitable threshold of the
leading maximum. Changing the amplitude and frequency of the rare
component, and sometimes selecting only a subset of its coordinates,
adapts this mechanism to the different moment assumptions.

To vary Gaussian width separately from dimension, the symmetric
construction pairs every leading coordinate with every rare-component
coordinate. Denote their respective counts by $r_n$ and $m_n$, and let
\begin{equation}
 1\le r_n\le m_n,\qquad m_n\to\infty,\qquad p_n:=r_nm_n,
 \qquad w_n:=\E\max_{j\le r_n}|G_j|,
 \label{lower:eq:dimensions}
\end{equation}
where $r_n,m_n$ are integers and the $G_j$ are independent standard
normal variables. The number $r_n$ controls the width and boundary
density of the leading maximum, while $m_n$ controls the maximum
of the rare components. The prescribed width $w_n$ lets us state
the growth conditions before constructing the observations.
Each symmetric construction will satisfy
\begin{equation}
 g_{\Sigma_n}=(1+o(1))w_n,\qquad
 w_n\asymp\sqrt{\log(er_n)}.
 \label{lower:eq:width-profile}
\end{equation}
Thus $r_n=1$ gives $g_{\Sigma_n}\to\sqrt{2/\pi}$, whereas
$r_n=m_n$ gives $g_{\Sigma_n}\asymp\sqrt{L_n}$.
The lower bounds retain logarithmic dependence on dimension even
when width stays bounded. They do not settle the necessity of the
additional logarithmic multipliers in Table~\ref{opt:tab:rates}.
These are existence results for the stated width regimes: we do
not assert a matching lower bound for every covariance matrix
with the same Gaussian width.

Appendix~\ref{app:sharpness-proofs} proves the scalar tail estimate
used for the independent-coordinate example and develops a common
shift comparison for the symmetric examples. The moment conditions
determine the parameter choices within that common construction.

\begin{table}[!htb]
\centering
\caption{Lower-bound rates and upper-bound multipliers for Gaussian
approximation over rectangles. Multiplying the rate in the second
column by the factor in the third gives an upper bound. The lower
bounds retain the Gaussian width of the constructed covariance.}
\label{opt:tab:rates}
\vspace{3pt}
\small
\setlength{\tabcolsep}{3pt}
\begin{tabular*}{\linewidth}{@{\extracolsep{\fill}}lcc@{}}
\toprule
Observation class and regime & Lower-bound rate & Upper-bound multiplier\\
\midrule
\shortstack[l]{\(\norm{X_{n,i}}_\infty\le B_n\)\\\(B_n\ge2\) may grow\\Theorem~\ref{opt:symmetric-lower}}
& \(\displaystyle B_n g_{\Sigma_n}L_n/\sqrt n\)
& \(\displaystyle C\sqrt{1+\log_+\frac{\sqrt n}{B_n\sqrt{L_n}}}\)\\
\midrule
\multicolumn{3}{@{}l@{}}{Envelope condition~\eqref{poly:envelope} alone: Theorem~\ref{thm:polynomial}\textup{(i)}}\\[3pt]
\shortstack[l]{\(q>2\), \(B_n\ge2\) may vary\\Theorem~\ref{lower:dense}\textup{(i)}}
& \(\displaystyle B_n g_{\Sigma_n}n^{-\alpha_q}L_n^{1-1/q}\)
& \(C_q\{\log(en)\}^{\alpha_q}\)\\
\midrule
\multicolumn{3}{@{}l@{}}{Envelope and coordinate conditions~\eqref{poly:envelope}--\eqref{poly:fourth}: Theorem~\ref{thm:polynomial}}\\[3pt]
\shortstack[l]{\(2<q<4\), fixed \(B\ge2\)\\\(\log n=o(L_n)\)\\Theorem~\ref{lower:sparse}}
& \(\displaystyle
 \frac{g_{\Sigma_n}n^{-\alpha_q}L_n^{1-1/q}}
 {\{\log(e+n/L_n)\}^{1-1/q}}\)
& \(C_{q,B}\)\\
\addlinespace[5pt]
\shortstack[l]{\(q=4\), fixed \(B\ge2\)\\Theorem~\ref{opt:q4-lower}}
& \(\displaystyle g_{\Sigma_n}n^{-1/4}L_n^{3/4}\)
& \(C_B\)\\
\addlinespace[5pt]
\shortstack[l]{\(q=4\), \(B_n\to\infty\)\\\(\log B_n=o(L_n)\)\\Theorem~\ref{opt:q4-growing}}
& \(\displaystyle
 \frac{B_n g_{\Sigma_n}n^{-1/4}L_n^{3/4}}
 {\{\log(e+B_n)\}^{3/4}}\)
& \(C\)\\
\addlinespace[5pt]
\shortstack[l]{\(q>4\), fixed \(B\ge2\)\\Theorem~\ref{lower:dense}\textup{(ii)}}
& \(\displaystyle B g_{\Sigma_n}n^{-\alpha_q}L_n^{1-1/q}\)
& \(C_q\{\log(en)\}^{\alpha_q}\)\\
\addlinespace[5pt]
\shortstack[l]{\(q>2\), \(B_n\ge2\) may vary\\Large \(\mathcal K_{n,q}\)\\Theorem~\ref{lower:sparse}}
& \(\displaystyle
 \frac{B_n g_{\Sigma_n}n^{-\alpha_q}L_n^{1-1/q}}
 {\{\log(e+\mathcal K_{n,q})\}^{1-1/q}}\)
& \(C_q\)\\
\bottomrule
\end{tabular*}
\par\medskip
\begin{minipage}{\linewidth}\small
Here \(\alpha_q=1/2-1/q\), and each displayed lower rate tends
to zero. The symmetric constructions use~\eqref{lower:eq:dimensions}
and satisfy~\eqref{lower:eq:width-profile}. The bounded rate also
holds with independent coordinates in Theorem~\ref{opt:skew-lower},
where \(g_{\Sigma_n}\asymp\sqrt{L_n}\). In the polynomial entries,
\(q\) is fixed, and constants may depend on \(q\) or a fixed
\(B\) as indicated. All upper bounds can be capped at \(1\).
The last entry assumes
\(\mathcal K_{n,q}:=B_n(n/L_n)^{2/q-1/2}\to\infty\) and
\(\log\mathcal K_{n,q}=o(L_n)\).
\end{minipage}
\end{table}

\subsection{Bounded observations}

If $\|X_{n,i}\|_\infty\le B_n$, then the normalized summands
have envelope $\delta_n:=B_n/\sqrt n$.
The envelope bound~\eqref{eq:intrinsic-envelope} therefore has scale
$B_ng_{\Sigma_n}L_n/\sqrt n$, multiplied by the additional
square-root logarithm in Table~\ref{opt:tab:rates}.
The first example attains this scale with independent coordinates.
Its skewed two-point law extends the construction
in~\cite[Proposition~2.1]{CCK} and makes the contribution of the
third moment explicit.

\begin{samepage}
\begin{theorem}[Bounded observations with independent coordinates]\label{opt:skew-lower}
Let \(p_n\to\infty\) be integers and \(B_n\ge2\). Set
\(\Sigma_n:=I_{p_n}\), and suppose that
\[
 e_n:=\frac{B_n g_{\Sigma_n}L_n}{\sqrt n}\longrightarrow0.
\]
There are i.i.d. centered observations with independent coordinates,
covariance \(\Sigma_n\), and \(\norm{X_{n,i}}_\infty\le B_n\)
almost surely, such that
\begin{equation}
 \liminf_{n\to\infty}\frac{\rh(W_n,Z_n)}{e_n}
 \ge\frac1{4e}.
 \label{opt:lower-constant}
\end{equation}
The envelope \(B_n\) is attained with positive probability.
\end{theorem}
\end{samepage}

The proof chooses a threshold where the Gaussian maximum has a
nondegenerate probability. At that threshold, each coordinate has
an exceedance probability of order $1/p_n$, and skewness produces
a relative error in its scalar tail. Multiplying the coordinate
probabilities turns these errors into the displayed discrepancy;
Subsection~\ref{app:skew-proof} gives the calculation.

The same rate can occur when every observation is centrally symmetric,
so that all mixed third moments vanish. For this result, we use the
common leading-component and rare-shift construction with a bounded
amplitude. Its two coordinate counts also allow the Gaussian width
to stay bounded.

\begin{theorem}[Centrally symmetric lower bound]\label{opt:symmetric-lower}
Let \(r_n,m_n,p_n,w_n\) be as in~\eqref{lower:eq:dimensions},
let \(B_n\ge2\), and suppose that
\[
 \frac{B_n w_nL_n}{\sqrt n}\longrightarrow0.
\]
For all sufficiently large \(n\), there are i.i.d. centered,
centrally symmetric observations with unit marginal variances,
\(\norm{X_{n,i}}_\infty\le B_n\), and width satisfying
\eqref{lower:eq:width-profile}, such that
\begin{equation}
 \rh(W_n,Z_n)\ge c\frac{B_n g_{\Sigma_n}L_n}{\sqrt n},
 \label{lower:symmetric-rate}
\end{equation}
where \(c>0\) is an absolute constant.
\end{theorem}

For $r_n=1$, the width converges to $\sqrt{2/\pi}$, and
\eqref{lower:symmetric-rate} still contains the factor $L_n$.
For comparison, if all coordinates of each observation coincide,
the width is exactly $\sqrt{2/\pi}$, and scalar Berry--Esseen
gives an error at most $CB_n/\sqrt n$. Thus bounded width
permits both scalar behavior and the larger errors constructed here.
The bounded symmetric calibration is proved in
Subsection~\ref{app:symmetric-lower}.

The symmetric covariance matrices have rank $r_n+m_n-1$.
They are singular when $r_n\ge2$; duplicating a coordinate
makes them singular also when $r_n=1$, without changing the
width, rectangle distance, or moment bounds.

\subsection{Fourth-moment envelopes}\label{opt:q4-comparison}

We next replace the almost-sure bound by moment conditions. For the
i.i.d. observations constructed here, membership in $\mathcal C_q(B,p)$
requires unit marginal variances and
\[
 \E\norm{X_{n,1}}_\infty^q\le B^q,\qquad
 \max_{j\le p}\E X_{n,1,j}^4\le B^2.
\]
At $q=4$, we keep the activation probability used for bounded
observations and increase the amplitude of the rare component.
Choosing its fourth power to have the same order as the reciprocal
activation probability keeps both fourth-moment bounds fixed.
This change yields a lower bound matching~\eqref{poly:q4}, including
its Gaussian-width factor.

\begin{theorem}[Fourth-moment lower bound]\label{opt:q4-lower}
Let \(r_n,m_n,p_n,w_n\) be as in~\eqref{lower:eq:dimensions},
and suppose that
\[
 w_n n^{-1/4}L_n^{3/4}\longrightarrow0.
\]
For all sufficiently large \(n\), there are i.i.d. centrally
symmetric observations in \(\mathcal C_4(2,p_n)\), with width
satisfying~\eqref{lower:eq:width-profile}, such that
\begin{equation}
 \rh(W_n,Z_n)\ge c g_{\Sigma_n}n^{-1/4}L_n^{3/4},
 \label{opt:eq:q4-lower}
\end{equation}
where \(c>0\) is absolute. For these observations, the lower bound
and~\eqref{poly:q4} give
\begin{equation}
 \rh(W_n,Z_n)\asymp g_{\Sigma_n}n^{-1/4}L_n^{3/4}.
 \label{lower:q4-fixed}
\end{equation}
\end{theorem}

The examples belong to $\mathcal C_4(B,p_n)$ for every fixed
$B\ge2$. They confirm the optimality of the quartic sample-size
rate conjectured for~\cite[Theorem~2.5]{CCKK2022} and show that
the logarithmic factor $L_n^{3/4}$ remains necessary when width
is bounded. Taking $r_n=m_n$ recovers the worst-case dimension
rate.

For growing $B_n$, increasing the amplitude alone can violate the
separate coordinate fourth-moment condition. We therefore add
\emph{coordinate masks}: each activated observation affects only a random
subset of the rare-component coordinates. Its maximum can remain
large while any fixed coordinate receives the large value less
often. The next calibration matches the dependence on $B_n$ in
\eqref{poly:q4}.

\begin{theorem}[Growing fourth-moment envelope]\label{opt:q4-growing}
Let \(r_n,m_n,p_n,w_n\) be as in~\eqref{lower:eq:dimensions},
and let \(B_n\ge2\) tend to infinity. Suppose that
\[
 \log B_n=o(L_n),\qquad
 \frac{B_n w_n n^{-1/4}L_n^{3/4}}
 {\{\log(e+B_n)\}^{3/4}}\longrightarrow0.
\]
For all sufficiently large \(n\), there are i.i.d. centrally
symmetric observations in \(\mathcal C_4(B_n,p_n)\), with width
satisfying~\eqref{lower:eq:width-profile}, such that
\begin{equation}
 \rh(W_n,Z_n)\ge
 c\frac{B_n g_{\Sigma_n}n^{-1/4}L_n^{3/4}}
 {\{\log(e+B_n)\}^{3/4}},
 \label{opt:eq:q4-growing}
\end{equation}
where \(c>0\) is absolute.
\end{theorem}

The upper bound~\eqref{poly:q4} and this lower bound agree in both
the growing-envelope factor and Gaussian width. The fixed and
growing-envelope results use the same shift comparison, with their
parameter choices verified in Subsections~\ref{app:symmetric-lower}
and~\ref{app:q4-growing}, respectively.

\subsection{Other polynomial-moment regimes}

The same rare-component construction also gives lower bounds for
every fixed $q>2$. Without coordinate masks, we choose the amplitude
to use the available $q$th-moment envelope. The resulting rate matches
Theorem~\ref{thm:polynomial}\textup{(i)} up to the factor
$\{\log(en)\}^{1/2-1/q}$.
When $q>4$ and $B$ is fixed, the rare component's contribution to
each coordinate fourth moment tends to zero, so the construction
also satisfies the additional coordinate condition.
Recall $\mathcal E_q,\mathcal C_q$, and $\alpha_q=1/2-1/q$
from Section~\ref{sec:poly}.

\begin{theorem}[Polynomial-envelope lower bounds]\label{lower:dense}
Fix \(q>2\), let \(B_n\ge2\), and let
\(r_n,m_n,p_n,w_n\) be as in~\eqref{lower:eq:dimensions}.
Suppose that
\[
 V_{n,q}:=B_n w_n n^{-\alpha_q}L_n^{1-1/q}\longrightarrow0.
\]
There is a constant \(c_q>0\), depending only on \(q\), such that:

\noindent\textup{(i)} For all sufficiently large \(n\), there are
i.i.d. centrally symmetric observations in \(\mathcal E_q(B_n,p_n)\),
with width satisfying~\eqref{lower:eq:width-profile}, for which
\begin{equation}
 \rh(W_n,Z_n)\ge c_q B_n g_{\Sigma_n}n^{-\alpha_q}L_n^{1-1/q}.
 \label{lower:dense-rate}
\end{equation}
\noindent\textup{(ii)} If \(q>4\) and \(B_n=B\ge2\) is fixed,
the observations in part~\textup{(i)} also belong to
\(\mathcal C_q(B,p_n)\) for all sufficiently large \(n\).
\end{theorem}

The upper bound in Theorem~\ref{thm:polynomial}\textup{(i)} is at
most $C_q\{\log(en)\}^{\alpha_q}$ times the rate
in~\eqref{lower:dense-rate}. This is the remaining gap for the
envelope class, and also for the class with both moment conditions
when $q>4$ and $B$ is fixed.

To match the Bennett bound when its envelope term dominates, we
again use coordinate masks to control the fourth moments. The
ratio $\mathcal K_{n,q}$ below is the sample-size-dependent version
of $\mathcal K_q$ in \eqref{poly:bennett-parameters}.

\begin{theorem}[Sparse polynomial-moment lower bound]\label{lower:sparse}
Fix \(q>2\), let \(B_n\ge2\), and let
\(r_n,m_n,p_n,w_n\) be as in~\eqref{lower:eq:dimensions}.
Define
\[
 \mathcal K_{n,q}:=B_n(n/L_n)^{2/q-1/2},\qquad
 R_{n,q}:=\frac{B_n w_n n^{-\alpha_q}L_n^{1-1/q}}
 {\{\log(e+\mathcal K_{n,q})\}^{1-1/q}},
\]
and suppose that
\begin{equation}
 \mathcal K_{n,q}\to\infty,\qquad
 \log\mathcal K_{n,q}=o(L_n),\qquad R_{n,q}\to0.
 \label{lower:sparse-regime}
\end{equation}
For all sufficiently large \(n\), there are i.i.d. centrally symmetric
observations in \(\mathcal C_q(B_n,p_n)\), with width
satisfying~\eqref{lower:eq:width-profile}, for which
\begin{equation}
 \rh(W_n,Z_n)\ge c_q
 \frac{B_n g_{\Sigma_n}n^{-\alpha_q}L_n^{1-1/q}}
 {\{\log(e+\mathcal K_{n,q})\}^{1-1/q}},
 \label{lower:sparse-rate}
\end{equation}
where \(c_q>0\) depends only on \(q\). The upper bound in
Theorem~\ref{thm:polynomial}\textup{(ii)} is at most a constant
depending only on \(q\) times this rate.
\end{theorem}

The three conditions in~\eqref{lower:sparse-regime} have distinct
roles. The divergence of $\mathcal K_{n,q}$ makes the envelope
term dominate the coordinate fourth-moment term in the upper bound.
The condition $\log\mathcal K_{n,q}=o(L_n)$ lets the mean number
of activated observations in the masked construction diverge, as
verified in Subsection~\ref{app:q4-growing}. Finally, $R_{n,q}\to0$
places the comparison in a regime of vanishing approximation error.

For fixed $B$ and $2<q<4$, the dimension condition holds whenever
$\log n=o(L_n)$. At $q=4$, the assumptions reduce to those of
Theorem~\ref{opt:q4-growing}. For $q>4$, the theorem requires
$B_n$ to grow sufficiently quickly that $\mathcal K_{n,q}\to\infty$.
The sharp rate for intermediate values of $\mathcal K_{n,q}$,
or dimensions outside these conditions, remains unresolved here.

In the symmetric results, $w_n$ can range from constant order to
order $\sqrt{L_n}$ through~\eqref{lower:eq:dimensions}.
The upper bounds hold for all arrays in the stated moment class
whose Gaussian width is at most a fixed multiple of $w_n$, and
the constructed examples attain the displayed lower scales.
Taking $r_n=m_n$ gives the worst-case comparisons over the full
moment classes, while $r_n=1$ gives their bounded-width counterparts.

\section{Remaining questions}\label{sec:discussion}

For bounded observations with unrestricted covariance, our upper and
lower bounds differ by an additional logarithmic factor. We conjecture
that this factor can be removed while retaining the Gaussian width
in the upper bound.
\begin{conjecture}[Sharp rectangle approximation under unrestricted covariance]
\label{conj:sharp-unrestricted}
Under~\eqref{eq:setup}, there is an absolute constant \(C\) such that
\[
 \rh(W,Z)\le C\min\{1,\delta g_\Sigma L\},\qquad L=\log(ep),
\]
where $g_\Sigma=\E\|Z\|_\infty$. The bound holds uniformly in
$n,p,\delta$ and $\Sigma$, including singular $\Sigma$.
\end{conjecture}

Theorem~\ref{opt:symmetric-lower} constructs centrally symmetric
observations attaining the scale $\delta g_\Sigma L$ in its stated
Gaussian-width regimes. These examples establish sharpness across
those regimes, with widths ranging from a constant to order $\sqrt L$.
Using $g_\Sigma\le C\sqrt L$, the conjecture also implies the
dimension-based bound $C\min\{1,\delta L^{3/2}\}$.

Theorem~\ref{thm:unrestricted} proves the conjectured rate for arrays with ${s_4D_L\le C\delta^2}$.

{
The conjectured rate holds when distinct coordinates are uniformly
separated from perfect positive or negative correlation. This condition
permits singular covariance matrices. The following corollary makes the
dependence on the correlation gap explicit while retaining the Gaussian
width $g_\Sigma$. The Supplemental Material~\cite[Appendix~H]{Supplement2026}
gives a proof using the smoothing
and interpolation estimates of Section~\ref{sec:smoothing}.

\begin{samepage}
\begin{corollary}[Approximation under a positive correlation gap]
\label{cor:positive-correlation-gap}
Under~\eqref{eq:setup}, define
\[
 c\coloneqq 1-\max_{j\ne k}|\Sigma_{jk}|,
\]
with \(c\coloneqq 1\) when \(p=1\). If \(c>0\), then
\[
 \rh(W,Z)\le C\min\left\{1,\,
 \delta g_\Sigma L\sqrt{1+\log(1/c)}\right\},
 \qquad L=\log(ep),
\]
where \(C\) is an absolute constant and $g_\Sigma=\E\|Z\|_\infty$.
\end{corollary}
\end{samepage}

The proof combines the fourth-order bound in
Proposition~\ref{an:prop:derivative} with an additional estimate that
retains the angles between pairs of faces. Taking the smaller bound
before integration gives the factor \(1+\log(1/c)\). The intrinsic
boundary coefficient is retained in both estimates, yielding the factor
$g_\Sigma L$ in the final bound. Since $g_\Sigma\le C\sqrt L$, the
dimension-based rate $C\delta L^{3/2}\sqrt{1+\log(1/c)}$ follows as a
consequence. The companion paper~\cite{CorrelationGap} develops further
moment-sensitive refinements under the same pairwise condition.

Consequently, for every fixed \(c_0>0\), the conjectured rate holds
uniformly over covariance matrices satisfying \(c\ge c_0\), including
singular matrices. The remaining question is whether the additional
factor can be removed uniformly as \(c\) approaches zero, including at
\(c=0\).
}

The uniform gap condition can be relaxed by measuring the total size of
the gaps below a chosen cutoff. The signed pair contribution vanishes
at perfect correlation, so a pair very close to duplication can be
less costly than a moderately separated pair. The following bound
retains this cancellation and the Gaussian-width factor.

\begin{corollary}[An exceptional correlation budget]
\label{cor:correlation-budget}
Under~\eqref{eq:setup}, fix a cutoff $c\in(0,1]$ and define
\[
 \eta_{jk}:=1-|\Sigma_{jk}|,\qquad
 B_c:=\sum_{\substack{1\le j<k\le p\\0<\eta_{jk}<c}}\sqrt{\eta_{jk}},
 \qquad J_c:=1+\log(1/c).
\]
The empty sum is zero. There is an absolute constant $C$, independent
of the cutoff, such that
\begin{equation}
 \rh(W,Z)\le C\min\left\{1,\,
      \delta g_\Sigma L\sqrt{J_c}
      +\delta^{2/3}(g_\Sigma L)^{2/3}B_c^{1/3}\right\}.
 \label{budget:eq:result}
\end{equation}
\end{corollary}

We call $B_c$ the \emph{exceptional correlation budget}. Each gap
below $c$ contributes its square root, and perfectly correlated pairs
contribute zero. The first term in~\eqref{budget:eq:result} controls
the remaining pairs. The second term measures the cost of the
exceptional pairs after choosing the smoothing width. Appendix~I of the
Supplemental Material~\cite{Supplement2026} proves the corollary using
the pair estimate from its Appendix~G and the intrinsic matrix estimates
from its Appendix~H.

For fixed $c>0$, the bound has order $\delta g_\Sigma L$ whenever
$B_c\le C_0\delta g_\Sigma L$, with the constant allowed to depend
on $c$ and $C_0$. More generally, the exceptional term is absorbed
by the first term when $B_c\le C_0\delta g_\Sigma LJ_c^{3/2}$.
These conditions constrain the total budget rather than the number
of exceptional pairs: $K$ gaps at most $a$ contribute at most
$K\sqrt a$. Using $g_\Sigma\le C\sqrt L$ recovers the dimension-based
bound $C\{\delta L^{3/2}\sqrt{J_c}+\delta^{2/3}LB_c^{1/3}\}$.
The corollary permits zero minimum gap, but it does not settle
Conjecture~\ref{conj:sharp-unrestricted}: many small positive gaps
can still have a large total budget.

A bound on the cumulative comparison error without the extra logarithm for all bounded arrays would give the full conjecture
and sharpen the polynomial transfer in Subsection~\ref{poly:transfer-sec}.

\clearpage
\appendix
\numberwithin{equation}{section}
\numberwithin{theorem}{section}
\numberwithin{lemma}{section}
\numberwithin{proposition}{section}
\numberwithin{corollary}{section}
\numberwithin{remark}{section}

\section{Proof of Lemma~\ref*{lem:endpoint} and logistic derivative estimates}\label{appx-sec:smoothing}
We first bound the error from replacing rectangle indicators by the
logistic approximation at the original endpoints. We then derive
the derivative formulas used along the comparison path and show how
boundary concentration improves their bounds after averaging.

\begin{proof}[Proof of Lemma~\ref{lem:endpoint}]
Fix a threshold vector $a\in\R^m$, and let $V$ denote either $S$ or
$G$. Recall that $m=2p$, $L=\log(ep)$, and $\kappa$ is defined
in~\eqref{eq:intrinsic-boundary-coefficient}. The exponential variables
in~\eqref{eq:mixture} are independent of $V$. To contain all their
threshold displacements within a single boundary layer, define
\[
 R_h:=h\max_{\alpha\le m}|\log(\omega_\alpha/\omega_0)|.
\]
For every realization of these variables, each threshold of
$A_{a,\omega}$ lies between $a_\alpha-R_h$ and $a_\alpha+R_h$.
Thus both $A_a$ and $A_{a,\omega}$ contain $A_a^{-R_h}$ and are
contained in $A_a^{R_h}$, so
\[
 |\ind_{A_a}(x)-\ind_{A_{a,\omega}}(x)|
 \le\ind_{A_a^{R_h}\setminus A_a^{-R_h}}(x),
 \qquad x\in\R^r.
\]
Condition on the exponential variables. The law of $V$ is unchanged
by independence, and the thresholds and $R_h$ are now fixed.
Lemma~\ref{lem:concentration}, with the coefficient $\kappa$ above,
therefore gives
\[
 \begin{aligned}
 &\E\bigl[
   |\ind_{A_a}(V)-\ind_{A_{a,\omega}}(V)|\mid\omega\bigr]\\
 &\hspace{2em}\le
 \Pp(V\in A_a^{R_h}\setminus A_a^{-R_h}\mid\omega)
 \le C(R_h+\delta)\kappa.
 \end{aligned}
\]
The mixture representation gives
$\E[\ind_{A_{a,\omega}}(V)\mid V]=F_{a,h}(V)$.
Taking expectations and using $|\E X|\le\E|X|$ now yields
\[
 |\Pp(V\in A_a)-\E F_{a,h}(V)|
 \le C(\E R_h+\delta)\kappa.
\]

It remains to bound the expected largest displacement. For $s\ge0$,
conditioning on $\omega_0$ and using the exponential tail gives
\[
 \Pp(\omega_\alpha>e^s\omega_0)
 =\int_0^\infty e^{-e^s t}e^{-t}\,dt
 =\frac1{1+e^s}.
\]
Exchanging $\omega_\alpha$ and $\omega_0$ gives the same probability for
the opposite logarithmic tail. Consequently,
\[
 \Pp\{|\log(\omega_\alpha/\omega_0)|>s\}
 =\frac2{1+e^s}\le2e^{-s}.
\]
Although the ratios share the denominator $\omega_0$, the union bound
does not require their independence. It gives
$\Pp(R_h>hs)\le\min\{1,2me^{-s}\}$, and integration of this tail yields
\[
 \E R_h
 \le h\int_0^\infty\min\{1,2me^{-s}\}\,ds
 =h\{\log(2m)+1\}\le ChL.
\]
Since $h\ge\delta$ and $L\ge1$, the endpoint error is therefore
\[
 |\Pp(V\in A_a)-\E F_{a,h}(V)|
 \le C(hL+\delta)\kappa\le Ch\kappa L.
\]
Apply this bound to $S$ and $G$ and use the triangle inequality:
\[
 |\Pp(S\in A_a)-\Pp(G\in A_a)|
 \le|\E F_{a,h}(S)-\E F_{a,h}(G)|+Ch\kappa L.
\]
Taking the supremum over $a$ proves~\eqref{eq:endpoint-min}.
\end{proof}

\subsection{\texorpdfstring{Mixture and derivative formulas}{Mixture and derivative formulas}}\label{appx-sec:mixture}

We now express each derivative as a sum of products of face
probabilities and directional projections. This form permits us to
sum the probability weights while retaining the factor $F(1-F)$,
whose expectation will be controlled by boundary concentration.

Fix $a\in\R^m$ and $h>0$, and abbreviate $F:=F_{a,h}$.
For $x\in\R^r$, define the unnormalized face weights and their sum by
\[
 w_\alpha(x):=e^{(u_\alpha\cdot x-a_\alpha)/h},
 \qquad T(x):=\sum_{\alpha=1}^m w_\alpha(x).
\]
Then $F=(1+T)^{-1}$, and the face probabilities
$\pi_\alpha=Fw_\alpha$ have total mass
\[
 \theta:=\sum_{\alpha=1}^m\pi_\alpha=1-F.
\]
We call $\theta$ the \emph{face mass}. All formulas below hold for
arbitrary finite thresholds, even when $A_a$ is empty. Deleting
an inequality is obtained by letting its threshold tend to infinity:
its weight tends to zero, and the finite sums below converge to the
corresponding formulas for the remaining faces.

For completeness, we first verify the mixture
representation~\eqref{eq:mixture} used in the smoothing proof.
Let $\omega_0,\ldots,\omega_m$ be independent exponential variables
with mean $1$, independent of every random vector at which $F$ is
evaluated. At a fixed $x$, the defining inequalities for the shifted
set are equivalent to
\[
 \begin{aligned}
 x\in A_{a,\omega}
 &\quad\Longleftrightarrow\quad
 u_\alpha\cdot x\le
 a_\alpha+h\log(\omega_\alpha/\omega_0)
       \quad\text{for every }\alpha\\
 &\quad\Longleftrightarrow\quad
 \omega_\alpha\ge\omega_0w_\alpha(x)
       \quad\text{for every }\alpha.
 \end{aligned}
\]
Conditional on $\omega_0$, independence of the remaining exponentials
therefore gives
\[
 \Pp_\omega(x\in A_{a,\omega}\mid\omega_0)
 =\prod_{\alpha=1}^m e^{-\omega_0w_\alpha(x)}
 =e^{-\omega_0T(x)}.
\]
Integrating over $\omega_0$ proves
\[
 \E_\omega\ind_{A_{a,\omega}}(x)
 =\int_0^\infty e^{-tT(x)}e^{-t}\,dt
 =\frac1{1+T(x)}=F(x).
\]
When $x$ is replaced by an independent random vector, conditioning
on that vector gives the same identity pointwise. All subsequent
averages of these indicators are justified by their boundedness.

Write $D^qf(x)[z_1,\ldots,z_q]$ for the $q$th derivative applied
to the directions $z_1,\ldots,z_q\in\R^r$. The first derivative
illustrates how the face probabilities enter:
\[
 \begin{aligned}
 DT(x)[z]&=h^{-1}\sum_\alpha
              w_\alpha(x)(u_\alpha\cdot z),\\
 DF(x)[z]&=-F(x)^2DT(x)[z]
          =-h^{-1}F(x)\sum_\alpha
              \pi_\alpha(x)(u_\alpha\cdot z).
 \end{aligned}
\]
In particular, $|DF(x)[z]|\le h^{-1}F(x)\theta(x)\norm z_U$,
where $\norm z_U=\max_\alpha|u_\alpha\cdot z|$ is the face norm.

For higher derivatives, every nonempty set $J$ of direction indices
satisfies
\[
 D^{|J|}T(x)[z_j:j\in J]
 =h^{-|J|}\sum_\alpha w_\alpha(x)
                      \prod_{j\in J}(u_\alpha\cdot z_j).
\]
Apply Fa\`a di Bruno's higher-order chain rule~\cite{Comtet1974} to
$F=f(T)$, where $f(t):=(1+t)^{-1}$ and
$f^{(k)}(t)=(-1)^k k!(1+t)^{-k-1}$.
Let $\Pi_q$ be the set of partitions of $\{1,\ldots,q\}$.
For each partition $\mathcal P$, choose an ordering
$J_1,\ldots,J_{|\mathcal P|}$ of its blocks. A block groups the
directions entering one derivative of $T$, and the number of blocks
determines the derivative of $f$. Substituting the two derivative
formulas into the chain rule gives
\begin{equation}
 \begin{split}
 D^qF(x)[z_1,\ldots,z_q]
 ={}&h^{-q}\sum_{\mathcal P\in\Pi_q}(-1)^{|\mathcal P|}|\mathcal P|!
 \sum_{\alpha_1,\ldots,\alpha_{|\mathcal P|}}
 F(x)\prod_{b=1}^{|\mathcal P|}\pi_{\alpha_b}(x)\\
 &\hspace{3em}\times
 \prod_{b=1}^{|\mathcal P|}\prod_{j\in J_b}
 (u_{\alpha_b}\cdot z_j).
 \end{split}
 \label{appx-eq:partition-derivative}
\end{equation}
Every face index ranges from $1$ to $m$, and different blocks may
select the same face. The factor $F\prod_b\pi_{\alpha_b}$ arises
because $F^{k+1}\prod_b w_{\alpha_b}=F\prod_b\pi_{\alpha_b}$
when $k=|\mathcal P|$.
See also~\cite[Lemma~13]{GaoZhang} for these derivative identities.

We next control the derivatives at nearby points, as required by
the Taylor remainders along the path. If $\norm v_U\le\varrho$, then
\[
 e^{-\varrho/h}\le
 \frac{w_\alpha(x+v)}{w_\alpha(x)}
 \le e^{\varrho/h},
 \qquad
 e^{-\varrho/h}\le
 \frac{1+T(x+v)}{1+T(x)}
 \le e^{\varrho/h}.
\]
The second comparison follows by summing the first and observing
that the constant term $1$ obeys the same inequalities.
It follows that $F(x+v)\le e^{\varrho/h}F(x)$ and
$\pi_\alpha(x+v)\le e^{2\varrho/h}\pi_\alpha(x)$.
Thus every positive weight product in
\eqref{appx-eq:partition-derivative}, with $k=|\mathcal P|\le q$,
satisfies
\[
 F(x+v)\prod_{b=1}^k\pi_{\alpha_b}(x+v)
 \le e^{(2k+1)\varrho/h}
        F(x)\prod_{b=1}^k\pi_{\alpha_b}(x).
\]
For $0\le\varrho\le h$, this multiplier is at most $e^{2q+1}$.
Each direction appears exactly once in the remaining projection
product, so its absolute value is at most
$\prod_{j=1}^q\norm{z_j}_U$. Finally,
\[
 \sum_{\alpha_1,\ldots,\alpha_k}
     F(x)\prod_{b=1}^k\pi_{\alpha_b}(x)
 =F(x)\theta(x)^k\le F(x)\theta(x),
 \qquad k\ge1.
\]
Taking absolute values in~\eqref{appx-eq:partition-derivative}
and summing over its finitely many partitions therefore proves,
for each fixed positive integer $q$,
\begin{equation}
 \sup_{\norm v_U\le\varrho}|D^qF(x+v)[z_1,\ldots,z_q]|
 \le C_qh^{-q}F(x)(1-F(x))\prod_{j=1}^q\norm{z_j}_U,
 \quad 0\le\varrho\le h.
 \label{appx-eq:local-envelope}
\end{equation}
Here $C_q$ depends only on $q$. The preceding comparison also holds
separately for each positive weight product, before the directional
projections are bounded. We use that form when random directions
must be summed through their covariance or higher moments.

To average~\eqref{appx-eq:local-envelope}, we need only a bound on
the probability of lying in a boundary layer. Such a bound for an
independent component suffices. Indeed, if $U=V+R$ with $V$ and $R$
independent, conditioning on $R$ replaces each threshold $a_\alpha$
by $a_\alpha-u_\alpha\cdot R$ while leaving the layer width unchanged.
A boundary bound uniform over thresholds for $V$ therefore also
holds for $U$.

\begin{lemma}[Averaging the derivative envelope]\label{appx-lem:envelope}
Suppose that $U\in\R^r$ satisfies, for all $a\in\R^m$ and $s>0$,
\[
 \Pp\{U\in A_a^s\setminus A_a^{-s}\}
 \le C_0\kappa(s+\delta),
\]
where $\kappa\ge1$ and $C_0$ is an absolute constant.
Uniformly over thresholds, for $h\ge4\delta$,
\begin{equation}
 \E F(U)(1-F(U))\le Ch\kappa.
 \label{appx-eq:small-envelope}
\end{equation}
Consequently, for each fixed positive integer $q$, fixed directions
$z_1,\ldots,z_q\in\R^r$, and $h\ge\max\{4\delta,\varrho\}$,
\begin{equation}
 \E\sup_{\norm v_U\le\varrho}|D^qF(U+v)[z_1,\ldots,z_q]|
 \le C_qh^{1-q}\kappa\prod_{j=1}^q\norm{z_j}_U.
 \label{appx-eq:averaged-derivatives}
\end{equation}
The constants depend only on $C_0$ and, for $C_q$, on $q$.
\end{lemma}

\begin{proof}
We compare $F(1-F)$ with the increase in the smoother when all
thresholds move outward across a layer of width $2h$. At a fixed
$x$, define
\[
 F_s(x):=F_{a+s\mathbf1,h}(x)
       =(1+e^{-s/h}T(x))^{-1},\qquad s\in\R.
\]
Differentiating with respect to the common threshold shift gives
\[
 \partial_sF_s(x)=h^{-1}F_s(x)(1-F_s(x)),
 \qquad
 \partial_s\log\{F_s(x)(1-F_s(x))\}
       =h^{-1}(1-2F_s(x)).
\]
The absolute value of the logarithmic derivative is at most $h^{-1}$.
Integrating this bound between $0$ and $s$ shows that
$F_s(x)(1-F_s(x))\ge e^{-1}F(x)(1-F(x))$ for $|s|\le h$.
A second integration, now of $\partial_sF_s$, yields
\[
 F_h(x)-F_{-h}(x)
 =\frac1h\int_{-h}^h F_s(x)(1-F_s(x))\,ds
 \ge\frac2e F(x)(1-F(x)).
\]
We have therefore proved the pointwise comparison
\begin{equation}\label{appx-eq:F(1-F)}
 F(1-F)\le\frac e2(F_h-F_{-h}).
\end{equation}

Use the same exponential variables for $F_h$ and $F_{-h}$.
The corresponding random sets are nested, so their indicator
difference is the indicator of a layer:
\[
 F_h(x)-F_{-h}(x)
 =\E_\omega\ind_{A_{a,\omega}^h\setminus
                         A_{a,\omega}^{-h}}(x).
\]
Here $A_{a,\omega}^s$ means that $s$ is added to each of the
already shifted thresholds. Evaluate at $U$, taking the exponentials
independent of $U$, and then condition on them. Their realized
thresholds are deterministic, so the boundary assumption gives
\[
 \E[F_h(U)-F_{-h}(U)]
 =\E_\omega\Pp\{U\in A_{a,\omega}^h\setminus
                        A_{a,\omega}^{-h}\mid\omega\}
 \le C_0\kappa(h+\delta).
\]
Combining this inequality with~\eqref{appx-eq:F(1-F)} and
$h\ge4\delta$ proves~\eqref{appx-eq:small-envelope}.
Notice that every pair of corresponding thresholds here differs
by exactly $2h$, regardless of the random displacements.
Thus this estimate uses a layer of fixed width and incurs no factor
$L$ from their maximum.

Finally, the directions in~\eqref{appx-eq:averaged-derivatives}
are fixed, so we may take expectations
in~\eqref{appx-eq:local-envelope} and apply
\eqref{appx-eq:small-envelope}. This replaces the factor $h^{-q}$
by $h^{1-q}\kappa$ and proves the claimed averaged derivative bound.
\end{proof}

\section{Proof of Proposition~\ref*{an:prop:derivative} (comparison along the path)}
\label{appx-sec:support}

We prove the three claims of Proposition~\ref{an:prop:derivative}
in the order in which they are used.
Subsection~\ref{appx-an:sec:initialization} proves the direct
derivative bounds~\eqref{eq:direct-initialization} and the
initialization bound~\eqref{eq:initialization-bound}.
Subsection~\ref{appx-an:sec:restoration} proves the seeded
derivative bounds~\eqref{eq:seeded-derivatives}, using the Gaussian
matrix and threshold estimates in
Subsections~\ref{appx-an:sec:matrix} and~\ref{appx-an:sec:threshold}.
We begin with the covariance and concentration estimates common
to these arguments, then derive the interpolation identity in
Subsection~\ref{appx-an:sec:path}.

\subsection{Covariance and concentration}
\label{appx-an:sec:coords}

The derivative estimates use two properties of the comparison path.
The aggregate covariance controls the sum over pairs of derivative
directions, while concentration near the boundary makes the average
derivative envelope small. We record these properties here, including
the forms needed after conditioning and after retaining only some
summands.

We use the support coordinates of
Subsection~\ref{sec:outline-smoothing}. Thus the vectors
\(\zeta_i^0\in\R^r\) are independent and centered, the face normals
\(u_1,\ldots,u_m\) are unit vectors, and
\begin{equation}
 \sum_i\E\zeta_i^0(\zeta_i^0)^\top=I_r,
 \qquad \norm{\zeta_i^0}_U\le\delta,
 \qquad \norm x_U=\max_{j\le m}|u_j\cdot x|,
 \quad m=2p.
 \label{appx-eq:projected-bound}
\end{equation}
In particular, the total second moment in every unit direction is
$1$. To use this fact when some of the randomness has already been
fixed, let \(\mathcal H\) be a \(\sigma\)-field and let \((Z_i)_i\)
satisfy
\[
 \sum_i\E[Z_iZ_i^\top\mid\mathcal H]\preceq I_r
 \quad\text{almost surely}.
\]
For fixed or \(\mathcal H\)-measurable unit vectors \(u,v\),
conditional Cauchy--Schwarz for each summand, followed by
Cauchy--Schwarz for the sum over \(i\), gives
\begin{equation}
 \begin{aligned}
 \sum_i\E[|(u\cdot Z_i)(v\cdot Z_i)|\mid\mathcal H]
 &\le\sum_i
   \left\{\E[(u\cdot Z_i)^2\mid\mathcal H]\,
          \E[(v\cdot Z_i)^2\mid\mathcal H]\right\}^{1/2}\\
 &\le\left\{\sum_i\E[(u\cdot Z_i)^2\mid\mathcal H]\,
             \sum_i\E[(v\cdot Z_i)^2\mid\mathcal H]\right\}^{1/2}
 \le1.
 \end{aligned}
 \label{appx-an:pair-contraction}
\end{equation}
The last inequality follows by applying the matrix bound to \(u\)
and \(v\) separately. This argument requires neither conditional
centering nor conditional independence of the \(Z_i\).
Taking \(Z_i=\zeta_i^0\) and the trivial \(\sigma\)-field recovers
the unconditional bound used in Section~\ref{sec:smoothing}.

We now explain how the pair estimate gives the factors $\delta$ and
$\delta^2$ in the third- and fourth-order bounds. For an array that
also satisfies the face-envelope bound $\norm{Z_i}_U\le\delta$
almost surely, each additional face component can be bounded by
$\delta$. Thus, for any face indices $\alpha_1,\ldots,\alpha_4$,
\[
 \begin{aligned}
 \sum_i\E\!\left[
   \left|\prod_{k=1}^3(u_{\alpha_k}\cdot Z_i)\right|
   \,\middle|\,\mathcal H\right]
 &\le\delta\sum_i\E[
   |(u_{\alpha_1}\cdot Z_i)(u_{\alpha_2}\cdot Z_i)|
   \mid\mathcal H]\le\delta,\\
 \sum_i\E\!\left[
   \left|\prod_{k=1}^4(u_{\alpha_k}\cdot Z_i)\right|
   \,\middle|\,\mathcal H\right]
 &\le\delta^2\sum_i\E[
   |(u_{\alpha_1}\cdot Z_i)(u_{\alpha_2}\cdot Z_i)|
   \mid\mathcal H]\le\delta^2.
 \end{aligned}
\]
The first inequality in each line bounds the remaining components
by $\delta$, and the second is~\eqref{appx-an:pair-contraction}.
The face indices need not be distinct.

The same second-moment assumption also bounds matrices with signed
coefficients. Let $c_i$ be real random variables with $|c_i|\le b$
almost surely, where $b\ge0$, and define
\[
 M:=\sum_i\E[c_iZ_iZ_i^\top\mid\mathcal H].
\]
For every unit vector $u$, we have
\[
 \begin{aligned}
 |u^\top Mu|
 &=\left|\sum_i\E[c_i(u\cdot Z_i)^2\mid\mathcal H]\right|\\
 &\le b\sum_i\E[(u\cdot Z_i)^2\mid\mathcal H]\le b.
 \end{aligned}
\]
The matrix $M$ is real and symmetric, so its operator norm equals
the supremum of $|u^\top Mu|$ over unit vectors. Consequently,
\[
 \left\|\sum_i\E[c_iZ_iZ_i^\top\mid\mathcal H]\right\|_{\mathrm{op}}
 \le b.
\]
This is the matrix form used when a repeated pair of derivatives
is moved onto the Gaussian density.

We next specify the boundary estimate needed to average the
derivative weights. Recall that
\[
 A_a:=\{x\in\R^r:u_j\cdot x\le a_j,\ j\le m\},
 \qquad A_a^s:=A_{a+s\mathbf1}.
\]
Thus \(A_a^s\) increases every face threshold by \(s\); the layer
\(A_a^s\setminus A_a^{-s}\) records the points whose membership can
change under a common threshold displacement of size \(s\).
Fix an absolute constant \(C_0>0\), large enough for the applications
below. We call a random vector \(V_0\) a \emph{concentrated base},
or simply a \emph{base}, with coefficient \(\kappa\ge1\), if
\begin{equation}
 \Pp\{V_0\in A_a^s\setminus A_a^{-s}\}
 \le C_0(s+\delta)\kappa,
 \qquad a\in\R^m,\quad s>0.
 \label{appx-an:concentration}
\end{equation}
The sets \(A_a\) are allowed to be empty. The estimate also holds
for any subsystem of the face inequalities: send the unused
thresholds to infinity and pass to the limit in the layer
indicators. Moreover, if \(R\) is independent of \(V_0\), then
\(V_0+R\) satisfies the same estimate. To see this, condition on
\(R\) and replace \(a_j\) by \(a_j-u_j\cdot R\). The uniformity
in \(a\) is what permits this random translation.

Lemma~\ref{lem:concentration} supplies such a bound from an independent
Gaussian or bounded-summand component whose common face variance is
bounded below by a positive absolute constant. For a bounded-summand
component, each summand's face norm must also be at most an absolute
constant times \(\delta\). Its intrinsic bound retains the expected
face norm of that component; its dimension-based bound permits
\(\kappa=\sqrt L\). In the application to
Theorem~\ref{thm:unrestricted}, we use the smaller coefficient in
\eqref{eq:intrinsic-boundary-coefficient}. The matrix and threshold
arguments below require only~\eqref{appx-an:concentration} from the
law of \(V_0\).

To pass from boundary probabilities to the smoother, write the
notation from Appendix~\ref{appx-sec:smoothing} as
\begin{equation}
 \label{appx-an:basic-notation}
 \begin{gathered}
 q_j(x)\coloneqq(u_j\cdot x-a_j)/h,\qquad
 N(x)\coloneqq1+\sum_{j=1}^m e^{q_j(x)},\\
 F(x)=F_{a,h}(x)=N(x)^{-1},\qquad
 \pi_j(x)=e^{q_j(x)}/N(x),\qquad \theta(x)=1-F(x).
 \end{gathered}
\end{equation}
Here \(q_j\) is the normalized violation of face \(j\), and
\(F\theta\) is the common envelope in the derivative bounds.
For \(F_s:=F_{a+s\mathbf1,h}\),
\eqref{appx-eq:F(1-F)} gives
\(F\theta\le(e/2)(F_h-F_{-h})\).
Fix $z\in\R^r$, and take the exponential variables in
\eqref{eq:mixture} independent of $V_0$. Using the same variables
for both smoothers, define the shifted thresholds
\[
 \widetilde a_j
 :=a_j-u_j\cdot z+h\log(\omega_j/\omega_0).
\]
The corresponding random sets are nested, so the mixture
representation gives, for every fixed $x\in\R^r$,
\[
 F_h(x+z)-F_{-h}(x+z)
 =\E_\omega\ind_{A_{\widetilde a}^{h}
                    \setminus A_{\widetilde a}^{-h}}(x).
\]
Evaluate this identity at $x=V_0$ and interchange the expectations.
Conditional on $\omega$, the thresholds $\widetilde a$ are fixed
and the law of $V_0$ is unchanged by independence. Hence
\eqref{appx-an:concentration} gives
\[
 \E\{F_h(V_0+z)-F_{-h}(V_0+z)\}
 =\E_\omega\Pp\{V_0\in
       A_{\widetilde a}^{h}\setminus A_{\widetilde a}^{-h}
       \mid\omega\}
 \le C_0(h+\delta)\kappa.
\]
Consequently, for \(h\ge\delta\),
\begin{equation}
 \sup_{z\in\R^r}\E F(V_0+z)\theta(V_0+z)\le Ch\kappa.
 \label{appx-an:weight-concentration}
\end{equation}
Each pair of corresponding random thresholds differs by exactly
$2h$. Thus the layer width is fixed, and this estimate incurs no
factor $L$ from the largest random threshold displacement.
Retaining a single face in the same argument gives
\[
 \sup_{z\in\R^r}\E
 \frac{e^{q_j(V_0+z)}}{(1+e^{q_j(V_0+z)})^2}
 \le Ch\kappa,\qquad j\le m.
\]
These estimates also apply after an independent Gaussian shift,
by conditioning on that shift. They use concentration of the
unconditional law of \(V_0\); no concentration assumption is made
on its law conditional on the original summands or masks.

Finally, the intrinsic coefficient requires control of the expected
face norm after thinning. Let \(I\subseteq\{1,\ldots,n\}\) be fixed.
Independence and centering of the omitted summands give
\(\E[W\mid(\xi_i)_{i\in I}]=\sum_{i\in I}\xi_i\).
Conditional Jensen's inequality for the convex function
\(x\mapsto\|x\|_\infty\) therefore yields
\begin{equation}\label{appx-an:retained-maximum}
 \E\left\|\sum_{i\in I}\xi_i\right\|_\infty
 =\E\left\|\E[W\mid(\xi_i)_{i\in I}]\right\|_\infty
 \le\E\|W\|_\infty.
\end{equation}
If \(I\) is random and independent of the summands, condition first
on \(I\) and apply the same inequality. Thus independent retention
masks do not increase the expected maximum, and multiplying the
retained sum by an amplitude of absolute value at most \(2\)
increases it by at most that factor. Since
\(\|\Gamma x\|_\infty=\|x\|_U\), the identical bound holds in support
coordinates. This supplies the expected-norm estimate for the bounded
components of the seeded bases; their face variances are checked
when Lemma~\ref{lem:concentration} is applied below.

\subsection{Interpolation identity with a Gaussian seed}
\label{appx-an:sec:path}

We now separate the change in a smoothed expectation into a term
involving the aggregate third moment and a fourth-order remainder.
The third derivatives must be evaluated at a common point so that
we can sum their signed coefficients before taking absolute values.
We first expand at a point with one summand removed, where
independence makes the Taylor expansion simple, and then move the
cubic derivative back to the full interpolation point.

The following identity covers both the Slepian and calibrated
thinning paths. The amplitude and retention probability are linked
so that the covariance remains constant. A fixed Gaussian seed
changes the variance from \(t\) to \(t+\varepsilon\), whose derivative
is still \(1\), and therefore leaves the scalar remainder kernels
unchanged.

\begin{lemma}[Interpolation identity with a Gaussian seed]
\label{appx-an:lem:identity}
Use the independent centered summands of
\eqref{appx-eq:projected-bound}, and take \(G\sim N(0,I_r)\)
independent of them. Let \(\varepsilon\ge0\), let
\(\gamma:(0,1)\to(0,\infty)\) be continuously differentiable, and
suppose
\(\varpi(t)\coloneqq(1-t)/\gamma(t)^2\in[0,1]\).
For each \(t\), take independent Bernoulli masks with success
probability \(\varpi(t)\), independent of the summands and \(G\),
and define
\begin{equation}
 \begin{aligned}
 Y_t&\coloneqq\gamma(t)\sum_i\mathbb B_{i,t}\zeta_i^0
                    +\sqrt{t+\varepsilon}\,G,\\
 H_{i,t}&\coloneqq Y_t-\gamma(t)\mathbb B_{i,t}\zeta_i^0.
 \end{aligned}
 \label{eq:interpolation-family}
\end{equation}
Thus \(\Cov(Y_t)=\{\varpi(t)\gamma(t)^2+t+\varepsilon\}I_r
=(1+\varepsilon)I_r\), and \(H_{i,t}\) is independent of the
removed summand and its mask. No coupling of the masks across
different times is required.

To express the Taylor remainders along segments that insert a
summand, let \(\zeta_i,\widetilde\zeta_i\) be copies of
\(\zeta_i^0\), with all copies mutually independent and independent
of the original summands, masks, and Gaussian vector. Recall the
aggregate third-moment tensor \(K_3\), and define the two scalar
functions used in the coefficients by
\[
 K_3=\sum_i\E(\zeta_i^0)^{\otimes3},\qquad
 q_3(t)\coloneqq(1-t)\gamma(t),\qquad
 \eta_t\coloneqq\frac{(1-t)\gamma'(t)}{\gamma(t)}.
\]
In particular, \(q_3(t)=\varpi(t)\gamma(t)^3\) is the multiplier
of each retained summand's third-moment tensor.

For a smooth function \(f:\R^r\to\R\) with bounded derivatives
through order four, the following identity holds for \(0<t<1\):
\begin{equation}
 \begin{aligned}
 \frac{d}{dt}\E f(Y_t)
 ={}&\frac{q_3'(t)}6\E D^3f(Y_t):K_3\\
 &+\int_0^1\sum_i\E\!\left[
       D^4f(H_{i,t}+s\gamma(t)\zeta_i):\mathcal T_{i,t}(s)
     \right]\dd s,
 \end{aligned}
 \label{appx-an:fourth-identity}
\end{equation}
where
\[
 \mathcal T_{i,t}(s)\coloneqq
 a_t(s)\zeta_i^{\otimes4}
 +b_t(s)\widetilde\zeta_i^{\otimes2}\otimes\zeta_i^{\otimes2}
 +c_t(s)\widetilde\zeta_i^{\otimes3}\otimes\zeta_i
\]
and
\begin{equation}
 \begin{aligned}
 a_t(s)&\coloneqq\frac{\gamma(t)^2}{6}(1-s)^2
                   \{\eta_t-1+(1+2\eta_t)s\},\\
 b_t(s)&\coloneqq\frac{1-t}{2}(1-s),\qquad
 c_t(s)\coloneqq\frac{1-t}{6}(1-\eta_t).
 \end{aligned}
 \label{appx-an:kernel}
\end{equation}
The colon contracts all corresponding indices. The symmetry of
\(D^4f\) makes explicit symmetrization of the tensors unnecessary.
\end{lemma}

The three kernels record different fourth-order contributions.
The kernel $a_t$ comes from expanding the segment Hessian at the
deleted-summand base, $b_t$ from restoring the removed summand in
the Hessian, and $c_t$ from moving the cubic derivative to the full
interpolation point. We derive them in this order.

\begin{proof}
We first differentiate the expectation while keeping track of the
three sources of change: the Gaussian variance, the retention
probabilities, and the amplitudes. Throughout the proof, write
\(\gamma=\gamma(t)\) and \(\varpi=\varpi(t)\).
For fixed mask values, the summands are bounded and the Gaussian
variance is positive. On a compact subinterval of \((0,1)\), the
amplitude, its derivative, and the derivative of the Gaussian scale
are bounded. The bounded derivatives of \(f\) therefore justify
differentiation under expectation by domination, as in
\cite[Section~2.3]{Folland1999}. Summing over the finitely many
mask configurations then permits differentiation of their
probabilities. This argument concerns the distributions at each
time and does not differentiate a sample path of the masks.

For a fixed vector \(x\), Gaussian integration by parts gives
\[
 \frac{d}{dt}\E_G f(x+\sqrt{t+\varepsilon}\,G)
 =\frac{1}{2\sqrt{t+\varepsilon}}\,
       \E_G Df(x+\sqrt{t+\varepsilon}\,G)[G]
 =\frac12\E_G\tr D^2f(x+\sqrt{t+\varepsilon}\,G).
\]
This is the heat identity obtained from
Stein's identity~\cite{Stein1981}.
To compute the retention-probability contribution of mask $i$,
vary only its success probability, keeping the other mask
probabilities, the amplitude, and the Gaussian variance fixed.
The law of $H_{i,t}$ then stays fixed, since this base excludes
mask $i$ and its summand. Conditional on $H_{i,t}$, averaging
over that mask and its summand gives
\[
 (1-\varpi)f(H_{i,t})
       +\varpi\,\E_{\zeta_i}f(H_{i,t}+\gamma\zeta_i),
\]
where \(\E_{\zeta_i}\) averages over the independent copy alone.
Differentiating this affine expression with respect to the mask
probability gives the difference
$\E_{\zeta_i}f(H_{i,t}+\gamma\zeta_i)-f(H_{i,t})$.
Multiplying by $\varpi'$ and summing over $i$ gives the second
term below.

For the amplitude contribution, keep the masks and Gaussian
variance fixed and differentiate the common multiplier $\gamma$.
The directional derivative is
$Df(Y_t)[\sum_i\mathbb B_{i,t}\zeta_i^0]$.
In its $i$th term, conditioning on $H_{i,t}$ and averaging over
the removed mask gives
$\varpi\,\E_{\zeta_i}Df(H_{i,t}+\gamma\zeta_i)[\zeta_i]$.
After multiplication by $\gamma'$, this gives the third term.
Together with the Gaussian contribution, the product rule
therefore yields
\begin{equation}
 \begin{aligned}
 \frac{d}{dt}\E f(Y_t)
 ={}&\frac12\E\tr D^2f(Y_t)\\
 &+\varpi'\sum_i\E\{f(H_{i,t}+\gamma\zeta_i)-f(H_{i,t})\}\\
 &+\varpi\gamma'\sum_i
       \E Df(H_{i,t}+\gamma\zeta_i)[\zeta_i].
 \end{aligned}
 \label{appx-an:three-contributions}
\end{equation}

The aggregate covariance turns the Gaussian term into the same
directional Hessians that appear in the other two terms:
\[
 \E\tr D^2f(Y_t)
 =\sum_i\E D^2f(Y_t)[\zeta_i,\zeta_i].
\]
Here the copies are independent of \(Y_t\) and
\(\sum_i\E\zeta_i\zeta_i^\top=I_r\).
For fixed $x$, the linear term vanishes after averaging, because
\[
 \E Df(x)[\zeta_i]=Df(x)[\E\zeta_i]=0.
\]
Second-order integral Taylor expansion of $f(x+\gamma\zeta_i)$
therefore gives the first formula below. Applying the fundamental
theorem of calculus to
$s\mapsto Df(x+s\gamma\zeta_i)[\zeta_i]$, and using the same
centering identity, gives the second:
\begin{align*}
 \E\{f(x+\gamma\zeta_i)-f(x)\}
 &=\gamma^2\int_0^1(1-s)
        \E D^2f(x+s\gamma\zeta_i)[\zeta_i,\zeta_i]\dd s,\\
 \E Df(x+\gamma\zeta_i)[\zeta_i]
 &=\gamma\int_0^1
        \E D^2f(x+s\gamma\zeta_i)[\zeta_i,\zeta_i]\dd s.
\end{align*}
Independence of \(\zeta_i\) and \(H_{i,t}\) permits these formulas
to be applied conditionally at \(x=H_{i,t}\).

To combine the two segment integrals, define the affine
\emph{signed density}
\[
 \nu_t(s)\coloneqq2\{1+\eta_t-(1+2\eta_t)s\}.
\]
Differentiating \(\varpi\gamma^2=1-t\) gives
\(\varpi'\gamma^2=-1-2\eta_t\) and
\(\varpi\gamma\gamma'=\eta_t\). Hence
\[
 \varpi'\gamma^2(1-s)+\varpi\gamma\gamma'
 =-\nu_t(s)/2.
\]
Substitution into~\eqref{appx-an:three-contributions} yields
\begin{equation}
 \frac{d}{dt}\E f(Y_t)
 =\frac12\sum_i\E\left[
    D^2f(Y_t)[\zeta_i,\zeta_i]
    -\int_0^1\nu_t(s)D^2f(H_{i,t}+s\gamma\zeta_i)
                       [\zeta_i,\zeta_i]\dd s\right].
 \label{appx-an:signed-hessian}
\end{equation}
The constant and linear terms in a Taylor expansion of the
segment Hessian will involve the mass and first moment of
$\nu_t$, respectively. Direct integration gives
\begin{equation}
 \int_0^1\nu_t(s)\dd s=1,\qquad
 \int_0^1s\nu_t(s)\dd s=(1-\eta_t)/3.
 \label{appx-an:signed-moments}
\end{equation}
The density may take negative values, so we use these as algebraic identities,
not as moments of a probability distribution.

We next expand the segment Hessian at \(H_{i,t}\), where the
inserted copy is independent of the evaluation point. For fixed
vectors \(x,z\), Taylor's formula reads
\[
 \begin{aligned}
 D^2f(x+s\gamma z)[z,z]
 ={}&D^2f(x)[z,z]+s\gamma D^3f(x)[z,z,z]\\
 &+\gamma^2\int_0^s(s-v)
           D^4f(x+v\gamma z)[z,z,z,z]\dd v.
 \end{aligned}
\]
Inserting this formula into~\eqref{appx-an:signed-hessian},
the mass identity in~\eqref{appx-an:signed-moments} leaves the
difference between the Hessians at \(Y_t\) and \(H_{i,t}\).
The first-moment identity gives the cubic coefficient
\[
 -\frac{\gamma(1-\eta_t)}6
 =\frac{(1-t)\gamma'-\gamma}{6}
 =\frac{q_3'(t)}6.
\]
To identify the fourth-order coefficient, we change the order of
integration over the triangle $0\le v\le s\le1$. For fixed $x,z$,
\[
 \begin{aligned}
 &\int_0^1\nu_t(s)\int_0^s(s-v)
       D^4f(x+v\gamma z)[z,z,z,z]\,\dd v\,\dd s\\
 &\quad=\int_0^1
       \left\{\int_v^1(s-v)\nu_t(s)\,\dd s\right\}
       D^4f(x+v\gamma z)[z,z,z,z]\,\dd v.
 \end{aligned}
\]
The bounded fourth derivatives and bounded summands justify this
interchange also after averaging at $x=H_{i,t}$ and $z=\zeta_i$.
Relabel the outer variable $v$ as $s$. Substituting
$v=s+(1-s)w$ in the inner integral then gives
\[
 \begin{aligned}
 \int_s^1(v-s)\nu_t(v)\,\dd v
 &=2(1-s)^2\int_0^1
       w\{1+\eta_t-(1+2\eta_t)[s+(1-s)w]\}\,\dd w\\
 &=\frac{(1-s)^2}{3}
       \{1-\eta_t-(1+2\eta_t)s\}.
 \end{aligned}
\]
Here we used $\int_0^1w\,\dd w=1/2$ and
$\int_0^1w^2\,\dd w=1/3$.
Multiplication by the factor $-\gamma^2/2$ in
\eqref{appx-an:signed-hessian} yields
\[
 -\frac{\gamma^2}{2}\int_s^1(v-s)\nu_t(v)\dd v
 =\frac{\gamma^2}{6}(1-s)^2
       \{\eta_t-1+(1+2\eta_t)s\}
 =a_t(s).
\]
For the terms evaluated at \(Y_t\) or \(H_{i,t}\), we may replace
\(\zeta_i\) by \(\widetilde\zeta_i\): the copies have the same law
and are independent of both evaluation points. This leaves
\(\zeta_i\) available to insert the removed summand in the next
expansions. In the segment remainder, $\zeta_i$ determines both
the evaluation point and the derivative directions, and we retain
the same copy in both roles. We have obtained
\begin{equation}
 \begin{aligned}
 \frac{d}{dt}\E f(Y_t)
 ={}&\frac12\sum_i\E
       \bigl[(D^2f(Y_t)-D^2f(H_{i,t}))
             [\widetilde\zeta_i,\widetilde\zeta_i]\bigr]\\
 &+\frac{q_3'(t)}6\sum_i
       \E D^3f(H_{i,t})
              [\widetilde\zeta_i,\widetilde\zeta_i,
               \widetilde\zeta_i]\\
 &+\sum_i\int_0^1a_t(s)
       \E D^4f(H_{i,t}+s\gamma\zeta_i)
              [\zeta_i,\zeta_i,\zeta_i,\zeta_i]\dd s.
 \end{aligned}
 \label{appx-an:deleted-base-expansion}
\end{equation}
The last line is already in the required form. We now express
the first line as another fourth-order remainder and move the
cubic derivative in the second line to \(Y_t\).

For the Hessian difference, condition on
\(H_{i,t},\widetilde\zeta_i\). The removed mask is zero with
probability \(1-\varpi\), in which case the difference vanishes.
When it is \(1\), its summand can be replaced in law by \(\zeta_i\).
The linear Taylor term is
$\gamma D^3f(H_{i,t})[\widetilde\zeta_i,\widetilde\zeta_i,\zeta_i]$.
Its conditional expectation is zero, since $\zeta_i$ is centered
and independent of the conditioned variables. The quadratic
remainder therefore gives
\[
 \begin{aligned}
 &\frac12\E
    \bigl[(D^2f(Y_t)-D^2f(H_{i,t}))
                 [\widetilde\zeta_i,\widetilde\zeta_i]\bigr]\\
 &\quad=\frac{\varpi\gamma^2}{2}\int_0^1(1-s)
       \E D^4f(H_{i,t}+s\gamma\zeta_i)
           [\widetilde\zeta_i,\widetilde\zeta_i,
            \zeta_i,\zeta_i]\dd s.
 \end{aligned}
\]
Since \(\varpi\gamma^2=1-t\), its coefficient is \(b_t(s)\).

For the cubic term, we write the derivative at $H_{i,t}$ as its
value at $Y_t$ minus the increment caused by restoring the removed
summand. This accounts for the negative sign in the following
first-order integral expansion, obtained under the same conditioning:
\[
 \begin{aligned}
 &\E D^3f(H_{i,t})
       [\widetilde\zeta_i,\widetilde\zeta_i,\widetilde\zeta_i]\\
 &\quad=\E D^3f(Y_t)
       [\widetilde\zeta_i,\widetilde\zeta_i,\widetilde\zeta_i]\\
 &\qquad\quad-\varpi\gamma\int_0^1
       \E D^4f(H_{i,t}+s\gamma\zeta_i)
       [\widetilde\zeta_i,\widetilde\zeta_i,
        \widetilde\zeta_i,\zeta_i]\dd s.
 \end{aligned}
\]
Multiplication by \(q_3'(t)/6\) makes the remainder coefficient
\[
 -\frac{q_3'(t)\varpi\gamma}{6}
 =\frac{(1-t)(1-\eta_t)}6=c_t(s).
\]
All cubic derivatives now share the full evaluation point, which
is independent of the copies. More explicitly, conditioning on
$Y_t$ gives
\[
 \E\!\left[
 D^3f(Y_t)[\widetilde\zeta_i,\widetilde\zeta_i,\widetilde\zeta_i]
 \,\middle|\,Y_t\right]
 =D^3f(Y_t):\E(\zeta_i^0)^{\otimes3}.
\]
Taking expectations and summing over $i$ therefore yields
\[
 \sum_i\E D^3f(Y_t)
       [\widetilde\zeta_i,\widetilde\zeta_i,\widetilde\zeta_i]
 =\E D^3f(Y_t):K_3.
\]
Substituting these last two expansions into
\eqref{appx-an:deleted-base-expansion} proves
\eqref{appx-an:fourth-identity}, with the three tensor terms
corresponding to \(a_t,b_t,c_t\), respectively.
\end{proof}

We apply the lemma with two choices of amplitude. For the Slepian
path, \(\gamma(t)=\sqrt{1-t}\) gives \(\varpi(t)=1\),
so every summand is retained. In this case
\(\eta_t=-1/2\), \(q_3'(t)/6=-\sqrt{1-t}/4\), and
\[
 \begin{aligned}
 a_t(s)&=-\frac{1-t}{4}(1-s)^2,\\
 b_t(s)&=\frac{1-t}{2}(1-s),\qquad
 c_t(s)=\frac{1-t}{4}.
 \end{aligned}
\]
For calibrated thinning, \(\gamma(t)=1+t\) gives
\(\varpi(t)=(1-t)/(1+t)^2\),
\(\eta_t=(1-t)/(1+t)\), and \(q_3'(t)/6=-t/3\). The kernels are
\[
 \begin{aligned}
 a_t(s)&=\frac{1+t}{6}(1-s)^2\{(3-t)s-2t\},\\
 b_t(s)&=\frac{1-t}{2}(1-s),\qquad
 c_t(s)=\frac{t(1-t)}{3(1+t)}.
 \end{aligned}
\]
In both cases the kernels are bounded on
\(0\le s\le1\), \(0<t<1\), and hence their absolute integrals
are bounded uniformly in time.
The first calibrated kernel agrees with~\cite[Lemma~4.1]{AL}, and
the common-point cubic form has the structure
of~\cite[Lemma~22]{GaoZhang}.

The later derivative estimates compare each remainder point with
the full path. Since both amplitudes are at most \(2\), the face
bounds on the original summands and their copies give
\begin{equation}
 H_{i,t}+s\gamma\zeta_i=Y_t+v_i,\qquad
 v_i\coloneqq s\gamma\zeta_i-\gamma\mathbb B_{i,t}\zeta_i^0,
 \qquad \norm{v_i}_U\le4\delta.
 \label{appx-an:translates}
\end{equation}
Indeed, the two terms in \(v_i\) have face norms at most
\(2\delta\) each.

It remains to connect the unnormalized path \(Y_t\) with the two
normalized stages in Section~\ref{sec:smoothing}. For the chosen
Gaussian seed, write
\begin{equation}
 \varepsilon=Lh^2,\qquad
 c_\varepsilon\coloneqq\sqrt{1+\varepsilon},\qquad
 v_\varepsilon=\frac{\varepsilon}{1+\varepsilon},
 \qquad T_v=\sqrt{1-v}\,S+\sqrt v\,G.
 \label{eq:seed-initialization}
\end{equation}
The Slepian choice with seed zero gives \(Y_v=T_v\). For the
calibrated choice with seed \(\varepsilon\),
\eqref{eq:path} gives \(Y_t=c_\varepsilon U_t\). The two stages
meet because
\[
 T_{v_\varepsilon}
 =\frac{S+\sqrt\varepsilon\,G}{\sqrt{1+\varepsilon}}
 =U_0.
\]
Recall that \(P_a^0(v)=\E F_{a,h}(T_v)\) and
\(P_a(t)=\E F_{a,h}(U_t)\). Thus
\(P_a^0(v_\varepsilon)=P_a(0)\), and
\[
 I_h=\sup_a|P_a^0(v_\varepsilon)-P_a^0(0)|
 \le\sup_a\int_0^{v_\varepsilon}|(P_a^0)'(v)|\,\dd v.
\]
Continuity at \(v=0\), bounded derivatives of \(F_{a,h}\), and the
bounded Slepian coefficients above justify integration from this
endpoint. The Slepian remainder \(R^0_{a,h}\) defined in
Subsection~\ref{sec:outline-seeded} is exactly the integral
remainder in~\eqref{appx-an:fourth-identity} with seed zero and
\(\gamma(v)=\sqrt{1-v}\).

For the calibrated path, incorporate its normalization into the
test function:
\begin{equation}
 f(x)\coloneqq F_{a,h}(x/c_\varepsilon)
       =F_{\widetilde a,\widetilde h}(x),\qquad
 \widetilde a\coloneqq c_\varepsilon a,\quad
 \widetilde h\coloneqq c_\varepsilon h.
 \label{eq:seeded-test}
\end{equation}
The equality follows by multiplying both the thresholds and the
smoothing width by \(c_\varepsilon\) in each face score. Since
\(f(Y_t)=F_{a,h}(U_t)\), the lemma gives
\[
 P_a'(t)=-\frac t3\E D^3f(Y_t):K_3+R_f(t),
\]
where \(R_f(t)\) denotes its integral remainder. The chain rule
gives \(D^3f(Y_t)=c_\varepsilon^{-3}D^3F_{a,h}(U_t)\).
We therefore recover~\eqref{eq:path-summary}, including its
factor \((1+\varepsilon)^{-3/2}\), with
\(R_{4,a,h}(t)=R_f(t)\). Bounds for the cubic contraction and the
integral remainder along \(Y_t\) thus give the normalized bounds
in Proposition~\ref{an:prop:derivative}.

\subsection{Direct derivatives and smooth initialization}
\label{appx-an:sec:initialization}

We prove the direct derivative bounds~\eqref{eq:direct-initialization}
and then integrate them to obtain the initialization
bound~\eqref{eq:initialization-bound}. In the cubic term, the common
evaluation point lets us sum the signed third moments before taking
absolute values. In the fourth-order remainder, the evaluation point
depends on an inserted summand, so we first compare its logistic
weights with those at the common point.

Fix \(a\in\R^m\), and write \(F=F_{a,h}\).
Along the Slepian path
\(T_v=\sqrt{1-v}\,S+\sqrt v\,G\), every face projection of the
bounded component has variance \(1-v\), and every face projection
of the Gaussian component has variance \(v\). At least one component
therefore has common face variance at least \(1/2\).
The bounded component has summand face norms at most \(\delta\)
and expected face norm at most \(\E\|W\|_\infty\).
The Gaussian component has expected face norm at most \(g_\Sigma\).
Apply Lemma~\ref{lem:concentration} to whichever component has
variance at least \(1/2\), treating the other as an independent
shift. The intrinsic and dimension-based bounds in that lemma give
the boundary coefficient \(\kappa\) from
\eqref{eq:intrinsic-boundary-coefficient}, uniformly in \(v\).
Lemma~\ref{appx-lem:envelope} consequently gives, for
\(h\ge\mathsf A\delta\) with \(\mathsf A\ge4\),
\begin{equation}
 \E F(T_v)(1-F(T_v))\le Ch\kappa,
 \qquad 0<v<1.
 \label{eq:slepian-weight}
\end{equation}

To bound the cubic term, expand \(D^3F(x):K_3\) by the partition
formula~\eqref{appx-eq:partition-derivative}. For fixed $x$, each
logistic weight is the same for every summand index $i$. We can
therefore sum the signed third-moment coefficients over $i$ before
taking absolute values. Since each face projection is an original
coordinate or its
negative, a coefficient with original coordinate indices \(j,k,l\)
equals, up to sign,
\[
 \sum_i\E[\xi_{ij}\xi_{ik}\xi_{il}]=e_k^\top T_j e_l.
\]
The third-moment assumption in~\eqref{eq:relative-moments} is stated
for quadratic forms. Polarization converts it to the mixed
coefficient needed here:
\[
 \begin{aligned}
 |e_k^\top T_j e_l|
 &=\frac14\left|
   (e_k+e_l)^\top T_j(e_k+e_l)
   -(e_k-e_l)^\top T_j(e_k-e_l)\right|\\
 &\le\frac{s_3}{4}\left\{
   (e_k+e_l)^\top\Sigma(e_k+e_l)
   +(e_k-e_l)^\top\Sigma(e_k-e_l)\right\}
 =s_3.
 \end{aligned}
\]
The last equality uses \(\Sigma_{kk}=\Sigma_{ll}=1\).
Thus this bound also holds when \(\Sigma\) is singular.

For a partition with \(k\) blocks, the sum of its positive logistic
weights is
\[
 \sum_{\alpha_1,\ldots,\alpha_k}
       F(x)\prod_{b=1}^k\pi_{\alpha_b}(x)
 =F(x)(1-F(x))^k
 \le F(x)(1-F(x)).
\]
There are only finitely many partitions of three derivative
directions, and each derivative contributes \(h^{-1}\).
Bounding each aggregate third-moment coefficient by \(s_3\)
therefore yields
\[
 |D^3F(x):K_3|\le Cs_3h^{-3}F(x)(1-F(x)).
\]
Averaging at \(x=T_v\) and using~\eqref{eq:slepian-weight} proves
\[
 |\E D^3F(T_v):K_3|\le Cs_3\kappa/h^2.
\]

We next bound the fourth-order remainder supplied by
Lemma~\ref{appx-an:lem:identity} with seed zero and the Slepian
amplitude. Its derivatives are evaluated at
\[
 x_{i,v,s}:=H_{i,v}+s\sqrt{1-v}\,\zeta_i,\qquad 0\le s\le1,
\]
and~\eqref{appx-an:translates} gives
\(\|x_{i,v,s}-T_v\|_U\le4\delta\).
We use the termwise weight comparison underlying
\eqref{appx-eq:local-envelope}, retaining each face label until
the corresponding directional moments have been bounded.
For each fixed list of face labels and each $1\le k\le4$, it gives
\[
 F(x_{i,v,s})\prod_{b=1}^k\pi_{\alpha_b}(x_{i,v,s})
 \le C F(T_v)\prod_{b=1}^k\pi_{\alpha_b}(T_v).
\]
Indeed, each face score changes by at most $4\delta/h$. The
multiplier from Appendix~\ref{appx-sec:smoothing} is at most
$e^{(2k+1)4\delta/h}\le e^9$, since $k\le4$ and
$h\ge\mathsf A\delta\ge4\delta$.
The weight on the right depends only on \(T_v\) and is
independent of all the copy variables in the derivative directions.

We first bound the directional products. Observe from
\eqref{eq:relative-moments} that
\[
 \sum_i\E\xi_{ij}^4=e_j^\top Q_j e_j
 \le s_4\Sigma_{jj}=s_4,\qquad j\le p.
\]
Let \(Z_{i,1},\ldots,Z_{i,4}\) denote the four directions in any
one of the three tensor terms of
\eqref{appx-an:fourth-identity}. Each is either \(\zeta_i\) or
\(\widetilde\zeta_i\), so each has the law of \(\zeta_i^0\).
For any fixed face labels $\beta_1,\ldots,\beta_4$, apply
H\"older's inequality first to the expectation for each $i$,
and then to the sum over $i$. These two steps give
\[
 \begin{aligned}
 \sum_i\E\prod_{\ell=1}^4
       |u_{\beta_\ell}\cdot Z_{i,\ell}|
 &\le\sum_i\prod_{\ell=1}^4
       \left\{\E
         |u_{\beta_\ell}\cdot Z_{i,\ell}|^4\right\}^{1/4}\\
 &\le\prod_{\ell=1}^4
       \left\{\sum_i\E
         |u_{\beta_\ell}\cdot Z_{i,\ell}|^4\right\}^{1/4}
 \le s_4.
 \end{aligned}
\]
Each fourth-moment sum is bounded by $s_4$, because the face
component has the law of an original coordinate or its negative.
This inequality does not require independence among the four
directions, and hence covers both repeated directions and products
involving the two copies.

To apply this bound to the fourth-derivative partition formula,
fix a partition with blocks $J_1,\ldots,J_k$ and face labels
$\alpha_1,\ldots,\alpha_k$. Each direction in block $J_b$ is paired
with the normal $u_{\alpha_b}$, so define
$\beta_\ell:=\alpha_b$ for $\ell\in J_b$.
The comparison above has moved the positive logistic weight to
$T_v$, which depends only on the original summands and the Gaussian
vector. Independence of the copies therefore gives the factorization
\[
 \begin{aligned}
 &\sum_i\E\!\left[
   F(T_v)\prod_{b=1}^k\pi_{\alpha_b}(T_v)
   \prod_{\ell=1}^4|u_{\beta_\ell}\cdot Z_{i,\ell}|\right]\\
 &\quad=\E\!\left[F(T_v)\prod_{b=1}^k\pi_{\alpha_b}(T_v)\right]
       \sum_i\E\prod_{\ell=1}^4|u_{\beta_\ell}\cdot Z_{i,\ell}|\\
 &\quad\le s_4\,\E\!\left[
       F(T_v)\prod_{b=1}^k\pi_{\alpha_b}(T_v)\right].
 \end{aligned}
\]
Sum over the face labels using the same identity as in the cubic
bound: the positive weights sum to $F(T_v)(1-F(T_v))^k$,
which is at most $F(T_v)(1-F(T_v))$. The finitely many partitions
contribute only an absolute constant, and the four derivatives
contribute $h^{-4}$. Applying~\eqref{eq:slepian-weight} then gives,
for each of the three direction patterns,
\[
 \begin{aligned}
 \sum_i\E\left|
   D^4F(x_{i,v,s})[Z_{i,1},Z_{i,2},Z_{i,3},Z_{i,4}]
          \right|
 &\le Cs_4h^{-4}\E F(T_v)(1-F(T_v))\\
 &\le Cs_4\kappa/h^3.
 \end{aligned}
\]
The absolute integrals of the three Slepian kernels in
Subsection~\ref{appx-an:sec:path} are bounded uniformly in \(v\).
Integrating the last estimate against those kernels therefore gives
\[
 |R^0_{a,h}(v)|\le Cs_4\kappa/h^3.
\]
Together with the cubic estimate, this proves
\eqref{eq:direct-initialization}.

Finally, the cubic coefficient in the Slepian identity is
\(-\sqrt{1-v}/4\), whose absolute value is at most \(1/4\).
The two derivative bounds thus imply
\[
 |(P_a^0)'(v)|
 \le C\kappa\left(\frac{s_3}{h^2}+\frac{s_4}{h^3}\right),
 \qquad 0<v<1.
\]
Take \(\varepsilon=Lh^2\) as in
Proposition~\ref{an:prop:derivative}, so that
\(v_\varepsilon=\varepsilon/(1+\varepsilon)\le Lh^2\).
Integrating over the initialization stage yields
\[
 \begin{aligned}
 I_h&\le\sup_a\int_0^{v_\varepsilon}|(P_a^0)'(v)|\,\dd v\\
 &\le C v_\varepsilon\kappa
       \left(\frac{s_3}{h^2}+\frac{s_4}{h^3}\right)
 \le C\kappa L(s_3+s_4/h).
 \end{aligned}
\]
To justify the lower endpoint, first integrate from a positive
time and then let it decrease to zero. The derivative bound is
uniform, and boundedness and continuity of \(F\) give
\(P_a^0(v)\to P_a^0(0)\). This proves
\eqref{eq:initialization-bound}.

\subsection{Paired derivatives and the seeded comparison}
\label{appx-an:sec:restoration}

We now prove the seeded derivative bounds in
\eqref{eq:seeded-derivatives}. The Gaussian component allows us to
replace two derivatives in a repeated direction by a centered
quadratic function of the Gaussian vector. This produces a factor
$\sigma^{-2}$ and leaves only first or second derivatives of the
logistic approximation. The remaining derivatives supply face
components, which we combine with the repeated direction into
coefficient matrices. Summing these matrices over the observations
before estimating their Gaussian averages will preserve the
aggregate moment bounds.

The analytic input is Lemma~\ref{appx-an:lem:matrix}, stated and
proved in Subsection~\ref{appx-an:sec:matrix}. It controls the sum of
the nuclear norms of averages of $GG^\top-I_r$ against
logistic product weights, without a factor depending on rank.
Here $\|\cdot\|_*$ denotes the \emph{nuclear norm}, the sum of
singular values, and $\|\cdot\|_{\op}$ denotes the
\emph{operator norm}, the largest singular value. The next lemma
reduces the derivative estimates to these matrix averages. Its
parameter $s$ counts the derivatives remaining after integration
by parts: $s=1$ treats the cubic term, and $s=2$ treats the
fourth-order remainder. The auxiliary vectors $Z_i,Z_{i,a}$ are
unrelated to the Gaussian target $Z$.

The conditioning in the lemma separates the Gaussian integration
from the other randomness. It freezes the base, shifts, and
directions, leaving a standard Gaussian vector for integration by
parts. The concentration assumption on the base refers throughout
to its unconditional law, as in~\eqref{appx-an:concentration}.

\begin{lemma}[Paired derivatives at translated bases]
\label{appx-an:lem:restoration}
Let \(\mathcal H\) be a \(\sigma\)-field, let \(V_0\) be an
\(\mathcal H\)-measurable concentrated base with coefficient
\(\kappa\), and let \(G\sim N(0,I_r)\) be independent of
\(\mathcal H\). Suppose that \(\delta\sqrt L\le\sigma\le1\),
and define \(U\coloneqq V_0+\sigma G\).
Fix \(s\in\{1,2\}\) and \(b\ge0\).
Let \((Z_i,Z_{i,1},\ldots,Z_{i,s},v_i)_i\) be random arrays
which, conditional on \(\mathcal H\), are independent of \(G\).
They may depend on \(\mathcal H\) and on one another.
Suppose that, almost surely,
\begin{equation}
 \sum_i\E[Z_iZ_i^\top\mid\mathcal H]\preceq I_r,
 \qquad \norm{Z_{i,a}}_U\le\delta\quad(a\le s),
 \qquad \norm{v_i}_U\le b\delta.
 \label{appx-an:conditional-budget}
\end{equation}
There are constants \(\mathsf A_b,C_b\), depending only on \(b\),
such that, for \(F=F_{a,h}\) and \(h\ge\mathsf A_b\delta\),
\begin{equation}
 \left|\sum_i\E D^{s+2}F(U+v_i)
         [Z_i,Z_i,Z_{i,1},\ldots,Z_{i,s}]\right|
 \le\frac{C_b}{\sigma^2}\,
                 \delta^s h^{1-s}\kappa L.
 \label{appx-an:restoration-bound}
\end{equation}
The constants are uniform over the threshold vector \(a\).

The factor \(\delta^s\) can be replaced by a constant \(M_s\ge0\)
under the following additional moment bounds.
For \(s=1\), suppose that \(v_i=0\) for every \(i\) and, almost
surely, for every face normal \(u_j\),
\begin{equation}
 \left\|\sum_i\E[Z_iZ_i^\top(u_j\cdot Z_{i,1})
                    \mid\mathcal H]\right\|_{\op}\le M_1.
 \label{appx-an:signed-cubic-budget}
\end{equation}
For \(s=2\), suppose that, almost surely, for each \(a=1,2\)
and each face normal \(u_j\),
\begin{equation}
 \sum_i\E[Z_iZ_i^\top(u_j\cdot Z_{i,a})^2
                    \mid\mathcal H]\preceq M_2I_r.
 \label{appx-an:weighted-fourth-budget}
\end{equation}
\end{lemma}

\begin{proof}
Condition on $\mathcal H$ and on the entire collection of arrays.
Since $G$ is independent of $\mathcal H$, and the arrays are
conditionally independent of $G$ given $\mathcal H$, the
conditional law of $G$ remains $N(0,I_r)$. The base $V_0$, shifts
$v_i$, and derivative directions are now fixed. We write $\E_G$
for expectation over the remaining Gaussian vector.

To see the two integrations by parts explicitly, define
\[
 \psi_i(g):=D^sF(V_0+v_i+\sigma g)
                  [Z_{i,1},\ldots,Z_{i,s}],\qquad g\in\R^r,
\]
and let $\partial_z$ denote differentiation in the Gaussian
argument $g$ along a fixed direction $z$. The chain rule gives
\[
 \partial_{Z_i}^2\psi_i(g)
 =\sigma^2D^{s+2}F(V_0+v_i+\sigma g)
                  [Z_i,Z_i,Z_{i,1},\ldots,Z_{i,s}],
\]
because each derivative in $g$ contributes a factor $\sigma$.
Apply Stein's identity~\cite{Stein1981} first to
$\partial_{Z_i}\psi_i$ and then to $(g\cdot Z_i)\psi_i(g)$:
\begin{align*}
 \E_G\partial_{Z_i}^2\psi_i(G)
 &=\E_G[(G\cdot Z_i)\partial_{Z_i}\psi_i(G)]\\
 &=\E_G[\{(G\cdot Z_i)^2-\|Z_i\|^2\}\psi_i(G)].
\end{align*}
The subtracted term comes from
$\partial_{Z_i}(g\cdot Z_i)=\|Z_i\|^2$. For fixed directions,
$\psi_i$ and its derivatives are bounded, so Gaussian
integrability justifies both integrations by parts.

The centered quadratic factor has the matrix representation
\[
 (G\cdot Z_i)^2-\|Z_i\|^2
 =\tr\{(GG^\top-I_r)Z_iZ_i^\top\}.
\]
Combining this identity with the chain rule yields
\begin{align*}
 &\E_G D^{s+2}F(U+v_i)[Z_i,Z_i,Z_{i,1},\ldots,Z_{i,s}]\\
 &\qquad=\sigma^{-2}\E_G\left[
       \tr\{(GG^\top-I_r)Z_iZ_i^\top\}
       D^sF(U+v_i)[Z_{i,1},\ldots,Z_{i,s}]\right].
\end{align*}
This accounts for the factor $\sigma^{-2}$ and introduces the
matrix $GG^\top-I_r$. Concentration has not been used at this
conditional stage. We will use it only after averaging over the
arrays and $\mathcal H$ again.

Only first and second derivatives remain. Define
$W_j(x):=F(x)\pi_j(x)$ and
$W_{jk}(x):=F(x)\pi_j(x)\pi_k(x)$. The first two cases of
\eqref{appx-eq:partition-derivative} give
\begin{align*}
 DF(x)[z]&=-h^{-1}\sum_jW_j(x)(u_j\cdot z),\\
 D^2F(x)[z,w]&=h^{-2}\left\{
       2\sum_{j,k}W_{jk}(x)(u_j\cdot z)(u_k\cdot w)
       -\sum_jW_j(x)(u_j\cdot z)(u_j\cdot w)\right\}.
\end{align*}
To organize these three sums, let
$\alpha=(\alpha_1,\ldots,\alpha_k)$ be an ordered tuple of face
labels, with $k=1$ or $2$, and define
\[
 W_\alpha(x):=F(x)\prod_{a=1}^k\pi_{\alpha_a}(x).
\]
For the first derivative, $\alpha=(j)$ and
$c_{i,(j)}:=u_j\cdot Z_{i,1}$. For the two second-derivative
sums, the respective coefficients are
\[
 c_{i,(j,k)}:=(u_j\cdot Z_{i,1})(u_k\cdot Z_{i,2}),
 \qquad
 c_{i,(j)}:=(u_j\cdot Z_{i,1})(u_j\cdot Z_{i,2}).
\]
Each coefficient contains $s$ face components, so
\eqref{appx-an:conditional-budget} gives
$|c_{i,\alpha}|\le\delta^s$.

We now average over the variables conditioned on and sum over
$i$. The first-derivative sum and the two second-derivative sums
have numerical coefficients $-1$, $2$, and $-1$, respectively.
After extracting that coefficient and the common factor
$\sigma^{-2}h^{-s}$, each contribution has the form
\begin{equation}
 \sum_\alpha\sum_i\E\left[
  \tr\{(GG^\top-I_r)Z_iZ_i^\top\}\,
       c_{i,\alpha}W_\alpha(U+v_i)\right].
 \label{appx-an:weighted-pair-form}
\end{equation}

The shifts $v_i$ can depend on the derivative directions, so the
translated weights in~\eqref{appx-an:weighted-pair-form} cannot
be separated from the coefficients by independence. We instead
express each translated weight as a convergent series of weights
at the common point $U$. In this representation, all dependence
on the shift remains in the coefficients.

We first derive the identity at deterministic points
$x,v\in\R^r$. Define
\[
 d_j(v):=e^{u_j\cdot v/h}-1,\qquad
 z_v(x):=\sum_j\pi_j(x)d_j(v),\qquad
 \eta_b:=e^{b\delta/h}-1.
\]
Here $d_j(v)$ is the relative change in the exponential weight
of face $j$. Recalling the notation
in~\eqref{appx-an:basic-notation}, we have
\[
 \begin{aligned}
 N(x+v)
 &=1+\sum_j e^{q_j(x)}\{1+d_j(v)\}\\
 &=N(x)+\sum_j e^{q_j(x)}d_j(v)
 =N(x)\{1+z_v(x)\}.
 \end{aligned}
\]
Thus $z_v(x)$ is the relative change in the common denominator.
In particular,
\[
 F(x+v)=\frac{F(x)}{1+z_v(x)},\qquad
 \pi_j(x+v)=\frac{\pi_j(x)e^{u_j\cdot v/h}}{1+z_v(x)}.
\]
The product $W_\alpha=F\prod_{a=1}^k\pi_{\alpha_a}$ contains
$k+1$ such denominator factors, which gives
\begin{equation}
 W_\alpha(x+v)=W_\alpha(x)
 e^{\sum_{a=1}^k u_{\alpha_a}\cdot v/h}
 (1+z_v(x))^{-k-1}.
 \label{eq:weight-translation}
\end{equation}
To expand the last factor, suppose that $\|v\|_U\le b\delta$.
The inequality $|e^t-1|\le e^{|t|}-1$ gives
$|d_j(v)|\le\eta_b$, and hence
\[
 |z_v(x)|\le\sum_j\pi_j(x)|d_j(v)|
             \le\eta_b\sum_j\pi_j(x)\le\eta_b.
\]
Choose $\mathsf A_b\ge1$ large enough that
$e^{b/\mathsf A_b}-1<1$. Since $h\ge\mathsf A_b\delta$,
we then have $\eta_b<1$, uniformly over $x$ and the allowed
shifts. The binomial series is therefore absolutely convergent:
\[
 \begin{aligned}
 (1+z_v(x))^{-k-1}
 &=\sum_{\ell\ge0}(-1)^\ell\binom{k+\ell}{\ell}z_v(x)^\ell\\
 &=\sum_{\ell\ge0}(-1)^\ell\binom{k+\ell}{\ell}
       \sum_{\beta_1,\ldots,\beta_\ell}
       \prod_{q=1}^\ell\pi_{\beta_q}(x)d_{\beta_q}(v).
 \end{aligned}
\]
The second equality expands the power of the finite sum defining
$z_v(x)$. Each $\beta_q$ ranges from $1$ to $m$, so the inner
sum runs over ordered face tuples. At $\ell=0$, it consists of
the empty product $1$. After substituting $x=U$ and $v=v_i$,
every logistic factor is evaluated at $U$, while the shift enters
only through $e^{\sum_a u_{\alpha_a}\cdot v_i/h}$ and the
factors $d_{\beta_q}(v_i)$.

We next justify interchanging this series with the expectations
in~\eqref{appx-an:weighted-pair-form}. First, Cauchy--Schwarz gives
\[
 \left|\tr\{(GG^\top-I_r)Z_iZ_i^\top\}\right|
 =\left|(G\cdot Z_i)^2-\|Z_i\|^2\right|
 \le(\|G\|^2+1)\|Z_i\|^2.
\]
For fixed $k$ and $\ell$, summing the nonnegative logistic
weights over both face tuples gives
\[
 \sum_{\alpha,\beta}F(U)
    \prod_{a=1}^k\pi_{\alpha_a}(U)
    \prod_{q=1}^\ell\pi_{\beta_q}(U)
 =F(U)\left\{\sum_j\pi_j(U)\right\}^{k+\ell}
 =F(U)(1-F(U))^{k+\ell}\le1.
\]
Together with $|c_{i,\alpha}|\le\delta^s$,
$e^{\sum_a u_{\alpha_a}\cdot v_i/h}\le e^{kb\delta/h}$,
and $|d_{\beta_q}(v_i)|\le\eta_b$, this bounds the sum of the
absolute values of all terms by
\[
 \delta^s e^{kb\delta/h}(1-\eta_b)^{-k-1}
       (\|G\|^2+1)\sum_i\|Z_i\|^2.
\]
Here we used
$\sum_{\ell\ge0}\binom{k+\ell}{\ell}\eta_b^\ell
=(1-\eta_b)^{-k-1}$.
Taking traces and expectations in the conditional covariance
bound~\eqref{appx-an:conditional-budget} yields
\[
 \sum_i\E\|Z_i\|^2
 =\E\tr\left\{\sum_i\E[Z_iZ_i^\top\mid\mathcal H]\right\}
 \le\tr(I_r)=r.
\]
Moreover, $G$ is independent of the joint collection of
$\mathcal H$ and the arrays, as established at the start of the
proof. Consequently,
\[
 \E\left[(\|G\|^2+1)\sum_i\|Z_i\|^2\right]
 =(r+1)\sum_i\E\|Z_i\|^2\le r(r+1)<\infty.
\]
The absolute series thus has an integrable bound, and Fubini's
theorem~\cite[Section~2.5]{Folland1999} permits the interchange.
The rank enters only this justification of integrability. The
bound on the resulting expectation below will not depend on it.

Every weight involving $G$ is now evaluated at $U$. Conditional
on $\mathcal H$ and $G$, these weights are fixed, because $V_0$
is $\mathcal H$-measurable. The arrays have the same conditional
law as they do given $\mathcal H$ alone, by their conditional
independence from $G$. We can therefore average the coefficient
matrices over the arrays while keeping the Gaussian weights
fixed. For each pair of ordered face tuples
$\alpha=(\alpha_1,\ldots,\alpha_k)$ and
$\beta=(\beta_1,\ldots,\beta_\ell)$, define
\[
 \Xi_{\alpha,\beta}:=\sum_i\E\left[
 Z_iZ_i^\top c_{i,\alpha}
 e^{\sum_{a=1}^k u_{\alpha_a}\cdot v_i/h}
 \prod_{q=1}^\ell d_{\beta_q}(v_i)
 \,\middle|\,\mathcal H\right].
\]
This matrix is $\mathcal H$-measurable and symmetric, since each
$Z_iZ_i^\top$ is symmetric and its multiplier is real. The
multiplier may have either sign, so we estimate the operator norm
through quadratic forms. We keep the sum over $i$ inside the
matrix being estimated, where the aggregate moment assumptions
can be applied.

For $z\in\R^r$, the identity
$z^\top Z_iZ_i^\top z=(z\cdot Z_i)^2$ and the coefficient bounds
from the preceding calculation give
\[
 \begin{aligned}
 |z^\top\Xi_{\alpha,\beta}z|
 &\le\delta^s e^{kb\delta/h}\eta_b^\ell
       \sum_i\E[(z\cdot Z_i)^2\mid\mathcal H]\\
 &=\delta^s e^{kb\delta/h}\eta_b^\ell
       z^\top\left\{\sum_i\E[Z_iZ_i^\top\mid\mathcal H]\right\}z\\
 &\le\delta^s e^{kb\delta/h}\eta_b^\ell\|z\|^2.
 \end{aligned}
\]
The last inequality is the conditional covariance
bound~\eqref{appx-an:conditional-budget}. For a real symmetric
matrix, the operator norm is the largest absolute eigenvalue,
which equals the supremum of the absolute quadratic form over
unit vectors. Hence, almost surely,
\[
 \|\Xi_{\alpha,\beta}\|_{\op}
 =\sup_{\|z\|=1}|z^\top\Xi_{\alpha,\beta}z|
 \le\delta^s e^{kb\delta/h}\eta_b^\ell.
\]
This is the use of the covariance bound described in
Subsection~\ref{appx-an:sec:coords}.

The additional moment assumptions sharpen the coefficient
estimate at this same step. For $s=1$, the derivative formula
has $k=1$, $\alpha=(j)$, and
$c_{i,(j)}=u_j\cdot Z_{i,1}$. If every shift is zero, then the
exponential multiplier is $1$ and $d_j(v_i)=0$. All terms with
$\ell\ge1$ vanish. For $\ell=0$, write $\beta=\varnothing$ for
the empty tuple. The remaining matrix is
\[
 \Xi_{(j),\varnothing}
 =\sum_i\E[Z_iZ_i^\top(u_j\cdot Z_{i,1})\mid\mathcal H],
 \qquad
 \|\Xi_{(j),\varnothing}\|_{\op}\le M_1,
\]
where the inequality is exactly
\eqref{appx-an:signed-cubic-budget}. Because the norm is taken
after summing over $i$, the cancellation permitted by that
assumption is retained.

For $s=2$, each coefficient is a product of two face components.
Write
\[
 x_i:=u_{j_1}\cdot Z_{i,1},\qquad
 y_i:=u_{j_2}\cdot Z_{i,2},\qquad c_{i,\alpha}=x_iy_i,
\]
where $j_1=j_2$ in the single-face sum and
$\alpha=(j_1,j_2)$ in the two-face sum. The moment
bound~\eqref{appx-an:weighted-fourth-budget}, applied to the
first slot and tested against $z\in\R^r$, gives
\[
 \begin{aligned}
 \sum_i\E[(z\cdot Z_i)^2x_i^2\mid\mathcal H]
 &=z^\top\left\{\sum_i\E[
       Z_iZ_i^\top(u_{j_1}\cdot Z_{i,1})^2
                         \mid\mathcal H]\right\}z\\
 &\le M_2\|z\|^2.
 \end{aligned}
\]
Applying the same assumption to the second slot gives the
identical bound with $y_i^2$ in place of $x_i^2$.
We can therefore control the mixed product by the pointwise
inequality $|x_iy_i|\le(x_i^2+y_i^2)/2$. Using also the
absolute bounds on the translation multipliers yields
\[
 \begin{aligned}
 |z^\top\Xi_{\alpha,\beta}z|
 &\le e^{kb\delta/h}\eta_b^\ell
       \sum_i\E[(z\cdot Z_i)^2|x_iy_i|\mid\mathcal H]\\
 &\le\frac12e^{kb\delta/h}\eta_b^\ell
       \sum_i\E[(z\cdot Z_i)^2(x_i^2+y_i^2)
                         \mid\mathcal H]\\
 &\le M_2e^{kb\delta/h}\eta_b^\ell\|z\|^2.
 \end{aligned}
\]
Taking the supremum over unit $z$ gives
\[
 \|\Xi_{\alpha,\beta}\|_{\op}
 \le M_2e^{kb\delta/h}\eta_b^\ell.
\]
The bounds on the translation factors and the mixed product are
pointwise, so they remain valid when the shifts and derivative
directions are dependent. Thus the coefficient bounds hold with
$\delta^s$ replaced by $M_s$ in the stated cases. Although
$\Xi_{\alpha,\beta}$ can depend on $\mathcal H$, all the
operator-norm bounds just obtained are deterministic.

It remains to pair these coefficient matrices with the Gaussian
averages and sum over the translation series. For a deterministic
base point $x\in\R^r$, define
\[
 M_{\alpha,\beta}(x):=
 \E_G\left[(GG^\top-I_r)F(x+\sigma G)
       \prod_{a=1}^k\pi_{\alpha_a}(x+\sigma G)
       \prod_{q=1}^\ell\pi_{\beta_q}(x+\sigma G)\right].
\]
Conditional on $\mathcal H$, the matrix $\Xi_{\alpha,\beta}$
and the base $V_0$ are fixed, while $G$ remains standard Gaussian.
We can therefore take $\Xi_{\alpha,\beta}$ outside the Gaussian
expectation and evaluate the remaining average as
$M_{\alpha,\beta}(V_0)$. Averaging over $\mathcal H$ then shows
that expression~\eqref{appx-an:weighted-pair-form} equals
\[
 \sum_{\ell\ge0}(-1)^\ell\binom{k+\ell}{\ell}
       \E\sum_{\alpha,\beta}
       \tr\{\Xi_{\alpha,\beta}M_{\alpha,\beta}(V_0)\}.
\]
Both matrices in the trace are symmetric. We use the
operator--nuclear norm inequality~\cite[Chapter~IV]{Bhatia1997},
which can be seen directly by writing
$M=\sum_{q=1}^r\lambda_q e_qe_q^\top$ in an orthonormal
eigenbasis:
\[
 |\tr(\Xi M)|
 \le\sum_{q=1}^r|\lambda_q|\,|e_q^\top\Xi e_q|
 \le\|\Xi\|_{\op}\sum_{q=1}^r|\lambda_q|
 =\|\Xi\|_{\op}\|M\|_*.
\]
Apply this inequality for each realization of $\mathcal H$
and use the deterministic bound on $\|\Xi_{\alpha,\beta}\|_{\op}$.
For each fixed $\ell$, we obtain
\[
 \begin{aligned}
 \left|\E\sum_{\alpha,\beta}
       \tr\{\Xi_{\alpha,\beta}M_{\alpha,\beta}(V_0)\}\right|
 &\le\E\sum_{\alpha,\beta}
       \|\Xi_{\alpha,\beta}\|_{\op}
       \|M_{\alpha,\beta}(V_0)\|_*\\
 &\le\delta^s e^{kb\delta/h}\eta_b^\ell
       \E_{V_0}\sum_{\alpha,\beta}
       \|M_{\alpha,\beta}(V_0)\|_*.
 \end{aligned}
\]
The last expectation depends only on the unconditional law of
$V_0$. Thus dependence between $\Xi_{\alpha,\beta}$ and $V_0$
causes no difficulty, and we can use the concentrated-base
assumption~\eqref{appx-an:concentration} in its stated,
unconditional form.

Concatenating $\alpha$ and $\beta$ gives an ordered tuple of
$k+\ell$ face labels, including repetitions. Hence the sum of
nuclear norms is exactly of the form covered by
Lemma~\ref{appx-an:lem:matrix}. Its hypotheses hold because
$G$ is independent of $V_0$, the Gaussian scale satisfies
$\delta\sqrt L\le\sigma\le1$, and
$h\ge\mathsf A_b\delta\ge\delta$. The lemma gives
\[
 \E_{V_0}\sum_{\alpha,\beta}\|M_{\alpha,\beta}(V_0)\|_*
 \le C_1^{k+\ell+1}h\kappa L.
\]

The factor $C_1^{k+\ell+1}$ grows with the number of logistic
factors. We therefore enlarge $\mathsf A_b$, if necessary, so that
\[
 C_1\bigl(e^{b/\mathsf A_b}-1\bigr)\le\tfrac12.
\]
For $h\ge\mathsf A_b\delta$, this ensures
$C_1\eta_b\le1/2$. Taking absolute values in the series
representation above and applying the two bounds shows that the
absolute value of~\eqref{appx-an:weighted-pair-form} is at most
\[
 \delta^s h\kappa L\,e^{kb\delta/h}C_1^{k+1}
 \sum_{\ell\ge0}\binom{k+\ell}{\ell}(C_1\eta_b)^\ell
 =\delta^s h\kappa L\,e^{kb\delta/h}C_1^{k+1}
       (1-C_1\eta_b)^{-k-1}.
\]
Here the same binomial-series identity used for the translation
now sums the bounds on its Gaussian averages. Since $k\le2$,
we have $(1-C_1\eta_b)^{-k-1}\le2^{k+1}$ and
$e^{kb\delta/h}\le e^{2b/\mathsf A_b}$. The remaining factors
are therefore bounded by a constant depending only on $b$.

Finally, restore the factor $\sigma^{-2}h^{-s}$ extracted before
\eqref{appx-an:weighted-pair-form}. Each derivative contribution,
before its numerical coefficient, is bounded by
\[
 \sigma^{-2}h^{-s}\,C_b\delta^s h\kappa L
 =\frac{C_b}{\sigma^2}\delta^s h^{1-s}\kappa L.
\]
For $s=1$ there is a single contribution with coefficient $-1$.
For $s=2$ there are two contributions with coefficients $2$ and
$-1$. Absorbing their absolute coefficients into $C_b$ proves
\eqref{appx-an:restoration-bound}.

For the cubic refinement, the zero shifts leave only $\ell=0$,
and the bound on $\Xi_{(j),\varnothing}$ replaces $\delta$ by
$M_1$ in the same trace estimate. For the fourth-order
refinement, the coefficient bound with $M_2$ replaces $\delta^2$
throughout the series calculation. These are the two asserted
improvements. Uniformity over the threshold vector follows from
the coefficient estimates and the uniform matrix bound in
Lemma~\ref{appx-an:lem:matrix}.
\end{proof}

We now apply the lemma to the calibrated path. The retained
summands lose variance as $t$ approaches $1$, so they cannot by
themselves provide a uniform concentration bound for the base.
We reserve half of the Gaussian variance for the base and use
the other half for integration by parts. We first verify the
base and independence assumptions, then check the sharper moment
bounds for the cubic term and the three fourth-order tensor terms.

\begin{proof}[Proof of the seeded estimates~\eqref{eq:seeded-derivatives}]
Fix $0<t<1$ and take $\varepsilon=Lh^2<1/4$. As
in~\eqref{eq:seeded-test}, work with the unnormalized path
$Y_t=\sqrt{1+\varepsilon}\,U_t$ and the rescaled test function
$f=F_{\widetilde a,\widetilde h}$, where
$\widetilde a=\sqrt{1+\varepsilon}\,a$ and
$\widetilde h=\sqrt{1+\varepsilon}\,h$. Thus
$f(Y_t)=F_{a,h}(U_t)$.
Let $G_0,G_1\sim N(0,I_r)$ be independent of each other and of
the original summands, masks, and copies. We realize the path as
\begin{equation}
 \begin{aligned}
 Y_t&=V_{0,t}+\sigma_tG_1,\\
 V_{0,t}&\coloneqq(1+t)\sum_i\mathbb B_{i,t}\zeta_i^0
                      +\sigma_tG_0,\qquad
 \sigma_t^2\coloneqq(t+\varepsilon)/2.
 \end{aligned}
 \label{eq:seeded-gaussian-split}
\end{equation}
Indeed, $\sigma_t(G_0+G_1)$ is centered Gaussian with covariance
$2\sigma_t^2I_r=(t+\varepsilon)I_r$, and it is independent of
all summands, masks, and copies. It therefore has the same joint
law with these variables as $\sqrt{t+\varepsilon}\,G$ does.
Using this same Gaussian sum in $Y_t$ and every removed-summand
base $H_{i,t}$ preserves all the joint laws in the interpolation
identity~\eqref{appx-an:fourth-identity}.

The Gaussian scale already satisfies the lemma's upper bound.
Its lower bound follows from the seed: if the constant
$\mathsf A$ in Proposition~\ref{an:prop:derivative} is at least
$\sqrt2$, then $h\ge\mathsf A\delta$ gives
\[
 \sigma_t^2\ge\varepsilon/2=Lh^2/2\ge\delta^2L,
 \qquad
 \sigma_t^2<(1+1/4)/2<1.
\]
Thus $\delta\sqrt L\le\sigma_t\le1$ uniformly in $t$.

We next verify concentration of $V_{0,t}$. The masks are
independent of the centered summands and have retention
probability $\varpi(t)=(1-t)/(1+t)^2$. Consequently,
\[
 \Cov\left((1+t)\sum_i\mathbb B_{i,t}\zeta_i^0\right)
 =(1+t)^2\varpi(t)\sum_i\E[\zeta_i^0(\zeta_i^0)^\top]
 =(1-t)I_r.
\]
The other component, $\sigma_tG_0$, has covariance
$\sigma_t^2I_r=(t+\varepsilon)I_r/2$ and is independent of this
sum. Since every face normal has unit length, the two components
have common face variances $1-t$ and $(t+\varepsilon)/2$,
respectively.

The retained summands remain independent and centered, and each
has face norm at most $(1+t)\delta\le2\delta$. Conditioning on
the masks and using~\eqref{appx-an:retained-maximum} gives
\[
 \E\left\|(1+t)\sum_i\mathbb B_{i,t}\zeta_i^0\right\|_U
 \le(1+t)\E\|W\|_\infty\le2\E\|W\|_\infty.
\]
Also, $\Gamma G_0$ has the law of the Gaussian target, so
$\E\|\sigma_tG_0\|_U=\sigma_tg_\Sigma\le g_\Sigma$.
If $t\le2/3$, apply the bounded-summand part of
Lemma~\ref{lem:concentration} to the retained sum, whose common
face variance is at least $1/3$. If $t>2/3$, apply its Gaussian
part to $\sigma_tG_0$, whose common face variance is greater
than $1/3$. In each case, the other component is an independent
shift, as allowed by the lemma. Its intrinsic and
dimension-based estimates therefore give, for every threshold
vector $a$ and layer width $w>0$,
\[
 \begin{aligned}
 \Pp\{V_{0,t}\in A_a^w\setminus A_a^{-w}\}
 &\le C(w+\delta)
       \min\{\sqrt L,\,1+\max\{\E\|W\|_\infty,g_\Sigma\}\}\\
 &=C(w+\delta)\kappa.
 \end{aligned}
\]
Here the factors $2$ in the summand envelope and $1/3$ in the
variance lower bound are absorbed into the absolute constant,
and $\kappa$ is defined in~\eqref{eq:intrinsic-boundary-coefficient}.
Thus $V_{0,t}$ is a concentrated base uniformly in $t$.

To apply the paired-derivative lemma, let $\mathcal H$ be
generated by the original summands, the masks at time $t$, and
$G_0$. The base $V_{0,t}$ is then $\mathcal H$-measurable, and
$G_1$ is independent of $\mathcal H$. Both copy arrays
$\zeta_i$ and $\widetilde\zeta_i$ are independent of
$\mathcal H$ and have the same marginal laws as the original
summands. Hence
\[
 \sum_i\E[\zeta_i\zeta_i^\top\mid\mathcal H]
 =\sum_i\E[\widetilde\zeta_i\widetilde\zeta_i^\top
                     \mid\mathcal H]
 =\sum_i\E[\zeta_i^0(\zeta_i^0)^\top]=I_r.
\]
This verifies the conditional second-moment requirement for
either choice of the repeated direction in the lemma.

The cubic derivative is evaluated at $Y_t$ and has zero shift.
At a fourth-order remainder point, 
\eqref{appx-an:translates} gives, for $0\le u\le1$,
\[
 H_{i,t}+u(1+t)\zeta_i=Y_t+v_i,\qquad
 v_i=(1+t)\{u\zeta_i-\mathbb B_{i,t}\zeta_i^0\}.
\]
These shifts are functions of the copies and variables in
$\mathcal H$. Thus the copies and shifts are jointly independent
of $G_1$ conditional on $\mathcal H$, even though a shift may
depend on a derivative direction. Each copy has face norm at
most $\delta$, and
\[
 \|v_i\|_U
 \le(1+t)\{u\|\zeta_i\|_U
                  +\mathbb B_{i,t}\|\zeta_i^0\|_U\}
 \le4\delta.
\]
The lemma therefore applies to $f$ at width
$\widetilde h\ge h$, with $b=4$, once we take
$\mathsf A\ge\max\{\sqrt2,\mathsf A_4\}$. Its concentration
assumption concerns the unconditional law of $V_{0,t}$ verified
above; the conditioning on $\mathcal H$ is used for the moment
bounds and the Gaussian integration.

It remains to verify the sharper moment assumptions with
$M_1=s_3$ and $M_2=s_4$. We first express the original moment
matrices in support coordinates. Recall the $r\times p$ map
\[
 P:=(\Gamma^\top\Gamma)^{-1}\Gamma^\top,
 \qquad \zeta_i^0=P\xi_i.
\]
Since $\Gamma$ has full column rank, $P\Gamma=I_r$, and hence
\[
 P\Sigma P^\top=P\Gamma\Gamma^\top P^\top=I_r.
\]
For $z\in\R^r$, applying the relative moment
bounds~\eqref{eq:relative-moments} to $P^\top z\in\R^p$ gives
\[
 \begin{aligned}
 |z^\top PT_jP^\top z|
 &\le s_3z^\top P\Sigma P^\top z=s_3\|z\|^2,\\
 z^\top PQP^\top z
 &\le s_4z^\top P\Sigma P^\top z=s_4\|z\|^2,
       \qquad Q\in\{Q_j,\widetilde Q_j\}.
 \end{aligned}
\]
The first matrix is symmetric, and the other two are positive
semidefinite. Taking the supremum over unit $z$ in the first
inequality and using the definition of positive-semidefinite
order in the other two yields
\[
 \|PT_jP^\top\|_{\op}\le s_3,\qquad
 PQ_jP^\top\preceq s_4I_r,\qquad
 P\widetilde Q_jP^\top\preceq s_4I_r.
\]
Only $\Gamma^\top\Gamma$ is inverted, so these comparisons
remain valid when $\Sigma$ is singular.

For the cubic term, take $Z_i=Z_{i,1}=\zeta_i$ and zero shift.
Each face normal has the form $u_\alpha=\pm\Gamma^\top e_j$
for some $j\le p$. Since $\Gamma\zeta_i^0=\xi_i$,
\[
 u_\alpha\cdot\zeta_i^0=\pm\xi_{ij}.
\]
The copies are independent of $\mathcal H$ and have the same
laws as the original support-coordinate summands. Thus the
conditional matrix in~\eqref{appx-an:signed-cubic-budget} is
\[
 \begin{aligned}
 \sum_i\E[\zeta_i\zeta_i^\top
                    (u_\alpha\cdot\zeta_i)\mid\mathcal H]
 &=\pm P\left\{\sum_i\E[\xi_i\xi_i^\top\xi_{ij}]\right\}P^\top\\
 &=\pm PT_jP^\top.
 \end{aligned}
\]
Its operator norm is at most $s_3$, so we may take $M_1=s_3$.
Moreover, the copies are independent of $Y_t$. Averaging their
cubic tensors therefore recovers
$K_3=\sum_i\E\zeta_i^{\otimes3}$. Applying
Lemma~\ref{appx-an:lem:restoration} gives
\[
 |\E D^3f(Y_t):K_3|
 =\left|\sum_i\E D^3f(Y_t)[\zeta_i,\zeta_i,\zeta_i]\right|
 \le\frac{Cs_3\kappa L}{\sigma_t^2}.
\]
To return to the normalized path, the chain rule for
$f(x)=F_{a,h}(x/\sqrt{1+\varepsilon})$ gives
\[
 D^3f(Y_t)=(1+\varepsilon)^{-3/2}D^3F_{a,h}(U_t).
\]
Multiplying the preceding bound by $t/3$ and using
$\sigma_t^2=(t+\varepsilon)/2$ now yields the cubic estimate
in~\eqref{eq:seeded-derivatives}:
\[
 \left|\frac{t}{3(1+\varepsilon)^{3/2}}
        \E D^3F_{a,h}(U_t):K_3\right|
 =\frac t3|\E D^3f(Y_t):K_3|
 \le Cs_3\kappa L\,\frac{t}{t+\varepsilon}.
\]

For the fourth-order remainder, the three tensors in
\eqref{appx-an:fourth-identity} correspond, respectively, to
\[
 (Z_i,Z_{i,1},Z_{i,2})
 =(\zeta_i,\zeta_i,\zeta_i),\quad
  (\widetilde\zeta_i,\zeta_i,\zeta_i),\quad
  (\widetilde\zeta_i,\widetilde\zeta_i,\zeta_i),
\]
where the first entry supplies the repeated pair in
$D^4f[Z_i,Z_i,Z_{i,1},Z_{i,2}]$. To verify
\eqref{appx-an:weighted-fourth-budget}, consider either remaining
slot and fix a face $u_\alpha=\pm\Gamma^\top e_j$.
When that slot uses the repeated copy, the projection is squared
and the sign disappears. For example,
\[
 \begin{aligned}
 \sum_i\E[\zeta_i\zeta_i^\top
                  (u_\alpha\cdot\zeta_i)^2\mid\mathcal H]
 &=P\left\{\sum_i\E[\xi_i\xi_i^\top\xi_{ij}^2]\right\}P^\top\\
 &=PQ_jP^\top\preceq s_4I_r.
 \end{aligned}
\]
The same calculation applies with $\widetilde\zeta_i$ in place
of $\zeta_i$. When the slot uses the independent copy, its
squared projection factors from the repeated pair:
\[
 \begin{aligned}
 \sum_i\E[\widetilde\zeta_i\widetilde\zeta_i^\top
                  (u_\alpha\cdot\zeta_i)^2\mid\mathcal H]
 &=\sum_i\E[\widetilde\zeta_i\widetilde\zeta_i^\top]
                   \E[(u_\alpha\cdot\zeta_i)^2]\\
 &=P\left\{\sum_i\E[\xi_i\xi_i^\top]\E[\xi_{ij}^2]\right\}P^\top\\
 &=P\widetilde Q_jP^\top\preceq s_4I_r.
 \end{aligned}
\]
Here conditioning on $\mathcal H$ leaves the copy laws and
their mutual independence unchanged. Thus the two slots give
$PQ_jP^\top$ in the first pattern,
$P\widetilde Q_jP^\top$ in the second pattern, and one of each
in the third. We may take $M_2=s_4$ in all three cases. The
shift can depend on these same copies, as allowed by the lemma.

Fix the remainder parameter $u\in[0,1]$. At the corresponding
points $Y_t+v_i$, the lemma with $s=2$ and width
$\widetilde h$ gives, for each of the three patterns,
\[
 \left|\sum_i\E D^4f(Y_t+v_i)
                [Z_i,Z_i,Z_{i,1},Z_{i,2}]\right|
 \le\frac{Cs_4\kappa L}{\widetilde h\,\sigma_t^2}.
\]
This estimate controls the sum over $i$ before taking its
absolute value. The scalar kernels $a_t(u),b_t(u),c_t(u)$ in
\eqref{appx-an:fourth-identity} are common to all indices, so
each can be factored out of its corresponding sum. Applying
the triangle inequality to these three sums and then integrating
over $u$ gives
\[
 \begin{aligned}
 |R_f(t)|
 &\le\frac{Cs_4\kappa L}{\widetilde h\,\sigma_t^2}
       \int_0^1\bigl(|a_t(u)|+|b_t(u)|+|c_t(u)|\bigr)\dd u\\
 &\le\frac{Cs_4\kappa L}{\widetilde h\,\sigma_t^2}.
 \end{aligned}
\]
The last step uses the uniform bounds on the absolute kernel
integrals established in Subsection~\ref{appx-an:sec:path}.
Finally,
\[
 \frac1{\widetilde h\,\sigma_t^2}
 =\frac{2}{\sqrt{1+\varepsilon}\,h(t+\varepsilon)}
 \le\frac{2}{h(t+\varepsilon)}.
\]
Since $f(Y_t)=F_{a,h}(U_t)$, the interpolation identity and
\eqref{eq:seeded-test} identify $R_f(t)$ with
$R_{4,a,h}(t)$. Consequently,
\[
 |R_{4,a,h}(t)|\le\frac{Cs_4\kappa L}{h(t+\varepsilon)},
\]
which is the fourth-order bound in~\eqref{eq:seeded-derivatives}
and completes the proof of the seeded estimates.
\end{proof}

\subsection{A matrix estimate for the logistic weights}
\label{appx-an:sec:matrix}

{To prove~\eqref{appx-an:matrix-bound} below, we use the covariance bound for}
a standard Gaussian law tilted by a log-concave weight. Pr\'ekopa's
theorem~\cite[Theorem~6]{Prekopa1973} implies that this covariance is
at most \(I_r\) in the positive-semidefinite order. The nuclear norm
can then be bounded by the squared length of the tilted mean and a
signed trace term. We control the means relative to a single face and sum the
weights before estimating the trace. A Gamma mixture expresses the
latter through a common-threshold derivative of a Gaussian distribution
function, bounded in Subsection~\ref{appx-an:sec:threshold}.
For a related application to convex cells,
see~\cite[Lemma~12]{GaoZhang}.

For the mean estimate, we compare each weight with a reference weight
depending on one normal coordinate. A second moment controls the
component along that normal, and the logarithm of the ratio of
reference mass to tilted mass controls the orthogonal component.
Using the constant weight \(1\) as the reference would instead introduce
the logarithm of the reciprocal tilted mass. The single-face reference
avoids a logarithm of the smoothing scale and refines the entropy argument
in~\cite[Lemma~12]{GaoZhang}.

\begin{lemma}[Gaussian mean relative to a single face]
\label{appx-an:lem:mean}
Let \(u\in\R^r\) be a unit vector and
\(\beta:\R\to[0,1]\) be a measurable function. Denote
\(b(g)\coloneqq\beta(u\cdot g)\) for all \(g\in\R^r\).
Also, let \(w:\R^r\to[0,1]\) be a measurable function
satisfying \(w(g)\le b(g)\) for all $g\in\R^r$.
In addition, suppose \(p_b\coloneqq\E b(G)>0\) and
\(p_w\coloneqq\E w(G)>0\), where \(G\sim N(0,I_r)\),
and let \(\mu_w\coloneqq\E[Gw(G)]/p_w\). Then
\begin{equation}
 p_w\norm{\mu_w}^2
 \le \E[w(G)(u\cdot G)^2]
        +2p_w\log(p_b/p_w).
 \label{appx-an:mean-entropy}
\end{equation}
\end{lemma}

\begin{proof}
Let $P_w$ be the probability measure on $\R^r$ with density
\[
\frac{d P_w(g)}{dg}\coloneqq \frac{w(g)\varphi_r(g)}{p_w},
\]
where $\varphi_r$ denotes the standard Gaussian density
on $\R^r$. Let $G_w\sim P_w$. By the definition of
$\mu_w$, we then have $\mu_w=\E[G_w]$.

We first bound the component of $\mu_w$ that is orthogonal to $u$. To do so, fix any \(t\in\R^r\) satisfying \(t\cdot u=0\).
By Jensen's inequality and the assumption \(w\le b\),
\begin{align*}
p_w\exp(t\cdot\mu_w)
&=p_w\exp\bigl(\E[t\cdot G_w]\bigr)
\le p_w\E[\exp(t\cdot G_w)]\\
&=\E[w(G)\exp(t\cdot G)]
\le\E[b(G)\exp(t\cdot G)].
\end{align*}
Also, since \(G\sim N(0,I_r)\),
the random variables \(u\cdot G\) and \(t\cdot G\)
are independent. In addition, \(t\cdot G\sim N(0,\norm{t}^2)\). Therefore,
\begin{align*}
\E[b(G)\exp(t\cdot G)]
&=\E[\beta(u\cdot G)\exp(t\cdot G)]\\
&=\E[\beta(u\cdot G)]\E[\exp(t\cdot G)]
=p_b\exp\bigl(\norm{t}^2/2\bigr).
\end{align*}
Combining the preceding inequalities and taking
logarithms yields
\begin{equation}
t\cdot\mu_w
\le \log(p_b/p_w)+\frac{\norm{t}^2}{2},
\qquad t\perp u.
\label{appx-an:mean-orthogonal-bound}
\end{equation}
Next, let $t=\mu_w-(u\cdot\mu_w)u$.
Since \(\norm{u}=1\), we have \(t\cdot u=0\), and so
$ t\cdot\mu_w =\norm{t}^2$.
Substituting this choice into
\eqref{appx-an:mean-orthogonal-bound} gives
$\norm{t}^2 \le \log(p_b/p_w)+\norm{t}^2/2$, and so $\norm{t}^2\leq 2\log(p_b/p_w)$.
Combining this inequality with the equality $\norm{\mu_w}^2 = (u\cdot\mu_w)^2 + \norm{t}^2$ yields
\begin{equation}
\norm{\mu_w}^2-(u\cdot\mu_w)^2
=\norm{t}^2
\le 2\log(p_b/p_w).
\label{appx-an:mean-orthogonal-result}
\end{equation}
Next, we bound the component of $\mu_w$ that is parallel to \(u\). By the definition of \(\mu_w\)
and the Cauchy--Schwarz inequality,
\begin{align*}
p_w(u\cdot\mu_w)^2
&=\frac{\bigl(\E[w(G)(u\cdot G)]\bigr)^2}{p_w}
\le
\frac{\E[w(G)]\E[w(G)(u\cdot G)^2]}{p_w}
=\E[w(G)(u\cdot G)^2].
\end{align*}
Combining this bound with \eqref{appx-an:mean-orthogonal-result},
we obtain
\begin{align*}
p_w\norm{\mu_w}^2
&=p_w(u\cdot\mu_w)^2
 +p_w\bigl(\norm{\mu_w}^2-(u\cdot\mu_w)^2\bigr)
 \le \E[w(G)(u\cdot G)^2]
 +2p_w\log(p_b/p_w),
\end{align*}
as claimed.
\end{proof}

We can now control the matrix sum without a factor depending on the
rank. We must sum the signed trace terms before taking an absolute value.
The resulting scalar weight admits a Gamma-mixture representation,
to which the next subsection's threshold estimate applies.

\begin{lemma}[Matrix estimate for logistic weights]
\label{appx-an:lem:matrix}
Let \(V_0\in\R^r\) satisfy the concentrated-base
bound~\eqref{appx-an:concentration}{ with coefficient \(\kappa\)}, and let \(G\sim N(0,I_r)\) be
independent of \(V_0\). Suppose that
\(\delta\sqrt L\le\sigma\le1\) and \(h\ge\delta\), and {define}
\(U\coloneqq V_0+\sigma G\). There is an absolute constant
\(C_1\ge1\) such that, for every threshold vector \(a\in\R^m\)
and every integer \(k\ge1\),
\begin{equation}
 \E_{V_0}\sum_{\alpha_1,\ldots,\alpha_k=1}^m
 \left\|\E_G\left[(GG^\top-I_r)F_{a,h}(U)
              \prod_{b=1}^k\pi_{\alpha_b}(U)\right]\right\|_*
 \le C_1^{k+1}h{\kappa L}.
 \label{appx-an:matrix-bound}
\end{equation}
Here \(\pi_j\) are the face probabilities associated with
\(F_{a,h}\), the sum runs over ordered face tuples, and
\(\norm{\cdot}_*\) is the nuclear norm.
\end{lemma}

\begin{proof}

Let \(w:\R^r\to[0,1]\) be a log-concave function
satisfying \(p_w\coloneqq\E[w(G)]>0\) and let
\(P_w\) be the probability measure on \(\R^r\) with density
\[
\frac{d P_w(g)}{dg}
\coloneqq\frac{w(g)\varphi_r(g)}{p_w},
\]
where \(\varphi_r\) denotes the standard Gaussian density.
Let \(G_w\sim P_w\), and denote its mean and covariance
matrix by
\[
\mu_w\coloneqq\E[G_w]=\frac{\E[Gw(G)]}{p_w},\qquad
K_w\coloneqq\E[(G_w-\mu_w)(G_w-\mu_w)^\top]=\frac{\E[GG^\top w(G)]}{p_w}-\mu_w\mu_w^\top.
\]

We first show that \(K_w\preceq I_r\), i.e. the matrix $I_r - K_w$ is positive semidefinite. To this end, observe that the function $(g, z)\mapsto e^{-\norm{g-z}^2/2}w(g)$
is log-concave. Therefore, the function \(H:\R^r\to(0,\infty)\) defined by
\[
H(z)\coloneqq
\int_{\R^r}e^{-\norm{g-z}^2/2}w(g)\dd g,\qquad z\in\R^r,
\]
is log-concave by Pr\'ekopa's theorem~\cite[Theorem~6]{Prekopa1973}.
In particular,
\begin{equation}\label{appx-an:prekopa-consequence}
\nabla^2\log H(0)\preceq0.
\end{equation}
We next calculate $\nabla^2\log H(0)$.
We have
\begin{align*}
H(z)
&=\int_{\R^r}
 e^{-\norm g^2/2+z\cdot g-\norm z^2/2}
 w(g)\dd g
 =(2\pi)^{r/2}e^{-\norm z^2/2}
 \E[w(G)e^{z\cdot G}].
\end{align*}
Denoting \(M(z)\coloneqq\E[w(G)e^{z\cdot G}]\), $z\in\R^r$, and taking logarithms, we obtain
\[
\log H(z)
=\frac r2\log(2\pi)-\frac{\norm z^2}{2}
 +\log M(z).
\]
Also, for all \(a, b = 1,\dots, r\),
\[
\partial_a M(z)
=\E[G_aw(G)e^{z\cdot G}],
\qquad
\partial_{ab}M(z)
=\E[G_aG_bw(G)e^{z\cdot G}].
\]
Therefore,
\begin{align*}
\partial_{ab}\log H(z)
&=-\mathbf 1\{a=b\}
 +\frac{\partial_{ab}M(z)}{M(z)}
 -\frac{\partial_aM(z)\partial_bM(z)}{M(z)^2}.
\end{align*}
Hence, given that \(M(0)=p_w\), we obtain
\begin{align*}
\partial_{ab}\log H(0)
&=-\mathbf 1\{a=b\}
 +\frac{\E[G_aG_bw(G)]}{p_w}
  -\frac{\E[G_aw(G)]\E[G_bw(G)]}{p_w^2}.
\end{align*}
Thus, $\nabla^2\log H(0)=K_w-I_r$, and so $K_w\preceq I_r$ by \eqref{appx-an:prekopa-consequence}, as required.

We now use this covariance matrix inequality to prove that
\begin{equation}
\left\|\E[(GG^\top-I_r)w(G)]\right\|_*
\le 2p_w\norm{\mu_w}^2
 -\E[H_2(G)w(G)],
\label{appx-an:matrix-trace-reduction}
\end{equation}
where we define \(H_2(g)\coloneqq\norm g^2-r\), $g\in\R^r$. To this end, observe that
\begin{align*}
\E[(GG^\top-I_r)w(G)]
&=p_w\bigl(K_w+\mu_w\mu_w^\top-I_r\bigr)
=p_w\mu_w\mu_w^\top-p_w(I_r-K_w).
\end{align*}
Also, define $A\coloneqq p_w\mu_w\mu_w^\top$ and $B\coloneqq p_w(I_r-K_w)$ and note that
both matrices are positive semidefinite, since
\(p_w>0\) and \(K_w\preceq I_r\).
Recall that the nuclear norm of a symmetric
matrix is the sum of the absolute values of its
eigenvalues. In particular, for a positive
semidefinite matrix it equals its trace.
The triangle inequality therefore gives
\begin{align*}
\norm{A-B}_*
&\le\norm A_*+\norm B_*
=\tr A+\tr B
=2\tr A-\tr(A-B).
\end{align*}
Here, $\tr A=p_w\norm{\mu_w}^2$ and
\begin{align*}
\tr(A-B)
&=\tr\E[(GG^\top-I_r)w(G)]
=\E[(\norm G^2-r)w(G)]
=\E[H_2(G)w(G)],
\end{align*}
by the definition of $H_2$ above. Combining these identities yields \eqref{appx-an:matrix-trace-reduction}, as required.

In the remainder of the proof, we use the general bound \eqref{appx-an:matrix-trace-reduction} to prove \eqref{appx-an:matrix-bound}. For each \(\alpha=(\alpha_1,\ldots,\alpha_k)\in\{1,\dots,m\}^k\), denote
\[
 w_\alpha(g)\coloneqq F(V_0+\sigma g)
                 \prod_{a=1}^k\pi_{\alpha_a}(V_0+\sigma g),\qquad g\in\R^r,
\]
\[
p_\alpha\coloneqq \E_G[w_\alpha(G)],
 \quad \mu_\alpha\coloneqq \frac{\E_G[Gw_\alpha(G)]}{p_\alpha}.
\]
Here, $\E_G[w_\alpha(G)]>0$. The definitions of the logistic weights give
\[
 \log w_\alpha(g)
 =\sum_{a=1}^k q_{\alpha_a}(V_0+\sigma g)
  -(k+1)\log N(V_0+\sigma g).
\]
Here, each \(q_j(V_0+\sigma g)\) is affine in \(g\) and it is straightforward to check that \(\log N(V_0+\sigma g)\) is convex in \(g\), e.g. see \cite[Section~3.1.5, pp.~72--74]{BoydVandenberghe2004}. Thus,
\(\log w_\alpha\) is concave, so \(w_\alpha\) is log-concave, as required. This implies that we can apply \eqref{appx-an:matrix-trace-reduction} with $w=w_\alpha$.

Now, use $\theta(x) = 1 - F(x) = \sum_{j=1}^m \pi_j(x)$ as in \eqref{appx-an:basic-notation} and observe that
\begin{equation}\label{appx-an:total-product-weight}
 \sum_\alpha w_\alpha(g)=F(V_0+\sigma g)\theta(V_0+\sigma g)^k,
\end{equation}
where the sum is over all $\alpha\in\{1,\dots,m\}^k$. Applying \eqref{appx-an:matrix-trace-reduction} with $w = w_{\alpha}$, summing over $\alpha$, and taking expectations over $V_0$ thus shows that
\begin{align}
& \E_{V_0}\sum_{\alpha_1,\ldots,\alpha_k}
 \left\|\E_G\left[(GG^\top-I_r)F(U)
              \prod_{a=1}^k\pi_{\alpha_a}(U)\right]\right\|_* \nonumber\\
& \qquad\qquad\qquad \le  2\sum_\alpha \E[p_\alpha\norm{\mu_\alpha}^2]
- \E[H_2(G)F(U)\theta(U)^k]. \label{appx-an:matrix-main-split}
\end{align}

We first bound the first term on the right-hand side of \eqref{appx-an:matrix-main-split}, i.e. the mean term.
For each $j=1,\dots, m$, set
\[
 b_j(g)\coloneqq
 \frac{e^{q_j(V_0+\sigma g)}}
 {\{1+e^{q_j(V_0+\sigma g)}\}^2},
 \qquad g\in\R^r,
 \]
 and \(p_{b_j}\coloneqq\E_G[b_j(G)]\).
Since $b_j$ depends only on $u_j\cdot g$ and
$w_\alpha\le F\pi_{\alpha_1}\le b_{\alpha_1}\leq 1$ for all $\alpha\in\{1,\dots, m\}^k$,
Lemma~\ref{appx-an:lem:mean} and
$|u_j\cdot G|\le\norm G_U$ give
\[
 \sum_\alpha p_\alpha\norm{\mu_\alpha}^2
 \le \sum_{\alpha}\E_G[w_{\alpha}(G)\norm G_U^2]
 +2\sum_\alpha p_\alpha
 \log(p_{b_{\alpha_1}}/p_\alpha).
\] 
Thus,
\begin{equation}
\sum_\alpha \E[p_\alpha\norm{\mu_\alpha}^2]
 \le \E[F(U)\theta(U)^k\norm G_U^2]
 +2\sum_\alpha \E[p_\alpha
 \log(p_{b_{\alpha_1}}/p_\alpha)],
 \label{appx-an:entropy-sum}
\end{equation}
where we used \eqref{appx-an:total-product-weight}.  The first term on the right-hand side of \eqref{appx-an:entropy-sum} is at most
$Ch{\kappa L}$: note that $\theta(U)^k \leq \theta(U)$, apply
\eqref{appx-an:weight-concentration} conditional on $G$, and then use $\E\norm G_U^2\le CL$.
To bound the second term on the right-hand side of \eqref{appx-an:entropy-sum}, 
define
\[
 P\coloneqq\sum_\alpha p_\alpha,\qquad
 B\coloneqq\sum_\alpha p_{b_{\alpha_1}}
 =m^{k-1}\sum_j p_{b_j}.
\]
Then
\[
 \sum_\alpha \E [p_\alpha
 \log(p_{b_{\alpha_1}}/p_\alpha)]
 \le \E[P\log(B/P)]
 \le \E[P]\log(\E[B]/\E[P]),
\]
where the first inequality follows from the log-sum inequality (e.g., see \cite[Theorem~2.7.1]{CoverThomas2006}) and the second from Jensen's inequality and concavity of $(x,y)\mapsto x\log(y/x)$ on $(0,\infty)^2$.
Here, by~\eqref{appx-an:weight-concentration} and \eqref{appx-an:total-product-weight}, $\E[P] \le Ch{\kappa}$. Also, taking the limit as $a_k\to\infty$ for all $k\neq j$ in \eqref{appx-an:weight-concentration} and applying Lebesgue's dominated convergence theorem, $\E[b_j(G)] \leq Ch{\kappa}$ for all $j=1,\dots, m$, so that  $\E[B] \le Cm^kh{\kappa}$.
Now, writing $\E[P] =h{\kappa}\,x$, where $0<x\le C$,
and using $\log m\le CL$, we obtain
\[
 \E[P]\log(\E[B]/\E[P])
 \le h{\kappa}\,x\log(Cm^k/x)
 \le C(k+1)h{\kappa L}.
\]
Substituting these bounds into \eqref{appx-an:entropy-sum} yields
\begin{equation}
  \sum_\alpha \E[p_\alpha\norm{\mu_\alpha}^2]
 \le C(k+1)h{\kappa L},
 \label{appx-an:tilted-mean-bound}
\end{equation}
yielding the required bound for the first term on the right-hand side of \eqref{appx-an:matrix-main-split}.

We next bound the second term on the right-hand side of \eqref{appx-an:matrix-main-split}, i.e. the signed trace term.
First, suppose that the normal vectors $u_1,\dots, u_m$ are linearly independent. For
\(s\in\R\), {define} \(F_s:=F_{a+s\mathbf1,h}\) and
\(\theta_s:=1-F_s\).
Since
\( \partial_s F_s= F_s\theta_s/h\) and \(\partial_s\theta_s^k=-kF_s\theta_s^k/h\),
we have
\begin{equation}\label{eq: F theta k trick}
 F\theta^k=-\frac hk
 \left.\partial_s\theta_s^k\right|_{s=0}.
\end{equation}
Also,
\begin{equation}\label{eq: theta expansion}
 \theta_s^k
 =\sum_{\ell=0}^k(-1)^\ell\binom{k}{\ell}F_s^\ell.
\end{equation}
Here, $\partial_s F_s^0 = 0$, and so we only need to consider $\partial_s F_s^{\ell}$ for $\ell\geq 1$. To do so, fix $\ell\ge1$ and let $T_\ell$ have the Gamma
distribution with shape $\ell$ and rate \(1\), independent of $V_0$ and $G$. Also, let $\omega_1,\ldots,\omega_m$ be independent exponential random variables with mean \(1\), independent of $T_\ell$, $V_0$, and $G$.
Define
\[
 Q_s :=\left\{x:
 u_j\cdot x\le a_j+s+h\log(\omega_j/T_\ell),
 \ j\le m\right\}.
\]
Then
\[
 \begin{aligned}
 \Pp(x\in Q_s\mid T_\ell)
 &=\prod_{j=1}^m
 \exp\left\{-T_\ell
 e^{(u_j\cdot x-a_j-s)/h}\right\}
 =\exp\{-T_\ell e^{-s/h}T(x)\},
 \end{aligned}
\]
where $T(x) = \sum_{j=1}^m e^{(u_j\cdot x-a_j)/h}$. Thus, by the law of iterated expectations,
\begin{equation}
\Pp(x\in Q_s) = \E[\exp\{-T_\ell e^{-s/h}T(x)\}] = \bigl(1+e^{-s/h}T(x)\bigr)^{-\ell} = F_s(x)^\ell,
 \label{appx-an:gamma-mixture}
\end{equation}
where the second equality follows from the Gamma Laplace transform formula.

\begingroup
Further, define the
Gaussian trace function and its common-threshold derivative by
\begin{equation}
 J(b):=\E_G[H_2(G)\ind\{u_j\cdot G\le b_j,\ j\le m\}],
 \quad \mathcal D(b):=\sum_j\partial_jJ(b),\quad b=(b_1,\dots,b_m)\in\R^m.
 \label{appx-an:threshold-functions}
\end{equation}
Also, define $b^0 = (b_1^0,\dots,b_m^0)\in\R^m$ by setting
\[
 b_j^0:=\{a_j+h\log(\omega_j/T_\ell)-u_j\cdot V_0\}/\sigma,\quad j=1,\dots, m.
\]
The event \(U\in Q_s\) is then equivalent to
\(u_j\cdot G\le b_j^0+s/\sigma\) for every \(j\) since $U = V_0 + \sigma G$. 
Then, by the law of iterated expectations,
\[
 \E[H_2(G)F_s(U)^\ell] = \E[H_2(G)\ind\{U\in Q_s\}]
 =\E[J(b^0+s\mathbf1/\sigma)],
\]
and hence
\[
 \frac{d}{ds}\E[H_2(G)F_s(U)^\ell]
 =\frac1\sigma\E[\mathcal D(b^0+s\mathbf1/\sigma)].
\]
To justify differentiation under the expectation, conditioning on
$u_j\cdot G=b_j$ gives
$|\partial_jJ(b)|\le\varphi(b_j)(b_j^2+2r)$, where $\varphi$ is the
standard Gaussian density. This bound has a finite supremum over $b$,
so dominated convergence applies for fixed $r$ and $m$. Moreover,
\begin{equation}\label{eq: key bound for surface integrals}
 \left|\frac{d}{ds}\E[H_2(G)F_s(U)^\ell]\right|
 \le\frac1\sigma\E[|\mathcal D(b^0+s\mathbf1/\sigma)|]
 \le C{\kappa\left(L+\frac\delta\sigma L^{3/2}\right)},
\end{equation}
where the last inequality follows by conditioning on $T_\ell$ and
$\omega$ and applying Lemma~\ref{appx-an:lem:threshold} below, whose
bound is uniform in the thresholds. Combining \eqref{eq: F theta k trick}, \eqref{eq: theta expansion}, and \eqref{eq: key bound for surface integrals} and using
\(\sum_{\ell=1}^k\binom{k}{\ell}\le2^k\), we obtain
\[
 \left|\E[H_2(G)F(U)\theta(U)^k]\right|
 \le\frac{C2^kh}{k}
       {\kappa\left(L+\frac\delta\sigma L^{3/2}\right)}.
\]
This gives the required bound for the second term on the right-hand side of \eqref{appx-an:matrix-main-split} when the normal vectors $u_1,\dots, u_m$ are linearly independent. Combining it with \eqref{appx-an:matrix-main-split} and
\eqref{appx-an:tilted-mean-bound} gives
\begin{equation}
 \E_{V_0}\sum_{\alpha_1,\ldots,\alpha_k=1}^m
 \left\|\E_G\left[(GG^\top-I_r)F_{a,h}(U)
              \prod_{b=1}^k\pi_{\alpha_b}(U)\right]\right\|_*
 \le C_1^{k+1}h{\kappa\left(L+\frac\delta\sigma L^{3/2}\right)}
 \label{appx-an:matrix-bound-unsimplified}
\end{equation}
for linearly independent normal vectors.

For arbitrary normal vectors, take \(\lambda\in(0,1)\) and define the following vectors in $\R^{r+m}$:
\[
 u_j^\lambda:=(\sqrt{1-\lambda^2}\,u_j,\lambda e_j),\qquad
 V_0^\lambda:=(V_0/\sqrt{1-\lambda^2},0),\qquad
 \widehat G:=(G,G'),
\]
where \(e_j\) is the \(j\)th coordinate vector in \(\R^m\), and
\(G'\sim N(0,I_m)\) is independent of $V_0$ and $G$. Also, define
\[
Q^\lambda(a):=\{z\in\mathbb R^{r+m}:u_j^\lambda\cdot z\le a_j,\ j\le m\},\quad a=(a_1,\dots,a_m)\in\R^m.
\]
Then the \(u_j^\lambda\) are linearly independent unit vectors in $\R^{r+m}$ and
since \(u_j^\lambda\cdot V_0^\lambda=u_j\cdot V_0\) for every \(j\), it follows from \eqref{appx-an:concentration} that for all \(a\in\mathbb R^m\) and \(s>0\),
\[
\begin{aligned}
\mathbb P\{V_0^\lambda\in Q^\lambda(a)^s\setminus Q^\lambda(a)^{-s}\}
&=\mathbb P\{V_0\in A_a^s\setminus A_a^{-s}\} \le C_0(s+\delta){\kappa}.
\end{aligned}
\]
Proceeding as before, we therefore obtain \eqref{appx-an:matrix-bound-unsimplified} with $V_0$, $G$, $I_r$, and $U=V_0+\sigma G$ replaced by $V_0^{\lambda}$, $\widehat G$, $I_{r+m}$, and $U^{\lambda} = V_0^{\lambda} + \sigma \widehat G$, respectively, and appropriately adjusted $F_{a, h}$ and $\pi_{\alpha}$. Also, as \(\lambda\downarrow0\),
\[
 u_j^\lambda\cdot(V_0^\lambda+\sigma\widehat G)
 =u_j\cdot V_0+\sigma\sqrt{1-\lambda^2}\,u_j\cdot G
                  +\sigma\lambda G'_j
 \longrightarrow u_j\cdot U.
\]
The weights therefore converge to their original values. Their
boundedness and the Gaussian second moments allow conditional dominated
convergence in each matrix entry. Continuity of the nuclear norm and
Fatou's lemma for the expectation over $V_0$ then give
\[
 \E_{V_0}\sum_{\alpha_1,\ldots,\alpha_k=1}^m
 \left\|\E_{\widehat G}\left[(\widehat G\widehat G^\top-I_{r+m})F_{a, h}(U)
              \prod_{b=1}^k\pi_{\alpha_b}(U)\right]\right\|_*
 \le C_1^{k+1}h{\kappa\left(L+\frac\delta\sigma L^{3/2}\right)}.
\]
Using \(\sigma\ge\delta\sqrt L\), we thus obtain \eqref{appx-an:matrix-bound} since replacing any $(r+m)\times(r+m)$ matrix by its $r\times r$ upper-left block cannot make the nuclear norm larger.
\par\endgroup
\end{proof}

\begin{remark}[Gaussian covariance contraction]\label{appx-rem:covariance-contraction}
For \(G\sim N(0,I_r)\) and a log-concave weight \(w:\R^r\to[0,1]\) with
\(p_w=\E w(G)>0\), let \(Q_w\) be the probability measure
\(dQ_w=(w/p_w)\,d\Phi_r\), and let \(K_w\) be its covariance.
The Pr\'ekopa argument in the proof above gives \(K_w\preceq I_r\)
even when \(w\) is a convex-cell indicator. For smooth positive
\(w\), the same bound follows from the Brascamp--Lieb variance
inequality~\cite{BrascampLieb1976}: the potential
\(V(g)=\norm g^2/2-\log w(g)\) has \(\nabla^2V\succeq I_r\),
so \(\operatorname{Var}_{Q_w}(a\cdot G)\le\norm a^2\) for every
\(a\in\R^r\).
\end{remark}

\subsection{An averaged Gaussian threshold estimate}
\label{appx-an:sec:threshold}
\begingroup
We now derive a bound for the common-threshold derivative in
\eqref{appx-an:threshold-functions} in the case when the normal vectors $u_1,\dots, u_m$ are linearly independent.

\begin{lemma}[Averaged Gaussian threshold derivative]
\label{appx-an:lem:threshold}
Suppose the unit normal vectors $u_1,\dots, u_m$ are linearly independent and let \(V_0\) be a
concentrated base{ with coefficient \(\kappa\)}. Take \(G\sim N(0,I_r)\) independent of \(V_0\). Fix \(\sigma>0\) and let $J(b)$ and $\mathcal D(b)$ be defined as in \eqref{appx-an:threshold-functions}.
Also, for every \(a\in\R^m\), {define}
\(b_j(V_0):=(a_j-u_j\cdot V_0)/\sigma\), $j=1,\dots, m$. Then
\begin{equation}
 \E[|\mathcal D(b(V_0))|]
{\le C\kappa\bigl(\sigma L+\delta L^{3/2}\bigr).}
 \label{appx-an:threshold-bound}
\end{equation}
\end{lemma}

\begin{proof}
Introduce the centered Gaussian random vector $X=(X_1,\dots,X_m)\in\R^m$ by setting $X_j := u_j\cdot G$, $j=1,\dots,m$ and let
\[
 C(b):=\Pp(X\leq b) = \Pp\{u_j\cdot G\le b_j,\ j\le m\},\qquad b\in\R^m,
\]
be its distribution function. Also, denote
\[
 C_j:=\partial_jC,\qquad C_{jl}:=\partial_j\partial_lC,
 \qquad \rho_{jl}:=u_j\cdot u_l,\qquad j, l=1,\dots, m.
\]
We first express \(\mathcal D\) through the first and mixed second
derivatives of \(C\). For $v>0$, let \(G_v\sim N(0,vI_r)\). Its density is
\[
\phi_{r, v}(g)
=(2\pi v)^{-r/2}\exp\left(-\frac{\|g\|^2}{2v}\right),\quad g\in\R^r,
\]
and so
\[
\frac{\partial}{\partial v}\phi_{r, v}(g)
=\frac12\left(\frac{\|g\|^2}{v^2}-\frac rv\right)
\phi_{r, v}(g).
\]
Since
\[
\Pp\{\sqrt v\,X\le b\}=\Pp\{u_j\cdot G_v\le b_j,\ j\le m\}
=\int_{\{g:u_j\cdot g\le b_j,\ j\le m\}}
\phi_{r,v}(g)\,dg,
\]
differentiating under the integral sign at \(v=1\) gives
\[
\left.\frac{d}{dv}\Pp\{\sqrt v\,X\le b\}\right|_{v=1}
=\frac12\E\left[
(\|G\|^2-r)\mathbf 1\{X\le b\}
\right]
=\frac12J(b).
\]
On the other hand, since the normal vectors $u_1,\dots, u_m$ are linearly independent, the covariance matrix of $X$ is positive definite, and so its distribution function $C$ is continuously differentiable. Thus, 
\[
\left.\frac{d}{dv}\Pp\{\sqrt v\,X\le b\}\right|_{v=1} = \left.\frac{d}{dv}C(b/\sqrt v)\right|_{v=1}
=-\frac12\sum_{j=1}^m b_jC_j(b).
\]
Combining the two identities gives
\begin{equation}\label{appx-an:radial-trace}
J(b)=-\sum_{j} b_jC_j(b).
\end{equation}

Further, we claim that
\begin{equation}
 \mathcal D(b)=\sum_j(b_j^2-1)C_j(b)
   -\sum_{j<l}(1-\rho_{jl})(b_j+b_l)C_{jl}(b).
 \label{appx-an:threshold-identity}
\end{equation}
To prove this claim,
fix \(j\) and recall that \(\rho_{jl}=\E[X_jX_l]=u_j\cdot u_l\).
Since \(X\) is Gaussian and \(\E[X_j^2]=1\), the residual
vector \((X_l-\rho_{jl}X_j)_{l\ne j}\) is independent of \(X_j\).
Thus,
\begin{equation}\label{appx-an:conditional-first-derivative}
C_j(b)
=\varphi(b_j)\Pp\{X_l-\rho_{jl}X_j \le b_l-\rho_{jl}b_j,\ l\ne j\},
\end{equation}
where \(\varphi\) is the standard normal density. To differentiate this expression further, let
\[
\Psi_j(c):=\Pp\{X_l-\rho_{jl}X_j\le c_l,\ l\ne j\},
\qquad c=(c_l)_{l\ne j}.
\]
The linear independence of \(u_1,\ldots,u_m\) implies that
the residual vector has a nondegenerate Gaussian distribution,
so \(\Psi_j\) is continuously differentiable. Substituting $\Psi_j$ into the expression for $C_j$, taking the derivative with respect to $b_j$, and using \(\varphi'(b_j)=-b_j\varphi(b_j)\) then gives
\begin{align*}
C_{jj}(b)
&=-b_jC_j(b)
 -\varphi(b_j)\sum_{l\ne j}\rho_{jl}
 \partial_l\Psi_j\big((b_q-\rho_{jq}b_j)_{q\ne j}\big)=-b_jC_j(b)-\sum_{l\ne j}\rho_{jl}C_{jl}(b),
\end{align*}
where the second equality follows by noting that differentiating the
expression for \(C_j(b)\) with respect to \(b_l\) yields
\[
C_{jl}(b)
=\varphi(b_j)
\partial_l\Psi_j\big((b_q-\rho_{jq}b_j)_{q\ne j}\big),
\qquad l\ne j.
\]
Thus,
\[
\begin{aligned}
\mathcal D(b)
&=\sum_{l=1}^m\partial_lJ(b)=-\sum_j C_j(b)
  -\sum_j b_j\sum_l C_{jl}(b)\\
&=-\sum_j C_j(b)
  -\sum_j b_j\left[
      -b_jC_j(b)
      +\sum_{l\ne j}(1-\rho_{jl})C_{jl}(b)
    \right]\\
&=\sum_j(b_j^2-1)C_j(b)
  -\sum_j\sum_{l\ne j}
       b_j(1-\rho_{jl})C_{jl}(b).
\end{aligned}
\]
This gives \eqref{appx-an:threshold-identity} by noting that \(\rho_{jl}=\rho_{lj}\) and \(C_{jl}(b)=C_{lj}(b)\).

Next, we bound the derivatives in \eqref{appx-an:threshold-identity} using the arguments related to those in~\cite{Nazarov2003,CCK17N}. For $j, l=1,\dots, m$, let \(\Pi_{j l}\) be the orthogonal projection onto \(\operatorname{span}(u_j, u_l)\). Define
\begin{equation}
 R(g):=1+\max\left\{\max_j|u_j\cdot g|,
                          \max_{j<l}\norm{\Pi_{jl}g}\right\},\qquad g\in\R^r,
 \label{appx-an:projection-radius}
\end{equation}
omitting the second maximum when \(m=1\).
For the record, note that this function is \(1\)-Lipschitz and since we are dealing with at most \(m+m^2\)
Gaussian projections of dimensions at most \(2\), it is standard to establish that
\begin{equation}
 \E R(G)^2\le CL,\qquad \E R(G)^3\le CL^{3/2}.
 \label{appx-an:projection-moments}
\end{equation}
We will prove that
\begin{equation}
 |\mathcal D(b)|
 \le C\E\left[
 R(G)^3\ind\left\{0<\max_j(u_j\cdot G-b_j)
 \le\frac1{R(G)}\right\}
 \right].
 \label{appx-an:adaptive-layer}
\end{equation}
To do so, fix \(j\) and note that
\(
H_j:=G-(u_j\cdot G)u_j
\)
is a standard Gaussian vector in \(u_j^\perp\), independent
of \(X_j=u_j\cdot G\). Moreover, for every \(l\ne j\),
\[
u_l\cdot H_j
=X_l-(u_l\cdot u_j)X_j
=X_l-\rho_{jl}X_j.
\]
Thus, substituting this identity into \eqref{appx-an:conditional-first-derivative} and denoting \(Y_j:=b_ju_j+H_j\), we obtain
\begin{equation}
\begin{aligned}
C_j(b)
&=\varphi(b_j)
\E\ind\{u_l\cdot Y_j\le b_l,\ l\ne j\}=\varphi(b_j)
\E\ind\{u_l\cdot Y_j<b_l,\ l\ne j\},
\end{aligned}
\label{appx-an:one-density}
\end{equation}
where the second equality follows because
\(\Pp(u_l\cdot Y_j=b_l)=0\) for every \(l\ne j\),
by the linear independence of the normal vectors. Further, define
\[
E_j:=\left\{
x+vu_j:
u_j\cdot x=b_j,\
u_l\cdot x<b_l\ \text{for all }l\ne j,\ 
0<v<\eta(x)
\right\},
\qquad
\eta(x):=\frac1{4R(x)},
\]
where $x$ varies over $\R^r$, and $v$ varies over $\R$.
In words, \(E_j\) consists of short outward normal segments
starting from the relative interior of the \(j\)-th facet. Also, for any $y\in E_j$, the decomposition $y = x+ vu_j$ is unique, and so, letting $Z$ be a $N(-b_j,1)$ random variable that is independent of $H_j$, we have that $G$ is equal in distribution to $Y_j + Zu_j$ and that
\[
Y_j + Zu_j \in E_j\quad\Longleftrightarrow\quad u_l\cdot Y_j < b_l\text{ for }{l\neq j}\text{ and }0<Z<\eta(Y_j)
\]
since $u_j\cdot Y_j = b_j$. Thus,
\[
\E[R(G)^3\ind_{E_j}(G)]
=\E\left[
\ind\{u_l\cdot Y_j<b_l,\ l\ne j\}
\int_0^{\eta(Y_j)}
R(Y_j+vu_j)^3\varphi(b_j+v)\,dv
\right].
\]
In addition, for $x+vu_j \in E_j$, we have $R(x) \geq 1+ |u_j\cdot x| = 1+ |b_j|$ and $0<v<\eta(x)\leq 1/4$, and so
\[
 \frac{\varphi(b_j+v)}{\varphi(b_j)}
 =e^{-b_jv-v^2/2}\ge e^{-1/4-1/32}.
\]
Also, the Lipschitz property of \(R\) implies
\(R(x+vu_j)\geq 3R(x)/4\). Therefore,
\begin{equation}
 \E[R(G)^3\ind_{E_j}(G)]
 \ge c(1+b_j^2)C_j(b),
 \label{appx-an:one-region}
\end{equation}
which is the bound on $C_j(b)$ we will use below.

Now fix \(j<l\) and denote
\(\alpha:=\arccos(\rho_{jl})\in(0,\pi)\) and
\(S_{jl}:=\operatorname{span}(u_j, u_l)\).
Since \(u_j\) and \(u_l\) are linearly independent,
there is a unique \(z\in S_{jl}\) satisfying
$u_j\cdot z=b_j$ and $u_l\cdot z=b_l$.
Let \(H_{jl}:=(I_r-\Pi_{jl})G\) and observe that \(G=\Pi_{jl}G+H_{jl}\), where the two components on the right-hand side are independent standard Gaussian vectors in \(S_{jl}\) and \(S_{jl}^\perp\), respectively. Also, denote \(Y_{jl}:=z+H_{jl}\) and let \(\varphi_2\) be the standard Gaussian density on \(S_{jl}\). In addition, let $f_{j l}$ be the joint density of the pair $(X_j,X_l)$.

Simple algebra shows that $f_{j l}(b_j, b_l) = \varphi_2(z)/\sin\alpha$. Thus, differentiating
\begin{align*}
C(b) & = \Pp(X_a\leq b_a, \ a=1,\dots,m) \\
&= \int_{-\infty}^{b_j}\int_{-\infty}^{b_l}f_{jl}(s,t)\Pp(X_a\leq b_a, \ a\neq j,l\mid X_j = s,X_l = t)dt ds,
\end{align*}
we have
\begin{align}
C_{jl}(b) & = \frac{\varphi_2(z)}{\sin\alpha}\E[\ind\{X_a \leq b_a, \ a\neq j,l\} \mid X_j = b_j, X_l = b_l] \nonumber\\
& = \frac{\varphi_2(z)}{\sin\alpha}
\E[\ind\{u_a\cdot Y_{jl}\leq b_a,\ a\ne j,l\}\mid X_j=b_j,X_l=b_l] \nonumber \\
& = \frac{\varphi_2(z)}{\sin\alpha}
\E[\ind\{u_a\cdot Y_{jl} <  b_a,\ a\ne j,l\}],\label{appx-an:two-density}
\end{align}
where the second line follows from noting that conditional on $X_j=b_j$ and $X_l = b_l$, we have 
\begin{align*}
\ind\{X_a\leq b_a, \ a\neq j,l\} 
&= \ind\{u_a\cdot G\leq b_a, \ a\neq j,l\} = \ind\{ u_a\cdot(\Pi_{jl} G + H_{jl}) \leq b_a,\ a\neq j,l \} \\
& = \ind\{u_a\cdot (z + H_{jl})\leq b_a, \ a\neq j,l\} = \ind\{u_a\cdot Y_{jl}\leq b_a, a\neq j,l\}
\end{align*}
and the third from noting that $Y_{jl}$ is independent of $(X_j,X_l)$ and using the fact that
\(\Pp(u_a\cdot Y_{jl}=b_a)=0\) for every \(a\ne j,l\).

Further, let
\[
K_{jl}:=\{s u_j+t u_l: s,t\ge0\},
\]
and
\[
E_{jl}:=\left\{
x+y:
\begin{array}{l}
u_j\cdot x=b_j,\quad u_l\cdot x=b_l,\\
u_a\cdot x<b_a\ \text{for all }a\ne j,l,\\
y\in\operatorname{int}_{S_{jl}}K_{jl},\
0<\|y\|<\eta(x)
\end{array}
\right\},
\qquad
\eta(x)=\frac1{4R(x)}.
\]
In words, \(E_{jl}\) consists of short outward sectors
starting from points where exactly the \(j\)-th and
\(l\)-th constraints are active. Also, for any $w\in E_{jl}$, the decomposition $w = x+y$ is unique, and so, letting \(Z_{jl}\) be a Gaussian vector in \(S_{jl}\)
with mean \(-z\) and identity covariance that is independent
of \(H_{jl}\), we have that $G$ is equal in distribution to \(Y_{jl}+Z_{jl}\) and that
\[
Y_{jl}+Z_{jl}\in E_{jl}
\quad\Longleftrightarrow\quad
u_a\cdot Y_{jl}<b_a\ (a\ne j,l),\
Z_{jl}\in \operatorname{int}_{S_{jl}}K_{jl},\
0<\|Z_{jl}\|<\eta(Y_{jl}),
\]
since $u_j\cdot Y_{jl} = b_j$ and $u_l\cdot Y_{jl} = b_l$. Thus,
\[
\begin{aligned}
\E[R(G)^3\ind_{E_{jl}}(G)]
=\E\bigg[
&\ind\{u_a\cdot Y_{jl}<b_a,\ a\ne j,l\}\int_{\substack{y\in K_{jl}\\
                 0<\|y\|<\eta(Y_{jl})}}
R(Y_{jl}+y)^3\varphi_2(z+y)\,dy
\bigg].
\end{aligned}
\]
In addition, for any
\(x+y\in E_{jl}\), the Lipschitz property of \(R\)
and \(\|y\|<1/(4R(x))\) imply
\[
R(x+y)\ge R(x)-\|y\|\ge\frac34R(x).
\]
Also, since \(z=\Pi_{jl}x\), the definition of \(R\)
gives \(\|z\|\le R(x)\), and so
\[
\begin{aligned}
\frac{\varphi_2(z+y)}{\varphi_2(z)}
&=\exp\left(-z\cdot y-\frac{\|y\|^2}{2}\right)\ge\exp\left(-\|z\|\|y\|-\frac{\|y\|^2}{2}\right)
\ge e^{-1/4-1/32}.
\end{aligned}
\]
Moreover, the sector \(K_{jl}\cap\{0<\|y\|<\eta(x)\}\)
has area \(\alpha\eta(x)^2/2\). Consequently,
\[
\begin{aligned}
&\int_{\substack{y\in K_{jl}\\
                 0<\|y\|<\eta(x)}}
R(x+y)^3\varphi_2(z+y)\,dy\\
&\qquad\ge
cR(x)^3\varphi_2(z)\frac{\alpha\eta(x)^2}{2}
=c\alpha R(x)\varphi_2(z)
\ge c\alpha\|z\|\varphi_2(z).
\end{aligned}
\]
Therefore,
\[
\E[R(G)^3\ind_{E_{jl}}(G)]
\ge c\alpha\|z\|\varphi_2(z)
\Pp(u_a\cdot Y_{jl}<b_a,\ a\ne j,l).
\]
Finally, since \(b_j+b_l=(u_j+u_l)\cdot z\), we have
\[
\begin{aligned}
\frac{(1-\rho_{jl})|b_j+b_l|}{\sin\alpha}
&\le
\frac{(1-\cos\alpha)\|u_j+u_l\|}{\sin\alpha}
\|z\| =2\sin(\alpha/2)\|z\|
\le\alpha\|z\|.
\end{aligned}
\]
Combining this inequality with
\eqref{appx-an:two-density} gives
\begin{equation}
\E[R(G)^3\ind_{E_{jl}}(G)]
\ge c(1-\rho_{jl})|b_j+b_l|C_{jl}(b),
\label{appx-an:two-region}
\end{equation}
which is the bound on $C_{jl}(b)$ we will use below.

We now show that the events \(E_j\) and
\(E_{jl}\) are pairwise disjoint and that their union is
contained in the event under the expectation sign on the right-hand side of
\eqref{appx-an:adaptive-layer}.
To do so, let
\[
P_b:=\{w\in\R^r:u_a\cdot w\le b_a,\ a=1,\dots,m\}.
\]
Consider any point \(g\) belonging to one of the events
\(E_j\) or \(E_{jl}\). By their definitions, we can write
\(g=x+y\), where \(x\in P_b\) and \(y\) is a nonzero
nonnegative linear combination of the normal vectors corresponding
to the constraints that are equalities at \(x\).
Thus, for every \(w\in P_b\), we have
\[
y\cdot(w-x)
=\sum_{a:u_a\cdot x=b_a}t_a(u_a\cdot w-b_a)
\le0,
\]
where the coefficients \(t_a\) are nonnegative. Consequently,
\[
\begin{aligned}
\|g-w\|^2
&=\|x+y-w\|^2=\|y\|^2+\|x-w\|^2+2y\cdot(x-w)\ge\|y\|^2+\|x-w\|^2
\ge\|g-x\|^2.
\end{aligned}
\]
Moreover, the last inequality is strict whenever \(w\ne x\).
Therefore, \(x\) is the unique closest point to \(g\) in \(P_b\).

This observation implies that all the events \(E_j\) and
\(E_{jl}\) are pairwise disjoint. Indeed, if \(g\) belonged
to two such events, the corresponding points \(x\) would
both be closest points to \(g\) in \(P_b\), and therefore
would coincide. However, the definition of \(E_j\) requires
that exactly the \(j\)-th constraint is an equality at \(x\),
whereas the definition of \(E_{jl}\) requires that exactly
the \(j\)-th and \(l\)-th constraints are equalities. Thus, the $E_j$ and the $E_{jl}$ are pairwise disjoint.

Next, we show that all these events are contained in the event under the expectation sign on the right-hand side of \eqref{appx-an:adaptive-layer}. Take any $g$ in either $E_j$ or $E_{jl}$.
Then \(g=x+y\) with \(x\in P_b\),
\(y\ne0\), and \(y\) being a nonnegative combination of active
normal vectors. Hence at least one constraint that is active at $x$ is violated
at \(g\). Indeed,
\[
y\cdot y
=\sum_{a:u_a\cdot x=b_a}t_a(u_a\cdot g-b_a)>0,
\]
and so \(u_a\cdot g>b_a\) for at least one \(a\).
On the other hand, since \(\|u_a\|=1\) and
\(u_a\cdot x\le b_a\), we have, for every \(a\),
\[
u_a\cdot g-b_a
=u_a\cdot x-b_a+u_a\cdot y
\le\|y\|.
\]
By construction, \(\|y\|<\eta(x)=1/(4R(x))\).
Since \(R\) is \(1\)-Lipschitz and \(R(x)\ge1\),
\[
R(g)\le R(x)+\|y\|
<R(x)+\frac1{4R(x)}
\le\frac54R(x)<4R(x).
\]
Therefore,
\[
0<\max_a(u_a\cdot g-b_a)
\le\|y\|<\frac1{4R(x)}
<\frac1{R(g)}.
\]
Thus every event \(E_j\) and \(E_{jl}\) is contained in
the outer-layer event in \eqref{appx-an:adaptive-layer}.

Finally, the disjointness of these events implies
\[
\begin{aligned}
&\sum_j\E[R(G)^3\ind_{E_j}(G)]
+\sum_{j<l}\E[R(G)^3\ind_{E_{jl}}(G)]\\
&\qquad\le
\E\left[
R(G)^3\ind\left\{
0<\max_a(u_a\cdot G-b_a)\le\frac1{R(G)}
\right\}\right].
\end{aligned}
\]
Combining this inequality with \eqref{appx-an:one-region}
and \eqref{appx-an:two-region}, we obtain
\[
\begin{aligned}
&\sum_j(1+b_j^2)C_j(b)
+\sum_{j<l}(1-\rho_{jl})|b_j+b_l|C_{jl}(b)\\
&\qquad\le C\E\left[
R(G)^3\ind\left\{
0<\max_a(u_a\cdot G-b_a)\le\frac1{R(G)}
\right\}\right].
\end{aligned}
\]
Since \(C_j(b)\ge0\) and \(C_{jl}(b)\ge0\) for \(j<l\), and
\(|b_j^2-1|\le1+b_j^2\), the representation
\eqref{appx-an:threshold-identity} now yields
\eqref{appx-an:adaptive-layer}, as required.

We are now ready to complete the proof of the lemma. We have
$$
0<\max_j(u_j\cdot G-b_j(V_0))
 \le\frac1{R(G)} \quad \Longleftrightarrow \quad V_0+\sigma G\in A_a^{\sigma/R(G)}\setminus A_a.
$$
Also,
$$
\Pp\left(V_0+\sigma G\in A_a^{\sigma/R(G)}\setminus A_a \mid G\right) {\leq C\kappa\{\sigma/R(G)+\delta\}}
$$
by \eqref{appx-an:concentration} with thresholds \(a_j-\sigma u_j\cdot G\). Therefore, \eqref{appx-an:projection-moments} and \eqref{appx-an:adaptive-layer}
give
\[
 \E[|\mathcal D(b(V_0))|]
{\le C\kappa\{\sigma\E[R(G)^2]+\delta\E[R(G)^3]\}
 \le C\kappa(\sigma L+\delta L^{3/2}),}
\]
which gives~\eqref{appx-an:threshold-bound} and completes the proof of the lemma.
\end{proof}
\par\endgroup

\section{Anti-concentration and comparison of expected maxima}\label{appx-app:concentration}
\label{appx-conc:conditioning}

\subsection{Proof of Lemma~\ref*{lem:concentration}}

We first prove a boundary estimate for a maximum of affine functions
of independent signs. Convexity along each edge of the discrete cube
relates the probability of a short interval to the expected maximum
of the underlying linear functions. Pairing each summand with an
independent copy then gives a conditional Rademacher representation
of a general bounded sum. Its conditional variances are random, so
we must also control the probability that any of them is too small.
The same expected maximum controls this exceptional event.
For the Gaussian assertion, integration by parts replaces the
averaging over cube edges.

This supplies a self-contained argument for the Rademacher estimate
used here, whose corresponding logarithmic bound is established
in~\cite{OST} and recorded in~\cite[Lemma~J.4]{CCKK2022}.
The logarithmic bounded-sum estimate also appears
in~\cite[Lemma~2.1]{AL}, where it is obtained using a bootstrap
argument and Kane's~\cite{Kane} noise-sensitivity inequality.

\begin{lemma}[Rademacher boundary estimate]\label{appx-lem:cube-boundary}
Let $\varepsilon$ be uniform on $\{-1,1\}^n$, and let
$b_1,\ldots,b_m\in\R^n$ satisfy
$\min_\alpha\norm{b_\alpha}_2^2\ge\sigma^2>0$ and
$\max_{\alpha,i}|b_{\alpha i}|\le b$.
For thresholds $a_1,\ldots,a_m$, define
\[
 F(x):=\max_{\alpha\le m}(b_\alpha\cdot x-a_\alpha),
 \qquad
 \mathcal M:=\E\max\{0,\max_{\alpha\le m}b_\alpha\cdot\varepsilon\}.
\]
Then, for every $t\in\R$ and $h\ge0$,
\begin{equation}
 \Pp\{t\le F(\varepsilon)\le t+h\}
 \le\frac{h+8b}{\sigma^2}\,\mathcal M.
 \label{appx-eq:cube-boundary}
\end{equation}
If also $\max_\alpha\norm{b_\alpha}_2^2\le1$, then
$\mathcal M\le\sqrt{2\log(m+1)}$.
\end{lemma}

To our knowledge, the intrinsic estimate in~\eqref{appx-eq:cube-boundary}
is new. It may be of independent interest because it retains the
expected positive maximum $\mathcal M$.
When all $b_\alpha$ have unit length, we can take $\sigma^2=1$,
and the final assertion recovers the logarithmic order of the
Rademacher bound in~\cite{OST,CCKK2022}. Keeping $\mathcal M$
gives an \emph{intrinsic refinement} that can be smaller for strongly
dependent linear forms. For example, if every $b_\alpha$ equals
the same unit vector, symmetry gives
$\mathcal M=\tfrac12\E|b_1\cdot\varepsilon|\le\tfrac12$,
regardless of $m$. Retaining this maximum also allows us to control
the random conditional variances in the bounded-sum argument below.

\begin{proof}
First suppose that the maximizing affine function is unique at every
cube vertex. Then $F$ is differentiable at each vertex, and its
gradient there equals the coefficient vector of the active branch.
We will bound the probability of
\[
 E:=\{x\in\{-1,1\}^n:t\le F(x)\le t+h\}
\]
by averaging changes of derivatives along cube edges.

Define
\[
 H:=h+8b,\qquad c:=t-6b,\qquad
 \psi(y):=\min\{1,\max\{0,(y-c)/H\}\}.
\]
Thus $\psi$ increases linearly on
$[c,c+H]=[t-6b,t+h+2b]$ and is constant outside that interval.
Let $\Psi(y):=\int_c^y\psi(z)\,dz$ and $g:=\Psi\circ F$.
Since $\Psi$ is convex and nondecreasing and $F$ is convex,
$g$ is convex. The extension of the ramp beyond $[t,t+h]$
ensures that an entire edge meeting $E$ lies in its linear region,
with enough distance from its lower endpoint to control the slopes.

To verify this, fix all signs except the $i$th and write $x_-$
and $x_+$ for the endpoints with $i$th coordinates $-1$ and $1$.
Define
\[
 f_\pm:=F(x_\pm),\qquad
 d_\pm:=\partial_iF(x_\pm),\qquad
 \bar d:=\frac{f_+-f_-}{2}.
\]
Convexity places the secant slope between the endpoint derivatives,
and the coefficient bound gives
\[
 -b\le d_-\le\bar d\le d_+\le b.
\]
The conditional average over the $i$th sign is
\[
 D_i:=\E_{\varepsilon_i}[
       \varepsilon_i\partial_i g(\varepsilon)]
 =\frac12\{\psi(f_+)d_+-\psi(f_-)d_-\}\ge0,
\]
where all other signs remain fixed. The inequality follows from
convexity of $g$ along the edge.

If at least one endpoint belongs to $E$, then
$|f_+-f_-|\le2b$ implies
$f_\pm\in[t-2b,t+h+2b]$. Hence
$\psi(f_\pm)=(f_\pm-c)/H$ and $f_\pm-c\ge4b$.
Writing $V:=d_+-d_-\ge0$, direct rearrangement gives
\[
 \begin{aligned}
 HD_i&=\bar d\,d_++\frac{f_--c}{2}V,\\
 HD_i&=\bar d\,d_-+\frac{f_+-c}{2}V.
 \end{aligned}
\]
For either sign, $|\bar d-d_\pm|\le V$ and $|d_\pm|\le b$,
so $\bar d\,d_\pm\ge d_\pm^2-bV$. Each remaining term is
at least $2bV$. Thus $HD_i\ge d_+^2$ and $HD_i\ge d_-^2$
whenever the edge meets $E$. Together with $D_i\ge0$ on all
other edges, this proves
\[
 D_i\ge\frac{
       \ind_E(x_+)d_+^2+\ind_E(x_-)d_-^2}{2H}
 =\frac1H\E_{\varepsilon_i}
       [\ind_E(\varepsilon)(\partial_iF(\varepsilon))^2].
\]
Now average over the remaining signs and sum over $i$:
\[
 \E[\varepsilon\cdot\nabla g(\varepsilon)]
 \ge\frac1H\E[\ind_E(\varepsilon)
                       \norm{\nabla F(\varepsilon)}_2^2]
 \ge\frac{\sigma^2}{H}\Pp(\varepsilon\in E).
\]
The last inequality uses the lower bound on the squared norm of
every active coefficient vector.

At a vertex whose active branch is $\alpha$,
$\nabla g(\varepsilon)=\psi(F(\varepsilon))b_\alpha$.
Because $0\le\psi\le1$,
\[
 \varepsilon\cdot\nabla g(\varepsilon)
 \le\max\{0,\max_\beta b_\beta\cdot\varepsilon\}.
\]
Its expectation is therefore at most $\mathcal M$. Comparing
this upper bound with the preceding lower bound proves
\eqref{appx-eq:cube-boundary} under the uniqueness assumption.

We remove that assumption by perturbing the thresholds, keeping all
coefficient vectors fixed. For a fixed vertex and two distinct
branch indices, equality of their values imposes a proper affine
hyperplane on the threshold vector. There are only finitely many
such hyperplanes. Thus, for every $\rho>0$, we can choose perturbed
thresholds within $\rho$ of the original ones so that every vertex
has a unique active branch. The resulting maximum $F_\rho$ satisfies
$|F_\rho-F|<\rho$, so
\[
 \{t\le F(\varepsilon)\le t+h\}
 \subseteq
 \{t-\rho\le F_\rho(\varepsilon)\le t+h+\rho\}.
\]
Apply the proved bound with interval width $h+2\rho$.
Neither $\sigma$, $b$, nor $\mathcal M$ changes under this
perturbation. Letting $\rho\downarrow0$ proves the assertion,
including when the original interval endpoints carry atoms.

Finally, suppose $\norm{b_\alpha}_2\le1$. Independence of the
signs and $\cosh x\le e^{x^2/2}$ give
\[
 \E e^{\lambda b_\alpha\cdot\varepsilon}
 =\prod_i\cosh(\lambda b_{\alpha i})
 \le e^{\lambda^2\norm{b_\alpha}_2^2/2}
 \le e^{\lambda^2/2}.
\]
For $M:=\max\{0,\max_\alpha b_\alpha\cdot\varepsilon\}$,
Jensen's inequality and $e^{\lambda M}\le
1+\sum_\alpha e^{\lambda b_\alpha\cdot\varepsilon}$ yield
\[
 \mathcal M\le\frac1\lambda\log\E e^{\lambda M}
 \le\frac{\log(m+1)}\lambda+\frac\lambda2.
\]
Taking $\lambda:=\sqrt{2\log(m+1)}$ proves the last assertion.
\end{proof}

\begin{proof}[Proof of Lemma~\ref{lem:concentration}]
We first prove the bounded-sum assertion at $v=1$ and $R=0$.
The goal is to apply the preceding lemma after conditioning.
Let $X_1',\ldots,X_n'$ be mutually independent copies of
$X_1,\ldots,X_n$, independent of the original array. Let
$\varepsilon_1,\ldots,\varepsilon_n$ be independent uniform signs,
independent of both arrays, and define
\[
 \bar X_i:=\frac{X_i+X_i'}2,\qquad
 \Delta_i:=\frac{X_i-X_i'}2,\qquad
 \mathcal G:=\sigma(X_i,X_i':1\le i\le n).
\]
For each $i$, the vector $\bar X_i+\varepsilon_i\Delta_i$ equals
$X_i$ or $X_i'$ with equal probability. These selections are
independent over $i$ and have the original marginal laws. Hence
\begin{equation}
 V\stackrel d=\widetilde V:=\sum_i\bar X_i+\sum_i\varepsilon_i\Delta_i.
 \label{appx-eq:pair-representation}
\end{equation}
Conditional on $\mathcal G$, only the signs remain random.
The coefficients $b_{\alpha i}:=u_\alpha\cdot\Delta_i$ satisfy
$|b_{\alpha i}|\le\eta$, and the conditional face variances are
\[
 \mathcal V_\alpha:=\sum_i b_{\alpha i}^2,\qquad
 \E\mathcal V_\alpha
 =\frac14\sum_i\E\{u_\alpha\cdot(X_i-X_i')\}^2
 =\frac12.
\]
The last equality uses independence and centering of each copy pair
and $\Var(u_\alpha\cdot V)=1$.

We retain the conditional expected maximum
\[
 \mathcal M_\Delta:=\E_\varepsilon
       \norm{\sum_i\varepsilon_i\Delta_i}_U,
 \qquad
 \mathcal R_V:=\frac12\E\norm{V-V'}_U,\qquad V':=\sum_iX_i'.
\]
The vectors $\Delta_i$ are independent and symmetric, so their
joint law is unchanged by independent sign changes. Averaging first
over the signs and then over $\mathcal G$ therefore gives
\[
 \E\mathcal M_\Delta
 =\E\norm{\sum_i\Delta_i}_U=\mathcal R_V
 \le\E\norm V_U,
\]
where the last inequality is the triangle inequality.

The Rademacher lemma needs a lower bound on every conditional
variance. On the event
\[
 \mathcal E:=\{\min_\alpha\mathcal V_\alpha\ge1/4\},
\]
we may use $\sigma^2=1/4$. To identify the relevant interval,
define the conditional thresholds
$a_\alpha':=a_\alpha-u_\alpha\cdot\sum_i\bar X_i$ and the maximum
\[
 F_{\mathcal G}(e):=
       \max_\alpha\left(\sum_i b_{\alpha i}e_i-a_\alpha'\right).
\]
Then
$\{\widetilde V\in A_a^s\setminus A_a^{-s}\}
=\{-s<F_{\mathcal G}(\varepsilon)\le s\}$.
This is contained in the closed interval event of width $2s$
used in Lemma~\ref{appx-lem:cube-boundary}.
The normals occur in opposite pairs, so the positive maximum in
that lemma is exactly the face norm defining $\mathcal M_\Delta$.
We obtain
\begin{equation}
 \Pp(\widetilde V\in A_a^s\setminus A_a^{-s}\mid\mathcal G)
 \le(8s+32\eta)\mathcal M_\Delta
 \qquad\text{on }\mathcal E.
 \label{appx-eq:conditional-intrinsic-concentration}
\end{equation}

We next bound $\Pp(\mathcal E^c)$. For this purpose, fluctuations
of the squared coefficients must be compared with the same
Rademacher maximum. We first establish the elementary
\emph{contraction inequality}
\begin{equation}
 \E_\varepsilon\max_\alpha
       \left|\sum_i\varepsilon_i z_{\alpha i}^2\right|
 \le4\eta\,\E_\varepsilon\max_\alpha
       \left|\sum_i\varepsilon_i z_{\alpha i}\right|,
 \qquad |z_{\alpha i}|\le\eta.
 \label{appx-eq:square-contraction}
\end{equation}
Fix real offsets $a_j$, numbers $z_j\in[-\eta,\eta]$, and a
$2\eta$-Lipschitz function $\phi$. Then
\[
 \max_j\{a_j+\phi(z_j)\}+\max_j\{a_j-\phi(z_j)\}
 \le\max_j\{a_j+2\eta z_j\}+\max_j\{a_j-2\eta z_j\}.
\]
Indeed, if $j$ and $k$ attain the two maxima on the left, their
sum is at most $a_j+a_k+2\eta|z_j-z_k|$.
The right-hand side is at least this quantity by assigning $j$
and $k$ to its two maxima in the appropriate order.
After division by $2$, this inequality compares expectations over
one sign. Conditioning on the other signs and repeating the
comparison for each coordinate replaces $\phi(z_{\alpha i})$
by $2\eta z_{\alpha i}$ throughout the expected maximum.

Apply this argument with $\phi(z)=z^2$, which is
$2\eta$-Lipschitz on $[-\eta,\eta]$, and include the zero vector
in the coefficient family. It follows that
\[
 \E_\varepsilon\max\left\{0,\max_\alpha
                     \sum_i\varepsilon_i z_{\alpha i}^2\right\}
 \le2\eta\,\E_\varepsilon\max_\alpha
                    \left|\sum_i\varepsilon_i z_{\alpha i}\right|.
\]
The absolute maximum on the left of
\eqref{appx-eq:square-contraction} is at most the sum of the positive
maximum and the corresponding negative maximum. Their expectations
are equal by sign symmetry. This supplies the additional factor
$2$ and proves~\eqref{appx-eq:square-contraction}.

Let $(\widetilde b_{\alpha i})$ be an independent copy of the
whole coefficient array. Conditional Jensen's inequality gives
\[
 \E\max_\alpha|\mathcal V_\alpha-\E\mathcal V_\alpha|
 \le\E\max_\alpha
       \left|\sum_i(b_{\alpha i}^2-\widetilde b_{\alpha i}^2)\right|.
\]
For each $i$, exchanging its two coefficient columns negates their
difference of squares. These differences are independent over $i$,
so their joint law is invariant under independent sign changes.
Consequently,
\[
 \begin{aligned}
 \E\max_\alpha|\mathcal V_\alpha-\E\mathcal V_\alpha|
 &\le\E\max_\alpha\left|\sum_i\varepsilon_i
                  (b_{\alpha i}^2-\widetilde b_{\alpha i}^2)\right|\\
 &\le2\E\max_\alpha\left|\sum_i\varepsilon_i b_{\alpha i}^2\right|\\
 &\le8\eta\,\E\mathcal M_\Delta=8\eta\mathcal R_V.
 \end{aligned}
\]
The second line uses the triangle inequality and identical
distributions of the two arrays. The third applies
\eqref{appx-eq:square-contraction} conditional on the coefficients.

On $\mathcal E^c$, at least one conditional variance is below
$1/4$, whereas its expectation is $1/2$. Markov's inequality gives
\begin{equation}
 \Pp(\mathcal E^c)
 \le4\E\max_\alpha|\mathcal V_\alpha-\E\mathcal V_\alpha|
 \le32\eta\mathcal R_V.
 \label{appx-eq:conditional-variance}
\end{equation}
We may now average~\eqref{appx-eq:conditional-intrinsic-concentration}
on $\mathcal E$ and bound the remaining conditional probability
by $1$. Using~\eqref{appx-eq:pair-representation}, we obtain
\begin{equation}
 \Pp(V\in A_a^s\setminus A_a^{-s})
 \le(8s+64\eta)\mathcal R_V
 \le64(s+\eta)\E\norm V_U.
 \label{appx-eq:concentration-pairs}
\end{equation}
This proves the intrinsic bound at variance $1$.

For its logarithmic consequence, we bound $\E\norm V_U$ using
exponential moments. If $0<\lambda\le1/(2\eta)$, the inequality
$e^y\le1+y+y^2$ for $|y|\le1/2$, together with centering, gives
\[
 \E e^{\lambda u_\alpha\cdot V}
 \le\prod_i\{1+\lambda^2\E(u_\alpha\cdot X_i)^2\}
 \le e^{\lambda^2}.
\]
The last step uses independence and the unit face variance.
Since the face directions occur in opposite pairs,
$\norm V_U=\max_\alpha u_\alpha\cdot V$.
Bounding the exponential of this maximum by the sum and applying
Jensen's inequality gives
\[
 \E\norm V_U\le\frac{\log m}{\lambda}+\lambda.
\]
Choose $\lambda:=\min\{\sqrt{\log m},1/(2\eta)\}$.
Since $m=2p$ and $L=\log(ep)$,
$\E\norm V_U\le C(\sqrt L+\eta L)$.
When $\eta\sqrt L\le1$, substitution
in~\eqref{appx-eq:concentration-pairs} gives $C(s+\eta)\sqrt L$.
When $\eta\sqrt L>1$, that bound follows directly from the
probability bound by $1$.

We turn to the Gaussian assertion, still with $v=1$ and $R=0$.
Here integration by parts supplies the analogue of the edge
identity, and we retain the expected maximum throughout,
as in~\cite{ChernozhukovAC2026}.
Write $V=AG$, where $G$ is standard Gaussian in $\R^d$,
$d:=\operatorname{rank}\Cov(V)$, and $AA^\top=\Cov(V)$.
Then
\[
 u_\alpha\cdot V=w_\alpha\cdot G,\qquad
 w_\alpha:=A^\top u_\alpha,\qquad
 \norm{w_\alpha}_2^2=\Var(u_\alpha\cdot V)=1.
\]
If two $w_\alpha$ coincide, retain only their smaller threshold.
This preserves the maximum of the affine functions, and the set of
distinct directions still occurs in opposite pairs. The number of
remaining directions is at most $m$.

Define the maximum and its smooth approximation by
\[
 F(x):=\max_\alpha(w_\alpha\cdot x-a_\alpha),\qquad
 F_\beta(x):=\frac1\beta\log\sum_\alpha
          e^{\beta(w_\alpha\cdot x-a_\alpha)},\qquad\beta>0.
\]
The associated weights
\[
 \lambda_{\beta,\alpha}(x):=
 \frac{e^{\beta(w_\alpha\cdot x-a_\alpha)}}
      {\sum_\gamma e^{\beta(w_\gamma\cdot x-a_\gamma)}}
\]
are nonnegative and sum to $1$. Differentiation yields
\[
 \nabla F_\beta=\sum_\alpha\lambda_{\beta,\alpha}w_\alpha,
 \qquad
 \Delta F_\beta
 =\beta\left\{\sum_\alpha\lambda_{\beta,\alpha}
                       \norm{w_\alpha}_2^2
                     -\norm{\nabla F_\beta}_2^2\right\}
 =\beta\{1-\norm{\nabla F_\beta}_2^2\}\ge0.
\]
Indeed, the gradient is a convex combination of unit vectors and
therefore has norm at most $1$.

Let $\psi:\R\to[0,1]$ be continuously differentiable and
nondecreasing, with bounded derivative. Apply the Gaussian identity
$\E[G_kH(G)]=\E[\partial_kH(G)]$ to
$H=\psi(F_\beta)\partial_kF_\beta$ and sum over $k$.
For fixed $\beta$, $H$ and its first derivatives are bounded,
which justifies integration by parts. We obtain
\[
 \begin{aligned}
 \E[\psi'(F_\beta(G))\norm{\nabla F_\beta(G)}_2^2]
 &=\E[\psi(F_\beta(G))
       \{G\cdot\nabla F_\beta(G)-\Delta F_\beta(G)\}]\\
 &\le\E\norm V_U.
 \end{aligned}
\]
To see the inequality, first drop the nonpositive term involving
$\Delta F_\beta$. The remaining integrand is at most
\[
 \max\{0,\max_\alpha w_\alpha\cdot G\}=\norm V_U,
\]
because $G\cdot\nabla F_\beta(G)$ is a convex combination of the
listed linear forms and $0\le\psi\le1$.

We now remove the smooth maximum. The bound
$F\le F_\beta\le F+\log(m)/\beta$ gives uniform convergence.
The directions $w_\alpha$ are distinct, so ties between two affine
functions lie on a proper hyperplane in $\R^d$, which has standard
Gaussian probability zero. At every point with a unique maximizing
branch, its weight tends to $1$ as $\beta\to\infty$.
Thus $\norm{\nabla F_\beta(G)}_2^2\to1$ almost surely.
The left-hand integrand is bounded by $\|\psi'\|_\infty$,
so dominated convergence gives
\[
 \E\psi'(F(G))\le\E\norm V_U.
\]

To extract the desired layer probability, let $\psi_0$ increase
linearly from $0$ to $1$ on $[-s,s]$, with constant extensions.
Convolve $\psi_0$ with a nonnegative smooth probability density
supported on $[-\tau,\tau]$, and call the result $\psi_\tau$.
Then $0\le\psi_\tau'\le1/(2s)$, and
$\psi_\tau'\to(2s)^{-1}\ind_{(-s,s)}$ away from the endpoints.
For every fixed $y$, the level set $\{F=y\}$ is contained in
the finite union of hyperplanes
$\{w_\alpha\cdot x=a_\alpha+y\}$. Each is proper because
$\norm{w_\alpha}_2=1$, so $F(G)$ has no atoms.
The event $V\in A_a^s\setminus A_a^{-s}$ is exactly
$-s<F(G)\le s$. Applying the preceding inequality to $\psi_\tau$
and letting $\tau\downarrow0$ therefore gives
\[
 \Pp(V\in A_a^s\setminus A_a^{-s})
 \le2s\E\norm V_U.
\]
Finally,
$\E e^{\lambda w_\alpha\cdot G}=e^{\lambda^2/2}$.
The exponential-maximum calculation in the Rademacher lemma
therefore gives $\E\norm V_U\le\sqrt{2\log(m+1)}$, and hence
\[
 \Pp(V\in A_a^s\setminus A_a^{-s})\le Cs\sqrt L.
\]
This logarithmic consequence is Nazarov's inequality
\cite{Nazarov2003,CCK17N} in the required form.

For either assertion at general variance $v>0$, apply the
variance-$1$ result to $V/\sqrt v$ with thresholds $a/\sqrt v$
and layer width $s/\sqrt v$. The bounded summands then have
face norm at most $\eta/\sqrt v$, and
$\E\norm{V/\sqrt v}_U=\E\norm V_U/\sqrt v$.
Thus the intrinsic coefficients become
$(s+\eta)\E\norm V_U/v$ and $2s\E\norm V_U/v$, respectively.
The logarithmic coefficients become
$(s+\eta)\sqrt{L/v}$ and $s\sqrt{L/v}$.

Finally, condition on the independent shift $R$.
In the event involving $V+R$, each threshold $a_\alpha$ is
replaced by $a_\alpha-u_\alpha\cdot R$.
The law of $V$ is unchanged, and every bound is uniform in these
thresholds, including when their inequalities are inconsistent.
Averaging over $R$ and taking the minimum with $1$ completes
the proof.
\end{proof}

\subsection{Comparison of expected maxima}

The intrinsic boundary estimates involve the expected maximum of
the bounded sum as well as its Gaussian counterpart. We now compare
these expectations without using a Gaussian approximation for
rectangle probabilities, so the argument does not rely on the
conclusion of Theorem~\ref{thm:unrestricted} in its own proof.
The bound separates the contribution of the covariance structure,
measured by $g_\Sigma$, from an additional term $\delta L$
coming from the summand envelope.

\begin{lemma}[Expected maximum of a bounded sum]
\label{intrinsic:lem:maximum}
Under~\eqref{eq:setup},
\begin{equation}\label{intrinsic:eq:maximum}
 \E\|W\|_\infty\le C(g_\Sigma+\delta L).
\end{equation}
\end{lemma}

\begin{proof}
Include both signs of every coordinate and a zero base point by
taking
\[
 T:=\{0,\pm e_1,\ldots,\pm e_p\},\qquad
 X_t:=t\cdot W,\qquad Y_t:=t\cdot Z,\qquad t\in T.
\]
Thus $X_0=Y_0=0$,
$\sup_{t\in T}|X_t|=\|W\|_\infty$, and
$\E\sup_{t\in T}|Y_t|=g_\Sigma$.
To compare the two processes, define
\[
 d_2(s,t):=\{\E(Y_t-Y_s)^2\}^{1/2},
 \qquad d_1(s,t):=\delta\ind\{s\ne t\}.
\]
The distance $d_2$ measures the standard deviation of a contrast.
It retains the dependence between coordinates: highly correlated
coordinates can be close in this distance. The discrete distance
$d_1$ records the uniform envelope used to control larger deviations.

Indeed, $X_t-X_s=\sum_i(t-s)\cdot\xi_i$ is a sum of independent
centered variables, and, for $s\ne t$,
\[
 |(t-s)\cdot\xi_i|\le2\delta,\qquad
 \sum_i\E\{(t-s)\cdot\xi_i\}^2
 =(t-s)^\top\Sigma(t-s)=d_2(s,t)^2.
\]
Bernstein's inequality therefore gives
\[
 \Pp\{|X_t-X_s|>C[\sqrt u\,d_2(s,t)+u\,d_1(s,t)]\}
 \le2e^{-u},\qquad u>0.
\]
The same bound is immediate for $s=t$. Its square-root and linear
terms will contribute separately to the expected maximum.

We use a chaining bound to preserve the distance $d_2$ while
controlling all increments. For a finite set $T$ with a semimetric
$d$ and for $\alpha>0$, define
\[
 \gamma_\alpha(T,d):=
 \inf_{(T_k)}\sup_{t\in T}
       \sum_{k\ge0}2^{k/\alpha}d(t,T_k),
 \qquad d(t,T_k):=\inf_{s\in T_k}d(t,s),
\]
where $T_k\subseteq T$, $|T_0|=1$, and
$|T_k|\le2^{2^k}$ for $k\ge1$.
At level $k$, the set $T_k$ supplies representative indices,
and $d(t,T_k)$ is the error in approximating $t$ by one of them.
The functional chooses these representatives to minimize the
largest weighted sum of approximation errors.
The two tail scales in Bernstein's inequality correspond to
$\gamma_2(T,d_2)$ and $\gamma_1(T,d_1)$.

More precisely, the mixed-tail bound
in~\cite[Theorem~3.5]{Dirksen2015}, applied at moment order $1$
with base point $0$, gives
\[
 \E\|W\|_\infty
 \le C\{\gamma_2(T,d_2)+\gamma_1(T,d_1)\}
       +2\sup_{t\in T}\E|X_t|.
\]
Absolute constants in the increment bound are absorbed into $C$
by homogeneity of the chaining functionals. The final term is at
most $2$, since each nonzero $t\in T$ selects a signed coordinate
of variance $1$.

We can bound the term involving $d_1$ directly.
Let $k_*$ be the smallest integer such that
$2^{2^{k_*}}\ge|T|=2p+1$.
Take $T_k=\{0\}$ for $k<k_*$ and $T_k=T$ thereafter.
This sequence is admissible, and its approximation errors vanish
from level $k_*$ onward. Consequently,
\[
 \gamma_1(T,d_1)
 \le\delta\sum_{k=0}^{k_*-1}2^k
 =\delta(2^{k_*}-1)
 \le C\delta\log(2p+1)\le C\delta L.
\]

For the remaining term, $d_2$ is exactly the canonical distance
of the centered Gaussian process $(Y_t)_{t\in T}$.
The Gaussian majorizing-measure theorem, in the form recalled
in~\cite[Remark~3.3(i)]{Dirksen2015}, gives
\[
 \gamma_2(T,d_2)
 \le C\E\sup_{t\in T}|Y_t-Y_0|=Cg_\Sigma.
\]
This is the step that replaces a bound based only on the number of
coordinates by their Gaussian expected maximum.
If $d_2(s,t)=0$ for distinct indices, then $Y_s=Y_t$ almost surely.
Identifying such indices leaves both the Gaussian maximum and
$\gamma_2$ unchanged. Thus the comparison also applies when
$\Sigma$ is singular.

Combining the two chaining estimates yields
$\E\|W\|_\infty\le C(g_\Sigma+\delta L)+2$.
Since any coordinate of $Z$ is standard normal,
$g_\Sigma\ge\E|Z_1|=\sqrt{2/\pi}$, so the constant term can be
absorbed into $Cg_\Sigma$. This proves
\eqref{intrinsic:eq:maximum}.
\end{proof}

\section{Proofs of the transfer corollaries}\label{app:transfers}

The two corollaries extend the rectangle bounds in different ways.
For sub-exponential observations, we cap the coordinates, apply the
bounded-summand theorem, and control the error from the discarded tails.
For polyhedra, we express the defining inequalities as a rectangle
event for standardized projections and check the assumptions on the
projected summands.

\subsection{Proof of Corollary~\ref*{cor:subexponential} (sub-exponential transfer)}\label{app:subexponential}
\begin{proof}[Proof of Corollary~\ref{cor:subexponential}]
We first construct a bounded sum whose marginal variances remain close
to $1$. We will compare it with its matching Gaussian vector, retaining
the Gaussian-width factor, and then transfer that comparison to the
original sum and target.

For brevity, write $\ell:=\log(en)$, $b:=K\ell/\sqrt n$, and $g:=g_n$.
If $b\ge1/16$, the claimed bound follows from $\rh\le1$, since
$g\ge\sqrt{2/\pi}$ and $L,D_n\ge1$. We therefore assume $b<1/16$.
This restriction will ensure that the capped variances stay bounded
away from zero.

The Orlicz-norm condition gives $\E e^{|X_{ij}|/K}\le2$.
Indeed, every scale larger than $K$ is admissible in the definition
of the norm, and monotone convergence allows those scales to decrease
to $K>0$. Choose the capping level $T:=4K\ell$ and define
\[
 C_T(x):=\operatorname{sgn}(x)\min\{|x|,T\},\qquad
 Y_i:=C_T(X_i)-\E C_T(X_i),\qquad D_i:=X_i-C_T(X_i),
\]
where $C_T$ acts coordinatewise on a vector.
The vector $Y_i$ is the centered capped observation, while $D_i$
is the discarded part. Since $\E X_i=0$, their normalized sums satisfy
\[
 W_T:=n^{-1/2}\sum_iY_i,\qquad
 W_D:=n^{-1/2}\sum_i(D_i-\E D_i),\qquad W_n=W_T+W_D.
\]

We need two controls on $W_D$: small marginal variances to justify
standardization, and a bound on its maximum to transfer rectangle
probabilities. Since $|D_{ij}|=(|X_{ij}|-T)_+$, the exponential-moment
assumption gives, for $s\ge0$,
\[
 \Pp(|D_{ij}|>s)\le2e^{-4\ell-s/K},\qquad
 \E|D_{ij}|^k
 =\int_0^\infty ks^{k-1}\Pp(|D_{ij}|>s)\,ds
\le2k!K^ke^{-4\ell}\quad(k=1,2,\ldots).
\]
Independence across $i$ therefore implies
\begin{equation}
 \max_j\Var(W_{D,j})
 \le\max_j\frac1n\sum_i\E D_{ij}^2\le4K^2e^{-4\ell}.
 \label{tail:residual-moments}
\end{equation}

To bound the maximum of the residual sum, use the pointwise inequality
$e^{|D_{ij}|/K}\le1+e^{-4\ell}e^{|X_{ij}|/K}$.
It gives $\E e^{|D_{ij}|/K}\le1+2e^{-4\ell}$.
Together with the first-moment bound above, the triangle inequality
and independence across $i$ yield
\begin{align*}
 \E\exp\!\left(\frac{\sqrt n\,|W_{D,j}|}{K}\right)
 &\le\exp\!\left(\frac1K\sum_i\E|D_{ij}|\right)
          \prod_i\E e^{|D_{ij}|/K}\\
 &\le e^{2ne^{-4\ell}}(1+2e^{-4\ell})^n
 \le e^{4ne^{-4\ell}}.
\end{align*}
Define $\bar L:=\log(4np)$. Markov's inequality gives
\[
 \Pp\!\left(\sqrt n\,|W_{D,j}|\ge K[\bar L+4ne^{-4\ell}]\right)
 \le e^{-\bar L}.
\]
Since $4ne^{-4\ell}=4e^{-4}n^{-3}\le1/4\le\bar L$, a union
bound over the $p$ coordinates gives
\begin{equation}
 \Pp(\|W_D\|_\infty>\varrho_1)\le\frac1{4n}\le\frac1{2n},
 \qquad \varrho_1:=\frac{2K\bar L}{\sqrt n}.
 \label{tail:sum-coupling}
\end{equation}
We have applied the union bound to the residual sums, rather than
requiring every observed coordinate to lie below $T$.
Consequently, $T$ depends on $K$ and $\log n$, while the dimension
enters through the allowed residual displacement $\varrho_1$.

We next check the variance normalization required by
Theorem~\ref{thm:unrestricted}. Define
$\tau:=2Ke^{-2\ell}$ and $a_j:=\operatorname{sd}(W_{T,j})$.
Since $W_n=W_T+W_D$ and $\operatorname{sd}(W_{n,j})=1$, the
triangle inequality in $L^2$ gives
\[
 |a_j-1|\le\operatorname{sd}(W_{D,j})\le\tau,
 \qquad \frac\tau b=\frac{2\sqrt n\,e^{-2\ell}}\ell
 \le\frac2{e^2}.
\]
Our restriction $b<1/16$ therefore ensures $a_j\ge1/2$.
Let $S:=\operatorname{diag}(a_1^{-1},\ldots,a_p^{-1})$.
The sum $SW_T$ has unit marginal variances, and its independent
centered summands satisfy
\[
 \left\|\frac{SY_i}{\sqrt n}\right\|_\infty
 \le\frac{2T}{\sqrt n\min_j a_j}\le16b.
\]

To use the width-sensitive bound, we also need to control the Gaussian
width after capping and standardization. Let $Z_T$ be centered Gaussian
with covariance $\Cov(W_T)$. The map $C_T$ is odd and $1$-Lipschitz,
so, for $s,t\in\{-1,1\}$,
\begin{align*}
 \E(sZ_{T,j}-tZ_{T,k})^2
 &=\frac1n\sum_i\Var\{C_T(sX_{ij})-C_T(tX_{ik})\}\\
 &\le\frac1n\sum_i\E(sX_{ij}-tX_{ik})^2
 =\E(sZ_{n,j}-tZ_{n,k})^2.
\end{align*}
The inequality first bounds a variance by a second moment and then
uses the Lipschitz property. The final equality uses independence
across observations and their zero means.
Apply the Sudakov--Fernique comparison for expected maxima
\cite[Theorem~2.6.5]{AdlerTaylorWorsley2015} to the two centered
Gaussian families indexed by $(j,s)\in\{1,\ldots,p\}\times\{-1,1\}$.
Their maxima are the maximum norms, and hence
\begin{equation}
 \E\|Z_T\|_\infty\le\E\|Z_n\|_\infty=g.
 \label{tail:width-contraction}
\end{equation}
This comparison requires the increment inequalities just verified,
but does not require equal marginal variances.
Since $a_j\ge1/2$, we also have
$\E\|SZ_T\|_\infty\le2\E\|Z_T\|_\infty\le2g$.

Positive diagonal scaling maps the class of rectangles bijectively
onto itself, so $\rh(W_T,Z_T)=\rh(SW_T,SZ_T)$.
Apply the envelope bound in Theorem~\ref{thm:unrestricted} to the
standardized sum, using envelope $16b$ and width at most $2g$.
It gives
\begin{equation}
 \rh(W_T,Z_T)\le C\min\left\{1,
 bgL\sqrt{1+\log_+\frac1{b\sqrt L}}\right\}.
 \label{tail:bounded-comparison}
\end{equation}
Here the numerical factors are absorbed into $C$; in particular,
$\log_+\{1/(16b\sqrt L)\}\le\log_+\{1/(b\sqrt L)\}$.

The comparison in~\eqref{tail:bounded-comparison} uses the capped
covariance. To return to the original Gaussian target, choose a jointly
centered Gaussian pair $(Z_T,Z_D)$ with the block covariance of
$(W_T,W_D)$. Such a pair exists because this block covariance is
positive semidefinite, even when singular. Its first marginal has the
law already specified for $Z_T$, and
\[
 \Cov(Z_T+Z_D)=\Cov(W_T+W_D)=\Cov(W_n).
\]
We may therefore realize the target as $Z_n=Z_T+Z_D$ on this coupling.
The components need not be independent. By~\eqref{tail:residual-moments},
$\max_j\Var(Z_{D,j})\le\tau^2$.
Gaussian tails and a coordinate union bound, at level
$\varrho_2:=\tau\sqrt{2\bar L}$, give
\begin{equation}
 \Pp(\|Z_D\|_\infty>\varrho_2)
 \le2p e^{-\bar L}=\frac1{2n},\qquad
 \varrho_2=\sqrt8\,Ke^{-2\ell}\sqrt{\bar L}.
 \label{tail:Gaussian-coupling}
\end{equation}

We now combine the two couplings through rectangle inclusions.
For a bounded rectangle $A=\prod_j[\ell_j,u_j]$ and $s\in\R$,
define $A^s:=\prod_j[\ell_j-s,u_j+s]$, with inconsistent bounds
interpreted as the empty set. If $\|W_D\|_\infty\le\varrho_1$,
then $W_n\in A$ implies $W_T\in A^{\varrho_1}$.
The analogous implication on the Gaussian coupling gives
\begin{align*}
 \Pp(W_n\in A)
 &\le\Pp(W_T\in A^{\varrho_1})+\frac1{2n}\\
 &\le\Pp(Z_T\in A^{\varrho_1})+\rh(W_T,Z_T)+\frac1{2n}\\
 &\le\Pp(Z_n\in A^{\varrho_1+\varrho_2})
                         +\rh(W_T,Z_T)+\frac1n.
\end{align*}
Using contracted rectangles in the same argument gives
\[
 \Pp(W_n\in A)
 \ge\Pp(Z_n\in A^{-(\varrho_1+\varrho_2)})
                         -\rh(W_T,Z_T)-\frac1n.
\]
Thus the remaining error is the probability of a Gaussian boundary
layer of thickness $\varrho_1+\varrho_2$.
The target $Z_n$ has unit marginal variances and width $g$, so the
Gaussian part of Lemma~\ref{lem:concentration}, applied in its support
coordinates, bounds this probability by $C(\varrho_1+\varrho_2)g$.
Taking the supremum over rectangles yields
\begin{equation}
 \rh(W_n,Z_n)\le\rh(W_T,Z_T)
       +C(\varrho_1+\varrho_2)g+\frac1n.
 \label{tail:transfer}
\end{equation}
Unbounded rectangles follow by monotone approximation. The argument
uses the error probability of each coupling separately, and does not
require independence between its components.

Finally, the transfer error in~\eqref{tail:transfer} has order $bgL$.
Indeed, $\bar L\le2\ell L$, $K\ge1$, and $g\ge\sqrt{2/\pi}$ imply
\[
 (\varrho_1+\varrho_2)g+\frac1n
 \le CKg\left[\frac{\bar L}{\sqrt n}
                   +e^{-2\ell}\sqrt{\bar L}\right]+\frac1n
 \le CbgL.
\]
For the last inequality, use $e^{-2\ell}=e^{-2}n^{-2}$ and
$\ell,L\ge1$ to bound the Gaussian displacement and $1/n$ by the
same scale as the first term. Since $D_n\ge1$, combining this bound
with~\eqref{tail:bounded-comparison} and $\rh\le1$ proves
\eqref{eq:subexponential}.
\end{proof}

\subsection{Proof of Corollary~\ref*{cor:polyhedral-transfer} (polyhedral transfer)}
\label{app:polyhedral-transfer}

\begin{proof}[Proof of Corollary~\ref{cor:polyhedral-transfer}]
We first express membership of the polyhedron as a rectangle event
for projections with unit variance. If a normal has variance
$\sigma_\alpha^2=0$, then the centered projections
$\nu_\alpha^\top W$ and $\nu_\alpha^\top Z$ are both zero almost surely.
For a fixed threshold $a_\alpha$, its inequality is therefore automatic
when $a_\alpha\ge0$ and excludes both vectors when $a_\alpha<0$.
In the first case we delete that inequality, and in the second case
the probability difference is zero. We may thus restrict attention
to the positive-variance inequalities.

Define the standardized projection matrix $M\in\R^{m\times p}$ by
$M_{\alpha j}:=\nu_{\alpha j}/\sigma_\alpha$.
For every $x\in\R^p$, dividing each defining inequality by its
positive standard deviation gives
\[
 x\in P_a
 \quad\Longleftrightarrow\quad
 \nu_\alpha^\top x\le a_\alpha\ \text{for all }\alpha
 \quad\Longleftrightarrow\quad
 (Mx)_\alpha\le a_\alpha/\sigma_\alpha\ \text{for all }\alpha.
\]
The last event is a lower orthant in $\R^m$ for the vector $Mx$.
Apply this identity to $W$ and $Z$, then take the supremum over $a$.
Since lower orthants are rectangles, we obtain
\begin{equation}
 \Delta_m(W,Z)\le\rh(MW,MZ).
 \label{eq:polyhedral-transfer}
\end{equation}
Both transformed vectors have covariance $M\Sigma M^\top$, and
\[
 (M\Sigma M^\top)_{\alpha\alpha}
 =\frac{\nu_\alpha^\top\Sigma\nu_\alpha}{\sigma_\alpha^2}=1,
 \qquad \E\|MZ\|_\infty=g_P.
\]
Thus the transformed Gaussian width is exactly the quantity in the
corollary. No invertibility of $M\Sigma M^\top$ is needed.

It remains to verify the moment assumptions after projection.
The normal-vector condition~\eqref{eq:polyhedral-normal-condition}
gives $\sum_j|M_{\alpha j}|\le J$.
The triangle inequality and weighted Cauchy--Schwarz therefore give,
for every $x\in\R^p$,
\[
 |(Mx)_\alpha|\le J\|x\|_\infty,
 \qquad
 (Mx)_\alpha^2
 \le\left(\sum_j|M_{\alpha j}|\right)
           \sum_j|M_{\alpha j}|x_j^2
 \le J\sum_j|M_{\alpha j}|x_j^2.
\]
Under the bounded-summand setup, the transformed summands $M\xi_i$
are independent and centered, and the first inequality gives their
envelope $J\delta$.

To check the relative fourth moments, fix a direction $v\in\R^m$
and define $u:=M^\top v\in\R^p$. Then
$v^\top M\xi_i=u^\top\xi_i$, so the second inequality above and
the bound for $Q_j$ in~\eqref{eq:relative-moments} give
\[
 \begin{aligned}
 \sum_i\E\bigl[(M\xi_i)_\alpha^2(v^\top M\xi_i)^2\bigr]
 &\le J\sum_j|M_{\alpha j}|
           \sum_i\E\bigl[\xi_{ij}^2(u^\top\xi_i)^2\bigr]\\
 &\le J^2s_4\,u^\top\Sigma u
 =J^2s_4\,v^\top M\Sigma M^\top v.
 \end{aligned}
\]
For the comparison involving independent copies, apply the same
pointwise bound inside the first expectation and then use the bound
for $\widetilde Q_j$:
\[
 \begin{aligned}
 \sum_i\E[(M\xi_i)_\alpha^2]\,\E[(v^\top M\xi_i)^2]
 &\le J\sum_j|M_{\alpha j}|
             \sum_i\E[\xi_{ij}^2]\,\E[(u^\top\xi_i)^2]\\
 &\le J^2s_4\,v^\top M\Sigma M^\top v.
 \end{aligned}
\]
These are precisely the two fourth-moment inequalities required by
Theorem~\ref{thm:unrestricted}, with relative bound $J^2s_4$.
Apply that theorem in dimension $m$, with envelope $J\delta$ and
width $g_P$, and use~\eqref{eq:polyhedral-transfer} to obtain the
bounded-summand conclusions.

For polynomial moment envelopes, the transformed observations $MX_i$
remain independent and centered, and the covariance of their normalized
sum again has unit diagonal. The envelope inequality gives
\[
 \max_{i\le n}\E\|MX_i\|_\infty^q
 \le J^q\max_{i\le n}\E\|X_i\|_\infty^q\le(JB)^q.
\]
Part~\textup{(i)} of Theorem~\ref{thm:polynomial} therefore applies
with envelope parameter $JB$.
If the coordinate fourth-moment condition also holds, weighted
H\"older's inequality gives
\[
 |(Mx)_\alpha|^4
 \le\left(\sum_j|M_{\alpha j}|\right)^3
          \sum_j|M_{\alpha j}|\,|x_j|^4
 \le J^3\sum_j|M_{\alpha j}|\,|x_j|^4.
\]
Average this inequality over the observations to obtain
\[
 \frac1n\sum_i\E(MX_i)_\alpha^4
 \le J^3\sum_j|M_{\alpha j}|
               \frac1n\sum_i\E X_{ij}^4
 \le J^4B^2.
\]
Part~\textup{(ii)} uses a single parameter for both moment conditions:
its $q$th power bounds the envelope moment and its square bounds the
coordinate fourth moments. The choice $B_*:=J^2B$ meets both
requirements, because $JB\le B_*$ when $J\ge1$ and
$J^4B^2=B_*^2$.
Applying that part with $B_*$ proves the asserted substitution
$B\mapsto J^2B$.

Finally, for sub-exponential observations, the triangle inequality
for the Orlicz norm gives
\[
 \|(MX_i)_\alpha\|_{\psi_1}
 \le\sum_j|M_{\alpha j}|\,\|X_{ij}\|_{\psi_1}\le JK.
\]
Thus Corollary~\ref{cor:subexponential} applies with $K$ replaced
by $JK$. In each application the new dimension is $m$, the width
is $g_P$, and the auxiliary quantities are evaluated at these
substituted parameters. Combining each resulting rectangle bound
with~\eqref{eq:polyhedral-transfer} proves the remaining claims.
All estimates use the normal vectors only through
$\sum_j|M_{\alpha j}|\le J$ and the stated width $g_P$, so the
constants are uniform over the permitted normal vectors.
\end{proof}

\section{\texorpdfstring{{Proof of Theorem~\ref*{thm:polynomial} (polynomial moment envelopes)}}{Proof of Theorem 2 (polynomial moment envelopes)}}\label{app:polynomial-proof}

The two parts of Theorem~\ref{thm:polynomial} use different information
about large observations. Under the envelope condition alone, we
condition on the large observations and apply the bounded theorem
to the remaining sum. With the additional coordinate fourth-moment
condition, we instead compare smooth expectations directly and use
Bennett's inequality to control the large increments. Throughout,
we retain the notation of Section~\ref{sec:poly}, including
$L=\log(ep)$, $g_\Sigma=\E\|Z\|_\infty$, and
$V_q=Bn^{-1/2+1/q}L^{3/2-1/q}$. Constants denoted by $C_q$ depend
only on the fixed moment order $q>2$.

\subsection{\texorpdfstring{{Part (i): polynomial envelope transfer}}{Part (i): polynomial envelope transfer}}\label{app:conditional}

We work under the envelope condition~\eqref{poly:envelope}, so that
$(X_i)_{i=1}^n\in\mathcal E_q(B,p)$. Our goal is
\eqref{poly:transfer-bound}, which is the conclusion
\eqref{poly:main-envelope} in this notation. The main issue is that
truncation changes both the sum and its covariance. We will control
these changes by adding shifts whose expected maximum norms are small.

First we record the version of the bounded theorem that will be
applied after conditioning. Let $Y_1,\ldots,Y_n$ be independent and
centered, with $\|Y_i\|_\infty\le M$, $M\ge1$, and
$n^{-1}\sum_i\E Y_{ij}^2\ge1/2$ for every $j$. If $G_Y$ is
Gaussian with the covariance of $n^{-1/2}\sum_iY_i$, define
$g_Y:=\E\|G_Y\|_\infty$. Then Theorem~\ref{thm:unrestricted} gives
\begin{equation}
 \rh\left(n^{-1/2}\sum_iY_i,G_Y\right)
 \le C\min\{1,Mg_YL\sqrt{\log(en)}/\sqrt n\}.
 \label{app:bounded-input}
\end{equation}
Indeed, write $\sigma_j^2:=n^{-1}\sum_i\E Y_{ij}^2$. Dividing
coordinate $j$ by $\sigma_j$ preserves the rectangle distance and
gives normalized summands with envelope $\sqrt2M/\sqrt n$.
The standardized Gaussian width is at most $\sqrt2g_Y$, because
$\max_j|(G_Y)_j/\sigma_j|\le\sqrt2\|G_Y\|_\infty$ pointwise.
For this envelope, the logarithmic factor $D_L$ in the bounded
theorem is at most $C\log(en)$, using $M\ge1$ and $L\ge1$.
This proves~\eqref{app:bounded-input} without an upper bound on
the marginal variances.

We also need to bound the expected norm of the part removed by
truncation. The following inequality separates its coordinate
variance from its envelope moment.

\begin{lemma}[An expected maximum under an envelope moment]\label{lem:maximum}
Let $Y_1,\ldots,Y_n$ be independent centered vectors in $\R^p$.
If
\[
 \max_j\sum_i\E Y_{ij}^2\le v,\qquad
 \sum_i\E\norm{Y_i}_\infty^q\le A^q,\qquad q>2,
\]
then
\begin{equation}
 \E\norm{\sum_iY_i}_\infty
 \le C_q\{\sqrt{vL}+AL^{1-1/q}\}.
 \label{app:maximal}
\end{equation}
\end{lemma}
\begin{proof}
If $A=0$, every $Y_i$ vanishes almost surely. Otherwise, choose
$u:=AL^{-1/q}$ and define the two centered parts
\[
 Y_i^{\le u}:=Y_i\ind\{\|Y_i\|_\infty\le u\}
       -\E[Y_i\ind\{\|Y_i\|_\infty\le u\}],
 \qquad Y_i^{>u}:=Y_i-Y_i^{\le u}.
\]
The first part has envelope at most $2u$, and
$\sum_i\Var(Y_{ij}^{\le u})\le\sum_i\E Y_{ij}^2\le v$.
Bernstein's inequality and a union bound over the $2p$ signed
coordinates therefore give, for $x\ge0$,
\[
 \Pp\left\{\left\|\sum_iY_i^{\le u}\right\|_\infty
       >C\bigl[\sqrt{v(L+x)}+u(L+x)\bigr]\right\}
 \le2e^{-x}.
\]
Integrating this bound, using $L\ge1$, yields
$\E\|\sum_iY_i^{\le u}\|_\infty\le C(\sqrt{vL}+uL)$.
The second part is the centered sum of the omitted vectors, so
the triangle inequality gives
\[
 \E\left\|\sum_iY_i^{>u}\right\|_\infty
 \le2\sum_i\E[\norm{Y_i}_\infty\ind\{\norm{Y_i}_\infty>u\}]
 \le2A^qu^{1-q}.
\]
Our choice of $u$ makes both $uL$ and $A^qu^{1-q}$ equal to
$AL^{1-1/q}$. Adding the two bounds proves~\eqref{app:maximal}.
\end{proof}

For an arbitrary threshold $T\ge2B$, we now prove
\begin{equation}
 \rh(W,Z)\le C_q\min\left\{1,\,
 g_\Sigma\left[
 \frac{T}{\sqrt n}L\sqrt{\log(en)}
 +B^{q/2}T^{1-q/2}\sqrt L+\frac{V_q}{\sqrt L}\right]\right\}.
 \label{poly:threshold}
\end{equation}
The first term is the approximation error for the bounded sum.
The remaining terms will bound the removed observations and the
Gaussian shifts needed to correct the covariance. The threshold
$T$ will be chosen only after these contributions have been combined.

To isolate the large observations, define
\[
 I_i:=\ind\{\norm{X_i}_\infty>T\},\qquad
 p_i:=\Pp(I_i=1),\qquad \mu_i:=\E[X_i\mid I_i=0].
\]
Markov's inequality and $q>2$ give
\begin{equation}
 p_i\le(B/T)^q\le2^{-q}<1/4.
 \label{app:p_i-bound}
\end{equation}
Thus each conditional mean is well defined. Subtracting it from
the small observations gives the decomposition
\begin{equation}
 Y_i:=(1-I_i)(X_i-\mu_i),\qquad
 R_i:=I_iX_i+(1-I_i)\mu_i,\qquad X_i=Y_i+R_i.
 \label{app:decomposition}
\end{equation}
Here $\E Y_i=0$, and hence $\E R_i=\E X_i-\E Y_i=0$.
The vectors $R_i$ are independent over $i$, as each is a function
of $X_i$ alone.

Let $\mathcal F:=\sigma(I_i,I_iX_i:1\le i\le n)$. This
information identifies the large observations and reveals their
values. Conditional on $\mathcal F$, a vector $Y_i$ is zero if
$I_i=1$; if $I_i=0$, its law is that of $X_i-\mu_i$ conditional
on $\|X_i\|_\infty\le T$. Since the conditioning is performed
separately for each independent observation, the $Y_i$ remain
conditionally independent and centered. Also,
$\|\mu_i\|_\infty\le T$, so $\|Y_i\|_\infty\le2T$.
The residual sum $R:=n^{-1/2}\sum_iR_i$ is
$\mathcal F$-measurable. Thus, after conditioning, $W$ is a
bounded centered sum plus a fixed shift.

We first bound that shift. Define $v_T:=B^qT^{2-q}$, the bound
on the second moment discarded from each observation. Centering
of $X_i$ implies $(1-p_i)\mu_i=-\E[X_iI_i]$, and therefore
\begin{equation}
 \norm{\mu_i}_\infty
 \le\frac{\E[\norm{X_i}_\infty I_i]}{1-p_i}
 \le2B^qT^{1-q},\qquad
 \E[\norm{X_i}_\infty^2 I_i]\le v_T.
 \label{app:tail-moments}
\end{equation}
The two terms in $R_i$ have disjoint supports. Consequently,
\[
 \E R_{ij}^2
 \le v_T+\|\mu_i\|_\infty^2\le Cv_T,
 \qquad
 \E\|R_i\|_\infty^q
 \le B^q+\|\mu_i\|_\infty^q\le C_qB^q.
\]
The last inequalities use $B/T\le1/2$ in
\eqref{app:tail-moments}. For the normalized residuals $R_i/\sqrt n$,
the variance budget is therefore $Cv_T$ and the envelope-moment
budget is $C_qB^qn^{1-q/2}$. Lemma~\ref{lem:maximum} gives
\begin{equation}
 g_\Sigma\,\E\norm R_\infty
 \le C_qg_\Sigma\{\sqrt{Lv_T}+V_q/\sqrt L\},
 \label{app:shift}
\end{equation}
where $Bn^{-1/2+1/q}L^{1-1/q}=V_q/\sqrt L$.

We next control the covariance of the bounded sum. Define
\[
 C_i:=\E X_iX_i^\top,\qquad C_i^0:=\Cov(X_i\mid I_i=0),
 \qquad
 Q:=\frac1n\sum_i(1-I_i)C_i^0,\qquad
 Q_0:=\frac1n\sum_iC_i^0.
\]
The matrix $Q$ is the covariance of $n^{-1/2}\sum_iY_i$
conditional on $\mathcal F$. The matrix $Q_0$ is deterministic
and includes every conditional covariance, even those omitted
when $I_i=1$. The identity
\[
 C_i^0=\frac{C_i-\E[X_iX_i^\top I_i]}{1-p_i}
                   -\mu_i\mu_i^\top
\]
shows that $0\preceq C_i^0\preceq C_i/(1-p_i)$.
Since $n^{-1}\sum_iC_i=\Sigma$, we obtain
\begin{equation}
 0\preceq Q\preceq Q_0\preceq\tfrac43\Sigma.
 \label{app:conditional-width-order}
\end{equation}
This order controls the Gaussian width without any rank assumption.
Indeed, for $0\preceq H\preceq\tfrac43\Sigma$, let
$G_H\sim N(0,H)$ and add an independent centered Gaussian with
covariance $\tfrac43\Sigma-H$. Conditional Jensen's inequality
for the convex function $\|\cdot\|_\infty$ gives
\[
 \E\|G_H\|_\infty\le\sqrt{4/3}\,g_\Sigma.
\]
In particular, this holds for every realization of $Q$.

To apply the bounded theorem, we must also exclude realizations
where a marginal variance of $Q$ is too small. Write
$\maxnorm M:=\max_{j,k}|M_{jk}|$. The envelope condition implies
$\maxnorm{C_i}\le B^2$, and the covariance identity gives
\[
 \maxnorm{C_i^0-C_i}
 \le\frac{p_iB^2+v_T}{1-p_i}+\|\mu_i\|_\infty^2
 \le Cv_T,
 \qquad \maxnorm{C_i^0}\le2B^2.
\]
Here $p_iB^2\le v_T$ follows from $B\le T$.
Since $Q_0-Q=n^{-1}\sum_iI_iC_i^0$, the triangle inequality
now yields, for $\Delta:=\maxnorm{Q-\Sigma}$,
\begin{equation}
 \E\Delta\le\maxnorm{Q_0-\Sigma}
             +\frac1n\sum_i p_i\maxnorm{C_i^0}\le Cv_T.
 \label{app:conditional-covariance}
\end{equation}
On $\mathcal E:=\{\Delta\le1/2\}$, every $Q_{jj}$ is at
least $1/2$. Markov's inequality gives
$\Pp(\mathcal E^c)\le2\E\Delta\le Cv_T$.

Conditional on $\mathcal F$, construct $G_Q\sim N(0,Q)$ using
additional Gaussian randomness. On $\mathcal E$, marginal
standardization changes its width by at most a factor $\sqrt2$.
Lemma~\ref{lem:concentration} therefore bounds its rectangle
boundary layers of width $s$ by $Csg_\Sigma$. In particular,
for a fixed vector $r$, the inclusions
$A^{-\|r\|_\infty}\subseteq A-r\subseteq A^{\|r\|_\infty}$
give
\[
 |\Pp(G_Q+r\in A\mid\mathcal F)
        -\Pp(G_Q\in A\mid\mathcal F)|
 \le Cg_\Sigma\|r\|_\infty.
\]
Apply~\eqref{app:bounded-input} to the conditional bounded sum,
whose envelope is $2T$, at the rectangle $A-R$. Comparing next
$G_Q+R$ with $G_Q$ gives, on $\mathcal E$,
\[
 \sup_A\bigl|\Pp(W\in A\mid\mathcal F)
             -\Pp(G_Q\in A\mid\mathcal F)\bigr|
 \le Cg_\Sigma\left\{
       \frac{T}{\sqrt n}L\sqrt{\log(en)}+\|R\|_\infty\right\}.
\]

It remains to compare the conditional Gaussian with $Z$. We may
assume that $v_T$ is smaller than a sufficiently small absolute
constant: otherwise the term $g_\Sigma\sqrt{Lv_T}$ in
\eqref{poly:threshold} bounds a positive absolute constant, so
$\rh\le1$ proves that inequality after increasing $C_q$.
We make the covariance comparison in two stages, using Gaussian
additions to retain the boundary coefficient $g_\Sigma$.

First, conditional on $\mathcal F$, let
$J_Q\sim N(0,Q_0-Q)$ be independent of $G_Q$. The covariance is
positive semidefinite by~\eqref{app:conditional-width-order}, and
$G_Q+J_Q$ has the fixed law $G_0\sim N(0,Q_0)$. On
$\mathcal E$, condition further on $J_Q$ and apply the shift
bound above. The expected Gaussian maximum with coordinate
variances bounded by $a$ is at most $C\sqrt{aL}$, so
\[
 \E\|J_Q\|_\infty
 \le C\sqrt L\,\E\sqrt{\max_j(Q_0-Q)_{jj}}
 \le C\sqrt{Lv_T}.
\]
The expectation here includes $\mathcal F$. The last inequality
uses Jensen's inequality and
\[
 \E\max_j(Q_0-Q)_{jj}
 \le\frac1n\sum_i p_i\max_j(C_i^0)_{jj}\le Cv_T.
\]
Thus
\[
 \E\!\left[\ind_{\mathcal E}\,
     \rh\{N(0,Q),N(0,Q_0)\}\right]
 \le Cg_\Sigma\sqrt{Lv_T}.
\]

Second, to compare $Q_0$ with $\Sigma$, define
$D_R:=\Cov(R)$ and $H_T:=n^{-1}\sum_i p_iC_i^0$.
In~\eqref{app:decomposition}, $R_i=\mu_i$ whenever $Y_i$ is
nonzero, and $\E[(1-I_i)(X_i-\mu_i)]=0$. It follows that
$\E Y_iR_i^\top=0$. Independence across observations then gives
\[
 \Sigma=Q_0-H_T+D_R,\qquad Q_0+D_R=\Sigma+H_T.
\]
Both $D_R$ and $H_T$ are positive semidefinite, with diagonal
entries at most $Cv_T$ by the preceding residual and covariance
bounds. Also $(Q_0)_{jj}\ge1-(D_R)_{jj}\ge1-Cv_T\ge1/2$.
Together with~\eqref{app:conditional-width-order}, this gives
the same boundary coefficient $Cg_\Sigma$ for $G_0$.

Let $G_D\sim N(0,D_R)$ be independent of $G_0$, and let
$G_H\sim N(0,H_T)$ be independent of $Z$. Their sums have the
same covariance, so $G_0+G_D\stackrel d=Z+G_H$. The triangle
inequality for rectangle distance and the two Gaussian shift
bounds therefore imply
\[
 \begin{split}
 \rh(G_0,Z)
 &\le\rh(G_0,G_0+G_D)+\rh(Z+G_H,Z)\\
 &\le Cg_\Sigma\{\E\|G_D\|_\infty+\E\|G_H\|_\infty\}
 \le Cg_\Sigma\sqrt{Lv_T}.
 \end{split}
\]
Neither covariance comparison requires an inverse covariance matrix.

We may now average the conditional comparison on $\mathcal E$,
use the two Gaussian comparisons, and bound the probability
discrepancy on $\mathcal E^c$ by $1$. This gives
\[
 \rh(W,Z)\le C_qg_\Sigma\left\{
 \frac{T}{\sqrt n}L\sqrt{\log(en)}
 +\E\|R\|_\infty+\sqrt{Lv_T}\right\}+Cv_T.
\]
Use~\eqref{app:shift}. Since $g_\Sigma\ge\sqrt{2/\pi}$,
$L\ge1$, and $v_T$ is small, the last term is absorbed by
$Cg_\Sigma\sqrt{Lv_T}$. Recalling
$\sqrt{v_T}=B^{q/2}T^{1-q/2}$ proves~\eqref{poly:threshold},
with the cap at $1$ supplied by $\rh\le1$.

To finish, write $\ell:=\log(en)$. When $n\ge L\ell$, choose
$T:=2B\{n/(L\ell)\}^{1/q}\ge2B$. This balances the two
threshold-dependent terms, since
\[
 \frac{T}{\sqrt n}L\sqrt\ell
   =2\frac{V_q}{\sqrt L}\ell^{\alpha_q},\qquad
 B^{q/2}T^{1-q/2}\sqrt L
   =2^{1-q/2}\frac{V_q}{\sqrt L}\ell^{\alpha_q},
 \qquad \alpha_q=\frac12-\frac1q>0.
\]
The remaining term $V_q/\sqrt L$ is no larger than this scale.
If $n<L\ell$, then instead
\[
 \frac{g_\Sigma}{\sqrt L}V_q\ell^{\alpha_q}
 =g_\Sigma B\sqrt L\left(\frac{L\ell}{n}\right)^{\alpha_q}
 \ge\sqrt{2/\pi}.
\]
Increasing the constant covers this case by $\rh\le1$ and proves
Theorem~\ref{thm:polynomial}\textup{(i)}.

\begin{remark}\label{rmk:E1}
If Conjecture~\ref{conj:sharp-unrestricted} holds, the same
conditional transfer retains Gaussian width and gives
$C_q\min\{1,(g_\Sigma/\sqrt L)V_q\}$. Choose
$T:=2B(n/L)^{1/q}$ when $n\ge L$. If $n<L$, then
$(g_\Sigma/\sqrt L)V_q
=g_\Sigma B\sqrt L(L/n)^{\alpha_q}\ge\sqrt{2/\pi}$,
so an enlarged constant covers this case by $\rh\le1$.
The inequality $g_\Sigma\le C\sqrt L$ then gives
$C_q\min\{1,V_q\}$ in the worst case over covariance matrices.
\end{remark}

\subsection{\texorpdfstring{{Part (ii): Bennett refinement}}{Part (ii): Bennett refinement}}\label{app:bennett}

We now prove~\eqref{poly:bennett-bound} under both moment
conditions~\eqref{poly:envelope} and~\eqref{poly:fourth}. The
coordinate fourth moments allow us to retain more information than
the conditional truncation argument used above. We first derive a
Gaussian comparison bound that keeps the variance and expected
maximum of the large increments separate. Bennett's inequality
will then bound that expected maximum using both moment conditions.

\begin{lemma}[Gaussian comparison with large-increment remainders]
\label{ben:comparison}
There is an absolute constant $C>0$ with the following property.
Let $n,p\ge1$, and let $X_1,\ldots,X_n\in\R^p$ be independent
centered random vectors with finite coordinate fourth moments and
$n^{-1}\sum_i\E X_{ij}^2=1$ for every $j\le p$. Define
\[
 W:=n^{-1/2}\sum_iX_i,\qquad \Sigma:=\Cov(W),\qquad
 Z\sim N(0,\Sigma),\qquad L:=\log(ep).
\]
Let $\widetilde X_i$ be mutually independent copies of $X_i$,
independent of the original array, and define
\[
 Y_i:=(\widetilde X_i-X_i)/\sqrt n,\qquad
 \Delta_1:=\frac1{n^2}\max_j\sum_i\E X_{ij}^4.
\]
For $t>0$, define
\[
 v_t:=\max_j\sum_i\E[Y_{ij}^2\ind\{\norm{Y_i}_\infty>t\}],
 \qquad M_t:=\E\norm{\sum_iY_i\ind\{\norm{Y_i}_\infty>t\}}_\infty.
\]
With $g_\Sigma:=\E\|Z\|_\infty$, for every $\phi>0$,
\begin{equation}
 \rh(W,Z)\le C\frac{g_\Sigma}{\sqrt L}\left\{
 \frac{\sqrt L}{\phi}+\phi L^2\sqrt{\Delta_1}
 +\phi^3L^{7/2}\Delta_1+\phi L^{3/2}v_t+\sqrt L M_t\right\},
 \quad t:=(\phi L)^{-1}.
 \label{ben:master}
\end{equation}
The covariance may be singular, and the observations need not be
identically distributed or symmetric.
\end{lemma}

The quantities $v_t$ and $M_t$ describe different effects of the
large increments: $v_t$ is their largest coordinate variance sum,
whereas $M_t$ measures the expected maximum norm of their sum.
Keeping the latter expectation intact permits the Bennett refinement.
The factor $g_\Sigma/\sqrt L$ records the improvement over the
dimension-based comparison.

Lemma~\ref{ben:comparison} is a version of the matching-covariance
case of Lemma~A.1 in the supplement to~\cite{CCKK2022}. We give
a self-contained proof using logistic smoothing and Slepian
interpolation. As in Appendix~H of that supplement, the proof uses
symmetry under interchanging a summand and its independent copy.
A trapezoidal Taylor identity makes the resulting fourth-order
remainder explicit.

Throughout this subsection, we use the original space $\R^p$ and
the coordinate face normals $u_j:=e_j$ and $u_{p+j}:=-e_j$,
for $j\le p$. Thus $m=2p$ and $\norm x_U=\norm x_\infty$.
For $a\in\R^{2p}$ and $h>0$, write
\[
 A_a:=\{x:u_\alpha\cdot x\le a_\alpha,\ \alpha\le2p\},\qquad
 F_{a,h}(x):=\left(1+\sum_{\alpha=1}^{2p}
          e^{(u_\alpha\cdot x-a_\alpha)/h}\right)^{-1}.
\]
These sets represent bounded closed rectangles and the empty set;
limits of thresholds cover unbounded rectangles. The mixture and
derivative formulas in Appendix~\ref{appx-sec:smoothing} apply to
these normals.

The observations are unbounded, so we first recover rectangle
probabilities from smooth expectations using anti-concentration
only for the Gaussian target. This avoids imposing a boundary
estimate on the original sum.

\begin{lemma}[Smoothing from the Gaussian endpoint]\label{ben:smoothing}
Let $V\in\R^p$ be any random vector, and let $Z\in\R^p$
be centered Gaussian with $\Var(Z_j)=1$ for every $j\le p$.
Define $g_\Sigma:=\E\|Z\|_\infty$. Then, for every $h>0$,
\begin{equation}\label{ben:gaussian-smoothing}
 \rh(V,Z)\le C\left\{
 \sup_{a\in\R^{2p}}|\E F_{a,h}(V)-\E F_{a,h}(Z)|
       +hg_\Sigma L\right\},\qquad L:=\log(ep),
\end{equation}
where $C$ is absolute.
\end{lemma}
\begin{proof}
Define the signed rectangle discrepancy and its supremum by
\[
 H(a):=\Pp(V\in A_a)-\Pp(Z\in A_a),\qquad
 \rho:=\sup_a|H(a)|,
\]
and let $D$ denote the supremum of the smooth discrepancies in
\eqref{ben:gaussian-smoothing}. In the mixture representation
\eqref{eq:mixture}, let $\omega_0,\ldots,\omega_{2p}$ be
independent standard exponential variables, independent of $V,Z$,
and define $\eta_\alpha:=h\log(\omega_\alpha/\omega_0)$.
Then $F_{a,h}(x)=\E_\omega\ind_{A_{a+\eta}}(x)$, so
\[
 \E F_{a,h}(V)-\E F_{a,h}(Z)=\E_\omega H(a+\eta).
\]
Write $R_h:=\|\eta\|_\infty$. The estimate at the start of
Appendix~\ref{appx-sec:smoothing} gives $\E R_h\le ChL$.
Choose $d:=KhL$, with an absolute $K$ large enough that Markov's
inequality gives $\Pp(R_h>d)\le1/4$.

Suppose first that $\sup_aH(a)=\rho>0$, and choose $a$ with
$H(a)\ge0$. On $\{R_h\le d\}$, all coordinates of
$d\mathbf1+\eta$ lie between $0$ and $d+R_h$. The probability
for $V$ can only increase when the rectangle expands, while the
increase for $Z$ is at most $C(d+R_h)g_\Sigma$ by
Lemma~\ref{lem:concentration}. Hence
\[
 H(a+d\mathbf1+\eta)\ge H(a)-C(d+R_h)g_\Sigma.
\]
On the complementary event the discrepancy is still at least
$-\rho$. Averaging and using the mixture representation gives
\begin{align*}
 D&\ge\E_\omega H(a+d\mathbf1+\eta)\\
  &\ge H(a)\Pp(R_h\le d)
      -C(d+\E R_h)g_\Sigma-\rho\Pp(R_h>d)\\
  &\ge\tfrac34H(a)-C(d+\E R_h)g_\Sigma-\tfrac14\rho.
\end{align*}
Letting $H(a)$ approach its supremum yields
$D\ge\rho/2-C(d+\E R_h)g_\Sigma$.

If instead $\sup_a(-H(a))=\rho>0$, use the inward shift
$a-d\mathbf1+\eta$. On $\{R_h\le d\}$, the probability for
$V$ can only decrease, and the decrease for $Z$ is at most
$C(d+R_h)g_\Sigma$. Thus
\[
 -H(a-d\mathbf1+\eta)\ge-H(a)-C(d+R_h)g_\Sigma.
\]
The same averaging argument gives the preceding bound for $\rho$.
If $\rho=0$, there is nothing to prove. Since
$d+\E R_h\le ChL$ and $\rho=\rh(V,Z)$, the lemma follows.
\end{proof}

We next derive the comparison identity for smooth functions.
Pairing each summand with its copy allows us to average the
Hessian at the two endpoints of a segment. The error in that
average involves a fourth derivative. This use of the copies
does not require the original observations to have zero third moments.

\begin{lemma}[Slepian interpolation with paired increments]
\label{ben:paired-interpolation}
Under the assumptions and notation of Lemma~\ref{ben:comparison},
take $Z$ independent of the original array and its copies. For
$h>0$, define $\mathcal X:=\sigma(X_1,\ldots,X_n)$ and
\[
 \bar Y_i:=Y_i\ind\{\norm{Y_i}_\infty\le h\},\qquad
 b_h:=\E\left[\sum_iY_i\ind\{\norm{Y_i}_\infty>h\}
                    \mid\mathcal X\right],\qquad
 T_h:=\frac12\sum_i\E[\bar Y_i\bar Y_i^\top\mid\mathcal X].
\]
Along the Slepian path
$W_s:=\sqrt s\,W+\sqrt{1-s}\,Z$, $0\le s\le1$,
every $C^4$ function $f:\R^p\to\R$ with bounded derivatives
of orders $1$ through $4$ satisfies, for $0<s<1$,
\begin{align}
 \frac d{ds}\E f(W_s)
 ={}&\frac12\E[(T_h-\Sigma):D^2f(W_s)]
       -\frac1{2\sqrt s}\E[b_h\cdot\nabla f(W_s)]\notag\\
 &-\frac{s}{8}\sum_i\E\int_0^1 r(1-r)
 D^4f(W_s+r\sqrt s\,\bar Y_i)
          [\bar Y_i,\bar Y_i,\bar Y_i,\bar Y_i]\,dr.
 \label{ben:slepian-identity}
\end{align}
Here $A:B:=\sum_{j,k}A_{jk}B_{jk}$ for matrices $A,B$.
\end{lemma}
\begin{proof}
For time $s$ in a compact subinterval of $(0,1)$, the derivative
along the path is dominated by a constant times $\|W\|_\infty+\|Z\|_\infty$,
which is integrable. We may therefore differentiate under the
expectation. Gaussian integration by parts then gives
\begin{equation}\label{ben:slepian-initial}
 \frac d{ds}\E f(W_s)
 =\frac1{2\sqrt s}\E[W\cdot\nabla f(W_s)]
       -\frac12\E[\Sigma:D^2f(W_s)].
\end{equation}
In fact,
$\E[Z\cdot\nabla f(W_s)]
=\sqrt{1-s}\,\E[\Sigma:D^2f(W_s)]$.
For singular $\Sigma$, write $Z=AG$ with $G$ standard Gaussian
in $\R^{\operatorname{rank}\Sigma}$ and $AA^\top=\Sigma$;
integration by parts in the coordinates of $G$ proves the same identity.

We transform the term involving $W$ by interchanging $X_i$ and
$\widetilde X_i$. This interchange preserves the joint law of
the arrays, sends $Y_i$ to $-Y_i$, and sends $W_s$ to
$W_s+\sqrt s\,Y_i$. The event $\{\|Y_i\|_\infty\le h\}$
is unchanged. On that event $Y_i=\bar Y_i$, and off it the
multiplier $\bar Y_i$ is zero. Consequently,
\[
 \E[\bar Y_i\cdot\nabla f(W_s)]
 =-\E[\bar Y_i\cdot\nabla f(W_s+\sqrt s\,\bar Y_i)].
\]
Also $\E[\sum_iY_i\mid\mathcal X]=-W$, because the copies
are centered. Conditioning additionally on the independent $Z$
does not change this expectation. Splitting the increments at $h$
and using the preceding interchange identity yields
\begin{align*}
 \E[W\cdot\nabla f(W_s)]
 &=-\sum_i\E[\bar Y_i\cdot\nabla f(W_s)]
       -\E[b_h\cdot\nabla f(W_s)]\\
 &=\frac12\sum_i\E\big[\bar Y_i\cdot
       \{\nabla f(W_s+\sqrt s\,\bar Y_i)-\nabla f(W_s)\}\big]
       -\E[b_h\cdot\nabla f(W_s)].
\end{align*}

To expand each gradient difference, recall the trapezoidal identity
\[
 \int_0^1g(r)\,dr
 =\frac{g(0)+g(1)}2-\frac12\int_0^1r(1-r)g''(r)\,dr
\]
for twice continuously differentiable $g$. It follows by two
integrations by parts, using $r(1-r)=0$ at both endpoints.
For a fixed $i$, take
\[
 g_i(r):=D^2f(W_s+r\sqrt s\,\bar Y_i)[\bar Y_i,\bar Y_i].
\]
The fundamental theorem of calculus expresses the inner product
above as $\sqrt s\int_0^1g_i(r)\,dr$.
Moreover,
\[
 g_i''(r)=sD^4f(W_s+r\sqrt s\,\bar Y_i)
                   [\bar Y_i,\bar Y_i,\bar Y_i,\bar Y_i].
\]
The copy interchange sends $g_i(1)$ to $g_i(0)$, since the two
directional factors change sign together. Thus
$\E g_i(1)=\E g_i(0)$, and the trapezoidal identity gives
\begin{align*}
 &\E\big[\bar Y_i\cdot
   \{\nabla f(W_s+\sqrt s\,\bar Y_i)-\nabla f(W_s)\}\big]\\
 &\quad=\sqrt s\,\E D^2f(W_s)[\bar Y_i,\bar Y_i]
 -\frac{s^{3/2}}2\E\int_0^1r(1-r)
       D^4f(W_s+r\sqrt s\,\bar Y_i)[\bar Y_i^{\otimes4}]\,dr.
\end{align*}
All these terms are integrable because the increments are truncated
and the relevant derivatives are bounded. Conditional on
$\mathcal X$ and $Z$, the Hessian at $W_s$ is fixed and
$\E[\bar Y_i\bar Y_i^\top\mid\mathcal X,Z]
=\E[\bar Y_i\bar Y_i^\top\mid\mathcal X]$.
Substitution in~\eqref{ben:slepian-initial} therefore gives the
coefficient $\tfrac14\sum_i\E[\bar Y_i\bar Y_i^\top\mid\mathcal X]
=T_h/2$ for the Hessian, and the remainder coefficient is
$(2\sqrt s)^{-1}\cdot\tfrac12\cdot(-s^{3/2}/2)=-s/8$.
This proves~\eqref{ben:slepian-identity}.
\end{proof}

\begin{proof}[Proof of Lemma~\ref{ben:comparison}]
Fix $\phi>0$ and set $h:=t=(\phi L)^{-1}$. We use
Lemma~\ref{ben:paired-interpolation} with $f=F_{a,h}$. Its three
terms require bounds on the conditional covariance error
$T_h-\Sigma$, the conditional drift $b_h$, and the fourth-order
remainder. The vectors $Y_i$ are independent and centrally
symmetric, but they need not be independent of $W$.

For the drift, conditional Jensen's inequality gives
$\E\|b_h\|_\infty\le M_h$. For the covariance bias, note that
$\tfrac12\sum_i\E Y_iY_i^\top=\Sigma$. Thus
\[
 \E T_h-\Sigma
 =-\frac12\sum_i\E[Y_iY_i^\top\ind\{\|Y_i\|_\infty>h\}].
\]
Applying Cauchy--Schwarz to the two coordinate factors and the sum
over $i$ bounds every entry in absolute value by $v_h/2$.
The fourth moments also satisfy
\[
 \max_j\sum_i\E Y_{ij}^4\le16\Delta_1,\qquad
 |\bar Y_{ij}\bar Y_{ik}|\le h^2,\qquad
 \max_{j,k}\sum_i\E\bar Y_{ij}^2\bar Y_{ik}^2\le16\Delta_1.
\]
The first bound follows from $|a-b|^4\le8(|a|^4+|b|^4)$;
the last follows by Cauchy--Schwarz and the first.

To bound the covariance fluctuation, conditional Jensen gives
\[
 \E\maxnorm{T_h-\E T_h}
 \le\frac12\E\max_{j,k}\left|
   \sum_i\{\bar Y_{ij}\bar Y_{ik}
                    -\E\bar Y_{ij}\bar Y_{ik}\}\right|.
\]
For fixed $j,k$, the summands on the right are independent and
centered, bounded by $2h^2$, and their variance sum is at most
$16\Delta_1$. To see the expected-maximum bound we need, let
$U_i$ be independent centered scalars with $|U_i|\le b$ and
$\sum_i\E U_i^2\le v$. Expansion of the exponential gives
\[
 \E e^{\lambda\sum_iU_i}
 \le\exp\!\left\{\frac{\lambda^2v}{2(1-|\lambda|b/3)}\right\},
 \qquad |\lambda|b<3.
\]
Sum this estimate over both signs of all $p^2$ entries. Jensen's
inequality for the logarithm bounds their expected maximum by
\[
 \frac{\log(2p^2)}{\lambda}
       +\frac{\lambda v}{2(1-\lambda b/3)},\qquad 0<\lambda b<3.
\]
Choosing $\lambda$ of order
$\min\{\sqrt{L/v},1/b\}$ gives $C(\sqrt{vL}+bL)$;
the case $v=0$ follows directly. Combining the fluctuation and
bias bounds proves
\begin{equation}\label{ben:conditional-matrix}
 \E\norm{T_h-\Sigma}_{\max}
 \le C\{\sqrt{L\Delta_1}+h^2L+v_h\}.
\end{equation}

For the fourth-order term, define
$Q_h:=\max_j\sum_i\bar Y_{ij}^4$.
If $0\le U\le b$, convexity of the exponential on $[0,1]$ gives
$\E e^{U/b}\le1+(e-1)\E U/b$.
Apply this with $U=\bar Y_{ij}^4$ and $b=h^4$, and multiply
over the independent observations. Since the exponential of a
maximum is bounded by the sum of the exponentials, Jensen gives
\begin{equation}\label{ben:fourth-maximum}
 \begin{split}
 \E Q_h
 &\le h^4\log p+(e-1)\max_j\sum_i\E\bar Y_{ij}^4\\
 &\le C(\Delta_1+h^4L).
 \end{split}
\end{equation}
This estimate keeps a maximum of coordinate sums, rather than a
sum of fourth moments of the envelope.

We next bound the derivatives, retaining the same coordinate sums.
Fix $a$ and write $F:=F_{a,h}$ and $\Psi:=F(1-F)$.
The positive weight formula~\eqref{appx-eq:partition-derivative}
with normals $\pm e_j$ gives
\[
 \sum_j|\partial_jF(x)|\le h^{-1}\Psi(x),\qquad
 \sum_{j,k}|\partial_{jk}F(x)|\le Ch^{-2}\Psi(x).
\]
For vectors $y_i$ and shifts $z_i$ satisfying
$\|z_i\|_\infty\le h$, the same formula gives
\begin{equation}\label{ben:fourth-contraction}
 \sum_i|D^4F(x+z_i)[y_i,y_i,y_i,y_i]|
 \le Ch^{-4}\Psi(x)\max_j\sum_i|y_{ij}|^4.
\end{equation}
To obtain this bound, we retain the positive weights until after
we have summed the directional products over $i$. For every partition in the
derivative formula, its weight at $x+z_i$ is bounded by an
absolute multiple of its weight at $x$, as established in
\eqref{appx-eq:local-envelope}. For each choice of four coordinate
indices, H\"older's inequality gives
\[
 \sum_i|y_{ij_1}y_{ij_2}y_{ij_3}y_{ij_4}|
 \le\prod_{\ell=1}^4
       \left(\sum_i|y_{ij_\ell}|^4\right)^{1/4}
 \le\max_j\sum_i|y_{ij}|^4.
\]
The sum of the positive weights for a partition with $k\ge1$
blocks is $F(x)(1-F(x))^k\le\Psi(x)$. There are only finitely
many partitions at derivative order $4$, proving
\eqref{ben:fourth-contraction}.

To average these derivative bounds, we use the Gaussian component
of the path. Uniformly over deterministic $x$ and $0<s<1$,
\begin{equation}\label{ben:gaussian-envelope}
 \E_Z\Psi(x+\sqrt{1-s}\,Z)
 \le C\frac{hg_\Sigma}{\sqrt{1-s}}.
\end{equation}
Indeed, let $F_r:=F_{a+r\mathbf1,h}$. The calculation in
Lemma~\ref{appx-lem:envelope} gives
$\partial_rF_r=F_r(1-F_r)/h$ and
$\Psi\le(e/2)(F_h-F_{-h})$. The mixture representation
\eqref{eq:mixture} expresses $F_h-F_{-h}$ as an average of
indicators of boundary layers of width $2h$. Conditional on the
mixture variables, the thresholds are fixed. Apply the Gaussian
part of Lemma~\ref{lem:concentration} to $\sqrt{1-s}\,Z$,
whose face variances are $1-s$ and whose expected maximum norm
is $\sqrt{1-s}\,g_\Sigma$. Its intrinsic boundary coefficient
is therefore $g_\Sigma/\sqrt{1-s}$, proving
\eqref{ben:gaussian-envelope}.

Apply~\eqref{ben:slepian-identity} to $F$. Conditional on
$\mathcal X$, both $T_h-\Sigma$ and $b_h$ are fixed, and
$W_s=\sqrt s\,W+\sqrt{1-s}\,Z$ has an independent Gaussian
component. Thus the first two derivative bounds can be averaged
using~\eqref{ben:gaussian-envelope}. For the remainder, condition
on both arrays. The increments $\bar Y_i$ are then fixed, and
the shifts $r\sqrt s\,\bar Y_i$ have maximum norm at most $h$.
Use~\eqref{ben:fourth-contraction} first to bring all weights to
$W_s$, and only then average over $Z$. We obtain
\begin{align*}
 \left|\frac d{ds}\E F(W_s)\right|
 \le{}&C\frac{g_\Sigma}{h\sqrt{1-s}}
                \E\norm{T_h-\Sigma}_{\max}
       +C\frac{g_\Sigma}{\sqrt{s(1-s)}}M_h\\
 &+C\frac{sg_\Sigma}{h^3\sqrt{1-s}}\E Q_h.
\end{align*}
All coefficient expectations are finite by the estimates above,
and the time factors are integrable:
\[
 \int_0^1\frac{ds}{\sqrt{1-s}}=2,\qquad
 \int_0^1\frac{ds}{\sqrt{s(1-s)}}=\pi,\qquad
 \int_0^1\frac{s\,ds}{\sqrt{1-s}}=\frac43.
\]
Integrate over $[\varepsilon,1-\varepsilon]$ and let
$\varepsilon\downarrow0$. The derivative bounds control the
integrals, while boundedness and continuity of $F$ give convergence
of the endpoint expectations. Uniformly in $a$,
\[
 |\E F_{a,h}(W)-\E F_{a,h}(Z)|
 \le Cg_\Sigma\left\{
 h^{-1}\E\norm{T_h-\Sigma}_{\max}+M_h+h^{-3}\E Q_h\right\}.
\]
Use~\eqref{ben:conditional-matrix} and~\eqref{ben:fourth-maximum},
then apply Lemma~\ref{ben:smoothing}. The terms $h^2L/h$ and
$h^4L/h^3$ both have order $hL$, as does the smoothing error.
Consequently,
\[
 \rh(W,Z)\le Cg_\Sigma\left\{
 hL+\frac{\sqrt{L\Delta_1}}h
 +\frac{\Delta_1}{h^3}+\frac{v_h}h+M_h\right\}.
\]
Substituting $h=t=(\phi L)^{-1}$ and factoring out
$g_\Sigma/\sqrt L$ gives~\eqref{ben:master}.
\end{proof}

The comparison bound leaves the expected maximum $M_t$ of the
large-increment sum to be estimated. The next lemma introduces
a second truncation level $u$: Bennett's inequality controls
increments up to $u$, and the envelope moment controls those
above $u$.

\begin{lemma}[A truncated Bennett maximum]\label{ben:maximum}
Suppose independent centrally symmetric vectors $T_i\in\R^p$
satisfy $\max_j\sum_i\E T_{ij}^2\le v$ and
$\sum_i\E\norm{T_i}_\infty^q\le A^q$, with $q>1$.
For every $u>0$ and $v>0$,
\begin{equation}
 \E\norm{\sum_iT_i}_\infty
 \le C\frac{uL}{\log(1+u^2L/v)}+\frac{A^q}{u^{q-1}}.
 \label{ben:maximum-bound}
\end{equation}
If $v=0$, the sum is zero almost surely.
\end{lemma}
\begin{proof}
Define $T_i^{(u)}:=T_i\ind\{\|T_i\|_\infty\le u\}$.
The truncation event is invariant under a change of sign, so
central symmetry makes these vectors centered. They remain
independent, with envelope $u$ and coordinate variance sums at
most $v$. Bennett's exponential bound
\cite[Theorem~2.9]{BLM2013} gives
\[
 \log\E\exp\left(\pm\lambda\sum_iT_{ij}^{(u)}\right)
 \le\frac v{u^2}(e^{\lambda u}-1-\lambda u),\qquad\lambda>0.
\]
This form also follows by expanding the exponential and using
$\E|T_{ij}^{(u)}|^k\le u^{k-2}\E(T_{ij}^{(u)})^2$ for
$k\ge2$. For $S^{(u)}:=\sum_iT_i^{(u)}$, the pointwise bound
\[
 e^{\lambda\|S^{(u)}\|_\infty}
 \le\sum_{j=1}^p\{e^{\lambda S_j^{(u)}}+e^{-\lambda S_j^{(u)}}\}
\]
and Jensen's inequality yield
\[
 \E\|S^{(u)}\|_\infty
 \le\frac{\log(2p)}\lambda
      +\frac v{\lambda u^2}(e^{\lambda u}-1-\lambda u).
\]
Choose $\lambda:=u^{-1}\log(1+u^2L/v)$. Then
$e^{\lambda u}-1=u^2L/v$, and $\log(2p)\le L$ gives
\[
 \E\|S^{(u)}\|_\infty
 \le\frac{2L}{\lambda}
 =\frac{2uL}{\log(1+u^2L/v)}.
\]
Finally, the expected norm of the omitted sum is at most
\[
 \sum_i\E[\|T_i\|_\infty\ind\{\|T_i\|_\infty>u\}]
 \le u^{1-q}\sum_i\E\|T_i\|_\infty^q
 \le A^qu^{1-q}.
\]
Adding the two estimates proves~\eqref{ben:maximum-bound}.
If $v=0$, every coordinate of every $T_i$ has second moment
zero, which proves the last assertion.
\end{proof}

We now apply Lemma~\ref{ben:comparison} under the assumptions of
Theorem~\ref{thm:polynomial}\textup{(ii)}. Keep the lemma's
notation $Y_i,v_t,M_t,\Delta_1$, and define
$A:=Bn^{-1/2+1/q}$ and $d:=B/\sqrt n$. The coordinate
fourth-moment condition and the envelope condition give, respectively,
\[
 \Delta_1\le d^2,\qquad
 \max_j\sum_i\E Y_{ij}^4\le16d^2,\qquad
 \sum_i\E\|Y_i\|_\infty^q\le C_qA^q.
\]
The last bound uses
$\|\widetilde X_i-X_i\|_\infty^q
\le2^{q-1}(\|\widetilde X_i\|_\infty^q+\|X_i\|_\infty^q)$.
For the tail variance, apply Cauchy--Schwarz first within each
expectation and then across observations. Markov's inequality
for the envelope gives
\begin{equation}
 \begin{split}
 v_t&\le
 \max_j\left(\sum_i\E Y_{ij}^4\right)^{1/2}
       \left(\sum_i\Pp(\|Y_i\|_\infty>t)\right)^{1/2}\\
 &\le C_qd(A/t)^{q/2}.
 \end{split}
 \label{ben:mixed}
\end{equation}
This is where the two moment conditions enter the same estimate.

The vectors $Y_i\ind\{\|Y_i\|_\infty>t\}$ are independent
and centrally symmetric, with coordinate variance sums bounded
by $v_t$ and envelope-moment budget at most $C_qA^q$.
Lemma~\ref{ben:maximum} applies to these vectors. The triangle
inequality gives an alternative bound by their first absolute
moments. Thus, for every $u>0$ and $v_t>0$,
\begin{equation}
 M_t\le C_q\left\{\frac{uL}{\log(1+u^2L/v_t)}
                      +\frac{A^q}{u^{q-1}}\right\},
 \qquad M_t\le C_qA^qt^{1-q}.
 \label{ben:tail}
\end{equation}
If $v_t=0$, these vectors vanish almost surely and $M_t=0$.

It remains to choose the smoothing bandwidth. Recall
\[
 U^2=dL^{5/2},\qquad V_q=AL^{3/2-1/q},\qquad
 \mathcal K_q=(V_q/U)^2,\qquad
 \mathcal L_q=\log(e+\mathcal K_q).
\]
The factor $g_\Sigma/\sqrt L$ in~\eqref{ben:master} does not
depend on the bandwidth. We estimate the expression in braces
and restore that factor at the end. If
$\phi^{-1}=r/\sqrt L$, then $t=r/L^{3/2}$, and the first
three terms in braces are bounded by
\[
 r,\qquad C\frac{U^2}{r},\qquad C\frac{U^4}{r^3}.
\]
The remaining two terms are $L^2v_t/r$ and $\sqrt L M_t$.
This representation shows directly which terms the bandwidth
must balance.

Suppose first that $\mathcal K_q\le1$. Choose $r:=U$, so
$\phi^{-1}=U/\sqrt L$ and $t=U/L^{3/2}$. The first three
terms are then bounded by $CU$. Since
$A/t=(V_q/U)L^{1/q}$, \eqref{ben:mixed} gives
\[
 \frac{L^2v_t}{U}\le C_qU\mathcal K_q^{q/4}.
\]
The direct estimate in~\eqref{ben:tail} similarly gives
\[
 \sqrt L M_t\le C_q\sqrt L A^qt^{1-q}
                 =C_qU\mathcal K_q^{q/2}.
\]
Both are at most $C_qU$, completing this case.

Suppose now that $\mathcal K_q>1$. Define $b:=1-1/q$ and choose
\[
 r:=V_q\mathcal L_q^{-b},\qquad
 \phi^{-1}:=r/\sqrt L,\qquad
 t:=AL^{-1/q}\mathcal L_q^{-b},\qquad
 u:=A(\mathcal L_q/L)^{1/q}=\mathcal L_q t.
\]
The value of $t$ equals $r/L^{3/2}$ as required, while the
larger truncation level $u$ is chosen to balance the two terms
in the Bennett estimate. Indeed, \eqref{ben:mixed} gives
\[
 v_t\le C_qdL^{1/2}\mathcal L_q^{(q-1)/2},\qquad
 \frac{u^2L}{v_t}
 \ge c_q\mathcal K_q\mathcal L_q^{2/q-(q-1)/2}
\]
when $v_t>0$, for a positive constant $c_q$.
For every fixed real $c$ and $c_0>0$, there is a positive
constant $c_1$ such that
\[
 \log\bigl(1+c_0K\{\log(e+K)\}^c\bigr)
 \ge c_1\log(e+K),\qquad K\ge1.
\]
To verify this, divide by $\log(e+K)$: the ratio tends to $1$
as $K\to\infty$, and is continuous and positive on every
compact subinterval of $[1,\infty)$. Applying this fact gives
$\log(1+u^2L/v_t)\ge c_q'\mathcal L_q$.
Both terms of~\eqref{ben:tail}, after multiplication by
$\sqrt L$, are therefore bounded by $C_qr$, because
\[
 \frac{uL^{3/2}}{\mathcal L_q}=r,
 \qquad \frac{\sqrt L A^q}{u^{q-1}}=r.
\]
If $v_t=0$, the same conclusion follows from $M_t=0$.

The other four terms in~\eqref{ben:master} are bounded, in order, by
\[
 r,\qquad C r\frac{\mathcal L_q^{2b}}{\mathcal K_q},\qquad
 C r\frac{\mathcal L_q^{4b}}{\mathcal K_q^2},\qquad
 C_q r\frac{\mathcal L_q^{2b+(q-1)/2}}{\mathcal K_q}.
\]
The middle two follow from $U^2=V_q^2/\mathcal K_q$ and
$r=V_q\mathcal L_q^{-b}$. The last follows by substituting
the bound on $v_t$ into $L^2v_t/r$. Each displayed ratio is
bounded for $\mathcal K_q\ge1$, with a constant depending
only on $q$, since any fixed power of $\log(e+K)$ grows more
slowly than $K$. Thus the expression in braces is at most
$C_qr$ in this case.

Combining the cases bounds that expression by
$C_q\{U+V_q/\mathcal L_q^{1-1/q}\}$. Restoring
$g_\Sigma/\sqrt L$ and using $\rh\le1$ proves
\eqref{poly:bennett-bound}, and hence
Theorem~\ref{thm:polynomial}\textup{(ii)}.

\section{Lower-bound constructions and proofs}\label{app:sharpness-proofs}

We prove the lower bounds by two mechanisms. For independent skewed
coordinates, a small relative error in each scalar tail accumulates
when we take the maximum. For the symmetric examples, the maximum
splits into a leading component and an independent random shift.
The difference between the mean shifts for the sum and its Gaussian
counterpart produces the lower bound. We first establish the scalar
tail estimate and use it for the skewed example. We then prove a
comparison for shifted maxima, introduce the common symmetric
construction in Subsection~\ref{app:lower-family}, and choose its
parameters for the remaining theorems.

We use $W_n,\Sigma_n,Z_n$, and $L_n$ as defined
in~\eqref{lower:eq:setup}. All limits are as $n\to\infty$ unless
otherwise specified. Positive absolute constants are denoted by
$C$ or $c_*$, and their values may change between inequalities.
Write $\Phi$ and $\varphi$ for the standard normal distribution
function and density, and define $\bar\Phi(x):=1-\Phi(x)$.
Each lower bound will be witnessed by a lower orthant. Intersecting
that orthant with $[-R,R]^p$ and letting $R\to\infty$ shows that
the same lower bound holds if the rectangle supremum is restricted
to bounded closed rectangles.

\subsection{A uniform scalar tail estimate}
\label{app:moderate-tail-proof}

The third moment determines the first correction to a Gaussian
tail. To use this correction for triangular arrays, we need an
estimate whose constants are uniform in the summand bound. We
also keep the tail strict, so that the estimate applies directly
to distribution functions even when the summands have atoms.

\begin{lemma}[Uniform scalar moderate-tail estimate]
\label{opt:moderate-tail}
Let $S:=\sum_{i=1}^n\eta_i$, where the real variables $\eta_i$ are
independent and centered, $\delta>0$, $|\eta_i|\le\delta$
almost surely, and $\sum_i\E\eta_i^2=1$. Write
$\gamma:=\sum_i\E\eta_i^3$. There are absolute constants $c_0,C>0$
such that, whenever $x\ge0$ and $\delta(1+x)\le c_0$,
\begin{equation}
 \left|\log\frac{\Pp(S>x)}{1-\Phi(x)}
       -\frac{\gamma x^3}{6}\right|
 \le C\{\delta(1+x)+\delta^2x^4\}.
 \label{opt:moderate-tail-formula}
\end{equation}
The tail is strict. The conclusion includes lattice distributions and
has constants independent of the lattice spacing or the position of
$x$ relative to its atoms.
\end{lemma}

\begin{proof}
We change the probability law so that $x$ becomes the mean of the
sum. Under the new law, a scalar Berry--Esseen bound controls the
fluctuations around $x$; the change of measure supplies the
exponential tail correction.

Define the \emph{cumulant-generating function}
$K(t):=\sum_i\log\E e^{t\eta_i}$. For $t\ge0$, define the
joint tilted law by
\[
 d\Pp_t:=e^{tS-K(t)}\,d\Pp.
\]
The normalizing identity $\E e^{tS}=e^{K(t)}$ follows from
independence. The tilted law is still a product law: the factor
for summand $i$ is
$e^{t\eta_i}/\E e^{t\eta_i}$. Write $\E_t$ and $\Var_t$
for expectation and variance under this law. Since
$|\eta_i|\le\delta$, each factor lies between
$e^{-2\delta t}$ and $e^{2\delta t}$.

To control the variance after tilting, let $\eta_i'$ be an
independent copy under the original law. The identity
$\Var(\eta_i)=\tfrac12\E(\eta_i-\eta_i')^2$ holds under
either product law. The density of the tilted pair relative to
the original pair lies between $e^{-4\delta t}$ and
$e^{4\delta t}$. Applying these bounds to the nonnegative
squared difference gives
\[
 e^{-4\delta t}\Var(\eta_i)
 \le\Var_t(\eta_i)\le e^{4\delta t}\Var(\eta_i).
\]
Differentiating the moment-generating functions is justified by
boundedness of the summands, and gives
$K'(t)=\sum_i\E_t\eta_i$ and
$K''(t)=\sum_i\Var_t(\eta_i)$. Thus, for a fixed small
absolute $c_1>0$,
\[
 e^{-4c_1}\le K''(t)\le e^{4c_1},\qquad
 0\le t\le c_1/\delta.
\]

We also need a uniform fourth-derivative bound. If
$\zeta_i:=\eta_i-\E_t\eta_i$, then $|\zeta_i|\le2\delta$
and $\Var_t(\eta_i)\le\delta^2$. The fourth derivative of
the $i$th log moment-generating function is its tilted fourth
cumulant, so
\[
 \left|\E_t\zeta_i^4-3\{\Var_t(\eta_i)\}^2\right|
 \le(4\delta^2+3\delta^2)\Var_t(\eta_i).
\]
Summing and using the bound on $K''$ yields
$\sup_{0\le t\le c_1/\delta}|K^{(4)}(t)|\le C\delta^2$.
At zero, $K(0)=K'(0)=0$, $K''(0)=1$, and $K'''(0)=\gamma$,
with $|\gamma|\le\delta\sum_i\E\eta_i^2=\delta$.
Taylor's theorem therefore gives, uniformly in this range,
\begin{align}
 K(t)&=t^2/2+\gamma t^3/6+O(\delta^2t^4),\notag\\
 K'(t)&=t+\gamma t^2/2+O(\delta^2t^3),\notag\\
 K''(t)&=1+\gamma t+O(\delta^2t^2).
 \label{opt:cumulant-expansions}
\end{align}

We now choose $t$ so that $K'(t)=x$. This choice must lie in
the range just controlled. With $t_0:=c_1/\delta$,
\[
 K'(t_0)=\int_0^{t_0}K''(u)\,du
 \ge e^{-4c_1}c_1/\delta.
\]
Choose $c_0<e^{-4c_1}c_1/2$. The hypothesis then implies
$x<K'(t_0)$. Since $K'(0)=0$ and $K''>0$, there is a
unique $t\in[0,t_0)$ with $K'(t)=x$, with $t=0$ if $x=0$.
For $x>0$, the upper and lower bounds on $K''$ give
$t\asymp x$.

The second expansion in~\eqref{opt:cumulant-expansions} first
gives $t-x=O(\delta x^2)$. Substitute this estimate back into
$x=t+\gamma t^2/2+O(\delta^2t^3)$ to replace $t^2$ by $x^2$
in the quadratic term. We obtain
\begin{equation}
 t=x-\gamma x^2/2+O(\delta^2x^3),\qquad
 K(t)-tx=-x^2/2+\gamma x^3/6+O(\delta^2x^4).
 \label{opt:tilt-root}
\end{equation}
For the second identity, one may first use $x=K'(t)$ to write
$K(t)-tx=-t^2/2-\gamma t^3/3+O(\delta^2t^4)$, and then
substitute the expansion of $t$.
Define $v:=\sqrt{K''(t)}>0$ and $s:=tv$. The square-root
expansion gives $v=1+\gamma x/2+O(\delta^2x^2)$.
In the product $tv$, the two terms of order $\gamma x^2$
cancel, leaving
\begin{equation}
 s=x+O(\delta^2x^3).
 \label{opt:tilted-scale}
\end{equation}

Under $\Pp_t$, the standardized sum $T:=(S-x)/v$ has mean
zero and variance $1$. Its summands remain independent, and
\[
 \sum_i\E_t|\eta_i-\E_t\eta_i|^3
 \le2\delta\sum_i\Var_t(\eta_i)=2\delta v^2.
\]
The scalar Berry--Esseen inequality~\cite{Berry1941}, applied
after division by $v$, gives
$\sup_y|\Pp_t(T\le y)-\Phi(y)|\le C\delta$, since $v$
is bounded above and away from zero.

The change of measure involves a weighted strict tail. To preserve
its endpoint convention, use, for $s\ge0$,
\[
 \E_t[e^{-sT}\ind_{\{T>0\}}]
 =\Pp_t(T>0)-s\int_0^\infty e^{-sy}\Pp_t(T>y)\,dy.
\]
For $s>0$, this follows by writing
$e^{-sT}=1-\int_0^T se^{-sy}\,dy$ on $\{T>0\}$ and using
Tonelli's theorem. At $s=0$, the integral term is defined to be
zero. Apply the same identity to a standard normal variable.
Berry--Esseen bounds the difference of the first terms by
$C\delta$, and bounds the difference of the integral terms by
$C\delta\int_0^\infty se^{-sy}\,dy=C\delta$. Thus
\begin{equation}
 \E_t[e^{-sT}\ind_{\{T>0\}}]=J(s)+O(\delta),\qquad
 J(u):=\int_0^\infty e^{-uy}\varphi(y)\,dy.
 \label{opt:strict-tail}
\end{equation}
No continuity of the law of $T$ has been used; an atom at zero
is excluded by the strict tail on both sides of the comparison.

For $u\ge0$, the integral satisfies
$J(u)\asymp(1+u)^{-1}$ and $|J'(u)|\le C(1+u)^{-2}$.
For example, the lower bound follows by integrating over
$0\le y\le(1+u)^{-1}$; for $u\ge1$ the upper bounds follow
by replacing $\varphi(y)$ with $\varphi(0)$ and integrating
$e^{-uy}$ or $ye^{-uy}$. The compact range $0\le u\le1$
is immediate from the defining integrals. Consequently,
$|(\log J)'(u)|\le C/(1+u)$.
The relative error in~\eqref{opt:strict-tail} is
$O(\delta(1+s))$. Decreasing $c_0$ if necessary makes it less
than $1/2$, so taking logarithms introduces an error of that
same order. Also, $s\asymp x$ for $x>0$ and
\eqref{opt:tilted-scale} give
$\log J(s)-\log J(x)=O(\delta^2x^2)$.

Finally, undoing the tilt and completing the square in the normal
integral give, respectively,
\[
 \Pp(S>x)=e^{K(t)-tx}\E_t[e^{-sT}\ind_{\{T>0\}}],\qquad
 1-\Phi(x)=e^{-x^2/2}J(x).
\]
Take logarithms and use~\eqref{opt:tilt-root}. The main term is
$\gamma x^3/6$, and the errors are bounded by
$C\{\delta(1+x)+\delta^2x^4+\delta^2x^2\}$.
For $x\le1$, the last term is absorbed by $C\delta(1+x)$;
for $x\ge1$, it is absorbed by $C\delta^2x^4$.
This proves~\eqref{opt:moderate-tail-formula}.
\end{proof}

\subsection{Proof of Theorem~\ref*{opt:skew-lower} (skewed bounded observations)}
\label{app:skew-proof}

Let $p_n$ and $B_n$ satisfy Theorem~\ref{opt:skew-lower}.
The scale to be attained is
\[
 e_n=\frac{B_n g_{I_{p_n}}L_n}{\sqrt n}\longrightarrow0.
\]
Since $g_{I_{p_n}}\sim\sqrt{2\log p_n}$, this is equivalent
to $B_n(\log p_n)^{3/2}/\sqrt n\to0$. We choose a threshold
at which the Gaussian maximum has distribution function exactly
$e^{-1}$. An exact choice avoids an additional error of order
$p_n^{-1}$, which need not be smaller than $e_n$.

\begin{proof}[Proof of Theorem~\ref{opt:skew-lower}]
Take all scalar entries $X_{n,i,j}$, $1\le i\le n$ and
$1\le j\le p_n$, to be mutually independent with the law
\begin{equation}
 \Pp(X=B_n)=\frac1{B_n^2+1},\qquad
 \Pp(X=-B_n^{-1})=\frac{B_n^2}{B_n^2+1}.
 \label{opt:two-point-law}
\end{equation}
Direct calculation gives $\E X=0$, $\E X^2=1$, and
$\mu_{3,n}:=\E X^3=B_n-B_n^{-1}\ge3B_n/4$.
Also $|X|\le B_n$, because $B_n\ge2$. Define
\[
 \delta_n:=B_n/\sqrt n,\qquad
 x_n:=\Phi^{-1}(e^{-1/p_n}),\qquad
 t_n:=1-\Phi(x_n),\qquad q_n:=\Pp(W_{n,1}>x_n).
\]
The threshold satisfies $\Phi(x_n)^{p_n}=e^{-1}$ exactly.
Normal tail estimates give $x_n\sim\sqrt{2\log p_n}$,
while $t_n=1-e^{-1/p_n}$ gives $p_nt_n\to1$.
In particular, $\delta_nx_n^3\to0$, so the condition of
Lemma~\ref{opt:moderate-tail} holds eventually.

Apply that lemma to the normalized scalar sum $W_{n,1}$.
Its aggregate third moment is $\mu_{3,n}/\sqrt n$.
The resulting logarithmic correction has leading term
$s_n:=\mu_{3,n}x_n^3/(6\sqrt n)$ and remainder
$O\{\delta_n(1+x_n)+\delta_n^2x_n^4\}$.
Since $e_n\asymp\delta_nx_n^3$, this remainder divided by
$e_n$ is $O(x_n^{-2}+\delta_nx_n)=o(1)$.
Also $s_n=O(e_n)\to0$, so exponentiation introduces only
$O(e_n^2)=o(e_n)$. Hence
\begin{equation}
 q_n=t_n\{1+s_n+o(e_n)\},\qquad
 s_n=\frac{\mu_{3,n}x_n^3}{6\sqrt n}.
 \label{opt:tail-linearization}
\end{equation}
The size of this correction relative to the target rate is
\[
 \frac{s_n}{e_n}
 =\frac{\mu_{3,n}}{B_n}\,
   \frac{x_n^3}{6g_{I_{p_n}}L_n},\qquad
 \frac{x_n^3}{g_{I_{p_n}}L_n}\longrightarrow2.
\]
Thus $\liminf s_n/e_n\ge1/4$.

Independence of the coordinates expresses the discrepancy on the
common-threshold orthant as
\[
 \Pp(\max_jW_{n,j}\le x_n)-\Pp(\max_jZ_{n,j}\le x_n)
 =(1-q_n)^{p_n}-(1-t_n)^{p_n}.
\]
The mean value theorem and~\eqref{opt:tail-linearization} write
this as
$-p_nt_n(1-u_n)^{p_n-1}\{s_n+o(e_n)\}$, where $u_n$
lies between $q_n$ and $t_n$. Since $q_n/t_n\to1$ and
$p_nt_n\to1$, we have $p_nu_n\to1$ and
$(1-u_n)^{p_n-1}\to e^{-1}$. Consequently,
\begin{equation}
 \Pp(\max_jW_{n,j}\le x_n)-e^{-1}
 =-\frac{\mu_{3,n}x_n^3}{6e\sqrt n}(1+o(1)).
 \label{opt:exact-product-asymptotic}
\end{equation}
The rectangle distance is at least the absolute value of this
orthant discrepancy. Dividing by $e_n$ proves
\eqref{opt:lower-constant}. Finally, a coordinate equals $B_n$
with positive probability, so
$\Pp(\|X_{n,i}\|_\infty=B_n)>0$ and the envelope is attained.
\end{proof}

\subsection{A common comparison for shifted maxima}
\label{app:lower-comparison}

The symmetric constructions will give maxima of the form $A+U$
and $G+V$, with independent components in each sum. If the
distribution functions of $A$ and $G$ are close, conditioning
reduces the comparison to a Taylor expansion of the distribution
function of $G$. Its first derivative multiplies the difference
between the mean shifts, while its second derivative controls
the error.

\begin{lemma}[Comparison of independently shifted variables]
\label{lower:shift-comparison}
Let $A$ and $G$ be real random variables such that the distribution
function $F$ of $G$ is twice continuously differentiable with
$\|F''\|_\infty<\infty$, and suppose that, for some $\epsilon\ge0$,
$\sup_x|\Pp(A\le x)-F(x)|\le\epsilon$.
Let $U$ be independent of $A$, and let $V$ be independent of $G$,
with $\E U^2+\E V^2<\infty$. Then, for every $t\in\R$,
\begin{equation}
\begin{split}
&\left|\Pp(A+U\le t)-\Pp(G+V\le t)
       -F'(t)(\E V-\E U)\right|\\
&\hspace{2em}\le \epsilon+
       \frac12\|F''\|_\infty(\E U^2+\E V^2).
\end{split}
\label{lower:eq:shift-comparison}
\end{equation}
\end{lemma}

\begin{proof}
Write $F_A(x):=\Pp(A\le x)$. Independence and conditioning give
\[
 \Pp(A+U\le t)=\E F_A(t-U),\qquad
 \Pp(G+V\le t)=\E F(t-V).
\]
The uniform bound on $F_A-F$ changes the first expectation to
$\E F(t-U)$ with error at most $\epsilon$. For every real $y$,
Taylor's theorem gives
\[
 |F(t-y)-F(t)+yF'(t)|\le\tfrac12\|F''\|_\infty y^2.
\]
Apply this formula to $U$ and $V$, take expectations, and subtract.
The constant terms cancel and the linear terms give
$F'(t)(\E V-\E U)$. The assumed second moments bound the two
remainders. Neither shift is required to be nonnegative, and no
joint independence between the two pairs is needed.
\end{proof}

We prepare two estimates for applying the lemma. The first
approximates the leading Rademacher maximum by a Gaussian maximum.
For an integer $n\ge1$, let
\[
 A_n:=n^{-1/2}\sum_{i=1}^n\eta_i,
\]
where the $\eta_i$ are independent Rademacher signs. For an integer
$m\ge1$, let $A_{n,1},\ldots,A_{n,m}$ be independent copies
of $A_n$ and define $\Lambda_m:=\log(em)$. There are absolute
constants $c_3,C>0$ such that, for all sufficiently large $n$
and $\Lambda_m\le c_3\sqrt n$,
\begin{equation}
 \epsilon_{n,m}:=\sup_x\left|\Pp\{\max_{j\le m}A_{n,j}\le x\}
                         -\Phi(x)^m\right|
 \le C\left(\frac{\sqrt{\Lambda_m}}{\sqrt n}
                 +\frac{\Lambda_m^2}{n}\right).
 \label{sym:eq:rademacher-max}
\end{equation}
For $m=1$, this is scalar Berry--Esseen. For $m\ge2$, applying
an absolute scalar error directly to the product of distribution
functions would introduce a factor $m$. We instead use the
relative tail estimate in Lemma~\ref{opt:moderate-tail}.

For $A_n$, the parameters of that lemma are $\delta=n^{-1/2}$
and $\gamma=0$. Hence
\[
 \left|\log\frac{\Pp(A_n>x)}{\bar\Phi(x)}\right|
 \le C\epsilon_n(x),\qquad
 \epsilon_n(x):=\frac{1+x}{\sqrt n}+\frac{x^4}{n},
 \qquad 0\le x\le c_2n^{1/4},
\]
for a sufficiently small absolute $c_2>0$ and all large $n$.
The lemma's hypothesis holds uniformly on this range. Choosing
$c_2$ small also bounds the displayed logarithm by a fixed
constant, so the two tails are comparable and
\[
 \left|\frac{\Pp(A_n>x)}{\bar\Phi(x)}-1\right|
 \le C\epsilon_n(x).
\]

We check the maximum over three ranges of $x$. If $x\le1$,
Berry--Esseen bounds the scalar distribution-function error by
$C/\sqrt n$. Both scalar distribution functions are at most a
fixed $r_0<1$ for large $n$, since $\Phi(1)<1$.
The mean value theorem for their $m$th powers gives an error at
most $Cm r_0^{m-1}/\sqrt n\le C/\sqrt n$.
If $1<x\le c_2n^{1/4}$, define
$u_x:=\Pp(A_n>x)$, $v_x:=\bar\Phi(x)$, and $w:=mv_x$.
The same mean value theorem, now using comparable tails, gives
\[
 |(1-u_x)^m-(1-v_x)^m|
 \le Cm|u_x-v_x|e^{-c_*mv_x}
 \le C\epsilon_n(x)w e^{-c_*w}.
\]
For $x\le2\sqrt{\Lambda_m}$, the boundedness of
$we^{-c_*w}$ gives the right side of~\eqref{sym:eq:rademacher-max}.
For $x>2\sqrt{\Lambda_m}$, the normal tail bound gives
$w\le me^{-x^2/2}\le e^{-x^2/4}$. The error is then at most
$C\epsilon_n(x)e^{-x^2/4}\le C/\sqrt n$.

Finally, choose $c_3<c_2^2/4$. For $x>c_2n^{1/4}$, both
scalar sums are sub-Gaussian, so a union bound controls the
difference of the maximum distribution functions by
\[
 2me^{-x^2/2}\le2e^{-c_2^2\sqrt n/4}=O(n^{-1/2}).
\]
Here $\log m\le c_3\sqrt n$ was used. These ranges cover all
real thresholds and prove~\eqref{sym:eq:rademacher-max}.

The second set of estimates concerns the density and moments of
the Gaussian maximum. For $m\ge2$, let $H_1,\ldots,H_m$ be
independent standard normal variables and define
\[
 H_*:=\max_{u\le m}H_u,\qquad F_m(x):=\Pp(H_*\le x)=\Phi(x)^m.
\]
Choose $u_m$ by $m\bar\Phi(u_m)=1$. This places the threshold
where the maximum has a nondegenerate distribution and a density
of order $\sqrt{\log m}$. As $m\to\infty$,
\begin{equation}
 u_m\sim\sqrt{2\log m},\qquad
 F_m'(u_m)\sim\frac{\sqrt{2\log m}}e,\qquad
 \E H_*\sim\sqrt{2\log m},\qquad \E H_*^2=O(\log m).
 \label{sym:eq:gaussian-max}
\end{equation}
We will also use the uniform bound
\begin{equation}
 \|F_m''\|_\infty\le C\log(em),\qquad m\ge2.
 \label{sym:eq:gaussian-smoothness}
\end{equation}

To verify~\eqref{sym:eq:gaussian-max}, use the normal tail estimates
$\bar\Phi(x)\asymp\varphi(x)/(1+x)$ for $x\ge0$ and
$\varphi(x)/\bar\Phi(x)\sim x$ as $x\to\infty$.
The equation defining $u_m$ gives its stated asymptotic, and
\[
 F_m'(u_m)=m\varphi(u_m)(1-1/m)^{m-1}
          \sim u_m/e.
\]
For any fixed $\varepsilon\in(0,1)$, those tail estimates
give $\Pp\{H_*>(1+\varepsilon)\sqrt{2\log m}\}\to0$ by a
union bound, and
\[
 \Pp\{H_*\le(1-\varepsilon)\sqrt{2\log m}\}
 \le\exp\{-m\bar\Phi((1-\varepsilon)\sqrt{2\log m})\}
 \longrightarrow0.
\]
Thus $H_*/\sqrt{2\log m}\to1$ in probability. Integrating
$\Pp(\max_u|H_u|>y)\le\min\{1,2me^{-y^2/2}\}$ yields
$\E H_*^2=O(\log m)$. This also gives uniform integrability
of the normalized maxima, proving the mean asymptotic.

For~\eqref{sym:eq:gaussian-smoothness}, differentiate twice:
\[
 F_m''(x)=m(m-1)\varphi(x)^2\Phi(x)^{m-2}
             -mx\varphi(x)\Phi(x)^{m-1}.
\]
The cases $m=2,3$ are bounded directly, since $\varphi$ and
$x\varphi(x)$ are bounded. For $m\ge4$ and $x<0$, the
inequality $\Phi(x)\le1/2$ makes the powers dominate the
polynomial factors in $m$, again giving an absolute bound.
For $0\le x\le2\sqrt{\log(em)}$, set $w:=m\bar\Phi(x)$.
Then $\varphi(x)\le C(1+x)\bar\Phi(x)$ and
$\Phi(x)^{m-2}\le e^{-w/2}$ show that the two terms are
bounded together by
\[
 C(1+x)^2(w+w^2)e^{-c_*w}\le C\log(em).
\]
For larger $x$, dropping the powers of $\Phi$ bounds them by
$C(m^2e^{-x^2}+mx e^{-x^2/2})\le C\log(em)$.
This proves the required uniform second-derivative estimate.

The leading maximum in our construction may have only a bounded
number of coordinates, including a single coordinate. Set $u_1:=0$;
for $r\ge2$, define $u_r$ by $r\bar\Phi(u_r)=1$.
Writing $F_r:=\Phi^r$ and
$g_r:=\E\max_{j\le r}|G_j|$ for independent standard normal
$G_j$, we have uniformly over $r\ge1$,
\begin{equation}
 F_r'(u_r)\asymp g_r\asymp\sqrt{\log(er)},\qquad
 \|F_r''\|_\infty\le C\log(er).
 \label{lower:eq:leading-density}
\end{equation}
For growing $r$, this follows from the preceding estimates and
$g_r\sim\sqrt{2\log r}$, obtained by the same tail argument
for the absolute maximum. The finitely many smaller values of
$r$ have positive densities at $u_r$. For $r=1$, all assertions
follow directly from $F_1=\Phi$ and $u_1=0$.

\subsection{The common symmetric construction}
\label{app:lower-family}

Each symmetric summand has a Rademacher component and an independent
component that is activated rarely. The activation follows the
construction in~\cite[Proposition~E.1]{FKLZ}. We use separate
indices $j\le r$ and $u\le m$ for the two components. This
allows the number of leading coordinates, which determines the
Gaussian width, to differ from the number of coordinates over
which the rare shifts are maximized.

Fix integers $n\ge1$, $1\le r\le m$, and $m\ge2$,
probabilities $0<\theta,\pi\le1$, and an amplitude $a>0$.
Define
\[
 p:=rm,\qquad R:=\log(er),\qquad \ell:=\log m,\qquad
 v:=\theta\pi a^2,\qquad s:=\sqrt{1-v},
\]
and require $0<v<1/2$. The activated component will contribute
variance $v$ to each coordinate, and the leading component will
contribute the remaining variance $s^2=1-v$. We suppress the
dependence of these parameters on $n$.

For $i\le n$, $j\le r$, and $u\le m$, let $\eta_{ij}$ and
$\varepsilon_{iu}$ be Rademacher signs, let $I_i$ have the
Bernoulli$(\theta)$ law, and let $J_{iu}$ have the
Bernoulli$(\pi)$ law, with all these variables mutually independent.
Define
\begin{equation}
 X_{i,ju}:=s\eta_{ij}+aI_iJ_{iu}\varepsilon_{iu},
 \qquad j\le r,\quad u\le m,\quad i\le n.
 \label{sym:eq:common-construction}
\end{equation}
The index $i$ labels observations, and the pair $(j,u)$ labels
coordinates. The first component is shared across all coordinates
with the same $j$, while the second is shared across those with
the same $u$. The indicator $I_i$ activates the second component
of observation $i$. The additional \emph{coordinate mask} $J_{iu}$
selects which of its $u$-indices contribute. Taking $\pi=1$
removes the masks.

We first verify the covariance, moments, and width of this vector.
These formulas will let us change the activation probabilities
and amplitude without changing the subsequent comparison argument.

\begin{lemma}[Properties of the common symmetric construction]
\label{lower:family-properties}
For the parameters and independent variables above, the observations
in~\eqref{sym:eq:common-construction} are i.i.d., centered, and
centrally symmetric, with unit marginal variances. Their covariance is
\begin{equation}
 \Sigma_{(j,u),(j',u')}=(1-v)\ind\{j=j'\}+v\ind\{u=u'\}.
 \label{sym:eq:common-covariance}
\end{equation}
It has rank $r+m-1$ and correlation gap $v$. For every
coordinate and every $q\ge1$,
\begin{align}
 \E X_{i,ju}^4
 &=1+4v-5v^2+\theta\pi a^4,
 \label{lower:eq:family-fourth}\\
 \norm{X_i}_\infty&\le s+aI_i\quad\text{almost surely},\qquad
 (\E\norm{X_i}_\infty^q)^{1/q}\le s+a\theta^{1/q}.
 \label{lower:eq:family-envelope}
\end{align}
For independent standard normal variables $G_j,H_u$, the vector
$Z_{ju}:=sG_j+\sqrt v\,H_u$ has covariance $\Sigma$.
Writing $g_t:=\E\max_{j\le t}|G_j|$, its width satisfies
\begin{equation}
 s g_r\le g_\Sigma\le s g_r+\sqrt v\,g_m.
 \label{lower:eq:family-width}
\end{equation}
\end{lemma}

\begin{proof}
The variables used for different observations are independent and
have the same laws. Averaging over the signs gives centering, and
reversing all signs within one observation preserves its law and
negates the whole vector. Thus the observations are i.i.d. and
centrally symmetric.

The cross terms between the two components have expectation zero.
For the first component,
$\E\eta_{ij}\eta_{ij'}=\ind\{j=j'\}$. For the second,
independence and centering of the signs give
\[
 \E[I_iJ_{iu}\varepsilon_{iu}J_{iu'}\varepsilon_{iu'}]
 =\theta\pi\ind\{u=u'\}.
\]
This proves~\eqref{sym:eq:common-covariance}, whose diagonal
entries are $s^2+v=1$. Its range consists of arrays
$(b_j+d_u)_{j,u}$, with $b\in\R^r$ and $d\in\R^m$.
Indeed, the covariance is the sum of the two positive covariance
operators associated with these component spaces, and both
coefficients $s^2,v$ are positive. The kernel of the map
$(b,d)\mapsto(b_j+d_u)_{j,u}$ consists exactly of
$b_j=c$, $d_u=-c$, so the range has dimension $r+m-1$.
Distinct coordinates have correlations $1-v$, $v$, or $0$;
the first value is attained by fixing $j$ and taking two different
$u$-indices. Because $m\ge2$ and $v<1/2$, the largest
correlation is $1-v$, giving gap $v$.

In the fourth-power expansion, all terms with an odd sign factor
have mean zero. The remaining terms give
\[
 \E X_{i,ju}^4=s^4+6s^2\theta\pi a^2+\theta\pi a^4
              =1+4v-5v^2+\theta\pi a^4.
\]
The bound $\|X_i\|_\infty\le s+aI_i$ is pointwise.
Minkowski's inequality and $\E I_i^q=\theta$ give its stated
moment bound. The Gaussian vector in the lemma has the same
covariance by independence of $G_j$ and $H_u$.
Conditional Jensen's inequality, with $G$ fixed, gives
\[
 \E_H\max_{j,u}|sG_j+\sqrt v H_u|
 \ge\max_{j,u}|sG_j+\sqrt v\,\E H_u|
 =s\max_j|G_j|.
\]
The triangle inequality gives the upper bound
$\max_{j,u}|sG_j+\sqrt v H_u|
\le s\max_j|G_j|+\sqrt v\max_u|H_u|$.
Taking expectations proves~\eqref{lower:eq:family-width}.
\end{proof}

We next identify the two shifts whose means will be compared.
Define
\[
 A_j:=n^{-1/2}\sum_i\eta_{ij},\qquad
 N:=\sum_iI_i,\qquad k:=n\theta,\qquad
 Q_u:=\sum_iI_iJ_{iu}\varepsilon_{iu},\qquad
 Q_*:=\max_{u\le m}Q_u.
\]
Here $N$ counts activated observations and $k=\E N$.
Let $A_*:=\max_{j\le r}A_j$, $G_*:=\max_{j\le r}G_j$,
and $H_*:=\max_{u\le m}H_u$, with the Gaussian variables
independent of the observations. Since the coordinate set contains
every pair $(j,u)$, maximization separates the two components:
\begin{equation}
 \max_{j,u}W_{ju}=sA_*+\frac a{\sqrt n}Q_*,\qquad
 \max_{j,u}Z_{ju}=sG_*+\sqrt v\,H_*.
 \label{sym:eq:common-maxima}
\end{equation}
For example, any maximizing $j$ for $A_j$ can be paired with
any maximizing $u$ for $Q_u$. The leading maximum and the shift
are independent in each sum, because they use disjoint independent
families of variables.

The maximum possible value of $Q_u$ is $N$. It is attained if
$J_{iu}\varepsilon_{iu}=1$ for every active observation.
Conditional on all $I_i$, this has probability $(\pi/2)^N$
for each $u$, independently across $u$. Hence the conditional
probability that no coordinate attains $N$ is
$(1-(\pi/2)^N)^m\le\exp\{-m(\pi/2)^N\}$.
On $N\le b$, the latter is at most $\exp\{-m(\pi/2)^b\}$.
Splitting off the event $N>b$ proves, for every real $b\ge0$,
\begin{equation}
 \Pp(Q_*\ne N)\le\Pp(N>b)+\exp\{-m(\pi/2)^b\}.
 \label{lower:eq:all-active}
\end{equation}
This explains the role of the large number $m$ of coordinates:
one coordinate can realize all active contributions even when
each individual coordinate has a small chance of doing so.

If $k\to\infty$ and the bound in~\eqref{lower:eq:all-active}
tends to zero, then
\begin{equation}
 \E Q_*=(1+o(1))k,\qquad \E Q_*^2\le k^2+k.
 \label{lower:eq:activation-moments}
\end{equation}
Indeed, $|Q_*|\le N$ and the binomial count satisfies
$\E N^2=k^2+k(1-\theta)\le k^2+k$. Cauchy--Schwarz gives
\[
 \E|N-Q_*|
 \le2\E[N\ind\{Q_*\ne N\}]
 \le2(\E N^2)^{1/2}\Pp(Q_*\ne N)^{1/2}=o(k).
\]
Thus $\E Q_*=\E N+o(k)$, proving the first assertion as well.

To apply the shift comparison, divide both maxima in
\eqref{sym:eq:common-maxima} by $s>0$ and define
\[
 V_W:=\frac{aQ_*}{s\sqrt n},\qquad V_G:=\frac{\sqrt v}{s}H_*.
\]
For $t_n:=su_r$, the events being compared become
$\{A_*+V_W\le u_r\}$ and $\{G_*+V_G\le u_r\}$.
Lemma~\ref{lower:shift-comparison}, with $F=F_r$, together with
\eqref{sym:eq:rademacher-max} and~\eqref{lower:eq:leading-density},
therefore gives, whenever $R\le c_3\sqrt n$,
\begin{equation}
\begin{split}
 &\Pp(\max_{j,u}W_{ju}\le t_n)-\Pp(\max_{j,u}Z_{ju}\le t_n)\\
 &\qquad=F_r'(u_r)(\E V_G-\E V_W)
       +O\{\epsilon_{n,r}+R\,\E(V_W^2+V_G^2)\}.
\end{split}
 \label{lower:eq:master-comparison}
\end{equation}
Because $F_r'(u_r)\asymp\sqrt R$, a difference between mean
shifts of order $ak/\sqrt n$ produces the scale
\begin{equation}
 d_n:=\frac{ak\sqrt R}{\sqrt n},\qquad k=n\theta.
 \label{lower:eq:proof-scale}
\end{equation}
For each choice of parameters below, we verify that this mean
difference has a fixed sign and the stated order, that $d_n\to0$,
and that the two remainder terms are $o(d_n)$.
We also verify $v\to0$ and $\sqrt v\,g_m\to0$.
Then~\eqref{lower:eq:family-width} and $g_r\ge\sqrt{2/\pi}$
give $g_\Sigma/g_r\to1$, proving~\eqref{lower:eq:width-profile}.

Finally, duplicating the first coordinate gives correlation gap
exactly zero without changing the comparison. The inverse image
of a rectangle under duplication is a rectangle or the empty set,
so the new rectangle distance is no larger than the original one.
Conversely, leaving the added coordinate unrestricted recovers
every original rectangle, proving equality of the distances.
Duplication also preserves Gaussian width, symmetry, envelope,
and coordinate moments. It changes the dimension from $rm$ to
$rm+1$, which leaves the asymptotic logarithmic factors unchanged.

\subsection{Proofs of Theorems~\ref*{opt:symmetric-lower}, \ref*{opt:q4-lower}, and~\ref*{lower:dense} (unmasked observations)}
\label{app:symmetric-lower}

For these three theorems, the same activation probability works
without coordinate masks. Only the amplitude changes: a bounded
amplitude gives the bounded-observation result, and a larger
amplitude uses the available polynomial moment budget.
Take $r=r_n$ and $m=m_n$ from the theorem being proved, with
the notation of Subsection~\ref{app:lower-family}. Since
$p_n=rm$ and $1\le r\le m$, we have
$\ell\asymp L_n$, $R\le1+\ell$, and
$w_n=g_r\asymp\sqrt R$. Choose
\[
 \pi:=1,\qquad \theta:=\frac{\ell}{2n},\qquad
 a:=\begin{cases}
 B_n-1,&\text{for Theorem~\ref{opt:symmetric-lower}},\\
 \theta^{-1/4},&\text{for Theorem~\ref{opt:q4-lower}},\\
 (B_n/2)\theta^{-1/q},&\text{for Theorem~\ref{lower:dense}}.
 \end{cases}
\]
We first check that these parameters define the required observations.
The mean number of active observations is $k=n\theta=\ell/2$,
so the comparison scale is
\[
 d_n=\frac{a\ell\sqrt R}{2\sqrt n}.
\]
For the three amplitudes, respectively, this scale is comparable to
\[
 B_n\sqrt R\,\ell/\sqrt n,\qquad
 \sqrt R\,n^{-1/4}\ell^{3/4},\qquad
 B_n\sqrt R\,n^{-\alpha_q}\ell^{1-1/q}.
\]
These are the scales assumed to vanish, since
$w_n\asymp\sqrt R$ and $L_n\asymp\ell$. Thus $d_n\to0$.

Each vanishing-scale condition also implies $\ell/n\to0$.
For the polynomial choice, write its scale as
$B_n\sqrt{R\ell}(\ell/n)^{\alpha_q}$ and use
$\alpha_q>0$ and $B_n\ge2$. For the bounded and fourth-moment
choices, the displayed scales give the conclusion directly.
Therefore $\theta\to0$ and $a\ge1$ eventually. In particular,
\[
 \frac{\ell^2R}{n}\le4d_n^2\to0,\qquad
 v=\theta a^2=\frac{2d_n^2}{\ell R}\to0.
\]
Thus $0<\theta\le1$ and $0<v<1/2$ for all sufficiently large
$n$. Also $R\le1+\ell$ and $\ell\to\infty$ imply
$R^2/n\le2\ell R/n\to0$, so $R\le c_3\sqrt n$ eventually.
The approximation for the leading maximum is therefore available.

We check the moment conditions using Lemma~\ref{lower:family-properties}.
For $a=B_n-1$, the pointwise envelope is $s+a\le B_n$.
For $a=\theta^{-1/4}$, we have $v=\sqrt\theta$ and
$\theta a^4=1$. Hence
\begin{align}
 \E X_{i,ju}^4&=2+4v-5v^2\le14/5<4,\notag\\
 \E\norm{X_i}_\infty^4&\le8(s^4+\theta a^4)\le16.
 \label{sym:eq:q4-moments}
\end{align}
The second inequality uses $(x+y)^4\le8(x^4+y^4)$ in the
pointwise bound $\|X_i\|_\infty\le s+aI_i$.
Thus these observations belong to $\mathcal C_4(2,p_n)$.
For $a=(B_n/2)\theta^{-1/q}$, the envelope bound gives
\[
 (\E\|X_i\|_\infty^q)^{1/q}\le s+B_n/2\le B_n,
\]
using $s\le1$ and $B_n\ge2$. Their coordinate fourth moment is
$1+4v-5v^2+(B_n^4/16)\theta^{1-4/q}$.
If $q>4$ and $B_n=B$ is fixed, this tends to $1$, since
$\theta\to0$ and $v\to0$. It is therefore at most $B^2$
eventually, as required for Theorem~\ref{lower:dense}\textup{(ii)}.

We next show that $Q_*$ nearly attains its maximal value $N$.
The count $N$ has mean and variance bounded by $\ell/2$.
Chebyshev's inequality gives
$\Pp(N>\ell)\le(\ell/2)/(\ell/2)^2=2/\ell$.
Since $m=e^\ell$ and $\pi=1$, taking $b=\ell$ in
\eqref{lower:eq:all-active} gives
\[
 \Pp(Q_*\ne N)\le2/\ell+
       \exp\{-e^{(1-\log2)\ell}\}\longrightarrow0.
\]
As $k=\ell/2\to\infty$, \eqref{lower:eq:activation-moments}
then yields $\E Q_*=(1/2+o(1))\ell$ and
$\E Q_*^2=O(\ell^2)$.

The mean shifts can now be compared explicitly. Since
$\sqrt v=a\sqrt{\ell/(2n)}$ and
$\E H_*=(1+o(1))\sqrt{2\ell}$,
\[
 \E V_G=(1+o(1))\frac{a\ell}{s\sqrt n},\qquad
 \E V_W=(1/2+o(1))\frac{a\ell}{s\sqrt n}.
\]
The Gaussian shift has the larger mean. Using $s\to1$, we obtain
\[
 \E V_G-\E V_W=(1/2+o(1))\frac{a\ell}{\sqrt n},\qquad
 \E(V_W^2+V_G^2)=O(a^2\ell^2/n).
\]
For the second assertion, use the bound on $\E Q_*^2$ for
$V_W$ and $\E H_*^2=O(\ell)$ from
\eqref{sym:eq:gaussian-max} for $V_G$.

Both errors in~\eqref{lower:eq:master-comparison} are smaller
than the resulting first-order term. For the leading maximum,
\eqref{sym:eq:rademacher-max} gives
\[
 \frac{\epsilon_{n,r}}{d_n}
 \le C\left(\frac1{a\ell}
       +\frac{R^{3/2}}{a\ell\sqrt n}\right)\longrightarrow0.
\]
Here $a\ge1$, $\ell\to\infty$, and the second term is at
most $C\sqrt{\ell/n}$ because $R\le1+\ell$.
The Taylor remainder satisfies
\[
 R\,\E(V_W^2+V_G^2)
 =O(Ra^2\ell^2/n)=O(d_n^2)=o(d_n).
\]
Substitution into~\eqref{lower:eq:master-comparison} therefore yields
\begin{equation}
\begin{split}
 &\Pp(\max_{j,u}W_{ju}\le t_n)-\Pp(\max_{j,u}Z_{ju}\le t_n)\\
 &\qquad=(1/2+o(1))F_r'(u_r)\frac{a\ell}{\sqrt n}
          +o(d_n)\ge c d_n
\end{split}
 \label{sym:eq:common-discrepancy}
\end{equation}
for all sufficiently large $n$. The uniform bound
$F_r'(u_r)\asymp\sqrt R$ ensures that the main term is
positive and bounded below by a fixed multiple of $d_n$, whether
$r$ stays bounded or grows.

We must finally express this scale in terms of the Gaussian width
of the constructed covariance. Since $g_m=O(\sqrt\ell)$,
\[
 \sqrt v\,g_m=O(a\ell/\sqrt n)=O(d_n/\sqrt R)\to0.
\]
Together with $v\to0$, the width bounds
\eqref{lower:eq:family-width} give $g_{\Sigma_n}/g_r\to1$.
Thus the construction has the prescribed width, and
\[
 d_n\asymp
 \begin{cases}
 B_n g_{\Sigma_n}L_n/\sqrt n,
   &\text{for Theorem~\ref{opt:symmetric-lower}},\\
 g_{\Sigma_n}n^{-1/4}L_n^{3/4},
   &\text{for Theorem~\ref{opt:q4-lower}},\\
 B_n g_{\Sigma_n}n^{-\alpha_q}L_n^{1-1/q},
   &\text{for Theorem~\ref{lower:dense}},
 \end{cases}
\]
with constants depending only on $q$ in the last case.
The orthant discrepancy in~\eqref{sym:eq:common-discrepancy}
proves all three lower bounds. Applying~\eqref{poly:q4} to the
fourth-moment example gives~\eqref{lower:q4-fixed}, and
Theorem~\ref{thm:polynomial}\textup{(i)} gives the stated
logarithmic comparison for the polynomial envelope.

For the bounded construction with $r=1$ and $B_n=2$, we have
$a=1$, $u_r=0$, and $F_r'(u_r)=\varphi(0)$.
Equation~\eqref{sym:eq:common-discrepancy} gives a discrepancy
of $(\varphi(0)/2+o(1))\ell/\sqrt n$. Since
$g_{\Sigma_n}\to\sqrt{2/\pi}=2\varphi(0)$ and $L_n\sim\ell$,
setting $\delta_n:=2/\sqrt n$ yields
\[
 g_{\Sigma_n}\to\sqrt{2/\pi},\qquad
 \liminf_{n\to\infty}
 \frac{\rh(W_n,Z_n)}{\delta_n g_{\Sigma_n}L_n}\ge\frac18.
\]
Thus the bounded-width example and its explicit constant follow
from the same calculation.

\subsection{Proofs of Theorems~\ref*{opt:q4-growing} and~\ref*{lower:sparse} (masked observations)}
\label{app:q4-growing}

For the remaining theorems, coordinate masks allow a large envelope
while keeping each coordinate's fourth moment small. We choose the
amplitude to use the envelope budget and then choose the mask
probability to satisfy the separate fourth-moment budget.
Fix $q>2$ and take the sequences in Theorem~\ref{lower:sparse}.
Write $B:=B_n$, $r:=r_n$, $m:=m_n$, $R:=\log(er)$,
$\ell:=\log m$, and $L:=L_n$. Define
\[
 \widehat K:=B(n/\ell)^{2/q-1/2},\qquad
 \Lambda:=\log(2\widehat K^2),\qquad
 k:=\frac{\ell}{4\Lambda},\qquad \theta:=k/n.
\]
Because $\ell\asymp L$, assumption~\eqref{lower:sparse-regime}
implies $\widehat K\asymp_q\mathcal K_{n,q}\to\infty$,
$\Lambda\asymp_q\log(e+\mathcal K_{n,q})=o(\ell)$,
and $k\to\infty$. Choose
\begin{equation}
 \pi:=B^{-2}\theta^{4/q-1},\qquad
 a:=\frac B2\theta^{-1/q},\qquad
 v:=\theta\pi a^2=\frac14\theta^{2/q},\qquad s:=\sqrt{1-v}.
 \label{q4:eq:sparse-parameters}
\end{equation}
These choices give $a\theta^{1/q}=B/2$ and
$\theta\pi a^4=B^2/16$. We check below that all probabilities
and variances are in their required ranges before applying the
construction.

First, substitution into~\eqref{lower:eq:proof-scale} gives
\[
 d_n=\frac{B n^{-\alpha_q}\ell^{1-1/q}\sqrt R}
 {2(4\Lambda)^{1-1/q}}\asymp_q R_{n,q}\longrightarrow0.
\]
To verify that the leading maximum can be approximated sufficiently
accurately, compare this with
$U_{\ell,R}:=B^{1/2}n^{-1/4}\ell^{3/4}\sqrt R$.
Direct division yields
\[
 \frac{d_n}{U_{\ell,R}}
 =\frac{\sqrt{\widehat K}}{2(4\Lambda)^{1-1/q}}
 \longrightarrow\infty.
\]
The limit holds because $\Lambda$ has logarithmic order in
$\widehat K$. Thus $U_{\ell,R}\to0$. Since
$U_{\ell,R}^4=B^2\ell^3R^2/n$ and $B\ge2$, it follows that
$\ell^3R^2/n\to0$, $\ell/n\to0$, and $R=o(\sqrt n)$.
In particular, $R\le c_3\sqrt n$ eventually.

Two further identities check the parameter ranges:
\[
 d_n=(B/2)\sqrt{kR}\,(k/n)^{\alpha_q},\qquad
 \pi=\widehat K^{-2}(4\Lambda)^{1-4/q}.
\]
In the first identity, $k\to\infty$, $R\ge1$, and
$\alpha_q>0$, so $d_n\to0$ forces $\theta=k/n\to0$.
The second identity gives $\pi\Lambda\to0$, since any fixed
power of $\Lambda$ is dominated by $\widehat K^2$.
Thus $\theta,\pi\in(0,1)$ and $v\in(0,1/2)$ for all
sufficiently large $n$, with $v\to0$ and $s\to1$.

We can now apply Lemma~\ref{lower:family-properties}. Its moment
formulas give
\begin{align}
 \E X_{i,ju}^4
 &=1+4v-5v^2+B^2/16\le9/5+B^2/16\le B^2,
 \label{q4:eq:sparse-coordinate-moments}\\
 (\E\norm{X_i}_\infty^q)^{1/q}
 &\le s+a\theta^{1/q}\le1+B/2\le B.
 \label{q4:eq:sparse-envelope-moment}
\end{align}
Here $1+4v-5v^2\le9/5$ and $B\ge2$ justify the first line.
Thus the observations belong to $\mathcal C_q(B,p_n)$.

We next verify that at least one coordinate receives all active
positive increments with high probability. The expression for $\pi$
gives
\[
 \log(2/\pi)=\Lambda-(1-4/q)\log(4\Lambda)\sim\Lambda.
\]
Since $k=\ell/(4\Lambda)$, we have
\[
 \log\{m(\pi/2)^{2k}\}
 =\ell-2k\log(2/\pi)=(1/2+o(1))\ell\ge\ell/4
\]
eventually. Also $\Var(N)\le k$, so
$\Pp(N>2k)\le1/k$. Taking $b=2k$ in
\eqref{lower:eq:all-active} gives
\[
 \Pp(Q_*\ne N)\le k^{-1}+\exp\{-e^{\ell/4}\}\to0.
\]
Therefore~\eqref{lower:eq:activation-moments} implies
\begin{equation}
 \E Q_*=(1+o(1))k,\qquad \E Q_*^2=O(k^2).
 \label{q4:eq:sparse-max-moments}
\end{equation}

The masks make the Gaussian shift smaller than the original shift
in this calibration. Indeed, $v=(k/n)\pi a^2$ and
$\E H_*\sim\sqrt{2\ell}$ give
\[
 \frac{\sqrt v\,\E H_*}{ak/\sqrt n}
 =\sqrt{\pi/k}\,\E H_*\sim\sqrt{8\pi\Lambda}\to0.
\]
The factor $1/s$ occurs in both shifts and cancels in their ratio.
Using~\eqref{q4:eq:sparse-max-moments}, we conclude that
\begin{equation}
 \E V_W-\E V_G=(1+o(1))\frac{ak}{s\sqrt n}.
 \label{q4:eq:sparse-mean-gap}
\end{equation}
The probability discrepancy therefore has the opposite sign from
the unmasked case, but its absolute value still has scale $d_n$.

To control the first error in~\eqref{lower:eq:master-comparison},
use~\eqref{sym:eq:rademacher-max} to obtain
\[
 \frac{\epsilon_{n,r}}{U_{\ell,R}}
 \le C\left(n^{-1/4}\ell^{-3/4}
       +n^{-3/4}R^{3/2}\ell^{-3/4}\right)\to0.
\]
The second term is at most $C(\ell/n)^{3/4}$ because
$R\le1+\ell$, and the first tends to zero directly.
Since $d_n/U_{\ell,R}\to\infty$, this proves
$\epsilon_{n,r}=o(d_n)$.
For the Taylor remainder, the second-moment bounds on $Q_*$
and $H_*$, together with $s$ bounded away from zero, give
\[
 R\E V_W^2=O(Ra^2k^2/n),\qquad
 R\E V_G^2=O(Rv\ell).
\]
Both are controlled by $d_n^2$, since
\[
 \frac{Ra^2k^2}{n}=d_n^2,\qquad
 Rv\ell=\frac{\pi\ell}{k}d_n^2=4\pi\Lambda d_n^2.
\]
Thus $R\E(V_W^2+V_G^2)=O(d_n^2)=o(d_n)$.
Using the sign in~\eqref{q4:eq:sparse-mean-gap}, the common
comparison~\eqref{lower:eq:master-comparison} yields
\begin{equation}
\begin{split}
 &\Pp(\max_{j,u}Z_{ju}\le t_n)-\Pp(\max_{j,u}W_{ju}\le t_n)\\
 &\qquad=(1+o(1))F_r'(u_r)\frac{ak}{s\sqrt n}
           +o(d_n)\ge c d_n
\end{split}
 \label{q4:eq:sparse-orthant}
\end{equation}
for all sufficiently large $n$.

The width is still determined by the leading coordinates. The
same calculation as for the Gaussian mean gives
\[
 \sqrt v\,g_m
 =O\left(\sqrt{\pi\Lambda}\,\frac{ak}{\sqrt n}\right)
 =O\left(\sqrt{\pi\Lambda}\,\frac{d_n}{\sqrt R}\right)\to0.
\]
Together with $v\to0$ and~\eqref{lower:eq:family-width}, this
proves~\eqref{lower:eq:width-profile}. Replacing $\sqrt R$ by
the comparable $g_{\Sigma_n}$ and $\ell$ by the comparable
$L_n$ in the expression for $d_n$ gives~\eqref{lower:sparse-rate}.
For the matching upper bound, the ratio of the envelope term to
the coordinate fourth-moment term in~\eqref{poly:main-fourth} is
\[
 \frac{\sqrt{\mathcal K_{n,q}}}
 {\{\log(e+\mathcal K_{n,q})\}^{1-1/q}}\to\infty.
\]
Both terms have the same Gaussian-width factor. Thus the envelope
term dominates, and the upper bound has the order of the lower
rate just proved.

Taking $q=4$ gives $\widehat K=B$,
$\Lambda=\log(2B^2)\asymp\log(e+B)$, and
$\mathcal K_{n,4}=B_n$. The assumptions and rate are then
exactly those of Theorem~\ref{opt:q4-growing}, proving that
result as well. The covariance rank is $r+m-1$, and its
correlation gap is
\[
 v=\frac14\left(\frac{\ell}{4n\Lambda}\right)^{2/q}\to0.
\]

\end{document}